\documentclass{emulateapj}
\usepackage{amssymb,amsmath,epsfig,graphicx, natbib, lscape}

\def\sles{\lower2pt\hbox{$\buildrel {\scriptstyle <}
   \over {\scriptstyle\sim}$}} 
\def\sgreat{\lower2pt\hbox{$\buildrel {\scriptstyle >}
    \over {\scriptstyle\sim}$}} 

\def\kms{\mbox {~km~s$^{-1}$}}

\def\cmv{\mbox {~cm$^{-3}$}}

\def\mpc3{\mbox {~Mpc$^{3}$}}

\def\asec{\mbox {\ifmmode {'' }\else $''~$\fi}}  
\def\amin{\mbox {\ifmmode {' }\else $'~$\fi}}    
\def\arcsper{\mbox {\ifmmode \rlap.{'' }\else $\rlap{.}'' $\fi}} 
\def\arcmper{\mbox {\ifmmode \rlap.{' }\else $\rlap{.}' $\fi}} 

\def\msun{\mbox {${\rm ~M_\odot}$}}

\def\Ha{\mbox {H$\alpha$}}
\def\Hb{\mbox {H$\beta$}}

\def\line{\mbox {~$\lambda$}}

\def\h0{\mbox {~H$_0$}}
\def\q0{\mbox {~q$_0$}}
 
\begin{document}
\title{Gas Excitation in ULIRGs: \\Maps of Diagnostic Emission-Line Ratios in Space and Velocity\footnote[1]{\lowercase{\uppercase{T}he data presented herein were obtained at the \uppercase{W.M. K}eck \uppercase{O}bservatory, which is operated as a scientific partnership among the \uppercase{C}alifornia \uppercase{I}nstitute of \uppercase{T}echnology, the \uppercase{U}niversity of \uppercase{C}alifornia and the \uppercase{N}ational \uppercase{A}eronautics and \uppercase{S}pace \uppercase{A}dministration. \uppercase{T}he \uppercase{O}bservatory was made possible by the generous financial support of the \uppercase{W.M. K}eck \uppercase{F}oundation.}}
}

\author{Kurt T. Soto, Crystal L. Martin}
\affil{Physics Department, University of California, Santa Barbara, CA 93106-9530}

\begin{abstract}
\label{sect:abst}

Emission-line spectra extracted at multiple locations across 39 ultraluminous infrared 
galaxies have been compiled into a spectrophotometric atlas. Line profiles of \Ha,
[\ion{N}{2}], [\ion{S}{2}], [\ion{O}{1}], \Hb, and [\ion{O}{3}] are resolved and fit 
jointly with common velocity components. 
Diagnostic ratios of these line fluxes are presented in a series of plots, showing 
how the Doppler shift, line width, gas excitation, and surface brightness change with 
velocity at fixed position and also with distance from the nucleus. One 
general characteristic of these spectra is the presence of shocked gas extending many 
kiloparsecs from the nucleus. 
In some systems, the shocked gas appears as part of a galactic gas disk
based on its rotation curve. These gas disks appear primarily
during the early stages of the merger. The general characteristics of the integrated
spectra are also presented.  

\end{abstract}

\subjectheadings{galaxies: starburst --- galaxies: evolution ---  galaxies: active --- galaxies: formation}

\section{Introduction}
\label{sect:intro}

The optical emission-line spectrum of a galaxy is a powerful diagnostic of recent
star formation \citep{Kennicutt:1998p3}, the gas-phase metallicity \citep{Kewley:2006p38},
and the excitation mechanism \citep{Kewley:2001p198,Baldwin:1981p7}. Fiber spectra of
low redshift galaxies have provided the first systematic examination of these properties 
in the central regions of low redshift galaxies 
\citep[e.g.][]{Kauffmann:2003p33,Kauffmann:2003p460,Tremonti:2004p1207}. 
Infrared spectra obtained with the new generation of multi-object spectrographs, including the 
James Webb Space Telescope, will make it possible to apply these diagnostics over the entire 
period of galaxy evolution. The interpretation of these spectra, however, is complicated
by measurement apertures that subtend a kiloparsec or more.

On scales larger than the narrow line region of an active galactic nucleus (AGN)
or a giant HII region, the physical process
that excites the gas is not necessarily uniform. 
The contribution to the excitation from shocks or the hard spectrum of an AGN
must be properly identified in order to use metallicity and
star formation diagnostics constructed under the assumption of photoionization by 
massive stars. 

To address this problem, we obtain spectra that resolve scales of $\sim$1 kpc and map
the excitation across a sample of ultraluminous infrared galaxies (ULIRGs).
Ratios of forbidden lines to Balmer lines 
are interpreted using grids of photoionization models and ``BPT diagrams'' \citep{Baldwin:1981p7}.
Flux ratios of lines that are close in wavelength, 
i.e. [\ion{N}{2}]/\Ha, [\ion{S}{2}]/\Ha, [\ion{O}{1}]/\Ha\ and [\ion{O}{3}]/\Hb, 
are used because
they are insensitive to reddening. While the line ratios produced by a power-law spectral energy
distribution are largely degenerate with those predicted by shock models 
\citep{Allen:2008p695}, spatial mapping has proven effective at distguishing 
shock excitation from photoionization.  
The large extranuclear extent of strong forbidden-line
emission relative to the Balmer lines \citep{Colina:2005p31,MonrealIbero:2010p590,Rich:2011p1044}
has uniquely identified shocks as the excitation mechanism in some ULIRGs. 
High forbidden-to-Balmer line ratios also appear to be common in the extranuclear regions
of $z \sim 2$ galaxies mapped  with infrared IFUs 
\citep{Genzel:2011p1007,Law:2007p1173,Wright:2010p1171}. 

Previous studies have shown that broad line profiles are often associated with shock 
excitation in mergers \citep{Rich:2011p1044}.
We improve upon these studies by resolving line profiles,
using multiple component fitting, and examining emission 
line ratios in velocity space. 
Since the next generation of optical IFUs such as 
MUSE \citep[Multi Unit Spectroscopic Explorer;][]{Henault:2003p1211} and 
KCWI \citep[Keck Cosmic Web Imager;][]{Martin:2010p1212}, will be able to 
map galaxies in the optical with large fields of view, 
our approach demonstrates how to analyze these data. We argue that
measuring variations in the forbidden-to-Balmer line flux ratio in the velocity coordinate 
may prove as useful as resolution in the spatial dimension for determining the excitation
mechanism.

The main parts of this paper are two sets of figures. 
The first set illustrates the variation of the forbidden-to-Balmer line flux ratio on scales
of approximately 60\kms\ spectrally and 1-2~kpc spatially; it also compares the lines ratios
to grids of photoionization models and shock models. 
The second set displays the emission-line profiles
across 39 galaxies, selected from the IRAS 2~Jy survey 
\citep{Murphy:1996p23,Strauss:1992p1215,Strauss:1990p1214} and known to be undergoing a
galaxy merger.  Section~2 presents the observations
and explains the measurement procedure. The discussion is mostly qualitative,
focusing on the trends observed during the progression of the merger. A companion paper
(Soto et al. 2012b) provides a quantitative analysis of the relationship between the 
gas excitation and gas kinematics. Throughout the paper we use a cosmology with 
$\Omega_m=0.27$, $\Omega_m=0.73$, and $H_0 = 71$\kms~Mpc$^{-1}$.

\section{Data}
\label{sect:data}

\subsection{Sample and Aperture Selection}
\label{sect:sample_ap}

Moderate-resolution spectra were obtained at Keck Observatory with the Echellete Spectrograph
and Imager \citep[ESI;][]{Sheinis:2002p9} under average seeing of 0\farcs8.
The observations and data reduction are described in \cite{Martin:2005p24, Martin:2006p5},
where the Na~I 5890, 96 interstellar absorption kinematics were measured previously and discussed
for the 18 objects with published CO velocities. In the remaining 21 objects in the sample,
redshifts were determined from the integrated \Ha\ emission line.

The broad spectral bandpass covers recombination lines from H and He and the following strong
forbidden lines: [\ion{S}{2}]$\lambda\lambda 6717,31$, [\ion{N}{2}]~$\lambda 6548, 6584$, 
[\ion{O}{1}] $\lambda 6300, 64$, [\ion{O}{3}]$\lambda 5007$, and 
sometimes [\ion{O}{2}]$\lambda
3727,29$ at lower S/N ratio. Dust-insensitive line ratios constructed from a subset of these line
flux measurements are presented in this paper for the full sample of 39 galaxies. 

Table \ref{tab:sample} lists the name, merger stage, infrared luminosity, and redshift of each
galaxy. The galaxies have IR luminosities L$_{IR} > 8.5\times10^{11}$ L$_\odot$, which identify
them as ultraluminous infrared galaxies (ULIRGs) and 
and 60$\mu$m flux $>$ 1.94 Jy \citep{Murphy:1996p23}. 
The redshift range from 0.043 to 0.15 corresponds to angular scales from 1.00 to 2.61~kpc/\arcsec.
The position angle of the 20\arcsec\ long slit is also listed for each galaxy.

\begin{deluxetable}{l l l l l}
\tablewidth{0pt}
\tablecaption{ \label{tab:sample} Sample and Morphology}
\tabletypesize{\scriptsize}
\tablehead{
\colhead{IRAS Name} 	& \colhead{Merger Class}       & \colhead{log($\frac{L_{IR}}{L_\odot}$)} & \colhead{$z$}& \colhead{PA}  \\
(1)			& (2)			& (3)			                 & (4)	        &  (5)	        	   \\
}
\startdata
00153$+$5454 &    IIIb\tablenotemark{a}   & 12.10 & 0.1116 &   -22.0 \\
00188$-$0856 &    V          				& 12.33 	& 0.1285 &    -3.3 \\
00262$+$4251 &    IVa\tablenotemark{a}    & 12.08 & 0.0972 &   -14.1 \\
01003$-$2238 &    V\tablenotemark{a}    	& 12.25 & 0.1177 &    10.0 \\
01298$-$0744 &    IVb        				& 12.29 & 0.1362 &   -89.3 \\
03158$+$4227 &    IVa\tablenotemark{a}    & 12.55 & 0.1344 &   -15.0 \\
03521$+$0028 &    IIIb       				& 12.48 & 0.1519 &    76.5 \\
05246$+$0103 &    IIIa\tablenotemark{a}   & 12.05 & 0.0971 &   109.5 \\
08030$+$5243 &    IVb\tablenotemark{a}    & 11.97 & 0.0835 &     0.0 \\
08311$-$2459 &    IVb\tablenotemark{a}    & 12.40 & 0.1006 &    67.5 \\
09111$-$1007 &    IVa\tablenotemark{a}    & 11.98 & 0.0542 &    73.5 \\
09583$+$4714 &    IIIa\tablenotemark{a}   & 11.98 & 0.0859 &   124.0 \\
10378$+$1109 &    IVb        				& 12.23 & 0.1362 &    11.3 \\
10494$+$4424 &    IVb        				& 12.15 & 0.0923 &    25.0 \\
10565$+$2448 &    M\tablenotemark{a}    	& 11.98 & 0.0431 &   109.0 \\
11095$-$0238 &    IVb        				& 12.20 & 0.1065 &    10.0 \\
11506$+$1331 &    V          				& 12.27 & 0.1273 &    80.3 \\
11598$-$0112 &    IVb         			& 12.40 & 0.1507 &   130.0 \\
12071$-$0444 &    IVb         			& 12.31 & 0.1284 &     0.0 \\
13451$+$1232 &    IIIb        			& 12.27 & 0.1212 &   105.7 \\
15130$-$1958 &    IVb         			& 12.03 & 0.1094 &   116.0 \\
15245$+$1019 &    IIIb\tablenotemark{a}	& 11.96 & 0.0755 &   127.8 \\
15462$-$0405 &    IVb         			& 12.16 & 0.1003 &   164.6 \\
16090$-$0139 &    IVa         			& 12.48 & 0.1336 &   107.6 \\
16474$+$3430 &    IIIb        			& 12.12 & 0.1115 &   161.8 \\
16487$+$5447 &    IIIb        			& 12.12 & 0.1038 &    66.2 \\
17028$+$5817 &    IIIa        			& 12.11 & 0.1061 &    94.5 \\
17208$-$0014 &    IIIb\tablenotemark{a}	& 12.38 & 0.0428 &   166.7 \\
17574$+$0629 &    IIIb\tablenotemark{a}	& 12.10 & 0.1096 &    51.2 \\
18368$+$3549 &    IVa\tablenotemark{a}	& 12.19 & 0.1162 &   -31.0 \\
18443$+$7433 &    V\tablenotemark{a}		& 12.23 & 0.1347 &    29.6 \\
18470$+$3233 &    IIIa\tablenotemark{a}	& 12.02 & 0.0785 &    67.3 \\
19297$-$0406 &    IVb\tablenotemark{a}	& 12.36 & 0.0857 &   149.5 \\
19458$+$0944 &    IVa\tablenotemark{a}	& 12.31 & 0.1000 &   118.1 \\
20046$-$0623 &    IIIb\tablenotemark{a}	& 12.02 & 0.0843 &    72.0 \\
20087$-$0308 &    IIIb\tablenotemark{a}	& 12.39 & 0.1057 &    85.5 \\
20414$-$1651 &    IVb         			& 12.19 & 0.0869 &    12.6 \\
23327$+$2913 &    IIIa       				& 12.03 & 0.1075 &    -4.3 \\
23365$+$3604 &    IIIb\tablenotemark{a}	& 12.13 & 0.0645 &   -19.5 \\
\enddata
\tablecomments{Column 1: Name, Column 2: Merger Classification \citep{Veilleux:2002p1039}. Column 3: IR luminosity \citep{Murphy:1996p23}. Column 4: Redshift. Column 6: Position angle of slit in degrees.
\tablenotetext{a} {Classifications estimated from $r$ band images and \cite{Murphy:1996p23}}
}

\end{deluxetable}

Fig.\ref{fig:ap_pos} shows two examples from the full figure set in Appendix A. 
The left panel shows an $r$ band image of each ULIRG from 
\cite{Murphy:1996p23}.
The slit position and the apertures used to extract spectra are marked.
In the middle panel, the same apertures are
marked and numbered on a cut-out of the two-dimensional spectrum near \Ha\ and [\ion{N}{2}].
All spectral orders were spatially registered with the order containing \Ha\ 
by cross correlation of the spatial continuum profile. 
The location of the brightest continuum emission defines the position of Aperture 0;
and the apertures are slightly separated ($\sim$ 0.1\arcsec) to reduce correlations between 
adjacent spectra.
Measurements of the Doppler shift, velocity dispersion, and excitation 
are summarized in the rightmost panel and described in Sect.\ref{sect:fitting}. 

In both galaxies included in Fig. \ref{fig:ap_pos}, line emission is 
detected over a much larger angle than is the continuum emission. In the first example, the
\Ha\ line profile of the low surface-brightness emission seen against the dark sky is noticeably
smoother than the double-peaked profiles of the extraplanar gas emanating from nearby 
starbursts \citep{Heckman:1990p710,Lehnert:1996p1025,Martin:1998p1216}. The second example illustrates
one of the 11 pre-mergers in the sample. Strong continuum emission is detected from two separate
galaxies as well as an extended, tidal feature.

\begin{figure*}
\centering
\includegraphics[keepaspectratio=true,trim = 0 0 0 20, clip=true,scale=0.9,width=0.9\linewidth]{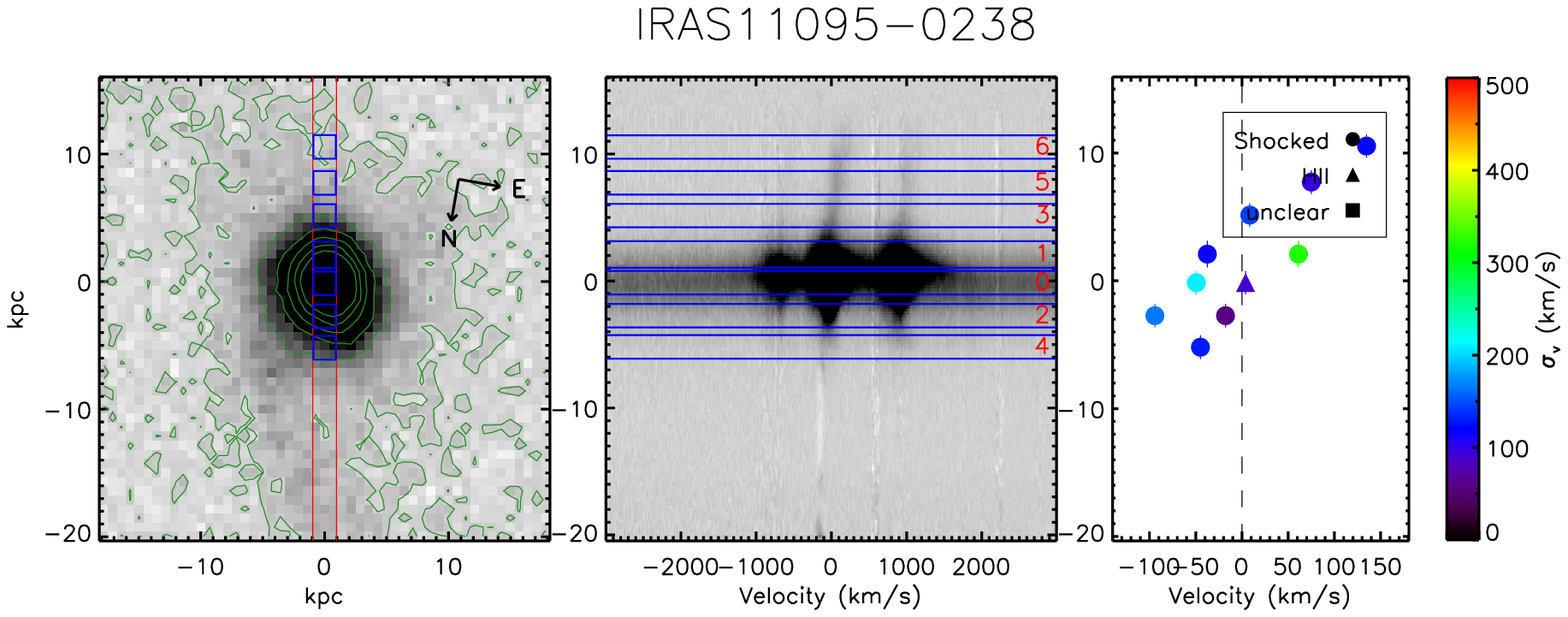}
\includegraphics[keepaspectratio=true,trim = 0 0 0 20, clip=true,scale=0.9,width=0.9\linewidth]{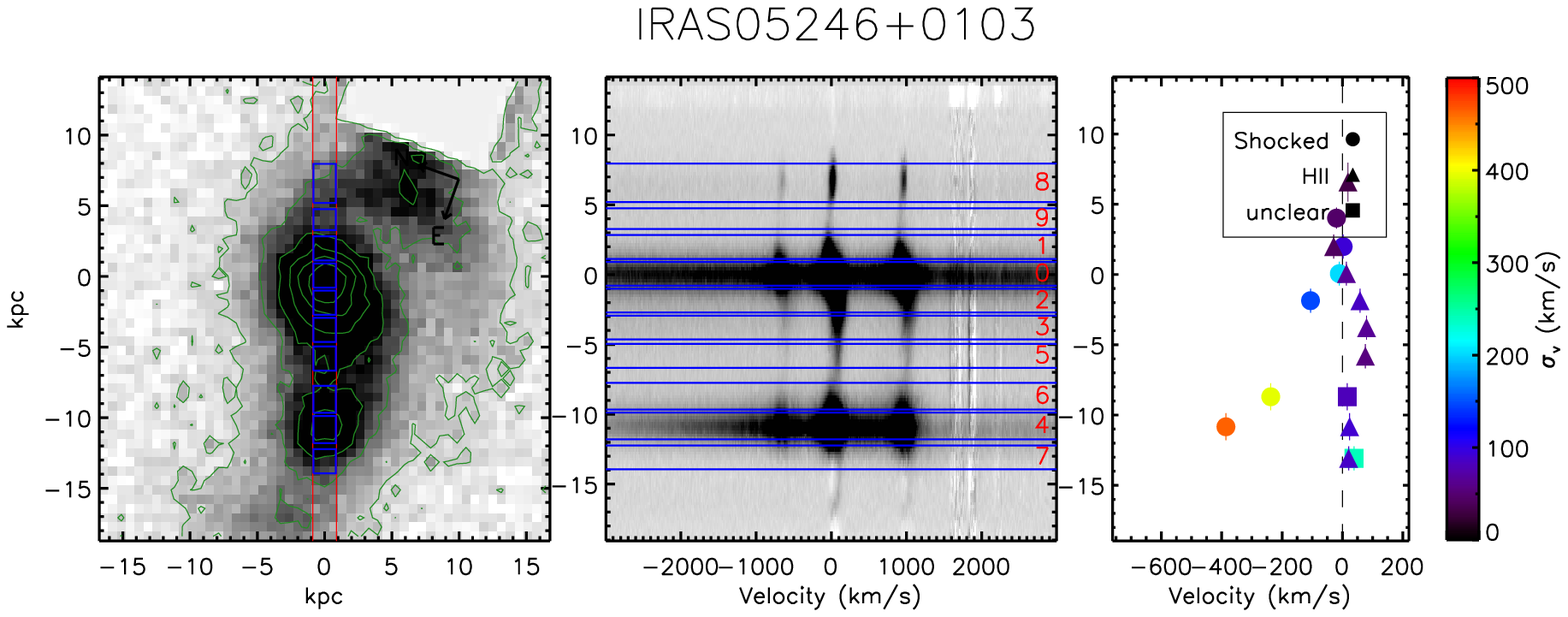}
\caption{\label{fig:ap_pos} 
\emph{Upper Row:} IRAS11095-0238 - a single nucleus ULIRG. 
\emph{Lower Row:} IRAS05246+0103 - a double nucleus ULIRG.
\emph{Left Column:} An $r$ band image from the Palomar 60 inch telescope \citep{Murphy:1996p23}. The green contours 
represent the median, and the 1, 2, 3, 4, and 5 $\sigma$ surface brightness levels in the image. 
The red lines 
show the spatial position of the 1\arcsec$\times$20\arcsec\ slit. The blue 
rectangles identify the positions of the selected apertures. 
\emph{Middle Column:} The 2D spectrum of \Ha\ + [\ion{N}{2}] centered in velocity space on \Ha. The 
blue rectangles identify the pixels that contribute to the individual apertures. The redshift 
range of the sample allows each object to be sampled by several apertures along the slit.  In 
IRAS~11095-0238 there is extended \Ha\ and [\ion{N}{2}] emission, 
allowing measurement of excitation mechanism out to 11 kpc from the nucleus. 
\emph{Right Column:} The position-velocity diagram for all of the fit components.  The line width 
of the components is indicated by the color of the plot point. The plot symbol represents the 
excitation category for that individual spectral component.
}
\end{figure*}

\begin{figure}
\centering
\includegraphics[keepaspectratio=true,trim = 0 0 0 10, clip=true, width=0.7\linewidth]{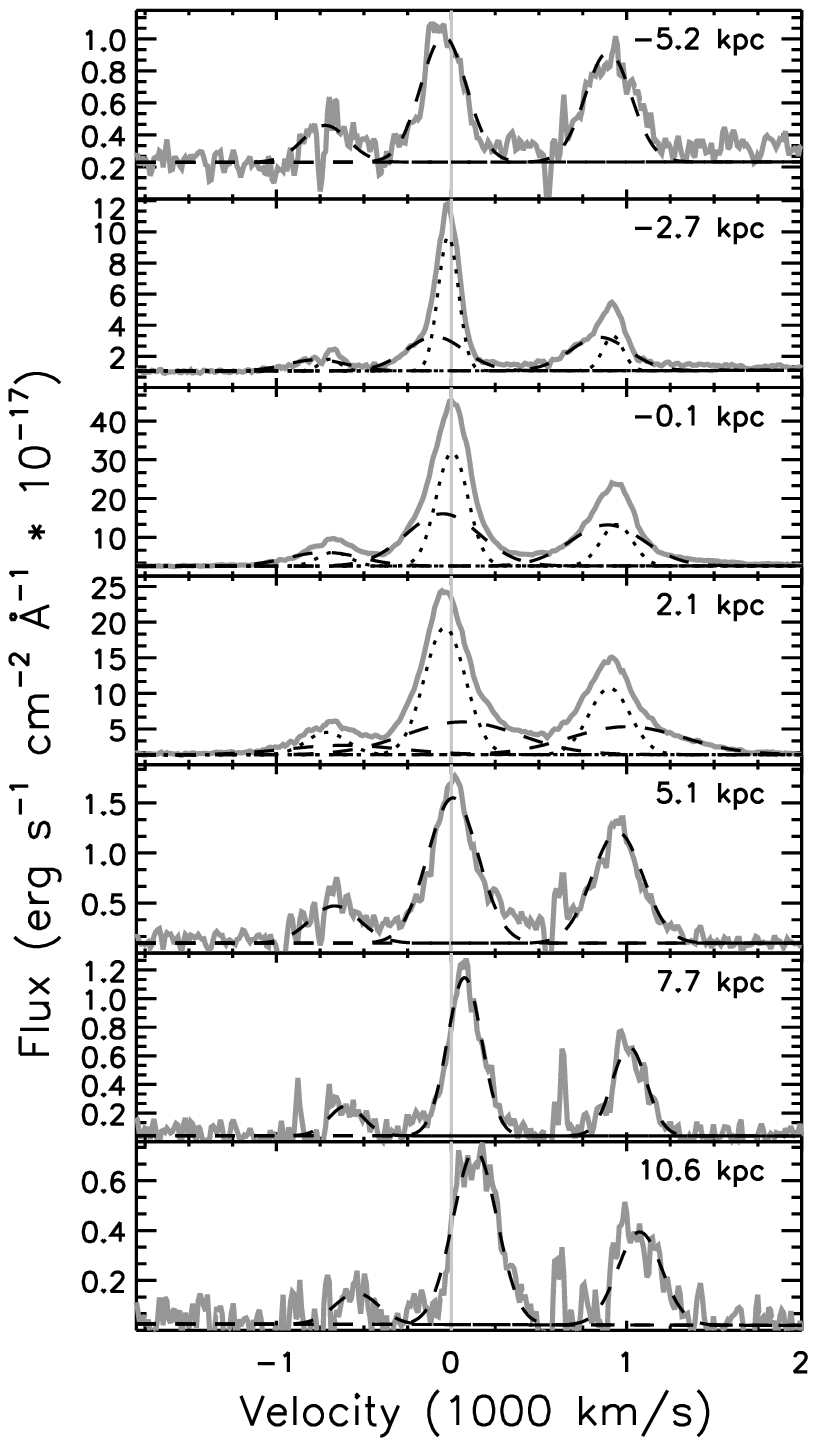}
\includegraphics[keepaspectratio=true,trim = 0 0 0 10, clip=true, width=0.7\linewidth]{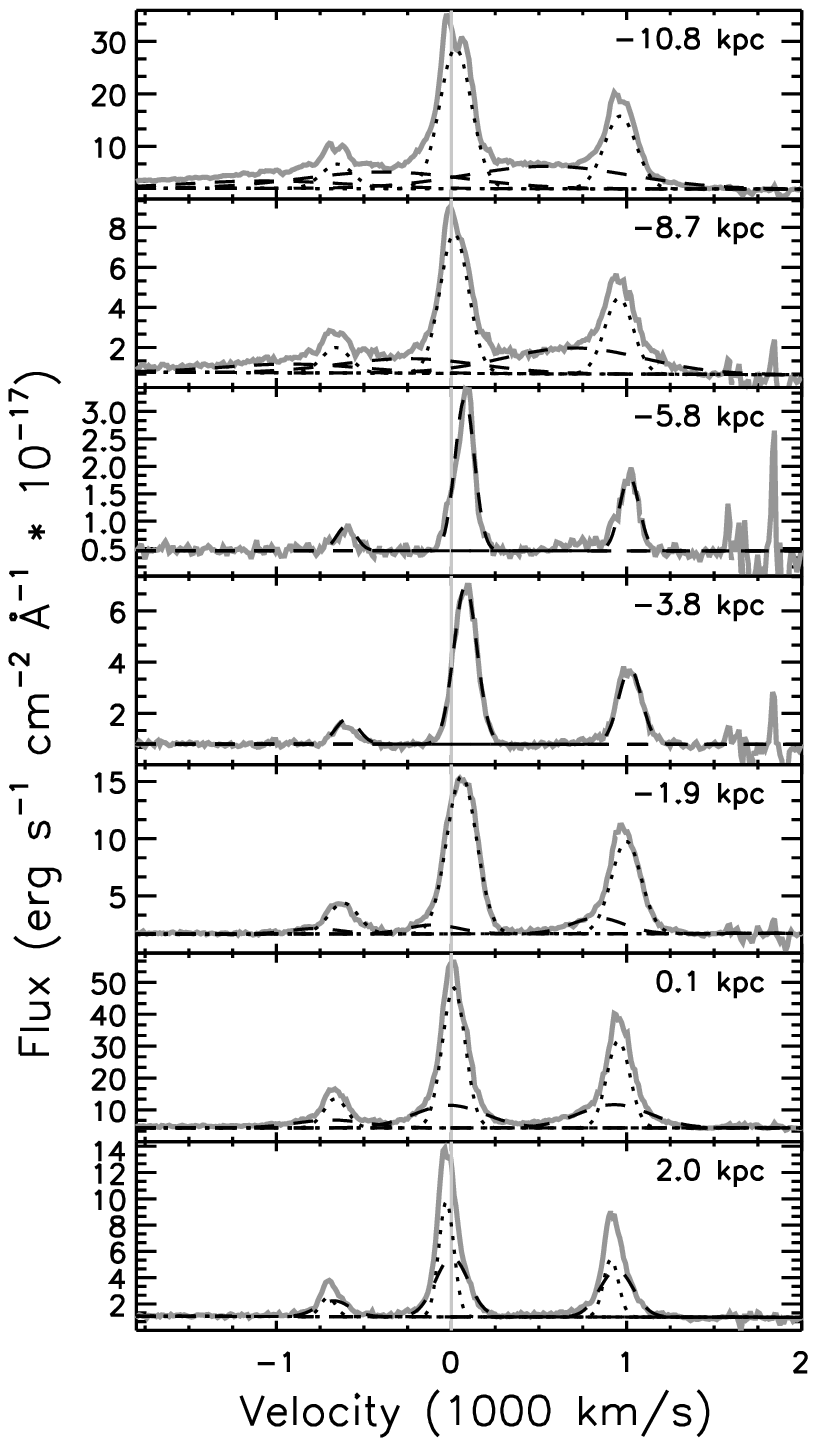}
\caption{\label{fig:allposprof} For the transitions \Ha\ + [\ion{N}{2}], we show the line profile for all positions in IRAS11095-0238 (above) and IRAS05246+0103 (below). {\it Above:} For IRAS11095-0238, the line profile shows a clear variation from aperture to aperture. In the aperture -2.7 kpc from the  $r$ band nucleus, the broad component (dashed) is blue-shifted relative to the narrow (dotted) component. On the other side of the nucleus at 2.1 kpc from the center, the broad component is slightly red-shifted relative to the narrow component. {\it Below:} In IRAS05246+0103 there is more than one nucleus (positions: 0.1 kpc and -10.8 kpc). A broad component can be seen in the line profile in each of them.}
\end{figure}

\subsection{Emission Line Fitting}
\label{sect:fitting}

In Appendix B, we show the observed emission line profiles as a function of aperture 
and of line transition for the full galaxy sample. In this section, we describe example 
figures for the same two galaxies as in Fig. \ref{fig:ap_pos}. 
First, Fig. \ref{fig:allposprof} shows the \Ha+[\ion{N}{2}] line profiles as a function 
of aperture position along the slit shown in Fig. \ref{fig:ap_pos}.
Prominent spatial gradients in the ratio of [\ion{N}{2}] to \Ha\ flux can be easily
seen by scanning up and down the first column for each galaxy.
Fig. \ref{fig:allposprof} illustrates the variation in the relative [\ion{N}{2}] to
\Ha\ strength along the slits shown in Fig. \ref{fig:ap_pos}. The ratio is very high in the 
extended, low-surface brightness emission surrounding IRAS11095-0238. IRAS05246+0103 shows 
similar variation at the various positions coinciding with the few kpc regions around
the nuclei.
The variations in the velocity
coordinate are more subtle but can be seen by comparing an unblended forbidden line, i.e., 
[\ion{O}{1}]\line 6300 or [\ion{O}{3}]\line 5007, with the \Hb\ profile from the same aperture.

\begin{figure}
\centering
\includegraphics[keepaspectratio=true, trim = 0 0 0 10, clip=true,width=0.7\linewidth]{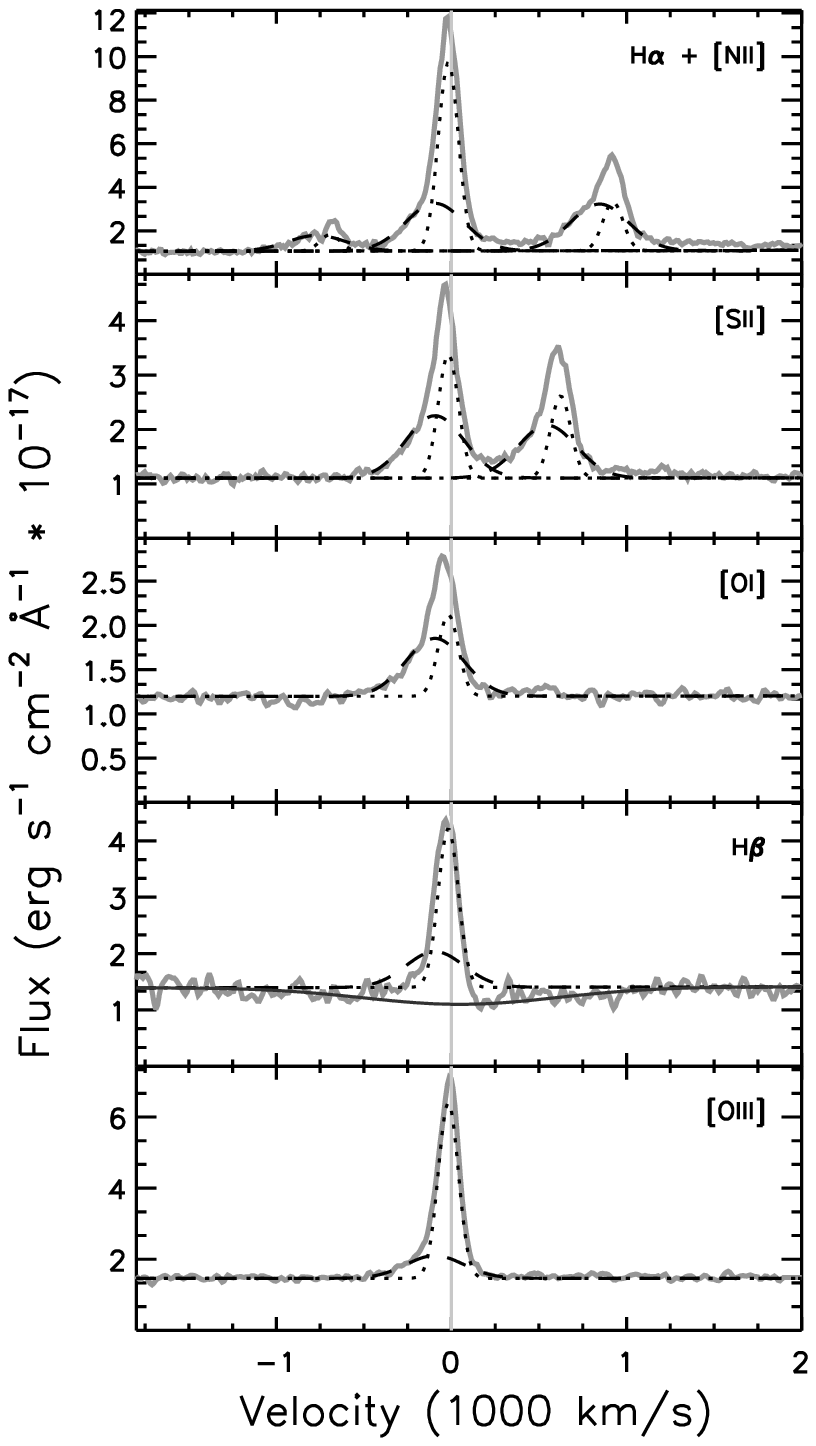}
\includegraphics[keepaspectratio=true, trim = 0 0 0 10, clip=true,width=0.7\linewidth]{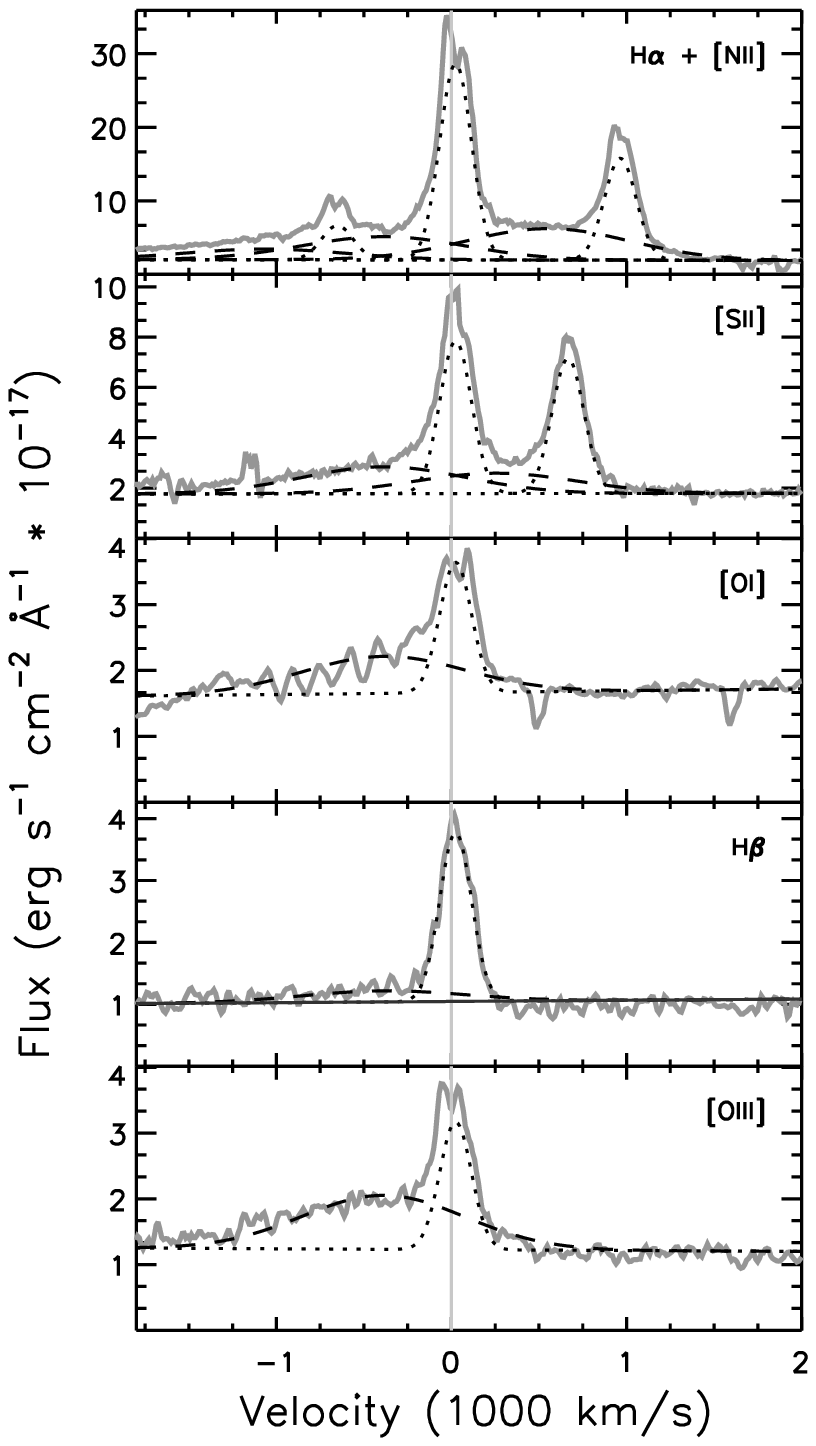}
\caption{\label{fig:alltransprof}  For a single aperture in IRAS11095-0238 (Above) and IRAS05246+0103 (Below), 
we present the line profiles for all analyzed transitions. 
Two components are required to fit the asymmetric line profiles evident in each transition.
The two components used in the fits have the same kinematics for each transition.}
\end{figure}

Next, in Fig. \ref{fig:alltransprof} we show the line profiles for a single aperture as a function
of line transition. 
In the IRAS11095-0238 spectrum, the broad, blue wing on the [\ion{O}{1}]\line 6300 profile 
is clearly stronger (relative to the total line flux) than the wing on the \Hb\ profile. 
IRAS05246+0103 shows this broad feature as well at in the measured transitions with \Hb\ being weaker. 
We note that in the spectral direction, quantifying these variations in the diagnostic line ratios is 
challenging. The line profiles must be moderately well resolved, e.g., FWHM of 60\kms\
for these echellete spectra. Because the \Ha, [\ion{N}{2}], and [\ion{S}{2}] lines are often
broad enough to blend with neighboring lines, coverage of one or more unblended 
lines like [\ion{O}{1}] \line 6300 is essential. 

Our line fitting procedure consists of fitting multiple Gaussian components simultaneously to 8 transitions (\Ha, [\ion{N}{2}]~$\lambda 6548,6583$, 
[\ion{S}{2}]$\lambda\lambda 6717,31$, [\ion{O}{1}] $\lambda 6300$, \Hb, 
and [\ion{O}{3}]$\lambda 5007$) with MPFIT in IDL.
The fitting method ties the individual 
kinematic components of the emission line profiles together, by requiring the same Doppler shift and 
velocity linewidth for all transitions. 
The amplitude of each component can vary independently. These requirements assume that the gas clouds 
that are identified by each kinematic component emit in all transitions. 
This method handles differing degrees of line blending by finding a solution suitable for 
all transitions. For some apertures, \Hb\ absorption contributed by the underlying stellar population 
significantly affects the line profile. We therefore include an \Hb\ absorption component in the
fitting as well, but allow it to vary independently from those of the emission components
\citep{Soto:2010p779}. 

We determined the number of components to fit by comparing fitting residuals to the 
flux measurement errors. 
The fitting for each aperture starts with a single component fit to all of the lines. 
If the residual flux exceeds the measurement error over a resolution element (FWHM $>$ 70) 
\kms, we include another fit component. We maintain spatial continuity in the fits by using the 
results from adjacent apertures as the initial guess for subsequent apertures. 

Two fitting components typically characterized the line profiles of all measured
transitions well, but we often see variations in amplitude and linewidth as a function of 
aperture and fitting component.
The fit of these 
multiple components allows us to identify separate kinematic components 
that vary in spatially different 
ways (Fig. \ref{fig:allposprof}). The position-velocity diagrams in the right column of 
Fig. \ref{fig:ap_pos} (and in Appendix A) further illustrate these 
variations in the cases of IRAS11095-0238 and IRAS05246+0103. 
All of this suggests that multiple kinematic components at a fixed position often arise from
physically distinct components of the galaxy and can be separated by this 
spatial and spectral deconstruction of the emission line profiles.
We include a full description of the fitted kinematic components in Appendix C, Table \ref{tab:flux}.

\clearpage
\subsection{Integrated Apertures}
\label{sect:integrated}

To allow examination of the impact of this spatial and kinematic structure seen in the spectral 
line ratios on the integrated spectrum, we extracted pseudo-integrated spectra. 
Using a large aperture that encompass all the apertures,
the flux per pixel is summed over this aperture.  
The resulting integrated line profiles were fit using the technique described in 
Section \ref{sect:fitting}. In Soto et al. (2012b) we compare the
spectral components identified in these integrated spectra to the components in sub-apertures. Our
analysis of the integrated spectra demonstrates when separate physical components, established
on the basis of the spatially resolved spectra, can be recognized in velocity space in the integrated
spectrum.

We use the integrated line fluxes for the individual galaxies in the mergers and
classify their overall excitation type using line ratio comparisons \citep{Kewley:2006p38}.
Table \ref{tab:int_spec_class} shows the result of these classifications. The HII class 
comprises the largest fraction of the sample (43\%) in the shows the classification of these 
integrated spectra, while LINER and Seyfert classes make up 18\% and 12\% of the sample.
Galaxies with mixed classifications in the different diagnostic diagram make up the 
remaining 27 \% of the sample. 

\begin{deluxetable}{lcccc}
\tablewidth{0pt}
\tabletypesize{\scriptsize}
\tablecaption{Integrated Spectral Classification \label{tab:int_spec_class}}
\tablehead{
IRAS Name &                
$\rm \left(\frac{[N~II]/{\rm H}\alpha}{[O~III]/{\rm H}\beta}\right)$ &
$\rm \left(\frac{[S~II]/{\rm H}\alpha}{[O~III]/{\rm H}\beta}\right)$ &
$\rm \left(\frac{[O~I]/{\rm H}\alpha}{[O~III]/{\rm H}\beta}\right)$ &
Total \\
(1)&(2)&(3)&(4)&(5)
}
\startdata
00153+5454    &  C    &   H/S   &   L/S &	M \\
00153+5454    &  C    &   L     &   L   &	L \\
00188$-$0856  &  A    &   T     &   S   &	S \\
00262+4251    &  A    &   L     &   T   &	L \\
01003$-$2238  &  A    &   S     &   S   &	S \\
01298$-$0744  &  C    &   T     &   L   &	L \\
03158+4227    &  A    &   T     &   L   &	L \\
03521+0028    &  X    &   X     &   X   &	X \\
05246+0103    &  C/A  &   H     &   T   &	H \\
05246+0103    &  A    &   L     &   T   &	L \\
08030+5243    &  C    &   H     &   H   &	H \\
08311$-$2459  &  X    &   X     &   X   &	X  \\
09111$-$1007  &  C    &   H     &   H   &	H \\
09583+4714    &  A    &   S     &   S   &	S \\
09583+4714    &  H    &   H     &   H   &	H \\
10378+1109    &  A    &   T     &   L   &	L \\
10494+4424    &  C/A  &   H     &   S/L &	M \\
10565+2448    &  C    &   H     &   H   &	H \\
11095$-$0238  &  C    &   L     &   L   &	L \\
11506+1331    &  C    &   H     &   H/S &	H \\
11506+1331    &  H    &   H     &   H/S &	H \\
11598$-$0112  &  C    &   H     &   H   &	H \\
12071$-$0444  &  A    &   T     &   S   &	S \\
13451+1232    &  A    &   S     &   S   &	S \\
15130$-$1958  &  A    &   S     &   S   &	S \\
15245+1019    &  C    &   H     &   H   &	H \\
15245+1019    &  A    &   H     &   S/L &	M \\
15462$-$0405  &  C    &   H     &   T   &	H \\
16090$-$0139  &  C/A  &   T     &   L   &	L \\
16474+3430    &  C    &   H     &   H   &	H \\
16474+3430    &  H    &   H     &   H   &	H \\
16487+5447    &  C    &   H     &   S/L &	M \\
16487+5447    &  A    &   S/L   &   S/L &	M \\
17028+5817    &  H    &   H     &   H   &	H \\
17028+5817    &  C    &   S/L   &   S/L &	S/L \\
17208$-$0014  &  C    &   H     &   H   &	H \\
17574+0629    &  H    &   H     &   H   &	H \\
18368+3549    &  A    &   S/L   &   S/L &	S/L \\
18443+7433    &  A    &   T     &   L   &	L \\
18470+3233    &  C    &   H     &   H   &	H \\
18470+3233    &  C    &   H     &   L   &	H/L \\
18470+3233    &  H    &   H     &   H   &	H \\
19297$-$0406  &  C    &   H     &   H   &	H \\
19458+0944    &  C    &   H     &   L   &	H/L \\
20046$-$0623  &  C    &   H     &   H   &	H \\
20087$-$0308  &  A    &   S/L   &   S/L &	S/L  \\
20414$-$1651  &  C    &   H     &   H/S &	H \\
23327+2913    &  A    &   S/L   &   S/L &	S/L\\
23365+3604    &  C    &   H     &   H   &  H 

\enddata
\tablecomments{ Column 1: IRAS Name.
Column 2: Spectral classifications as defined in \cite{Kewley:2006p38} with fluxes from the sum of 
both kinematic components in the spatially integrated measurements along the ESI longslit. These 
diagnostics come from the comparision of [\ion{N}{2}]/\Ha\ vs. [\ion{O}{3}]/\Hb. A refers to 
galaxies that exceed the maximum excitation possible from star formation alone 
suggesting a possible AGN, H refers to the 
regions below the empirical limit to the excitation by HII regions \citep{Kauffmann:2003p33}, C refers 
to the region between these, where the integrated emission line ratio is expected to be a combination 
of HII and AGN contributions. 
Column 3: Spectral classifications as defined in \cite{Kewley:2006p38} for the line ratios 
[\ion{S}{2}]/\Ha\ vs. [\ion{O}{3}]/\Hb. L refers to line ratios indicating LINER, 
S refers to line ratios indicating a Seyfert galaxy. 
Column 4: Spectral classifications as defined in \cite{Kewley:2006p38} for the line 
ratios [\ion{O}{1}]/\Ha\ vs. [\ion{O}{3}]/\Hb.
Column 5: The total classification determined by combination of the 3 diagnostic diagrams. 
The additional classification of M is included where a mix of all three classifications makes
the class designation ambiguous.
}
\end{deluxetable}

\section{Results}
\label{sect:results}

\subsection{Line Ratios}
\label{sect:ratios}

Deblending the line profiles allows us to investigate the underlying excitation mechanism that is relevant for each kinematic component.
In the 
standard BPT diagram, ratios of forbidden transitions to Balmer transitions ([\ion{N}{2}]/\Ha, 
[\ion{S}{2}]/\Ha, [\ion{O}{1}]/\Ha\ and [\ion{O}{3}]/\Hb) probe the energy of the associated ionizing 
radiation \citep{Baldwin:1981p7}. High forbidden to Balmer line ratios exceeding the range possible 
for star formation can imply the presence of either AGN \citep{Kewley:2006p38} or shocks \citep{Allen:2008p695}. 

In the right column of Fig. \ref{fig:ap_pos} (and in Appendix B), we present the kinematic
components along with an indicator for the type of excitation that is most likely relevant. 
The top right panel of Fig. \ref{fig:ap_pos} is an example of one of the more irregular kinematic patterns for the galaxy 
IRAS11095-0238.  In this case the entire galaxy has shock-like excitation.  The lower right panel 
of Fig. \ref{fig:ap_pos}
shows a different kinematic pattern the narrow HII-like emission in IRAS05246+0103 and broad 
shock-like emission closer to the nuclei.

\subsection{Shock Models}
\label{sect:shock_models}

Shock models predict the line ratios created by the shocks for a 
range of magnetic field strengths, densities, and shock velocities 
\citep[$v_{sh}$;][]{Allen:2008p695}. Here, we focus on magnetic field ranging from 1 to 100 
$\mu$G for starburst galaxies \citep{Thompson:2006p833}. 
The shock-only model with solar metallicity and electron 
density $n_e = 10~{\rm cm}^{-3}$, overlapped with a large fraction of the measured line ratios. 
The line ratios fall into the region of the BPT diagram that more often suggests LINER-like 
excitation. If these measured line ratios can be attributed to shocks, the models suggests that 
the precursor to the shock is not a significant contributor to the ionization 
\citep{Dopita:1995p1120}. Some of the measured line ratios fall just below the shock grid, 
but the effect of averaging over an $\sim 1{\rm kpc}^2$ can include contributions by HII regions
as well, moving the line ratio toward the HII region of the diagram. The spread in the 
[\ion{O}{1}]/\Ha\ ratio with $v_{sh}$ in these models (Fig. \ref{fig:shock_grid}) indicates 
that the [\ion{O}{1}]/\Ha\ vs [\ion{O}{3}]/\Hb\ diagram is a more sensitive identifier of 
shocks and shock velocity. The [\ion{S}{2}]/\Ha\ and [\ion{N}{2}]/\Ha\ gridlines, on the 
other hand, pile-up for gridlines with $v_{sh} > 250$ \kms. 

In a different class of objects, line emitting red galaxies, extended LINER-like emission is 
detected at extended radii as well, but post-AGB stars are the suspected source of 
ionization \citep{Yan:2012p1226}.
These post-AGB stars would make the largest {\it relative contribution} to the ULIRG spectra at 
large galactocentric radii because there is an increase in stellar age with 
radius \citep{Soto:2010p779}; the central regions are dominated by ongoing star formation.
However, the ages of the stellar populations necessary to create this excitation are few Gyr, 
much larger than the $\sim 0.4$ Gyr implied by the measured stellar population ages.
Furthermore, the scaling between \Ha\ luminosity and stellar mass from this form of 
excitation \citep{Yan:2012p1226} implies an order of magnitude more stellar mass 
($2~\times~10^{11}\msun$) than is found in these galaxies \citep{Tacconi:2002p1236}.
We therefore conclude that LINER-like excitation in our ULIRG sample most likely results from shocks.

Having identified components of the emission beyond the range of ionization by \ion{H}{2} regions, 
we estimate $v_{sh}$ from the position of the component line ratios on the shock grids. We 
estimate errors in $v_{sh}$ from the position of the end of the error bar in each measurement, and 
using the difference in these velocities as the $v_{sh}$ error. We present these estimates on a 
per component basis in Appendix C, Table \ref{tab:vshock}. The error in $v_{sh}$ is dominated by systematic 
errors from the uncertainty in model selection and the influence of a 
radiative precursor region on the emitted flux.

As with shock-only models,
shock + precursor grids with $n_e = 10~{\rm cm}^{-3}$ cover a reasonable fraction of the 
measured line ratios, but we find that there is a small systematic offset of 25 \kms.
Similarly, in models with  
$n_e = 0.1~{\rm cm}^{-3}$, we find a similar systematic offset with a 
slightly larger scatter.

\subsection{Excitation Categories}
\label{sect:categories}

\begin{figure*}
\centering
\epsfig{file=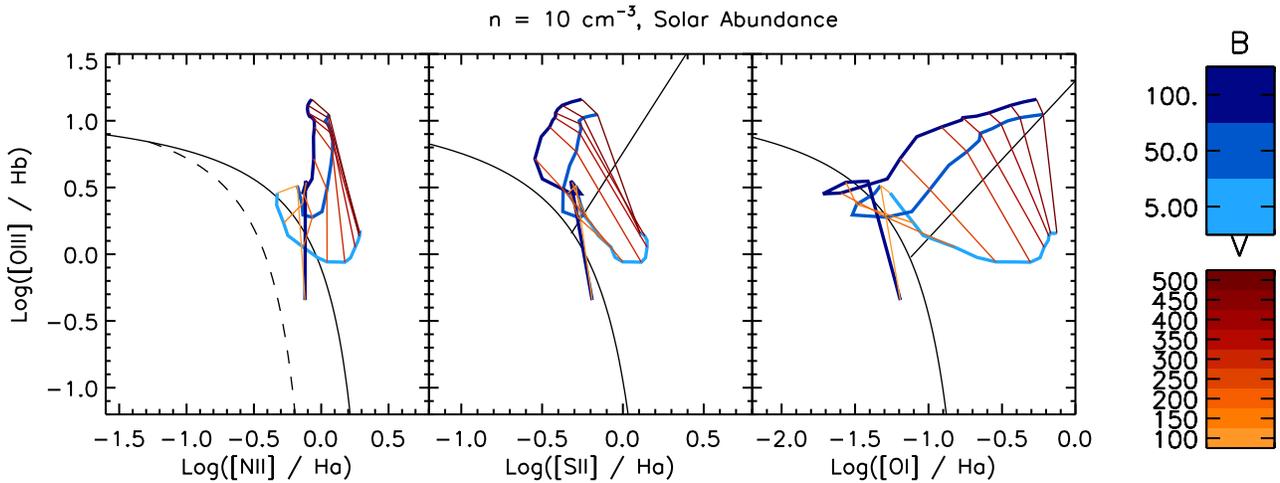, keepaspectratio=true, width=\linewidth}
\caption{\label{fig:shock_grid} Shock grids \citep{Allen:2008p695} are presented for a solar 
abundance with $n_e$ = 10~\cmv.  The yellow - orange lines are contours of constant shock 
velocity in \kms. The cyan - blue lines are lines of constant magnetic field in $\mu$G. Also 
included in the plots are the ``extreme starburst lines'' and the lines that indicate the 
distinction between Seyferts and LINERs.  Larger forbidden to Balmer line ratios can indicate 
either a faster shock, or a larger contribution by AGN. In Appendix A we present these 
diagnostics along with the individual line ratios for each component in each galaxy. }
\end{figure*}

When trying to characterize the underlying physical processes present in merging galaxies, many 
mechanisms are at work exciting the gas in the object which leads to the measured optical emission 
lines. The measured emission line ratios span a range of values within the diagnostic diagrams 
making the interpretation difficult. We attempt to simplify the analysis by categorizing the 
measured components into two categories; ``HII-like'' and ``shock-like''. We make this distinction 
by comparing a component's emission line ratios to the maximum ionization in the extreme 
starburst case \citep{Kewley:2006p38}. Below this line, we define the line ratio as ``HII-like''; 
above it, we define the line ratio as ``shock-like'', since the spectral energy 
distribution of a starforming region is not sufficient to create these emission ratios. Figure 
\ref{fig:shock_grid} displays where the shock models and diagnostic diagrams intersect, showing 
the tendency of these fast shock models to exhibit emission line ratios in the region of the 
diagnostic diagram typically associated with photoionization by active galactic nuclei.

The clearest cases for grouping a component into ``shock-like'' or ``HII-like'' is when all three 
diagnostic measurements (log([\ion{O}{3}]/\Hb) versus log([\ion{N}{2}]/\Ha), 
log([\ion{S}{2}]/\Ha) and log([\ion{O}{1}]/\Ha))
agree, i.e., are below or above this extreme starburst line. In cases 
where the measures do not agree, we generally rely upon the sensitivity of the [\ion{O}{1}] 
transition to determine the classification for a particular component. For cases where the errors 
in flux ratio are consistent with either case, we classify the component as ``unclear''. For each 
galaxy, we present the emission line ratios for each component on these diagrams in Appendix A.

\section{Location of Shocked Gas}
\label{sect:shock_location}

The categorization of emission into different excitation categories allows the identification of 
``shock-like'' ratios beyond the nuclei of the galaxies. For the 38 galaxies that exhibit 
``shock-like'' emission line ratios in at least one of the components, the spatial position 
of the ``shock-like'' component lies 3 kpc from the nucleus. 
For 14 of the objects, ``shock-like'' emission also appears within 2 kpc of the nucleus.
Only one galaxy shows exclusively ``HII-like'' emission -- IRAS16474+3430. 
 
The kinematic spatial variations along the slit suggest that some of these galaxies host gas disks, 
despite disturbance caused by the merger interaction. 
The smooth transition of Doppler shift from red to blue identifies the candidate disks, 
while the excitation categories indicate that gas with ``shock-like'' 
excitation share the same kinematics.
We define a ``disk'' in these cases as when the contiguous components with $\sigma_v < 150$ \kms\ 
cross the line $v = 0$ \kms. We make an exception in the cases where two nuclei are present, as 
indicated by the presence of a second continuum peak. In these cases, the objects have motions 
associated with their interaction, so are allowed to have a velocity offset. Using these 
parameters, we find 12 systems with clear disks, presented in Figure \ref{fig:best_disk} and 
described in Section \ref{sect:individuals}. We include the measured rotation gradients 
in Table \ref{tab:disk}. 

\subsection{Evidence for Gas Disks}

\subsubsection{Individual Objects}
\label{sect:individuals}

\begin{figure*}
\centering
\epsfig{file=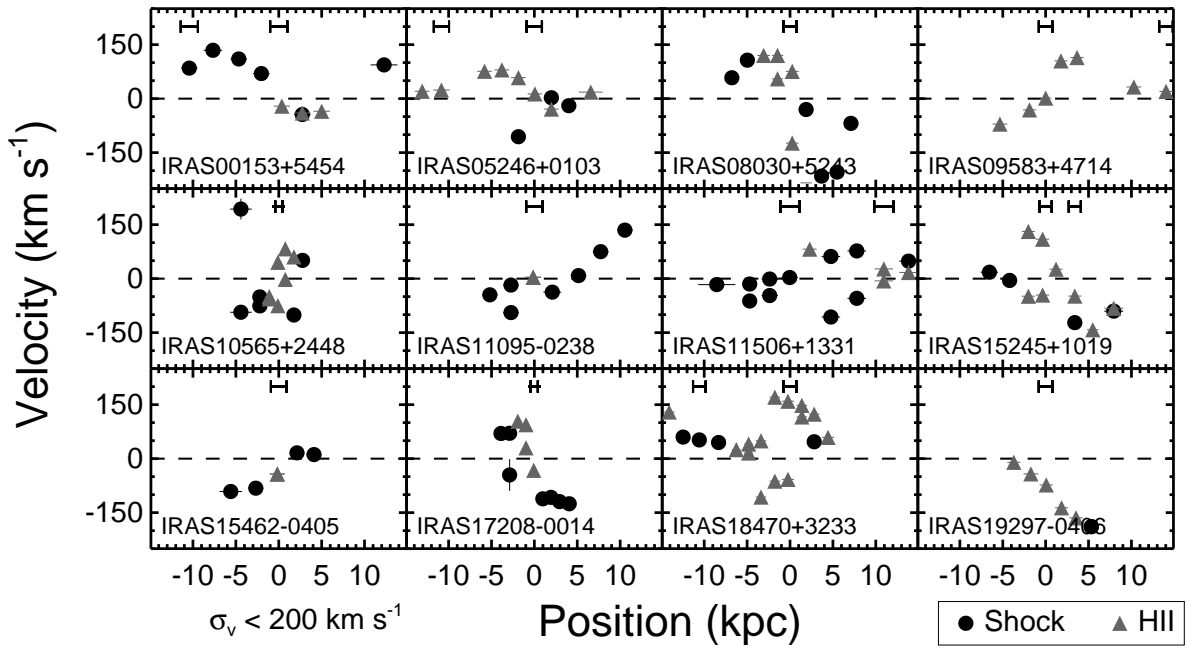, width=0.9\linewidth, keepaspectratio=true}
\caption{\label{fig:best_disk} In the above plot we show position-velocity diagrams for components 
with $\sigma_v < 200$~\kms\ for 12 objects with profiles most indicative of rotation. We identify 
nuclei based on the position of continuum sources in the 2D spectra, which are marked above the 
data points. In IRAS00153+5454,  IRAS11095-0238,  and IRAS15462-0450 the shocked components appear 
to be part of a rotating gas disk. IRAS05246+0103, IRAS08030+5243, IRAS09583+4714, and 
IRAS19297-0406 on the other hand, present predominantly HII-like emission in its gas disk. 
IRAS17208-0014 and IRAS18470+3233 show a mixture of HII and shock, with the shock in the outer 
edges of the disk. IRAS11506+1331, IRAS18470+3233, and IRAS 15245+1019, are double nucleus objects 
with possibly multiple disks, and a large mixture of HII and shock excitation.}
\vspace{0.4cm}
\end{figure*}

\emph{IRAS00153+5454} - Fig. A\ref{fig:i00153_bbc} shows two continuum sources, identifying it 
as a double nucleus object with a separation of $\sim$ 10 kpc. The emission for the northern 
nucleus is dominated by ``shock-like'' emission, while a smaller shock region exists in the 
southern nucleus. The velocities in the northern nucleus may incdicate rotation in a disk.

\emph{IRAS01298-0744} - The emission in this object extends to $\sim 12$ kpc from the continuum 
source.   Rotation is evident in the central and northern regions of the disk (Fig. 
A\ref{fig:i01298_bbc}), while the emission in the far south presents more of a flat rotation 
profile. Apertures 4 and 5 in Fig. B\ref{fig:i01298_prof} show a peculiar emission component on 
the red side of the \Ha+[\ion{N}{2}] line profile. This emission is not fit, since it does not 
clearly show up in any of the other transisitions.

\emph{IRAS03158+4227} - The narrow components are primarily ``shock-like'', and only shows a 
shallow rotation gradient in Fig. Fig. A\ref{fig:i03158_bbc}. The morphology of the $r$ band image 
is consistent with a face on disk-like objects, which could explain the flat rotation curve.

\emph{IRAS05246+0103} - This double nucleus object is spatially extended by approximately 11 kpc, 
however they both appear to have extended gas disks. The extended emission is mostly dominated by 
``HII-like'' emission. In the position-velocity plot (Fig. A\ref{fig:i05246_bbc}) shows two 
disk-like rotation curves with broader emission closer to the nuclei.  The narrow emission lines 
in these cases is more often HII-like, but some components at the East side of the eastern nucleus 
have shock-like line ratios. In the line profile plots (Fig. B\ref{fig:i05246_prof}), apertures 
4 and 6 show a good example of the simultaneous fitting decomposing a blended \Ha+[\ion{N}{2}] 
profile.

\emph{IRAS08030+5243} - one of the clearer examples of disk associated rotation, which includes 
narrow ``shock-like'' emission. At the far ends of this rotation curve (Fig. 
A\ref{fig:i08030_bbc}) in this object the Doppler shift turns around and  approaches the systemic 
velocity. The central aperture in this case shows a double peaked narrow emission profile (Fig. 
B\ref{fig:i08030_prof}), where the apertures outside of this have more clear broad profiles.

\emph{IRAS08311-2459} - The gas disk in this object emits both ``HII-like'' and shock-like 
emission at distances up to 4 kpc from the continuum source. In this case the rotation (Fig. 
A\ref{fig:i08311_bbc}) in narrower emission lines is clear compared to the broad emission line 
profiles, which are at a constant Doppler shift of $\sim 50$\kms along the slit. The diagnostic 
diagrams in this case show strong [\ion{O}{3}]/\Hb\ ratios, however the [\ion{O}{1}] in this case 
is unmeasured due to flaws in the data files.

\emph{IRAS09111-1007} - The rotation in this object shifts from -50 \kms\ to 100 \kms\ showing 
that this object may host a disk (Fig. A\ref{fig:i09111_bbc}). The disk components are comprised 
equally of ``HII-like'' and ``shock-like'' emission components.  In the center few apertures there 
are strongly blueshifted components relative to the velocities of the disk.

\emph{IRAS09583+4714} - One of the clearest examples of disk associated rotation appears in Fig. 
A\ref{fig:i09583_bbc}. Two nuclei are present, however the eastern nucleus exhibits strong 
rotation, while the western nucleus is mostly a flat rotation curve. The rotation curves are a 
comprised of ``HII-like'' emission components. For this object we do not have an $r$ band image to 
compare to the 2D spectroscopy, which makes the interpretation slightly more difficult.

\emph{IRAS10565+2448} - This lower redshift object shows that the disk rotation (Fig. 
A\ref{fig:i10565_bbc}.) is complex, however there is again a trend from red to blue ($\pm 100$
\kms) indicating possible rotation in a disk. the further extended components are ``shock-like'' 
in this case.

\emph{IRAS11095-0238} - The rotation in this object has a larger extent on one side of this object 
(Fig. A\ref{fig:i11095_bbc}). The emission lines are dominated by ``shock-like'' emission 
through out the object, mostly appearing in the LINER region of the diagnostic diagrams.

\emph{IRAS11506+1331} - Two disks are resolved in Fig. A\ref{fig:i11506_bbc} for this object, 
where the gas between the two nuclei are separated in velocity space by $\sim 150$ \kms. Most of 
the gas in the object emits ``shock-like'' emission, again in the LINER region of the diagnostic 
diagram.

\emph{IRAS11598-0112} - The emission line profiles in this object are generally more chaotic, 
however, the narrow part of the line profile presents a general red to blue trend in 
Fig. A\ref{fig:i11598_bbc} over a wide spatial ($\pm 8$ kpc) and velocity. The fits in Fig. 
B\ref{fig:i11598_prof} had some difficulty capturing the line profile in aperture 0.

\emph{IRAS15245+1019} - This ULIRG has double with nuclei separated by $\sim$ 4kpc, as shown in 
Fig. A\ref{fig:i15245_bbc}. It is difficult to distinguish which galaxy each of the components 
belongs to, since they span the same spatial range, however, in either interpretation there 
appears to be two disks involved.

\emph{IRAS15462-0405} - Narrow emission shows a disk like rotation profile in Fig. 
A\ref{fig:i15245_bbc}, though the broad part of the line profile behaves much differently than 
the rest of the objects. 

\emph{IRAS16487+5447} - This object is a double separated by 5 kpc with a complicated position-
velocity diagram (Fig. A\ref{fig:i16487_bbc}). There are two extended emission components, 
however, and the narrower emission component presents more disk like rotation. 

\emph{IRAS17208-0014} - This object is a double, separated by $\sim$ 2 kpc. In Fig.
A\ref{fig:i17208_bbc} The rotation seems to cross the two nuclei with the north side having a 
flat rotation profile at -100 \kms, and the south side more chaotic.

\emph{IRAS17574+0629} - The position-velocity diagram (Fig. A\ref{fig:i17574_bbc}) separates 
the two galaxies in velocity space, revealing two separate disks. The morphology of the $r$ band 
image gives some indication that the merging disks may be nearly orthogonal or a multiple merger. 
The emission in this collision is dominated by ``HII-like'' emission.

\emph{IRAS18470+3233} - This object is a multiple merger. In Fig. A\ref{fig:i18470_bbc} the 
western galaxy in this merger hosts two emission features that present ``disk-like'' line profiles 
separated by a 150 \kms offset. The eastern galaxy has a spatially large rotation profile, which 
includes shock inoization in the central 4 kpc.

\emph{IRAS19297-0406} - This object presents rotation profiles in Fig. A\ref{fig:i19297_bbc} 
that extend to $\pm 5$ kpc and are mostly ``HII-like'' in ionization. The whole rotation curve is 
offset from systemic, implying that there is a possible slight error in the used redshift.

\emph{IRAS20046-0623} - The continuum profile in this object is flat along the spatial axis and 
extended over 8 kpc as can be seen in the 2D spectrum of Fig. A\ref{fig:i20046_bbc}. There are 
no clear peaks in the continuum profile. The $r$ band image indicates that the two intersecting 
galaxies are nearly perpendicular.  A rotation gradient is evident along the major axis of the 
sampled objects, while the minor axis shows little rotation.

\begin{deluxetable}{l r r r r r r}
\tablewidth{0pt}
\tabletypesize{\scriptsize}
\tablecaption{ \label{tab:disk} Disks}
\tablehead{
IRAS Name	& \multicolumn{2}{c}{Edge Positions}	&	\multicolumn{2}{c}{$v_{rad}$}	&	rotation grad	\\
			& \multicolumn{2}{c}{kpc}	&	\multicolumn{2}{c}{\kms}	&	\kms~kpc$^{-1}$	\\
(1)			& (2)	& (3)	& (4)	& (5)	& (6)
}
\startdata
00153$+$5454(a)  &  -10.3  &  -2.2  &    79  &  133  &   2.1 \\
00153$+$5454(b)  &    0.2  &  11.9  &   -45  &   93  &   10.5 \\ 
01298$-$0744     &  -10.6  &   3.9  &   -34  &   28  &    3.1 \\
03158$+$4227     &   -6.7  &   9.6  &   -17  &   35  &    2.0 \\
05246$+$0103(a)  &   -5.8  &   6.6  &  -105  &   79  &   5.6 \\
05246$+$0103(b)  &  -13.1  &  -8.7  &    16  &   23  &   1.0 \\ 
08030$+$5243     &   -3.6  &   3.7  &  -348  &  461  &  73.8 \\
08311$-$2459     &   -4.1  &   4.3  &  -131  &   56  &   24.9 \\
09111$-$1007     &   -4.2  &   7.4 \tablenotemark{a}  &  -115  &  113  &   11.6 \\
09583$+$4714     &   -5.3  &   3.7 \tablenotemark{a}  &   -70  &  114  &   22.6 \\ 
10565$+$2448   &   -4.4  &   2.8  &  -101  &  193  &    0.7 \\
11095$-$0238     &   -5.2  &  10.6  &   -94  &  134  &   11.3 \\
11506$+$1331(a)  &   -8.5  &   7.8  &   -62  &  283  &    9.1 \\
11506$+$1331(b)  &    4.8  &  13.8  &  -106  &   48  &   15.6 \\
11598$-$0112     &   -5.2  &  10.4 \tablenotemark{a}  &  -192  &  336  &  -23.6 \\
15245$+$1019     &   -6.6  &   7.9  &  -142  &  131  &  11.3 \\
15462$-$0405     &   -5.6  &   4.1  &   -91  &   16  &   12.5 \\
16487$+$5447     &   -8.3  &   7.2  &  -294  &   28  &   15.8 \\
17208$-$0014     &   -4.6  &   4.6  &  -127  &  143  &  32.4 \\ 
18470$+$3233(a)  &   -4.8  &   4.4  &  -106  &  170  &   11.1 \\
18470$+$3233(b)  &  -15.7  &  -6.3  &    25  &  147  &  12.8 \\
19297$-$0406     &   -6.4  &   5.8  &  -192  &   43  &  19.8\\

\enddata
\tablenotetext{a} {08311-2459, 09583+4714 and 11598-0112 have a highly displaced 
HII region not included in the description of the possible disk.}
\tablecomments{Col.(1):IRAS name; Cols.(2\&3): Distance of the rotation profile edge 
from the continuum source in kpc; Cols.(4\&5): Minimum and maximum velocities of 
rotation profile in \kms. Col.(6): Rotation gradient in \kms\ kpc$^{-1}$.}
\end{deluxetable}

\subsubsection{Conclusions}
\label{sect:disk_conclusion}

Out of the entire sample of 39 ULIRGs, the twenty objects in Sect. \ref{sect:individuals} are the 
ULIRGs with candidate gas disks. 
	These candidate gas disks have narrow emission lines with $\sigma \le 150 \kms$ that vary smoothly
		in Doppler shift along the slit.
	The remaining objects in the sample have either little evidence of an extended gas, 
		 broader emission features, or unclear spatial trends in the gas kinematics.
	Four objects out of the full sample (IRAS03521+0028, IRAS10378+1109, IRAS16090-0139,IRAS18368+3549)
		exhibit only less ordered motion that is not clearly part of a disk.

The sample included in this study chooses galaxies at various merger phases, which we can compare 
to the presence of a rotation profile in the position velocity diagrams.
	Twelve objects with the clearest rotation profiles presented in Fig. \ref{fig:best_disk} 
		show strong gradients in Doppler shift along the slit.
	In this representative set of twelve ULIRGs, seven are in binaries, one is a multiple merger, 
		and the remaining four are at a later single nucleus phase with extended diffuse emission. 
	The number of early merger phase objects with clear disks suggests that a gas disk appears
		at early stages of the merger, then removed as the galaxies coalesce.  

Spatially resolved emission line diagnostics allow us to examine the relationship between the 
presence of rotation and the spatial distribution of gas excitation. Two disks appear in some of 
the double nuclei objects, increasing the number of disks to consider to 15. IRAS05246+0103, 
IRAS19297-0406, IRAS09583+4714, IRAS15245+1019 and IRAS18470+3233 show disks that are strongly 
dominated by HII-like excitation. In IRAS11095-0238, both disks in IRAS11506+1331, IRAS15462-0405, 
and IRAS00153+5454 the rotating gas is dominated by shock-like excitation. The remaining 5 disks 
(IRAS08030+5243, IRAS10565+2448, IRAS15245, IRAS17208-0014, and IRAS18470+3233) are evenly mixed 
in both HII-like and shock-like excitation. The distribution of excitation in these objects places 
the HII-like regions closer to nuclei, with shock-like excitation in the outer few kpc, with an 
exception for IRAS18470+3233. This HII-centered distribution also appears in the shock dominated 
disks, where just the central aperture is HII-like, but the rest of the disk is dominated by 
shock-like excitation. The distribution of excitation however is not clearly related to the merger 
class in this subsample.

There are at least 4 different possible origins of narrow shocked components 
of the emission line profiles. (1) In merger models \citep{Cox:2004p791}, shocks occur
as the gas disks collide. (2) Shocks also occur as gas that was previously removed via tidal
stripping falls back into the galaxies. (3) Shocks can also be produced by the dissipation of 
energy injected by massive stars created in the burst of star formation. (4) Shock-like emission 
can also be produced by photoionization via aging post-AGB stars.
The gas disks in this study, however, appear more frequently in the earlier, 
binary stages of the merger. This suggests that the earlier stages of the 
merger can trigger shocks, rather than only appearing after the galaxies have coalesced.

We stress, however, that the 
complicated gas kinematics present at all stages of the merger are often difficult to discern 
with long slit measurements and imaging data. 
One situation that leads to confusion is that some velocity 
profiles do not reach the same maximum velocity in both directions along the slit from the 
nucleus. Additionally, the orientation of this gas disk with respect to the line of sight is not 
clear, meaning that objects with a shallow rotation gradient may simply be close to face-on, 
thereby decreasing the line of sight velocity. Furthermore, the slit position angle was selected 
to either sample multiple nuclei or sample an elongation in the $r$ band image of the object, so a 
possible misalignment with the major axis would interfere with the detection of a disk.

Future observations that employ integral field 
spectroscopy will allow a better analysis of the gas kinematics and excitation in concert with its 
spatial distribution along the merger sequence. Spatial continuity in these types of measurements 
allow a better understanding of larger scale features such as disk inclination, as well as 
isolating regions of varying excitation. Better measurement of the kinematic features can further 
better estimates of merger phase and lead better understanding of the processes that 
create shocked gas disks.

\section{Summary}
\label{sect:summary}
In this analysis of longslit ESI data, we observed complex line profiles in the various line 
species used for the examination of emission line excitation. 
	The spectral resolution and signal to noise obtained with ESI allow detailed 
		investigation of the underlying emission mechanisms.
	Spatial resolution along the long slit allows us to further understand the extended 
		structure of these sources. 
	By performing a simultaneous multicomponent fits to the Balmer and forbidden lines, 
		we were able to identify shocked gas disks at early merger stages. 
	These data are relevant for many investigative purposes such as informing model creation and
		examination of gas processes in the merging galaxies and strong star formation environments.

\acknowledgments{
The authors thank Moire Prescott and Lee Armus for many stimulating discussions and helpful suggestions.
	The authors also wish to recognize and acknowledge the very significant cultural role and 
		reverence that the summit of Mauna Kea has always had within the indigenous 
		Hawaiian community.  
	We are most fortunate to have the opportunity to conduct observations from this mountain. 
	This work was supported by the National Science Foundation under 
		contract AST-0909182 and AST-1109288.
}
{\it Facilities:}  \facility{Keck}

\bibliographystyle{apj}
\bibliography{Soto_Atlas_emulateapj}

\clearpage
\appendix 
\section{Appendix A}
\setcounter{figure}{0}
\setcounter{table}{0}

In this Appendix we include the figures that represent the spatial distribution, kinematics and 
	excitation from the entire suite of measurements in the full sample.
	The combination of these plots in the figure set help to illustrate the variations throughout
		the ULIRGs.
	The emission line profiles for the measured transitions are presented in Appendix B.
	
\begin{figure}[h]
\centering
\includegraphics[keepaspectratio=true, width=\linewidth,trim = 0 0 0 0, clip=true,scale=1]{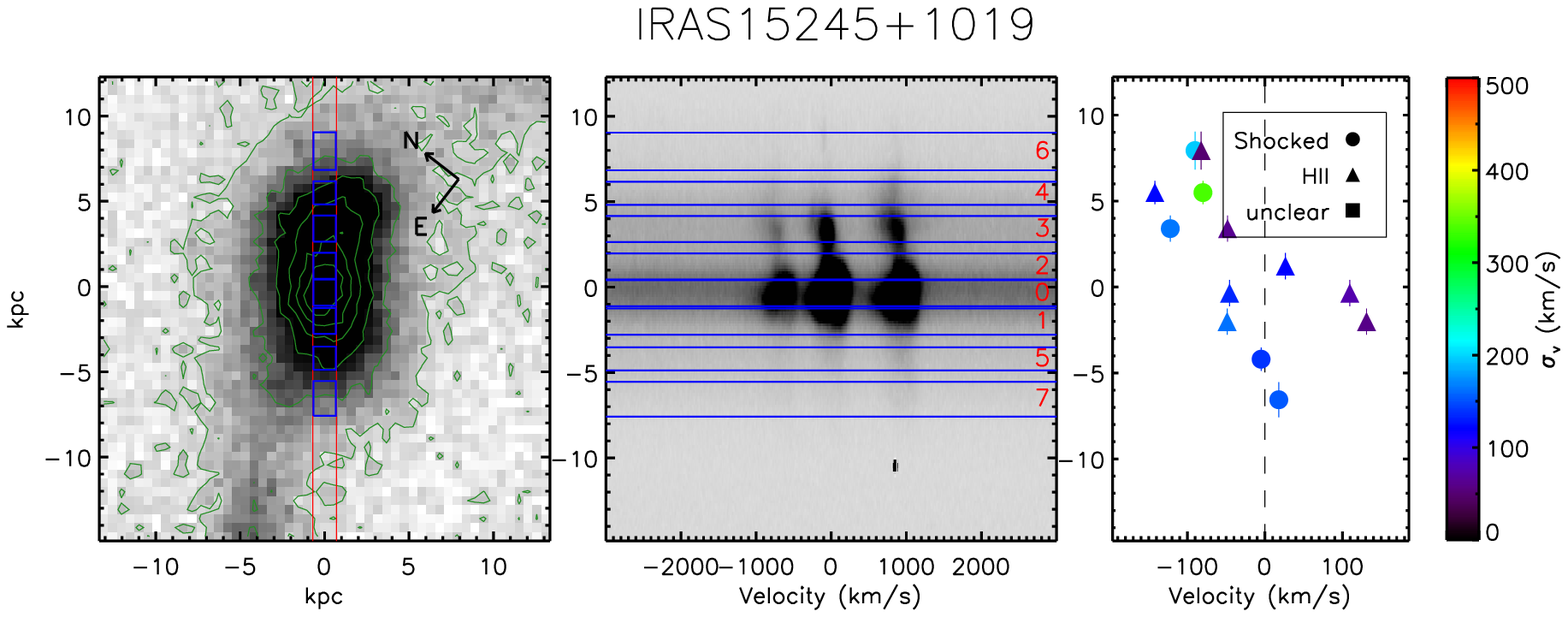}
\includegraphics[keepaspectratio=true, width=\linewidth,trim = 0 0 0 20, clip=true,scale=1]{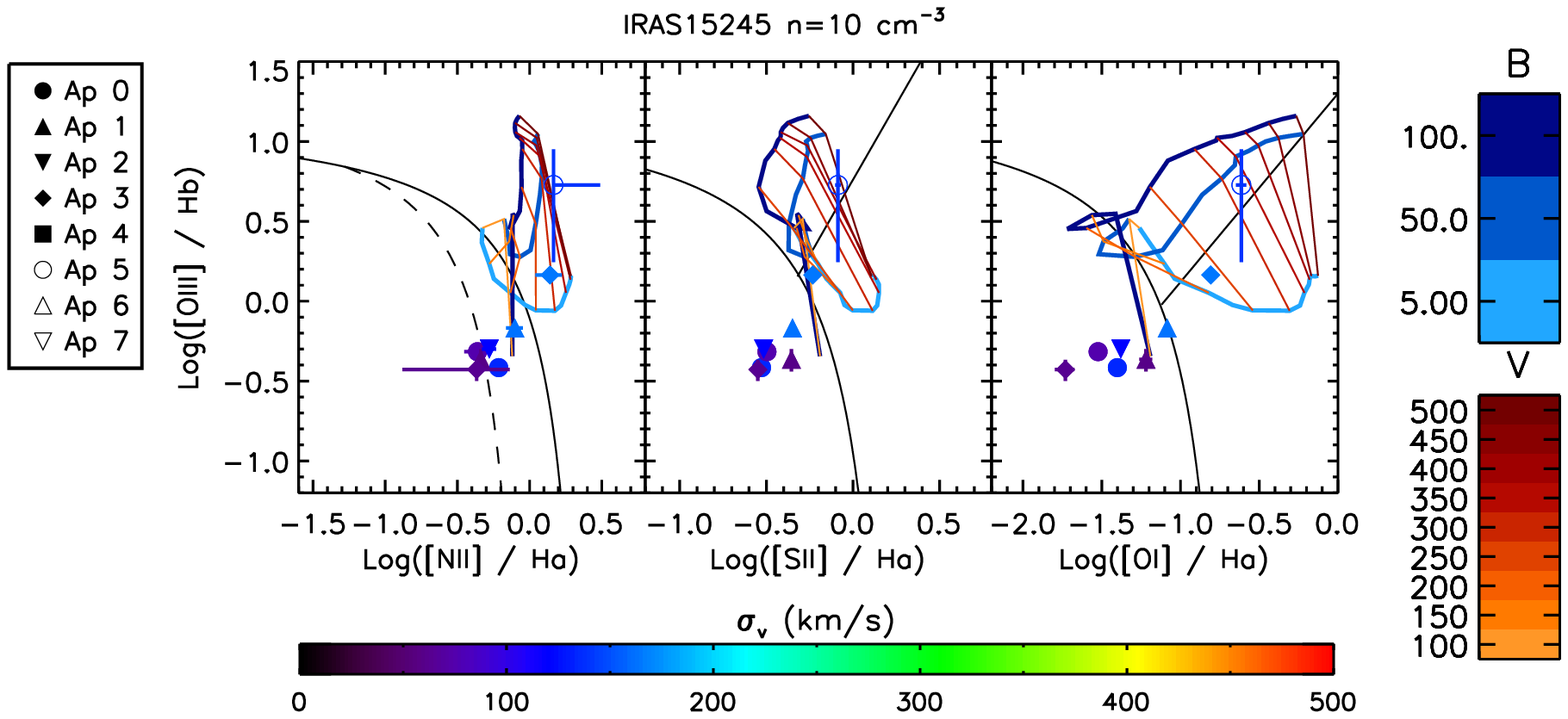}
\caption{\label{fig:i15245_bbc}
\scriptsize
\emph{Top Left:}
An $r$ band image from the Palomar 60 inch telescope. The green contours 
represent the median, and the 1, 2, 3, 4, and 5 $\sigma$ surface brightness levels in the image. 
The red lines 
show the spatial position of the 1\arcsec$\times$20\arcsec\ slit. The blue 
rectangles identify the positions of the selected apertures.
\emph{Top Middle:} 
The 2D spectrum of \Ha+[\ion{N}{2}] centered in velocity space on \Ha. The 
blue rectangles identify the pixels that contribute to the individual apertures. The redshift 
range of the sample allows each object to be sampled by several apertures along the slit.
\emph{Top Right:} 
The position-velocity diagram for all of the fit components.  The linewidth 
of the components is indicated by the color of the plot point. The plot symbol represents the 
excitation category for that individual spectral component.
\emph{Bottom Row:} Line ratio diagrams showing the line ratios for the emission components in 
these selected apertures. The black solid line represents the extreme star formation limit, above which 
the excitation mechanism in not expected to be from star formation \citep{Kewley:2006p38}. The 
dashed line in the [\ion{O}{3}]/\Hb\ vs [\ion{N}{2}]/\Ha\ represents the empirical star formation 
limit \citep{Kauffmann:2003p33}. 
Shock grids \citep{Allen:2008p695} are presented for a solar 
abundance with (blue and orange lines) with 
$n_e = 10~{\rm cm}^{-3}$, with a velocity range of 100 to 500 \kms, and magnetic field strength 
5 to 100 $\mu$G expected for star forming galaxies. Overplotted on these diagnostics are 
flux ratios from the individual components in the apertures identified in the above plots. 
The full sample is included in the online version of the article.
}
\end{figure}

\begin{figure}[h]
\centering
\epsfig{file=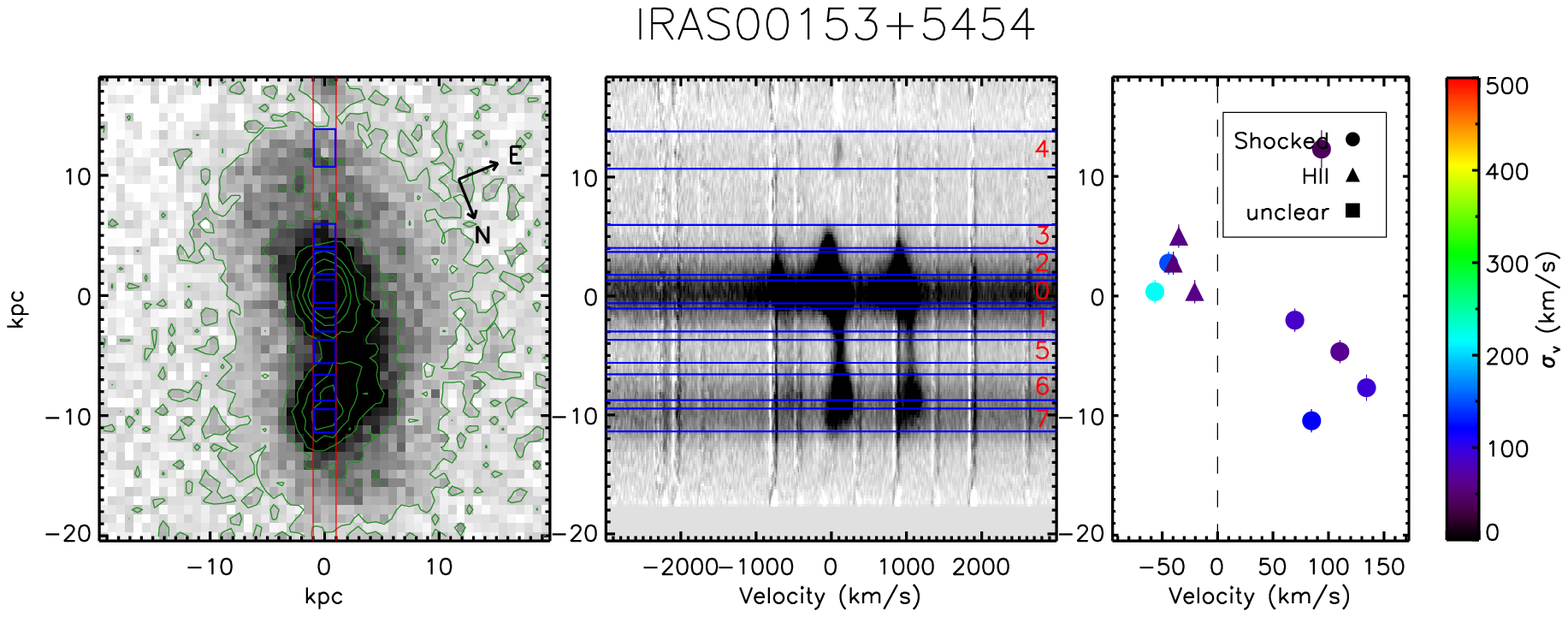, keepaspectratio=true, width=\linewidth}
\epsfig{file=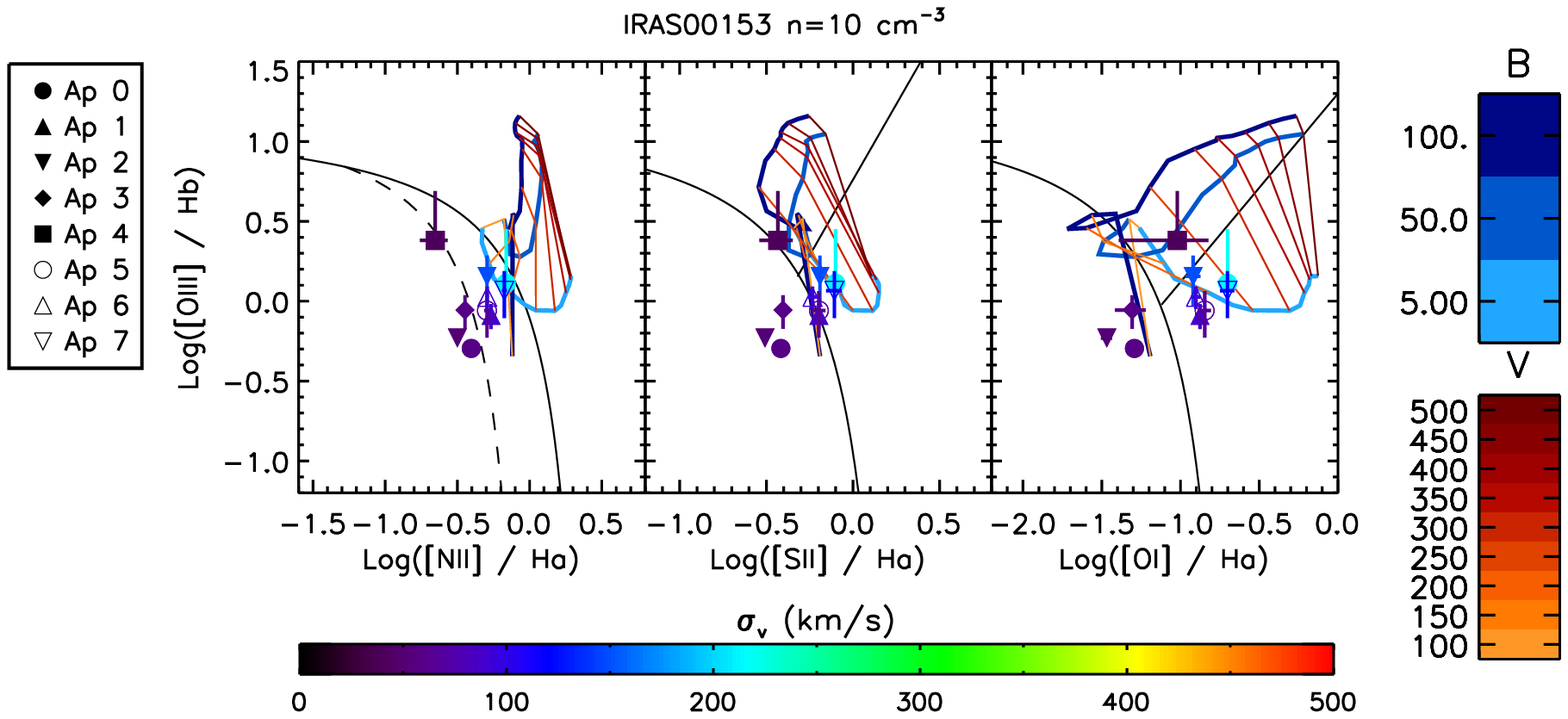, keepaspectratio=true, width=\linewidth}
\caption{\label{fig:i00153_bbc}
}
\end{figure}
\clearpage

\begin{figure}[h]
\centering
\epsfig{file=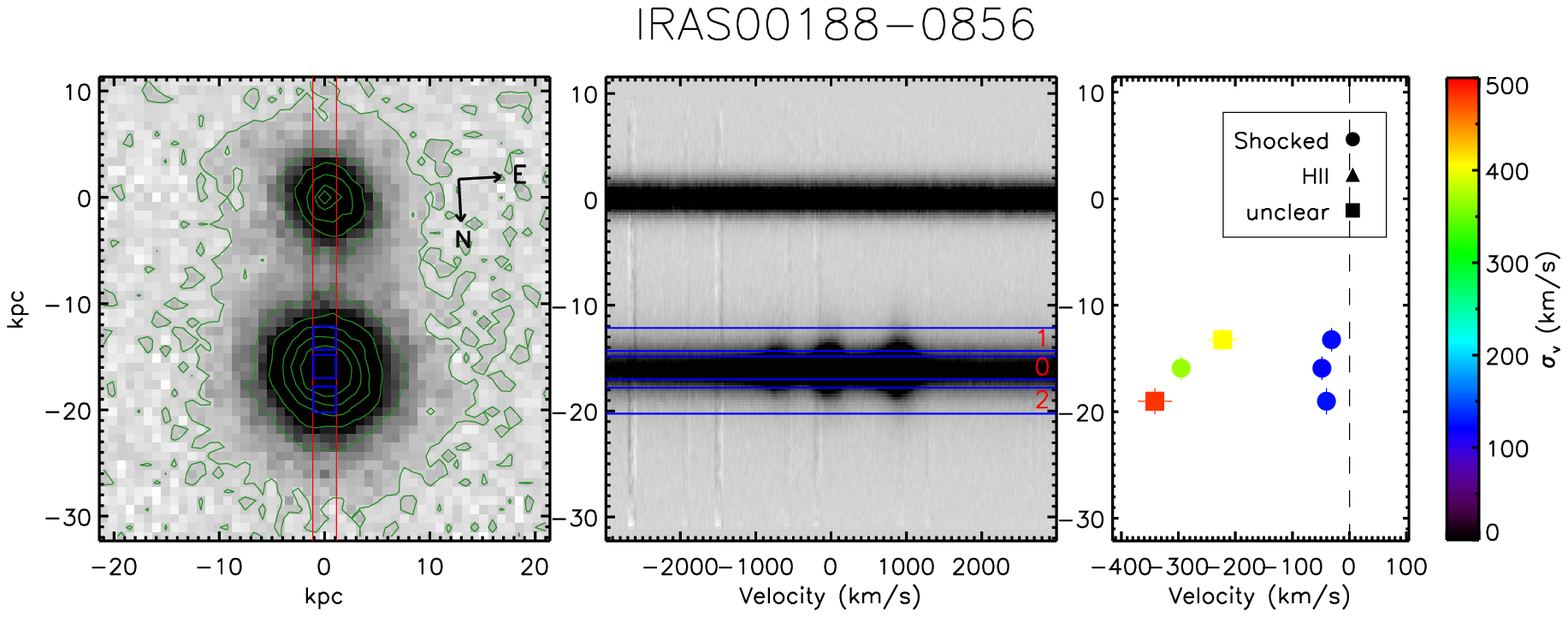, keepaspectratio=true, width=\linewidth}
\epsfig{file=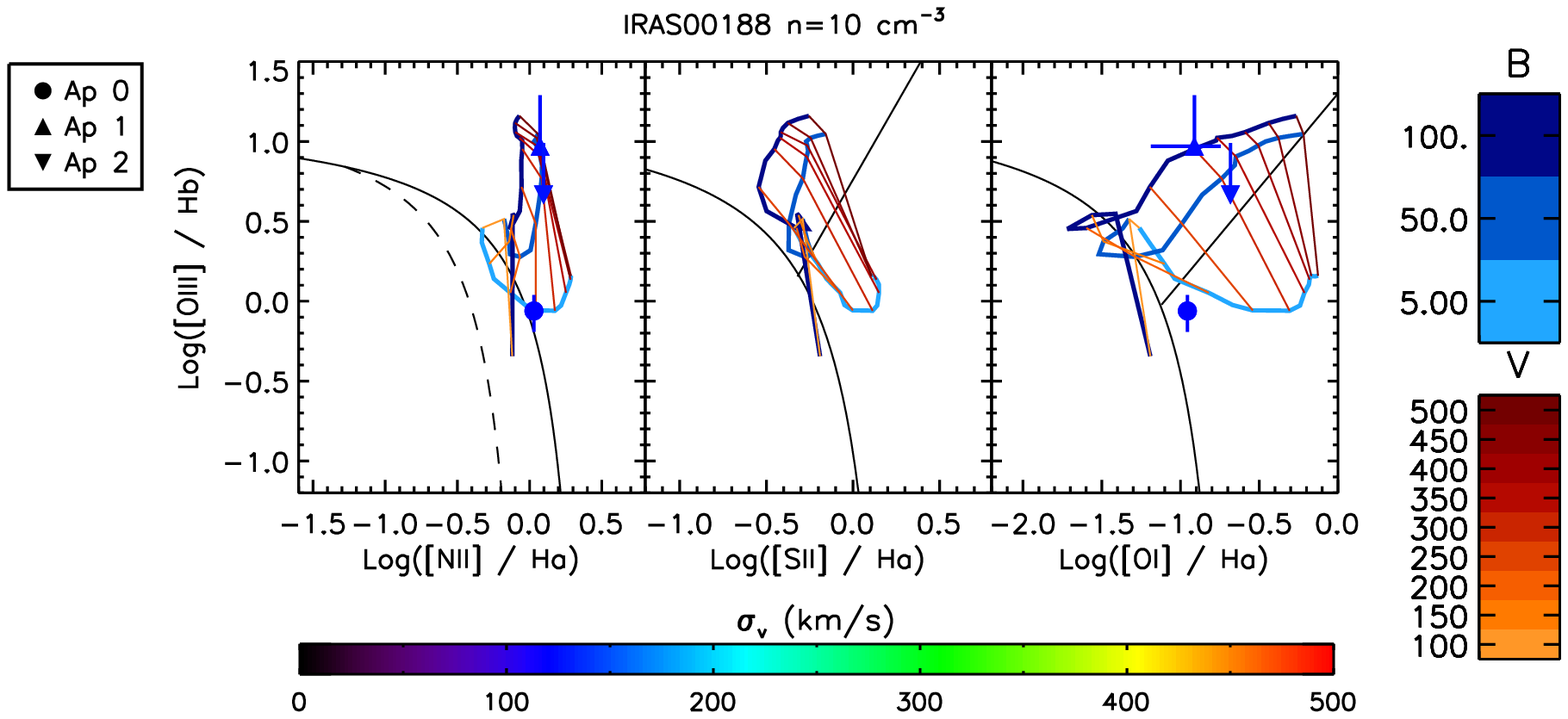, keepaspectratio=true, width=\linewidth}
\caption{\label{fig:i00188_bbc}
}
\end{figure}
\clearpage

\begin{figure}[h]
\centering
\epsfig{file=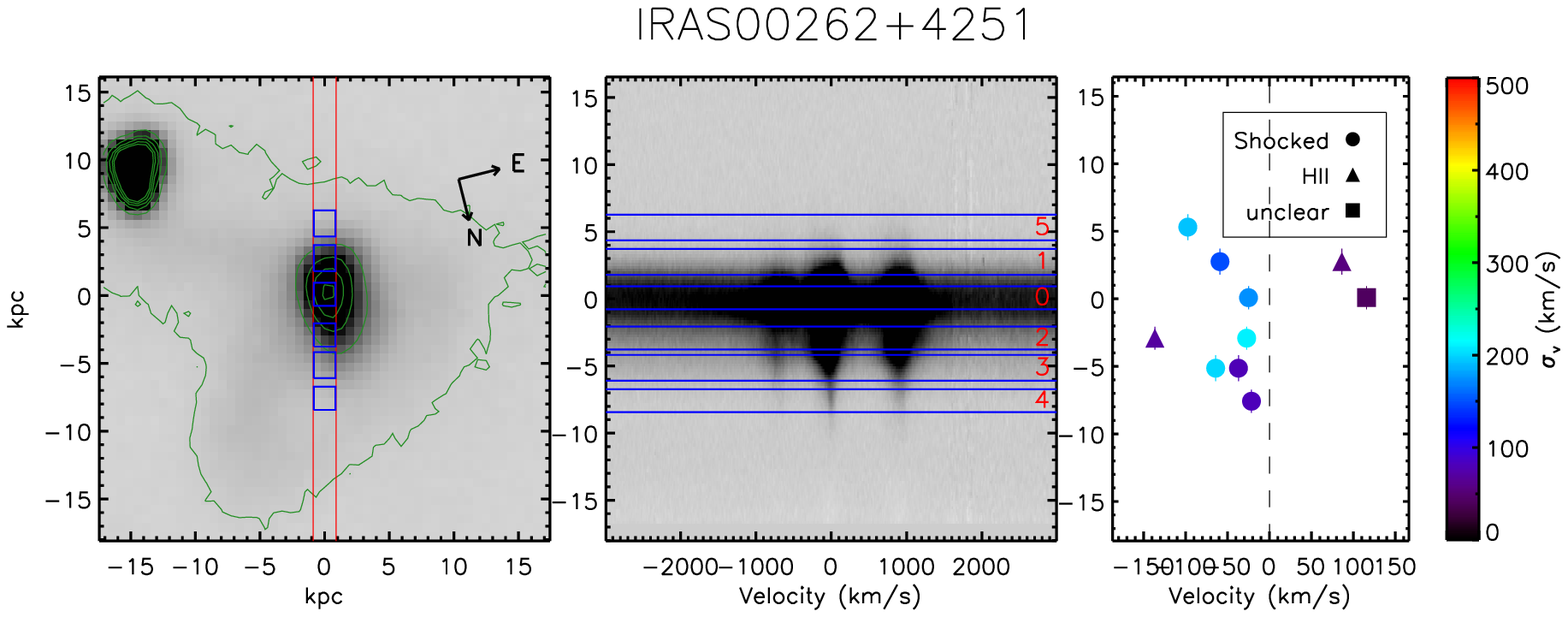, keepaspectratio=true, width=\linewidth}
\epsfig{file=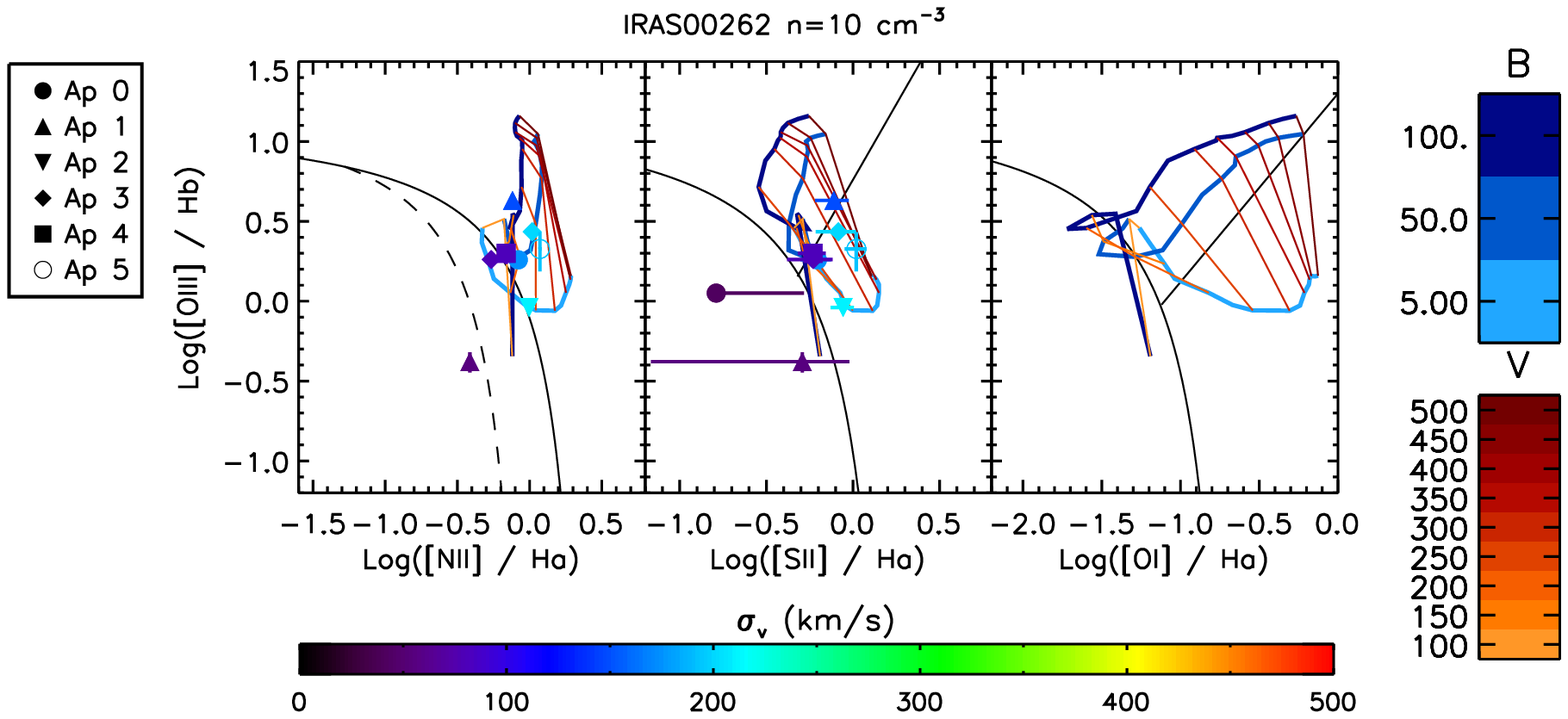, keepaspectratio=true, width=\linewidth}
\caption{\label{fig:i00262_bbc}
}
\end{figure}
\clearpage

\begin{figure}[h]
\centering
\epsfig{file=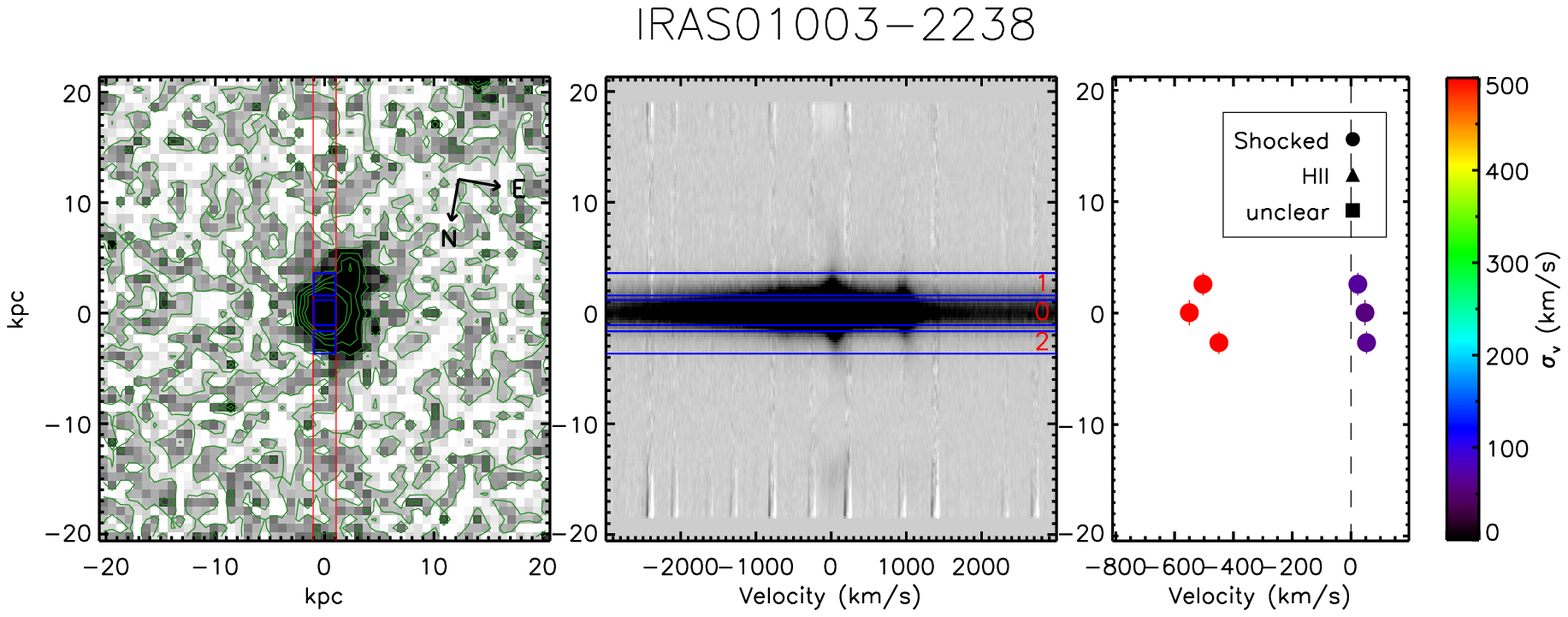, keepaspectratio=true, width=\linewidth}
\epsfig{file=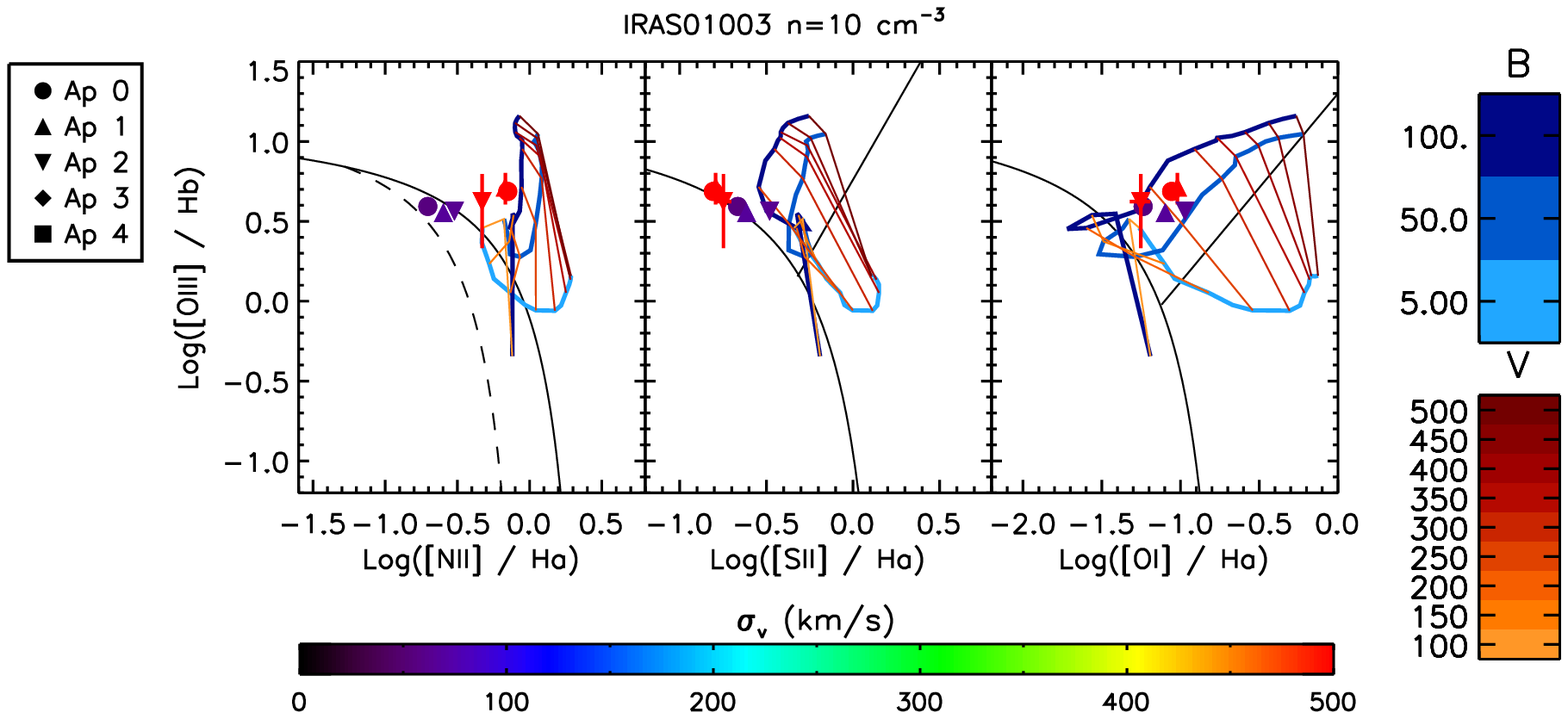, keepaspectratio=true, width=\linewidth}
\caption{\label{fig:i01003_bbc}
}
\end{figure}
\clearpage

\begin{figure}[h]
\centering
\epsfig{file=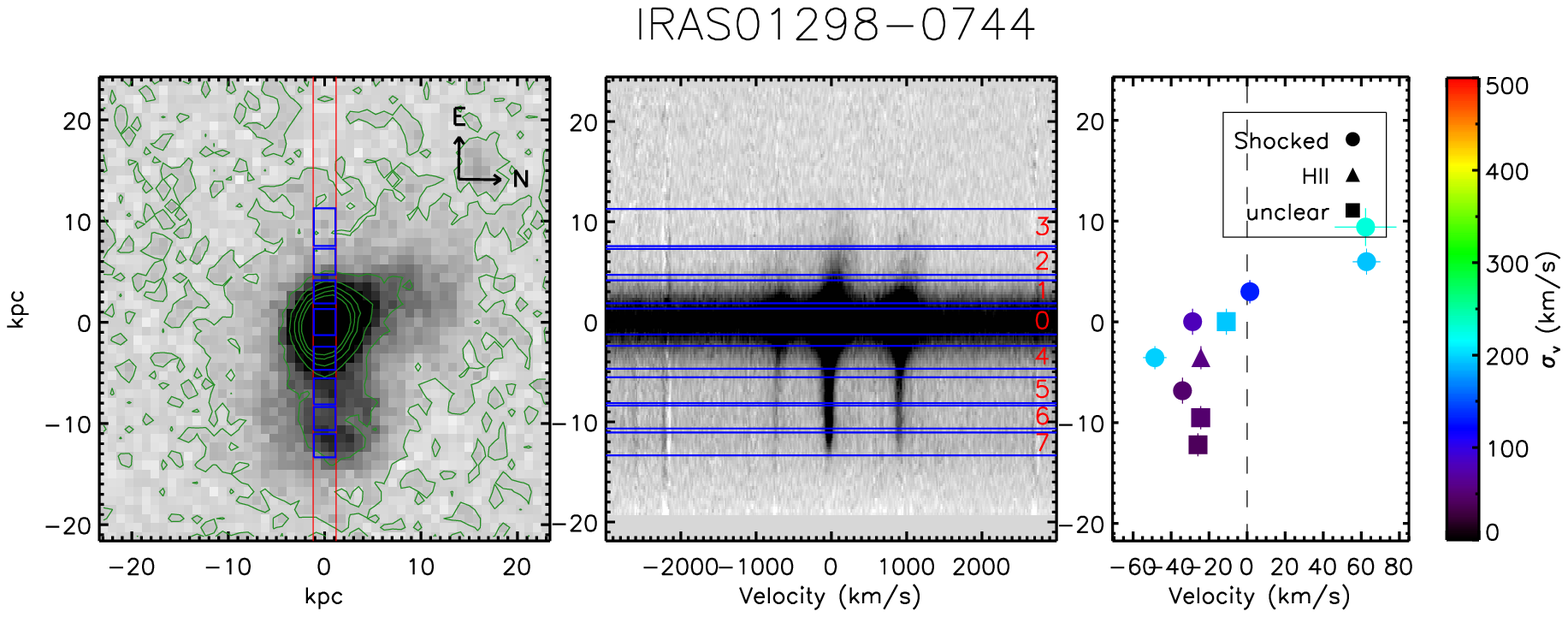, keepaspectratio=true, width=\linewidth}
\epsfig{file=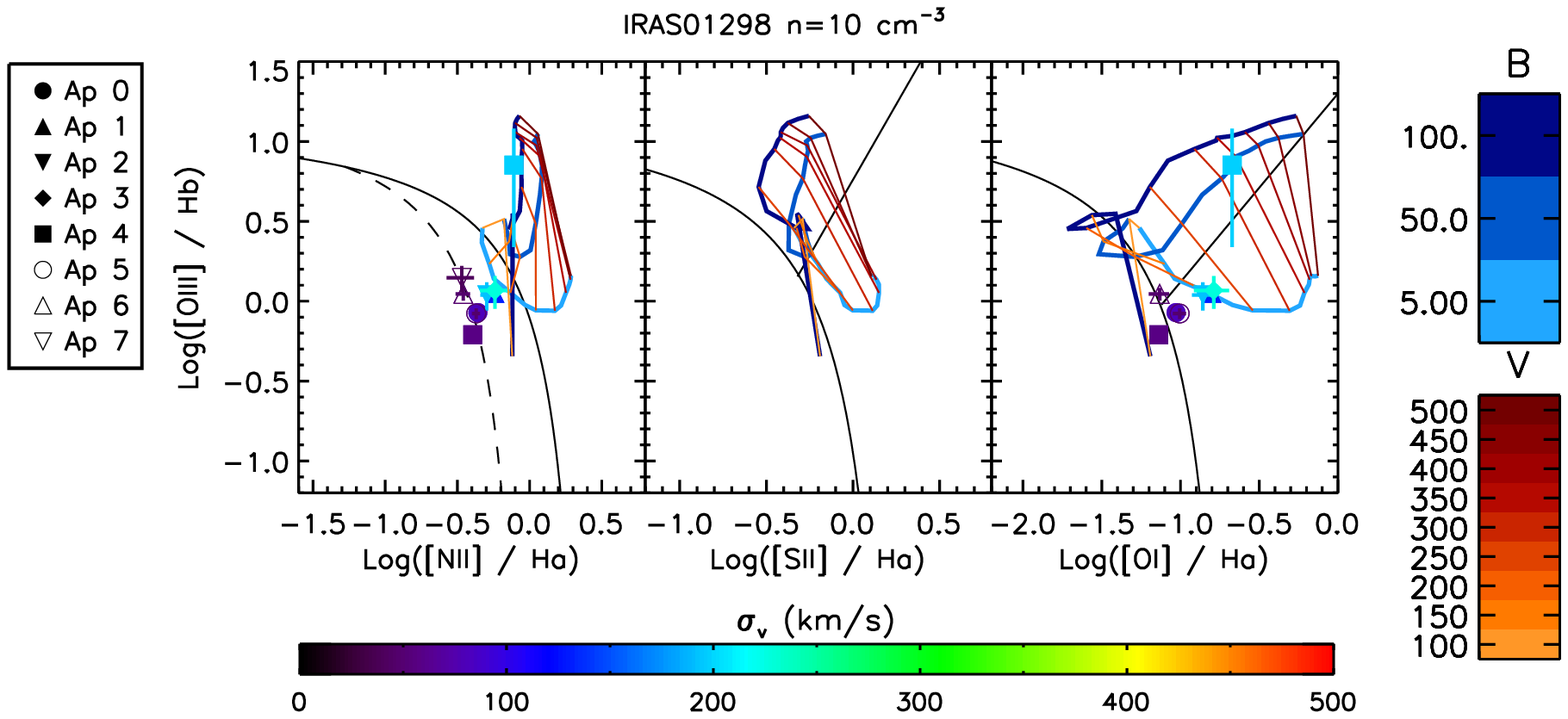, keepaspectratio=true, width=\linewidth}
\caption{\label{fig:i01298_bbc}
}
\end{figure}
\clearpage

\begin{figure}[h]
\centering
\epsfig{file=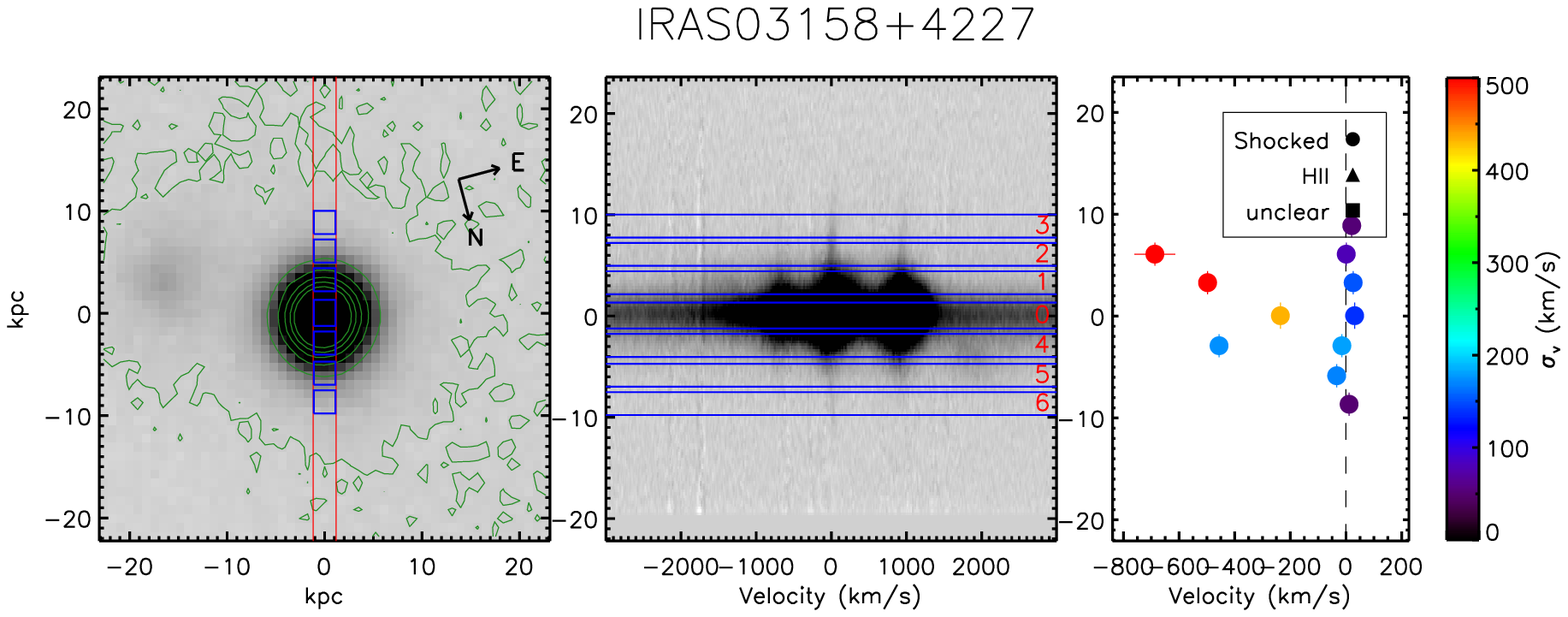, keepaspectratio=true, width=\linewidth}
\epsfig{file=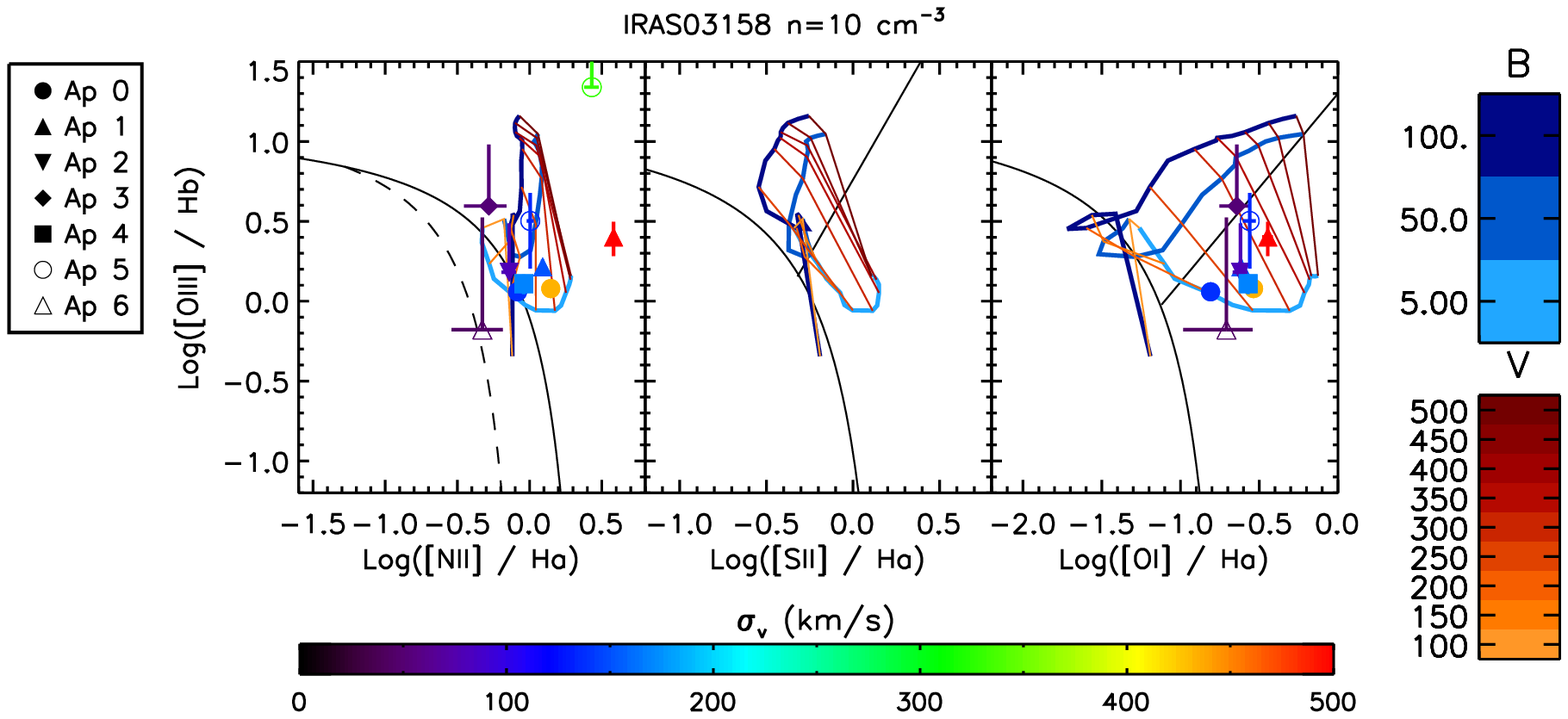, keepaspectratio=true, width=\linewidth}
\caption{\label{fig:i03158_bbc}
}
\end{figure}
\clearpage

\begin{figure}[h]
\centering
\epsfig{file=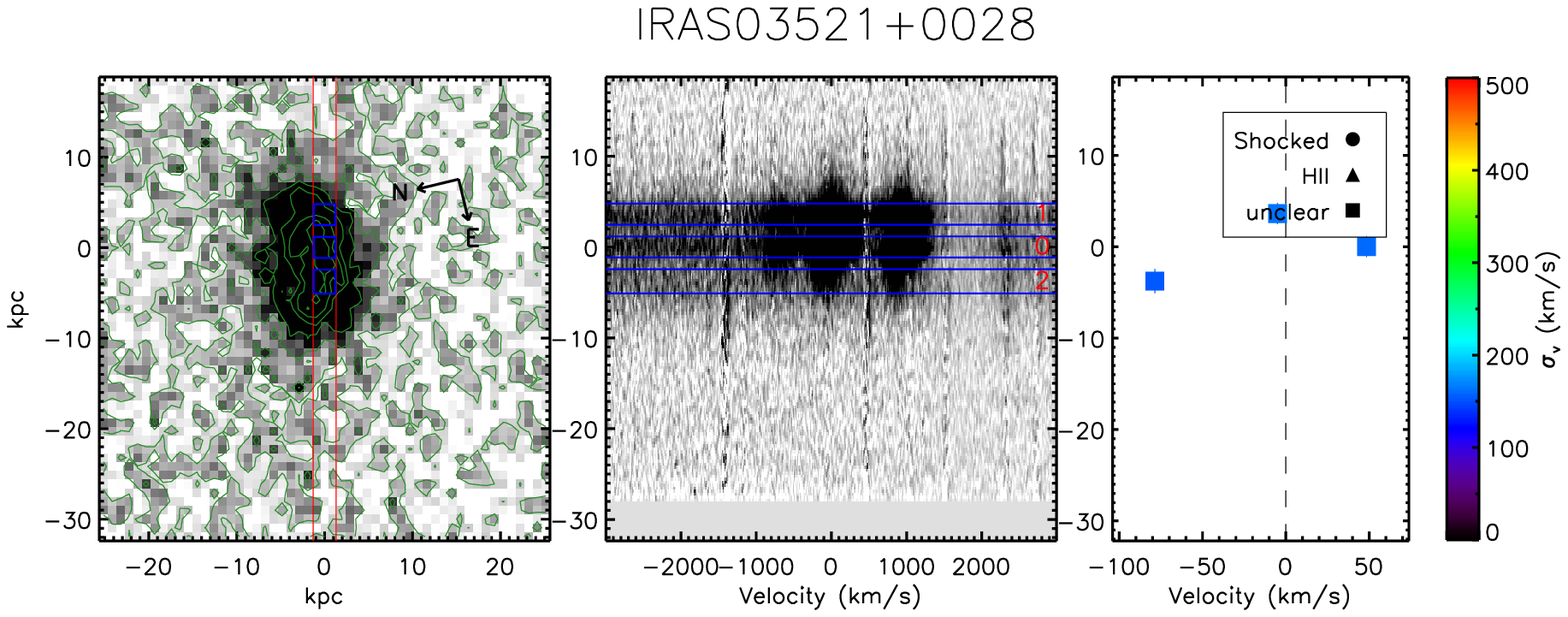, keepaspectratio=true, width=\linewidth}
\epsfig{file=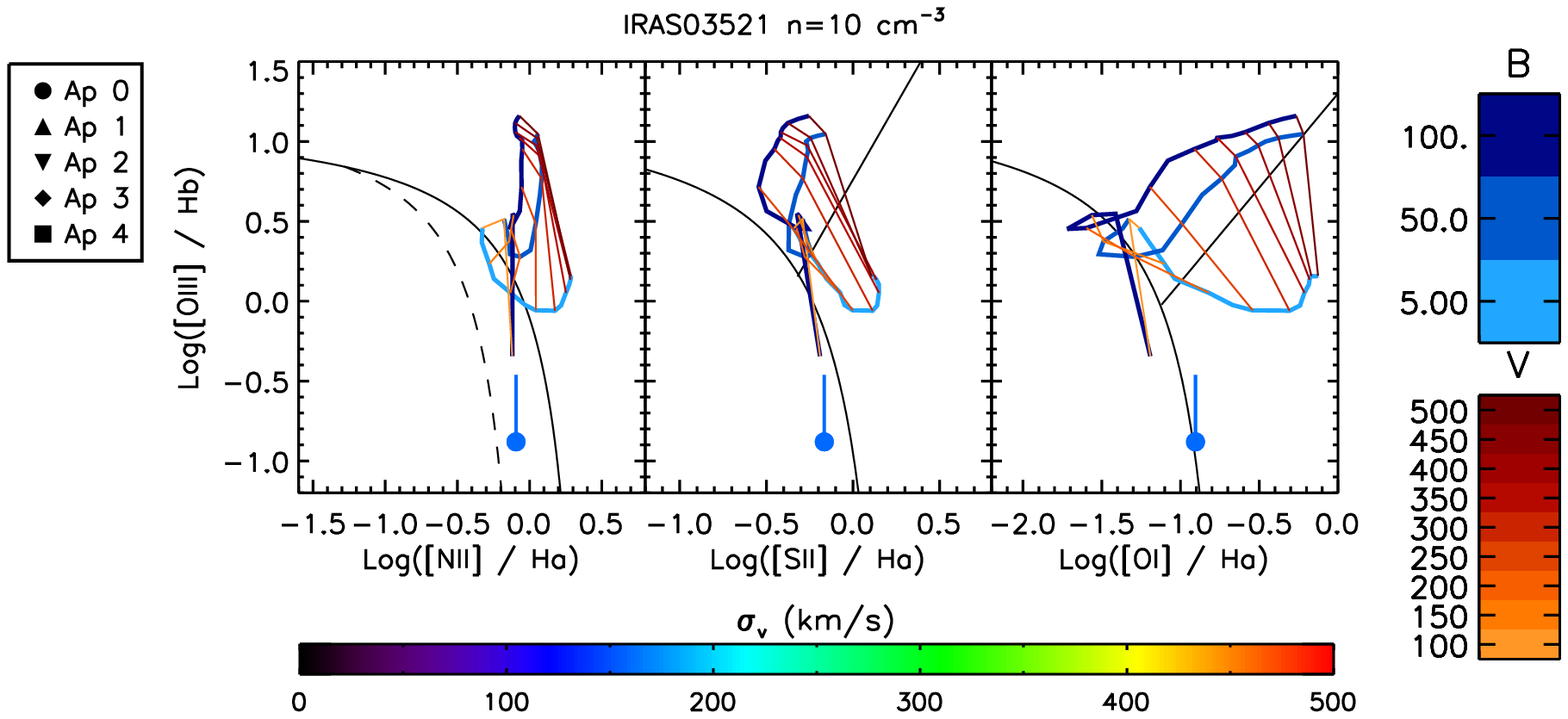, keepaspectratio=true, width=\linewidth}
\caption{\label{fig:i03521_bbc}
}
\end{figure}
\clearpage

\begin{figure}[h]
\centering
\epsfig{file=i05246pos3.eps, keepaspectratio=true, width=\linewidth}
\epsfig{file=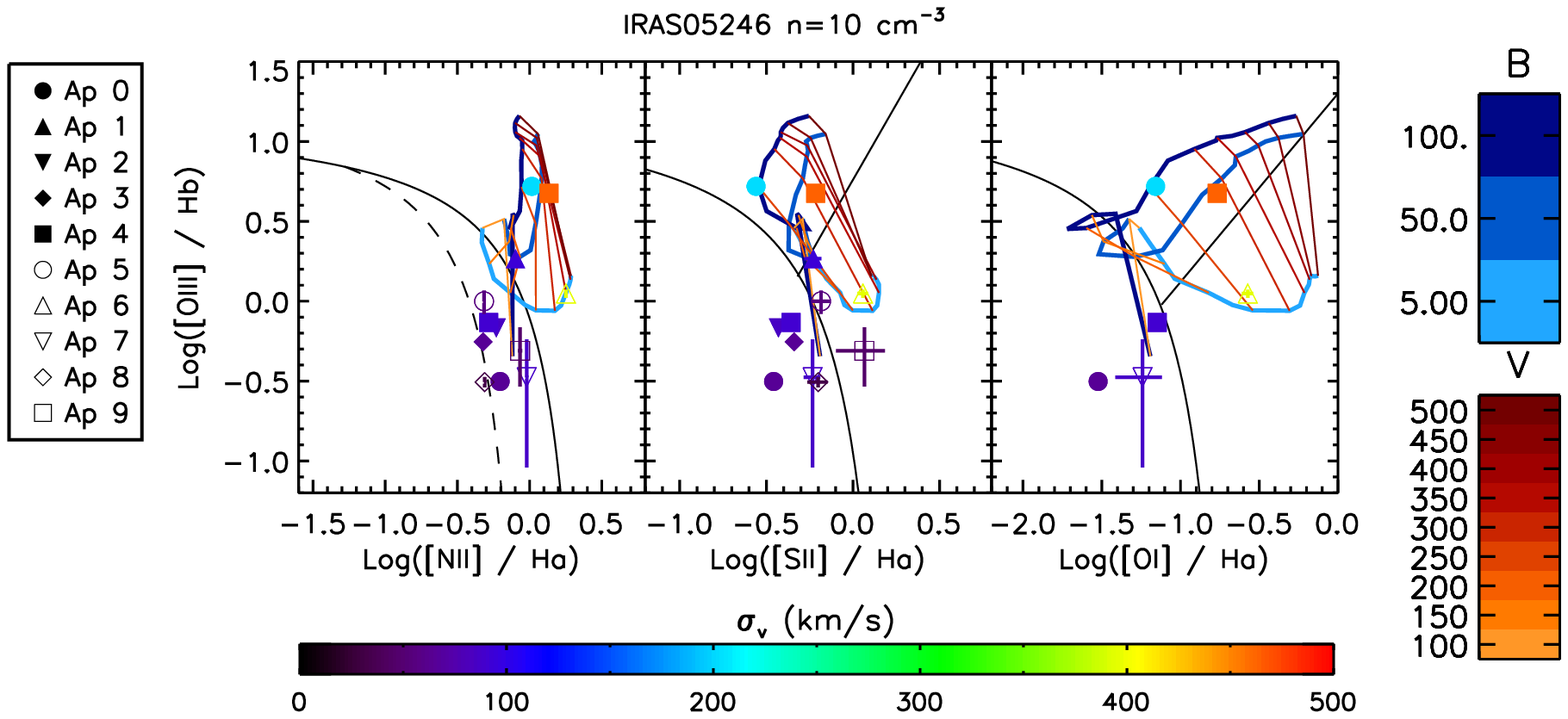, keepaspectratio=true, width=\linewidth}
\caption{\label{fig:i05246_bbc}
}
\end{figure}
\clearpage

\begin{figure}[h]
\centering
\epsfig{file=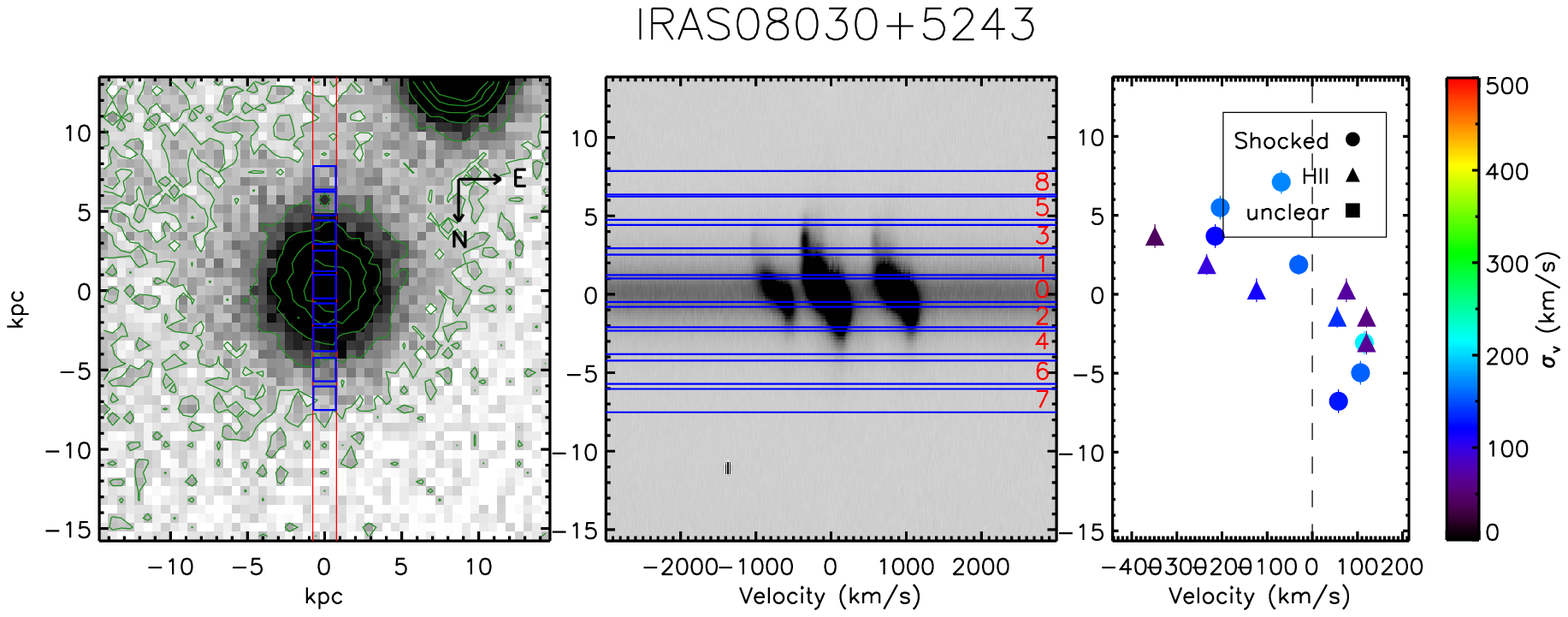, keepaspectratio=true, width=\linewidth}
\epsfig{file=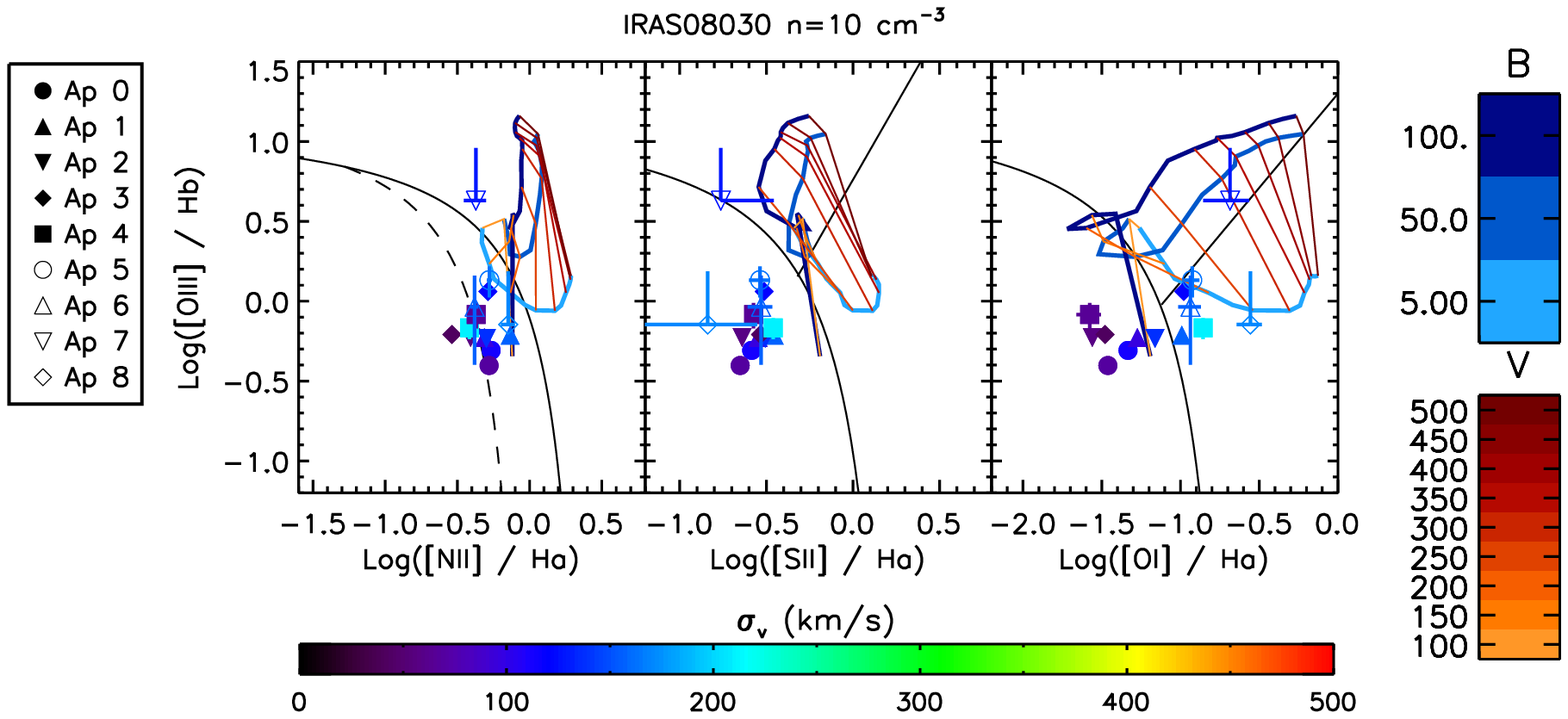, keepaspectratio=true, width=\linewidth}
\caption{\label{fig:i08030_bbc}
}
\end{figure}
\clearpage

\begin{figure}[h]
\centering
\epsfig{file=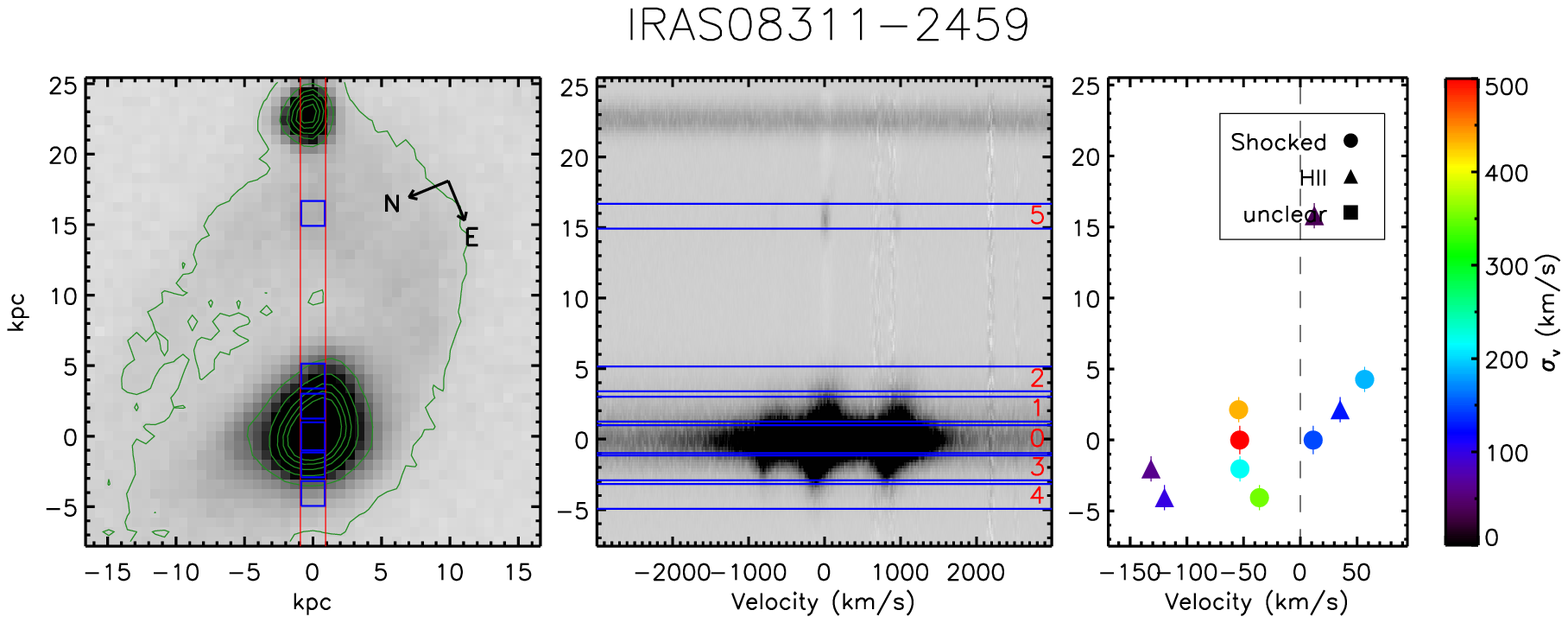, keepaspectratio=true, width=\linewidth}
\epsfig{file=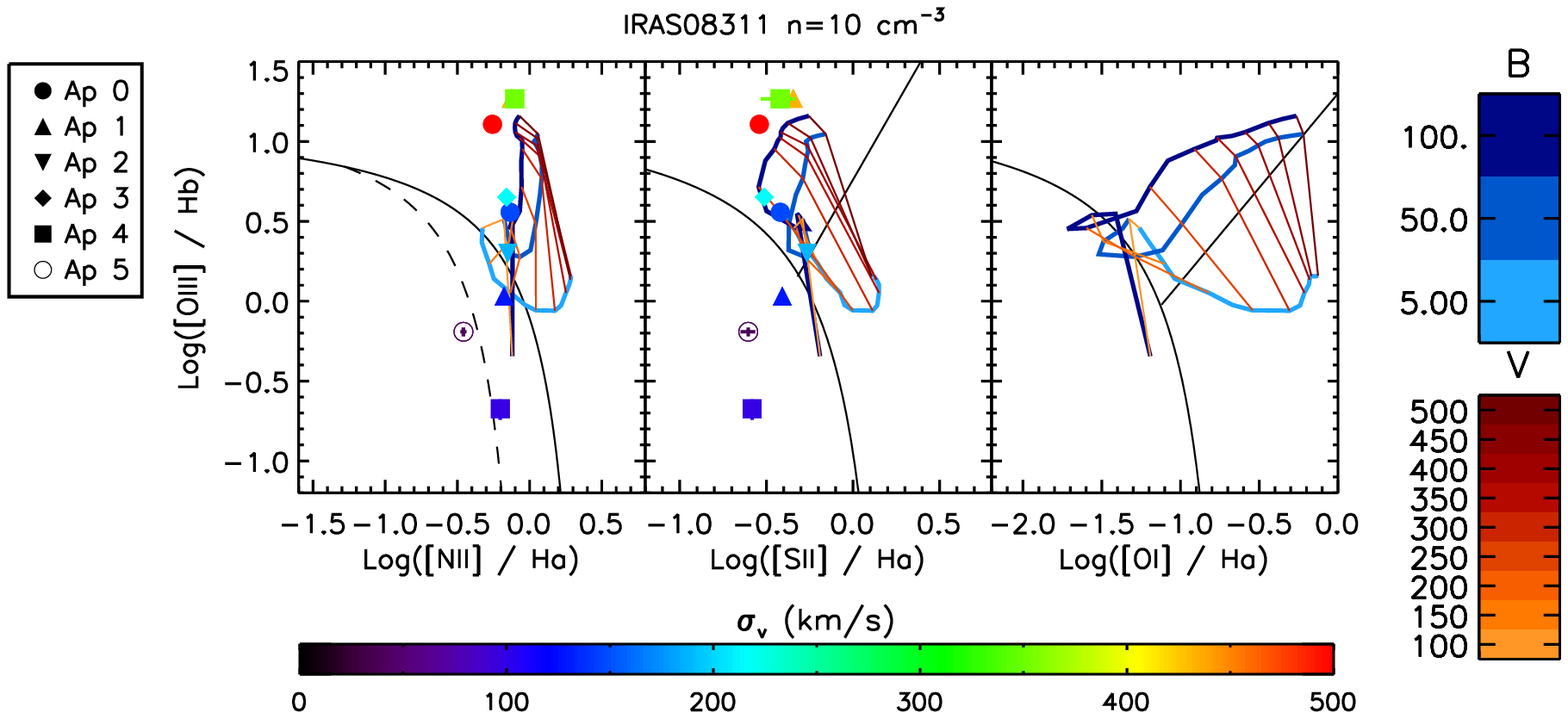, keepaspectratio=true, width=\linewidth}
\caption{\label{fig:i08311_bbc}
}
\end{figure}
\clearpage

\begin{figure}[h]
\centering
\epsfig{file=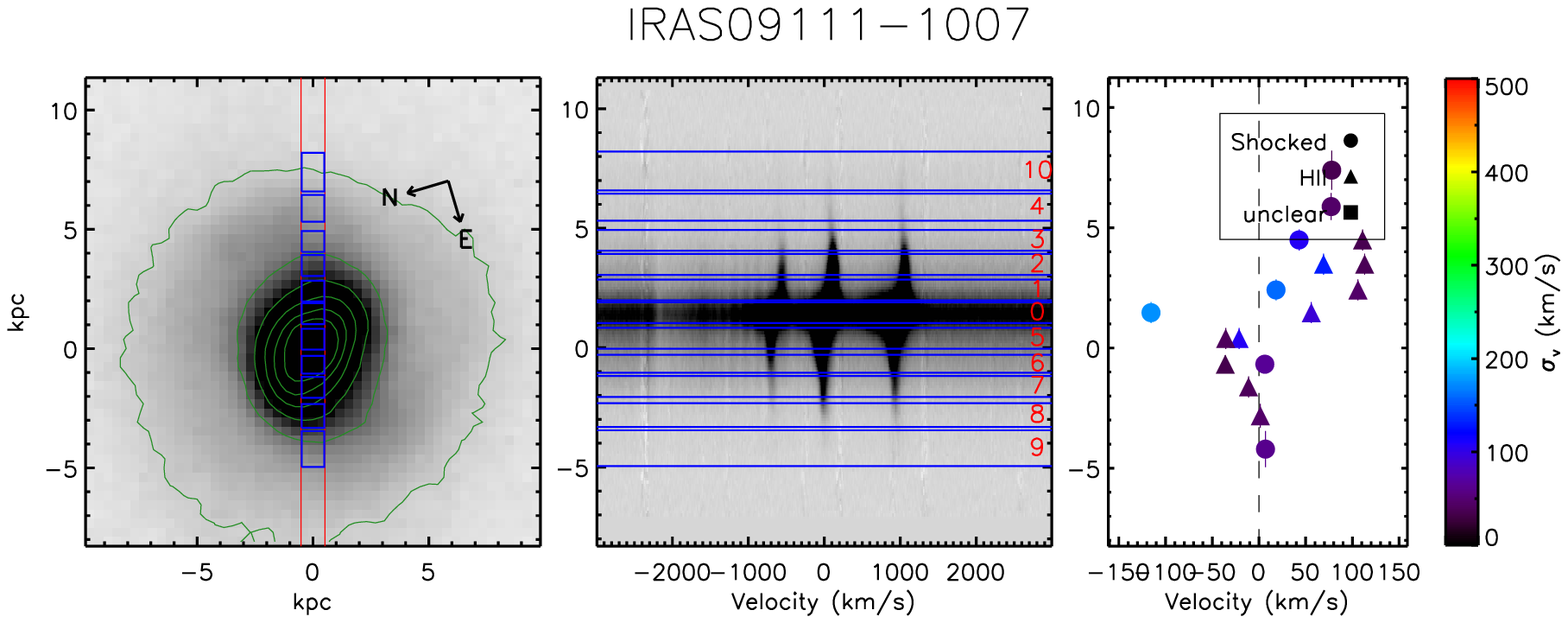, keepaspectratio=true, width=\linewidth}
\epsfig{file=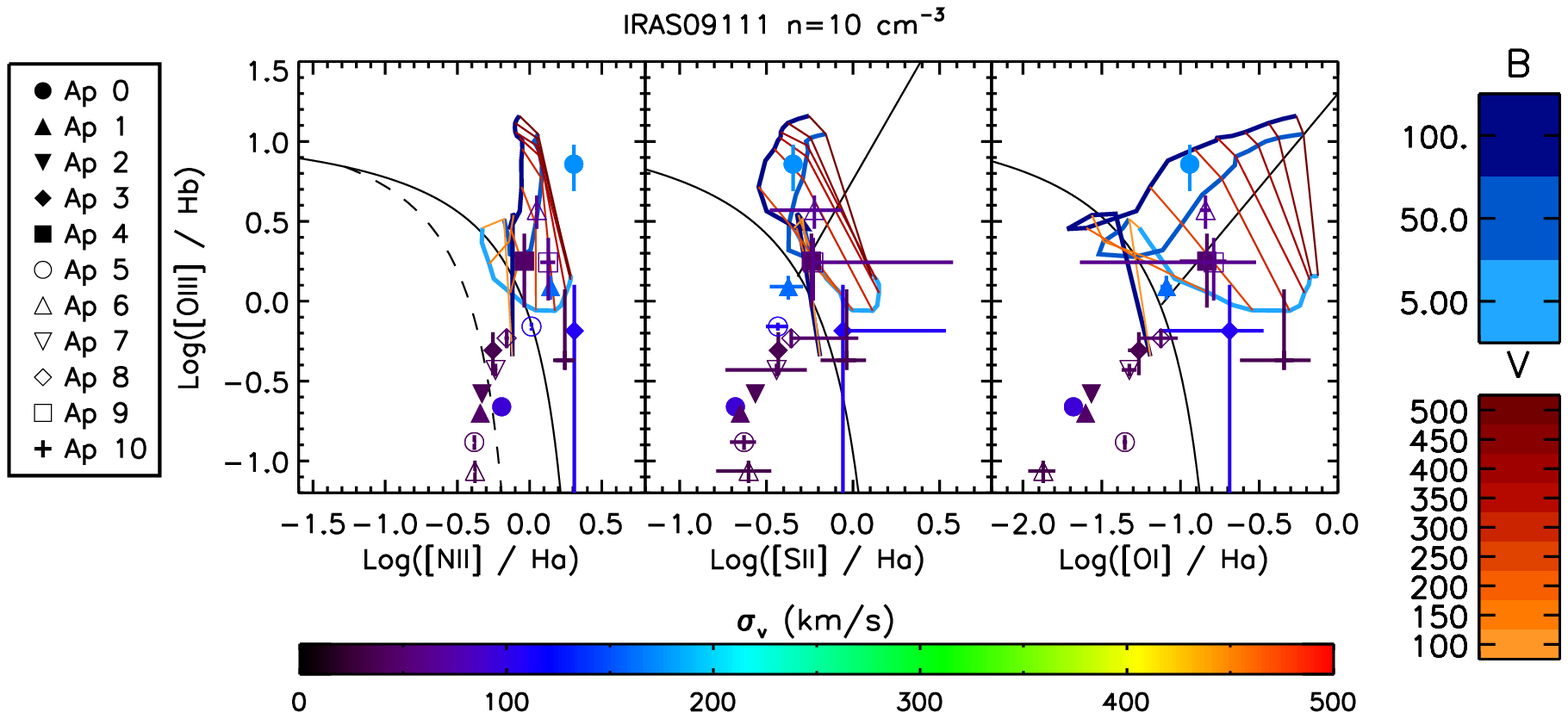, keepaspectratio=true, width=\linewidth}
\caption{\label{fig:i09111_bbc}
}
\end{figure}
\clearpage

\begin{figure}[h]
\centering
\epsfig{file=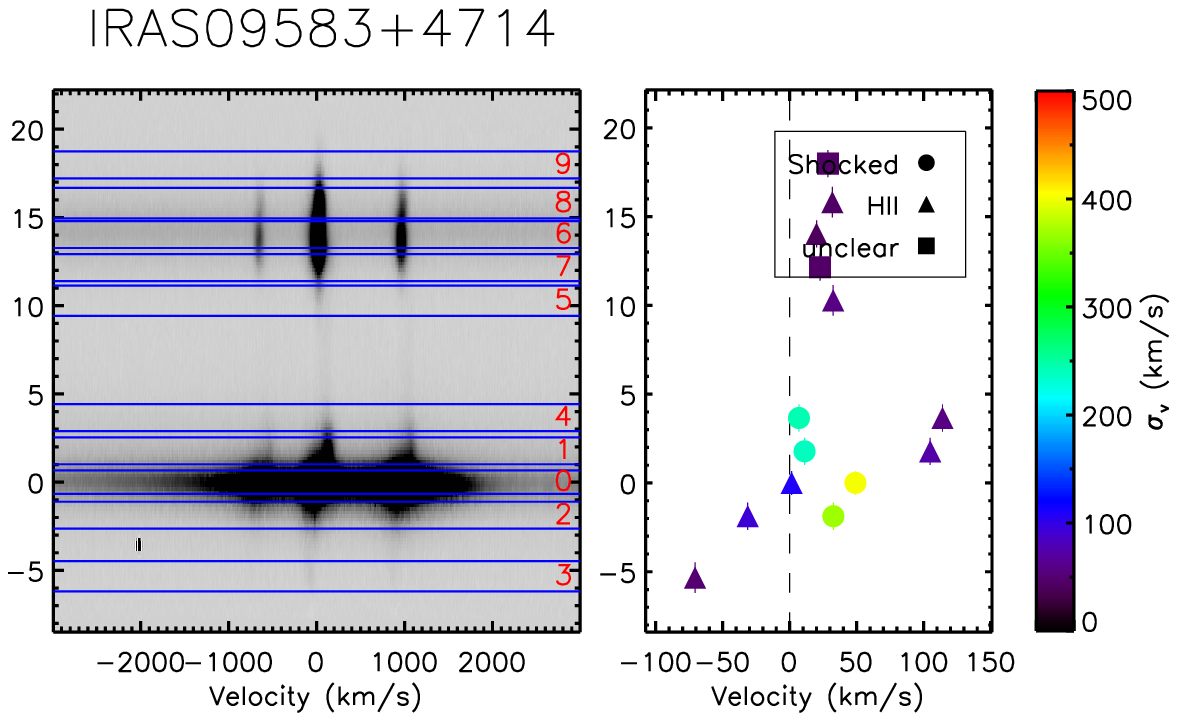, keepaspectratio=true, width=\linewidth}
\epsfig{file=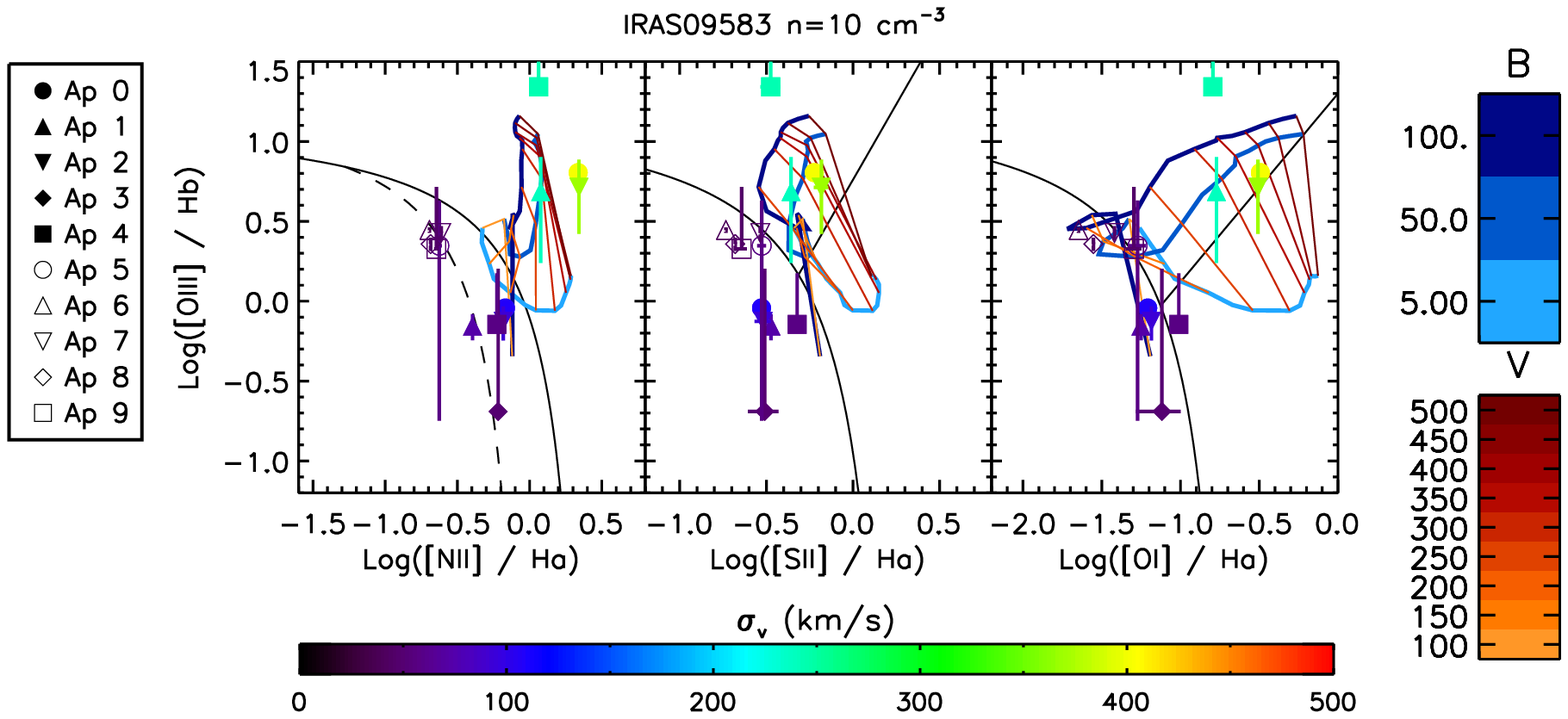, keepaspectratio=true, width=\linewidth}
\caption{\label{fig:i09583_bbc}
}
\end{figure}
\clearpage

\begin{figure}[h]
\centering
\epsfig{file=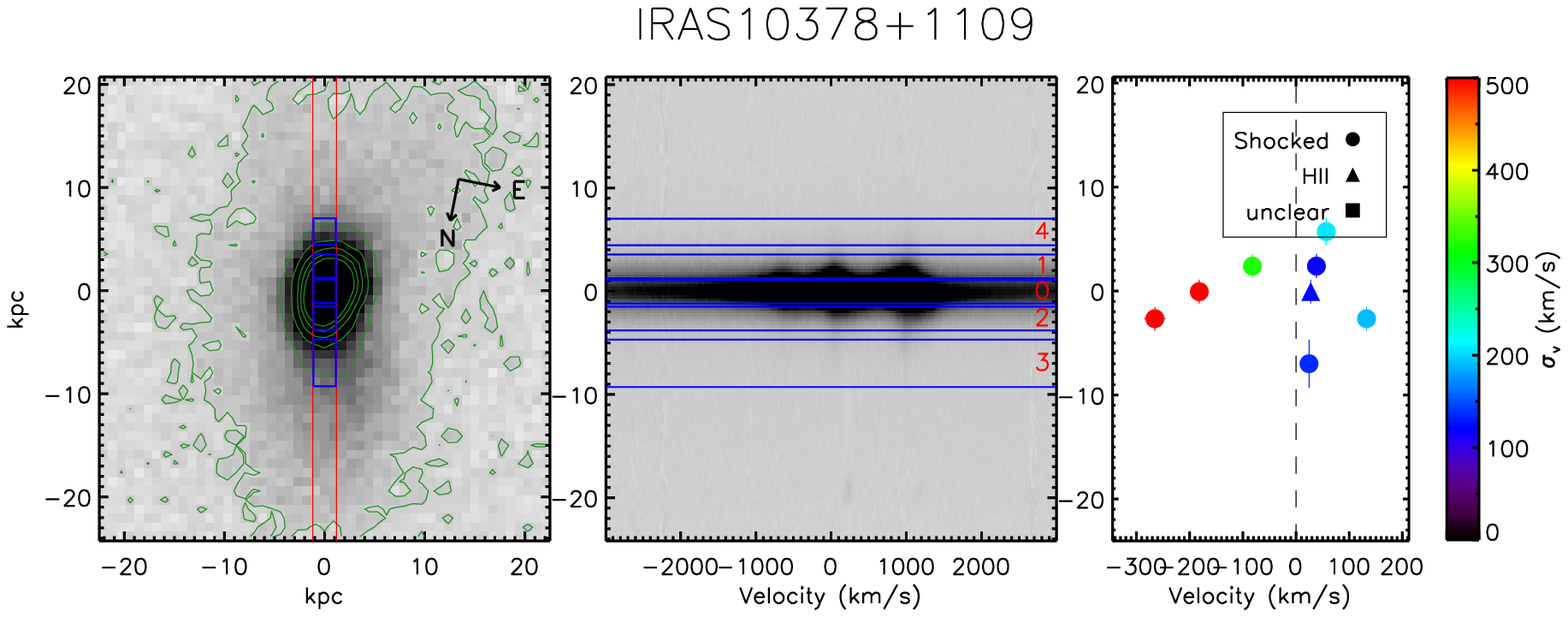, keepaspectratio=true, width=\linewidth}
\epsfig{file=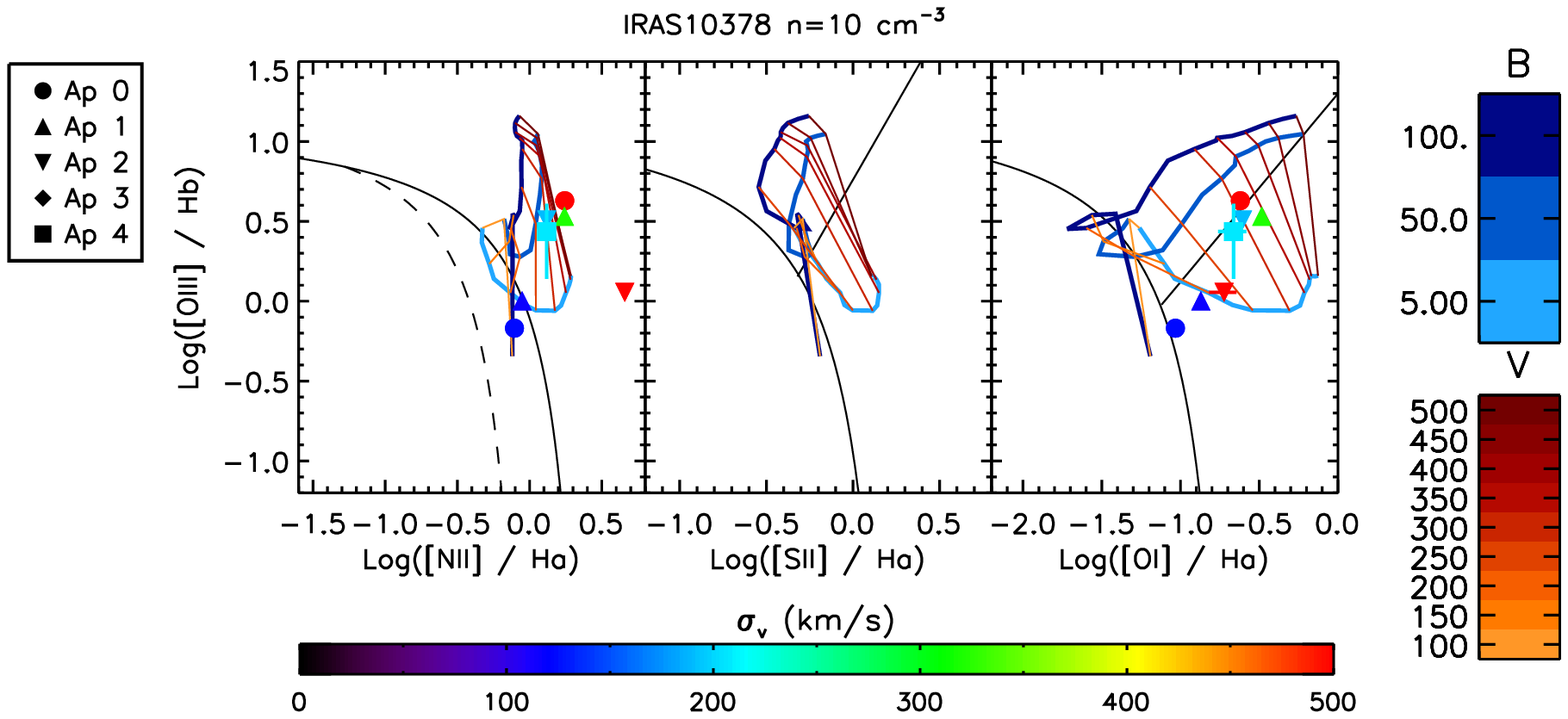, keepaspectratio=true, width=\linewidth}
\caption{\label{fig:i10378_bbc}
}
\end{figure}
\clearpage

\begin{figure}[h]
\centering
\epsfig{file=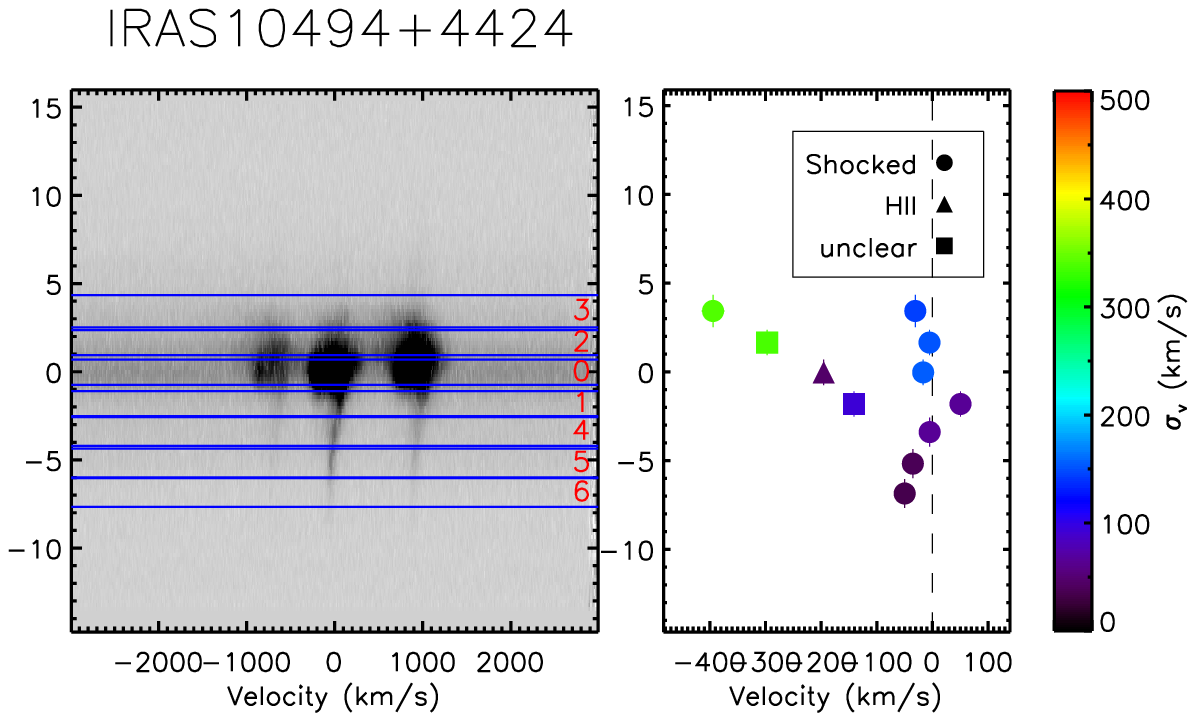, keepaspectratio=true, width=\linewidth}
\epsfig{file=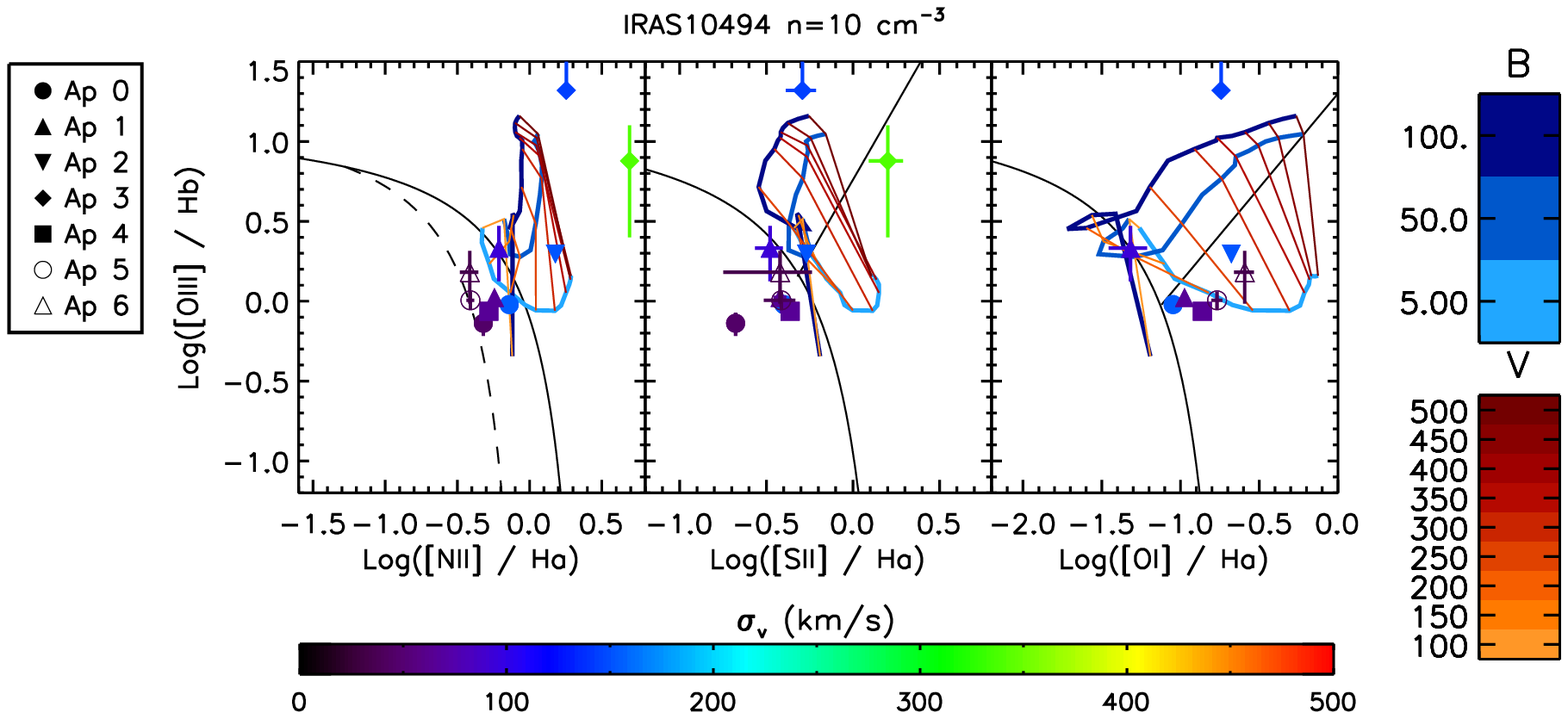, keepaspectratio=true, width=\linewidth}
\caption{\label{fig:i10494_bbc}
}
\end{figure}
\clearpage

\begin{figure}[h]
\centering
\epsfig{file=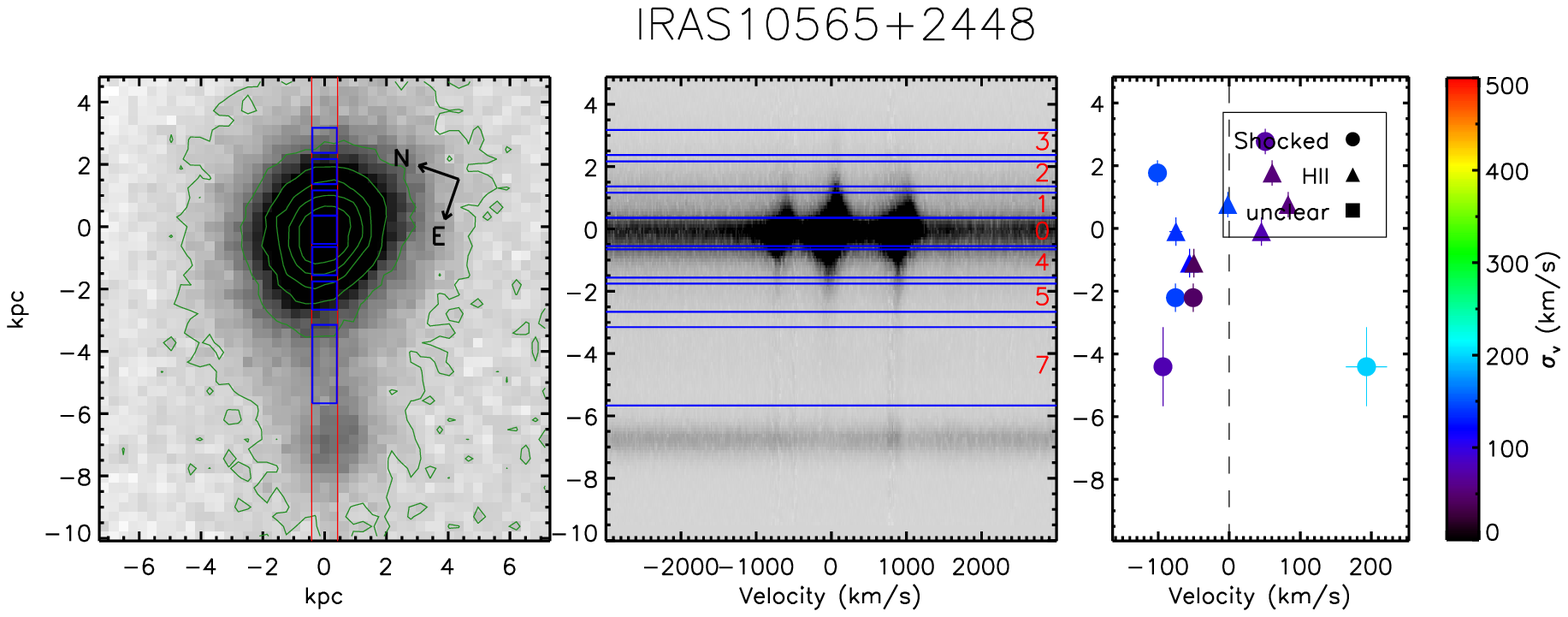, keepaspectratio=true, width=\linewidth}
\epsfig{file=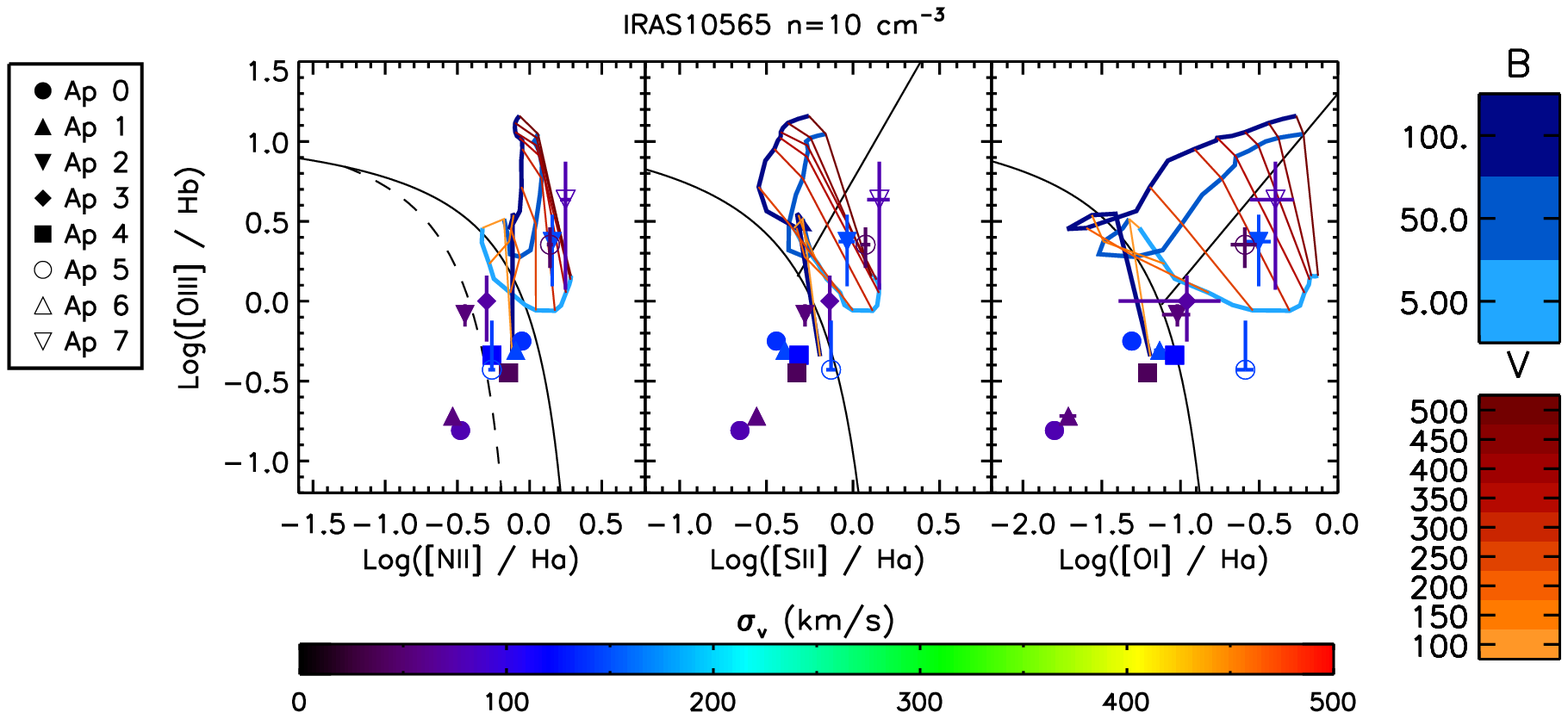, keepaspectratio=true, width=\linewidth}
\caption{\label{fig:i10565_bbc}
}
\end{figure}
\clearpage

\begin{figure}[h]
\centering
\epsfig{file=i11095pos3.eps, keepaspectratio=true, width=\linewidth}
\epsfig{file=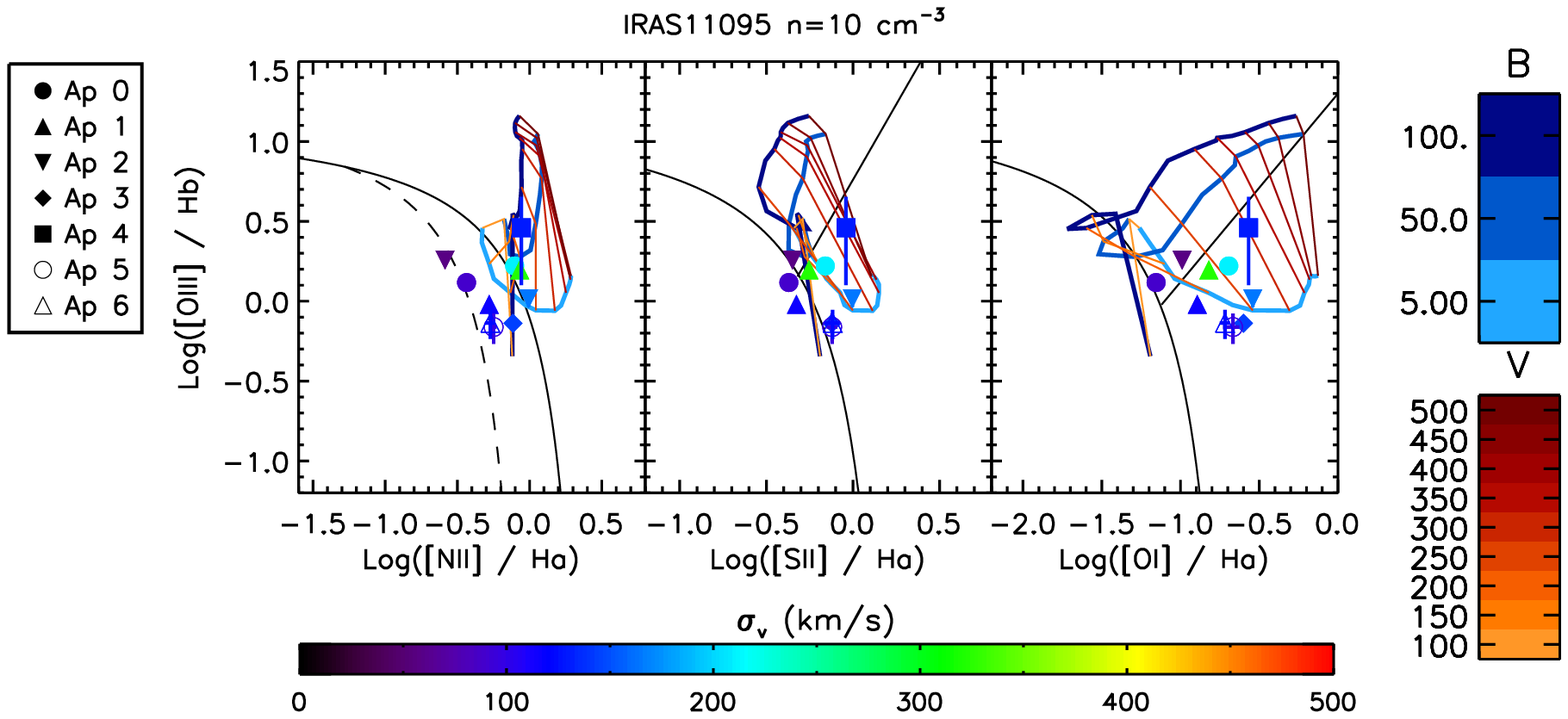, keepaspectratio=true, width=\linewidth}
\caption{\label{fig:i11095_bbc}
}
\end{figure}
\clearpage

\begin{figure}[h]
\centering
\epsfig{file=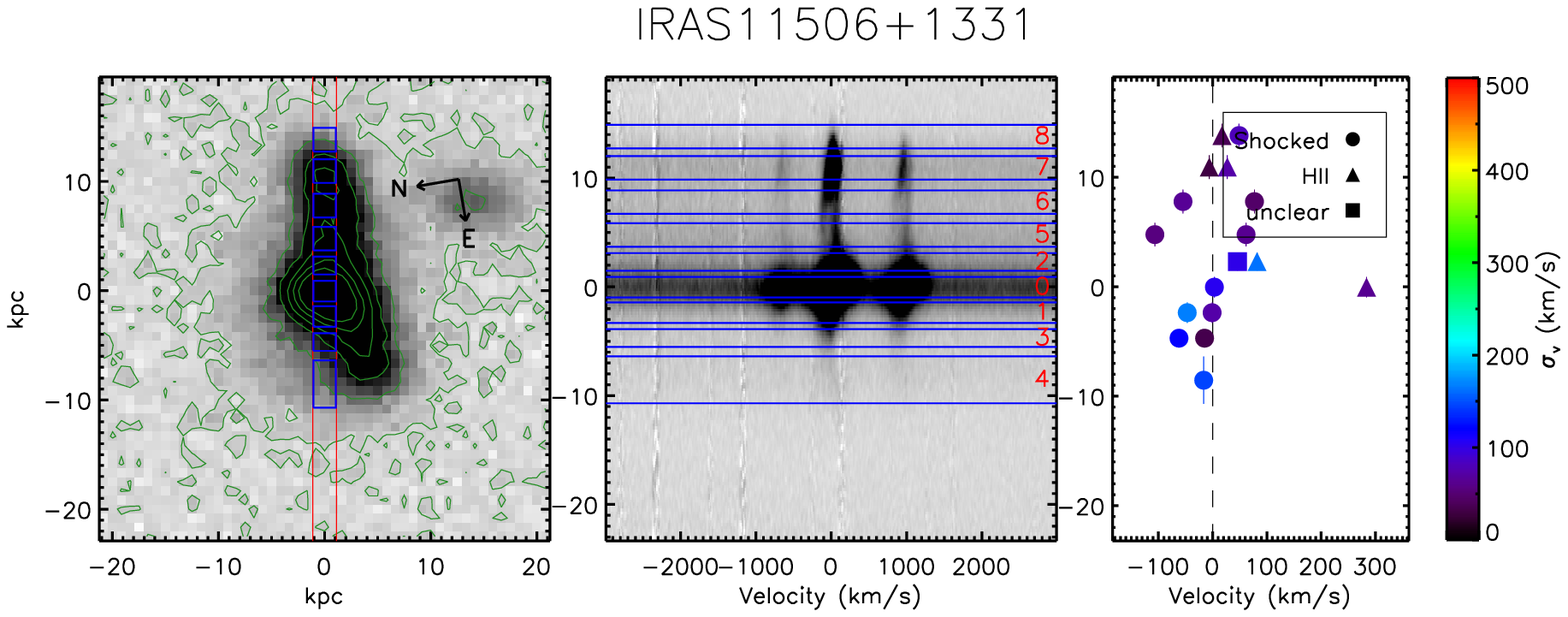, keepaspectratio=true, width=\linewidth}
\epsfig{file=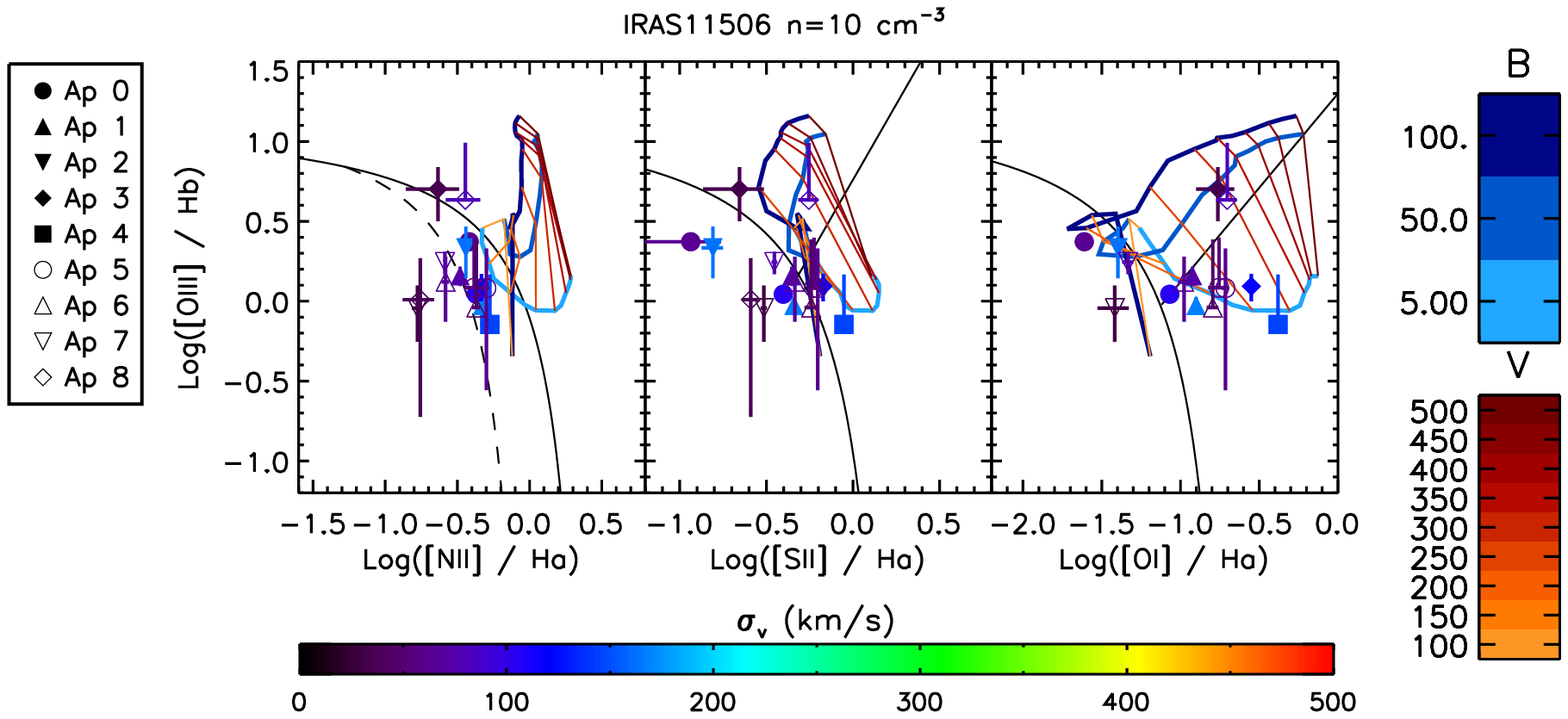, keepaspectratio=true, width=\linewidth}
\caption{\label{fig:i11506_bbc}
}
\end{figure}
\clearpage

\begin{figure}[h]
\centering
\epsfig{file=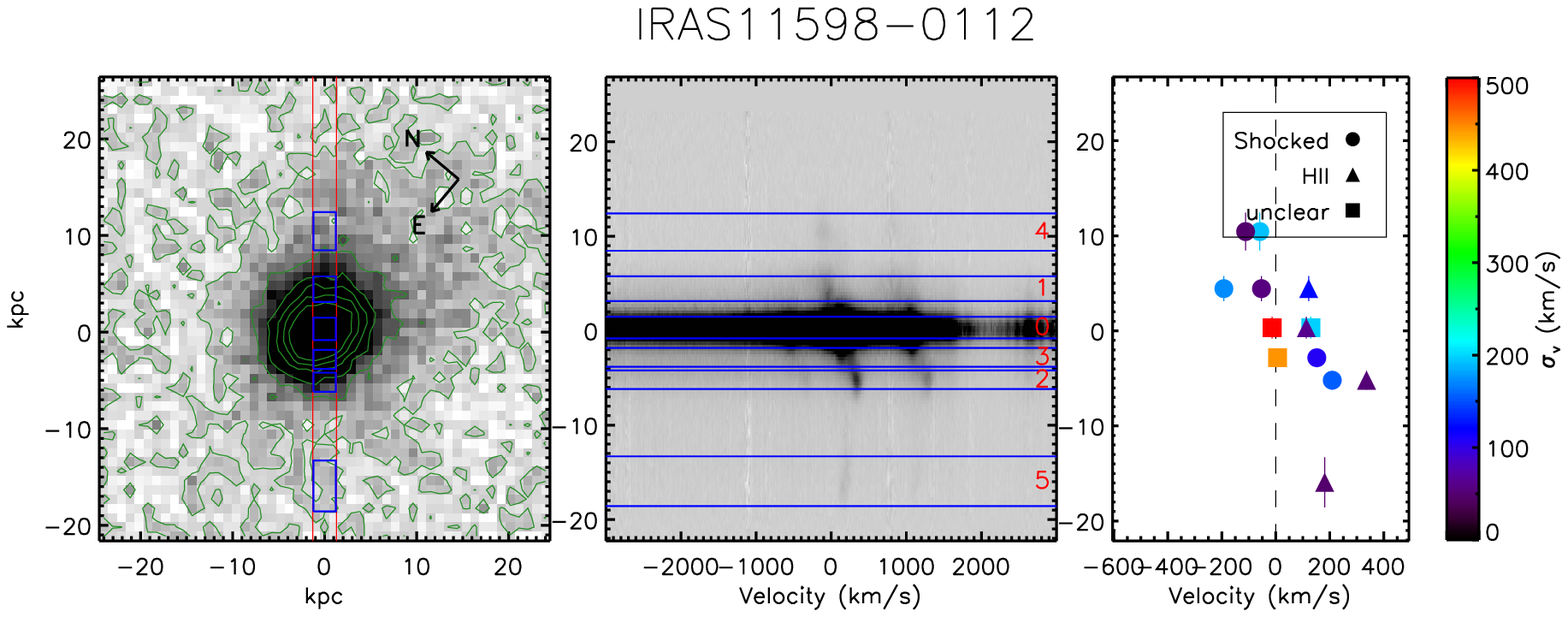, keepaspectratio=true, width=\linewidth}
\epsfig{file=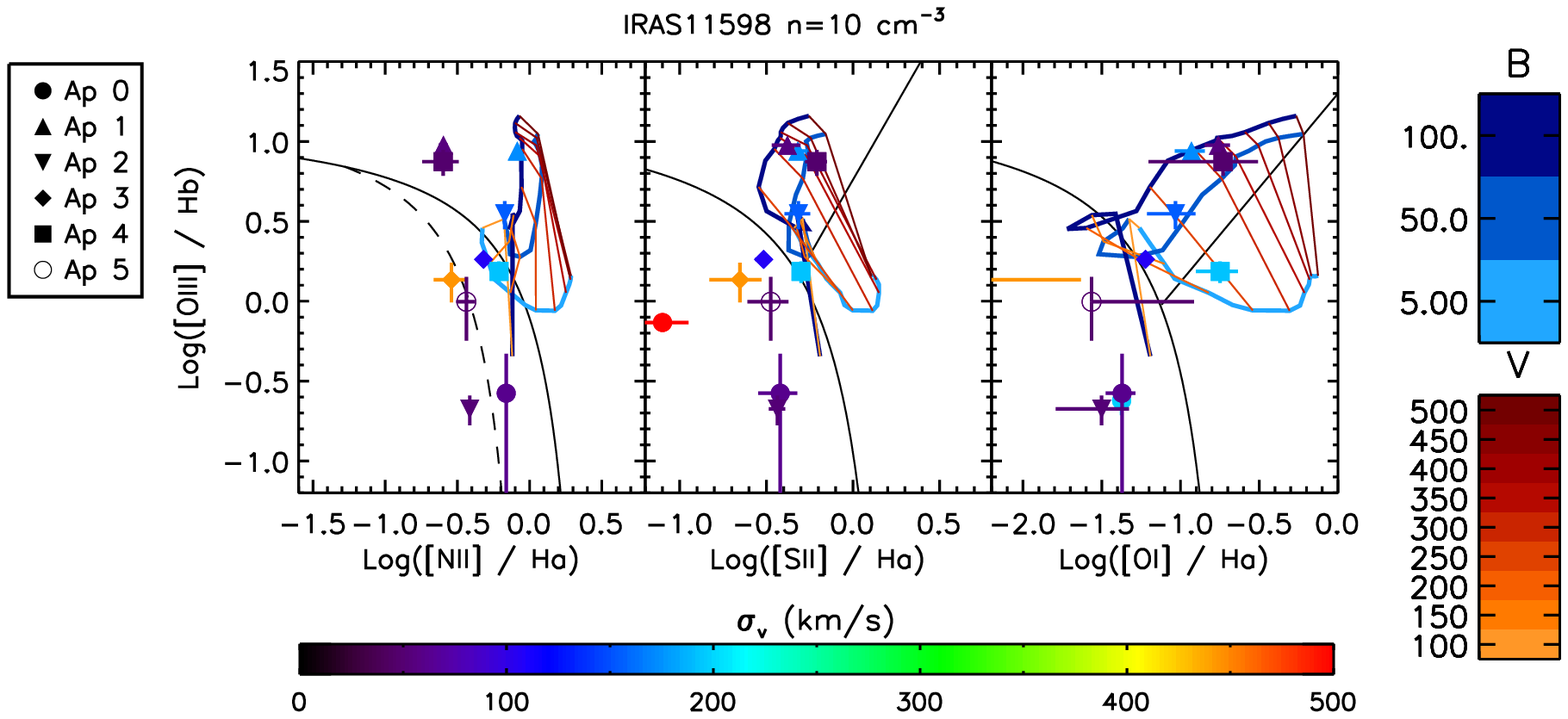, keepaspectratio=true, width=\linewidth}
\caption{\label{fig:i11598_bbc}
}
\end{figure}
\clearpage

\begin{figure}[h]
\centering
\epsfig{file=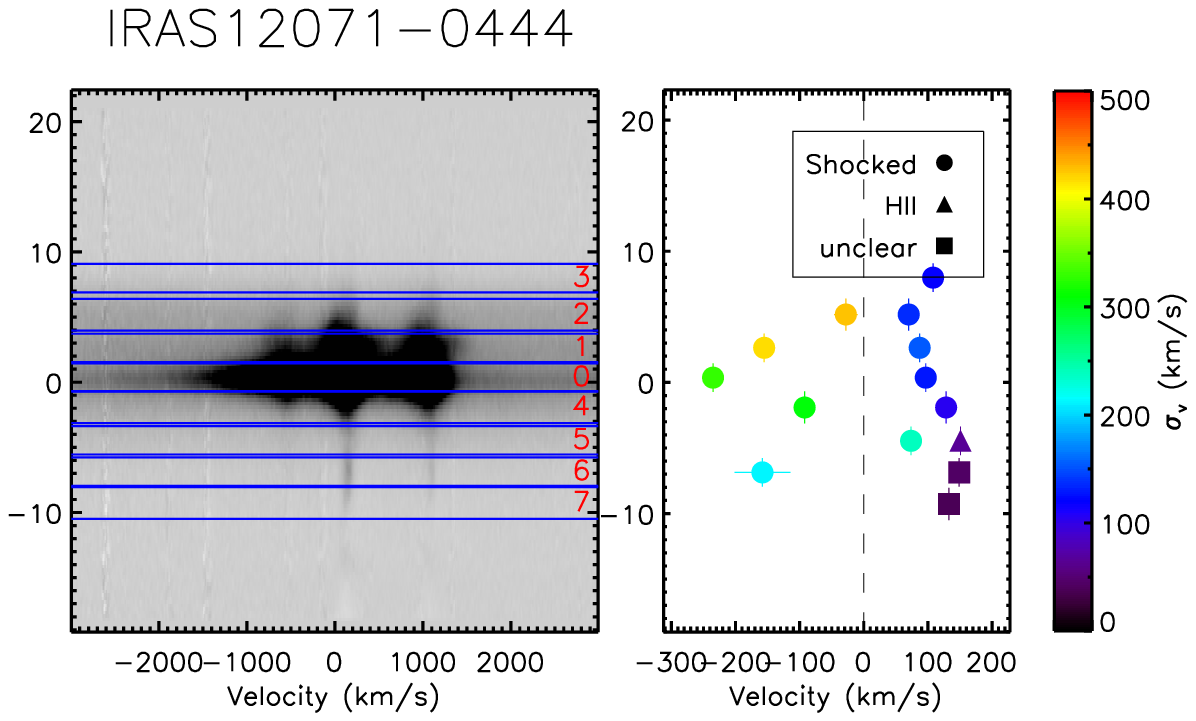, keepaspectratio=true, width=\linewidth}
\epsfig{file=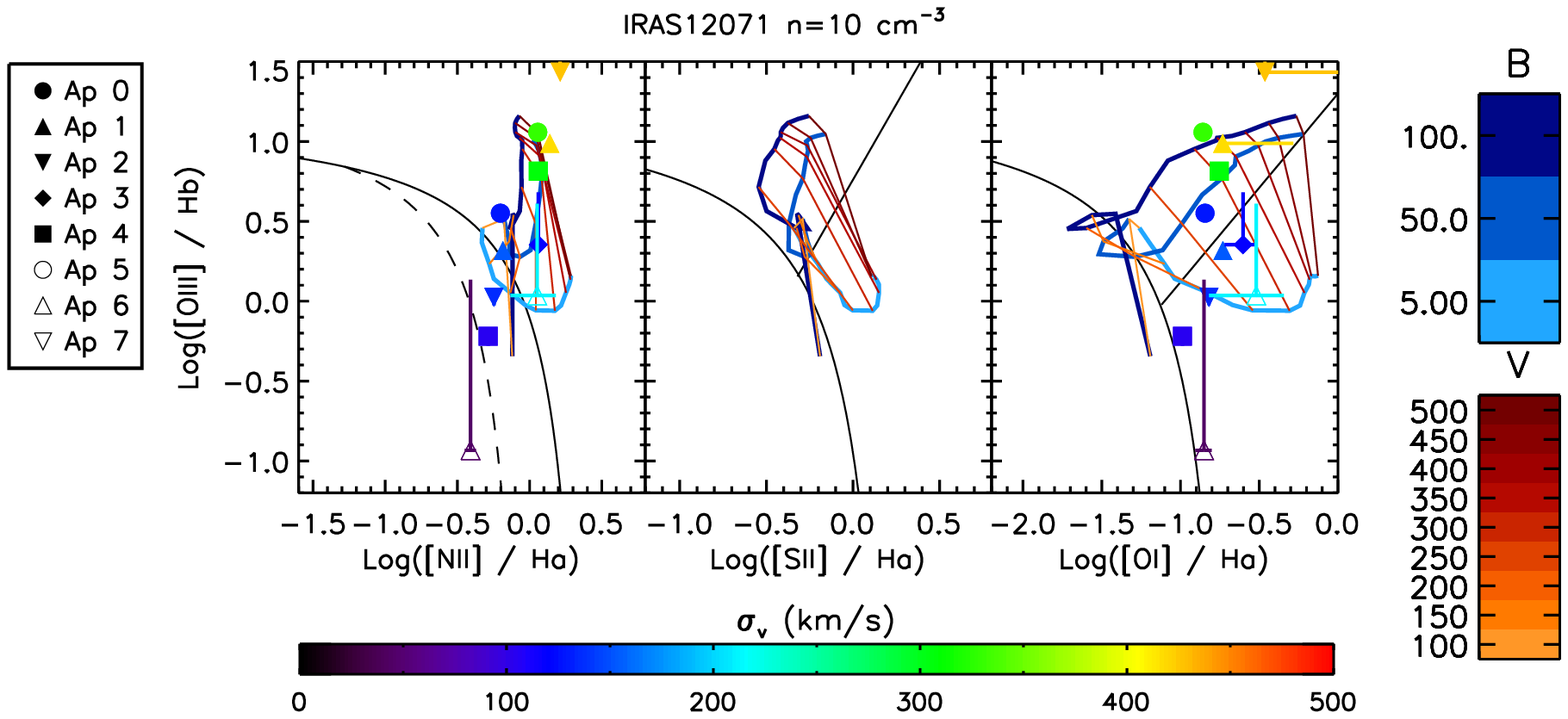, keepaspectratio=true, width=\linewidth}
\caption{\label{fig:i12071_bbc}
}
\end{figure}
\clearpage

\begin{figure}[h]
\centering
\epsfig{file=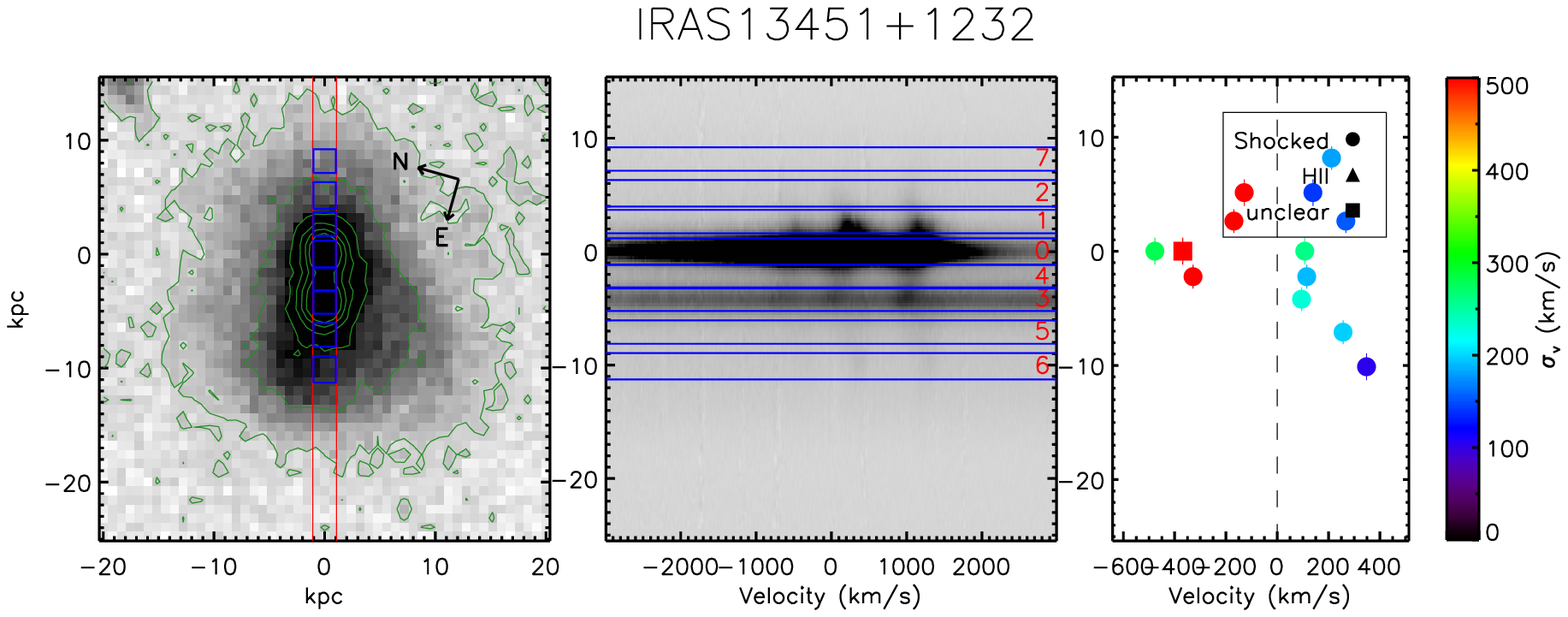, keepaspectratio=true, width=\linewidth}
\epsfig{file=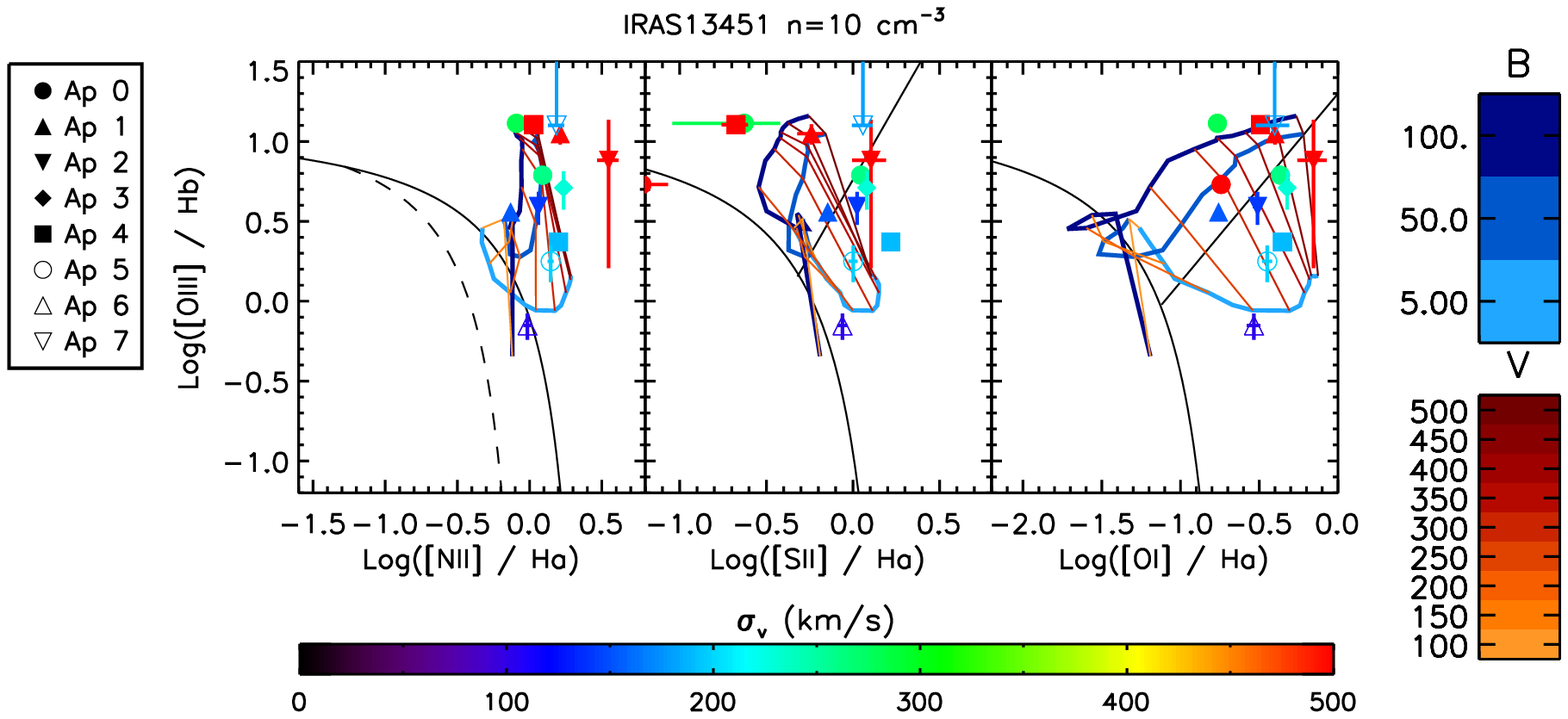, keepaspectratio=true, width=\linewidth}
\caption{\label{fig:i13451_bbc}
}
\end{figure}
\clearpage

\begin{figure}[h]
\centering
\epsfig{file=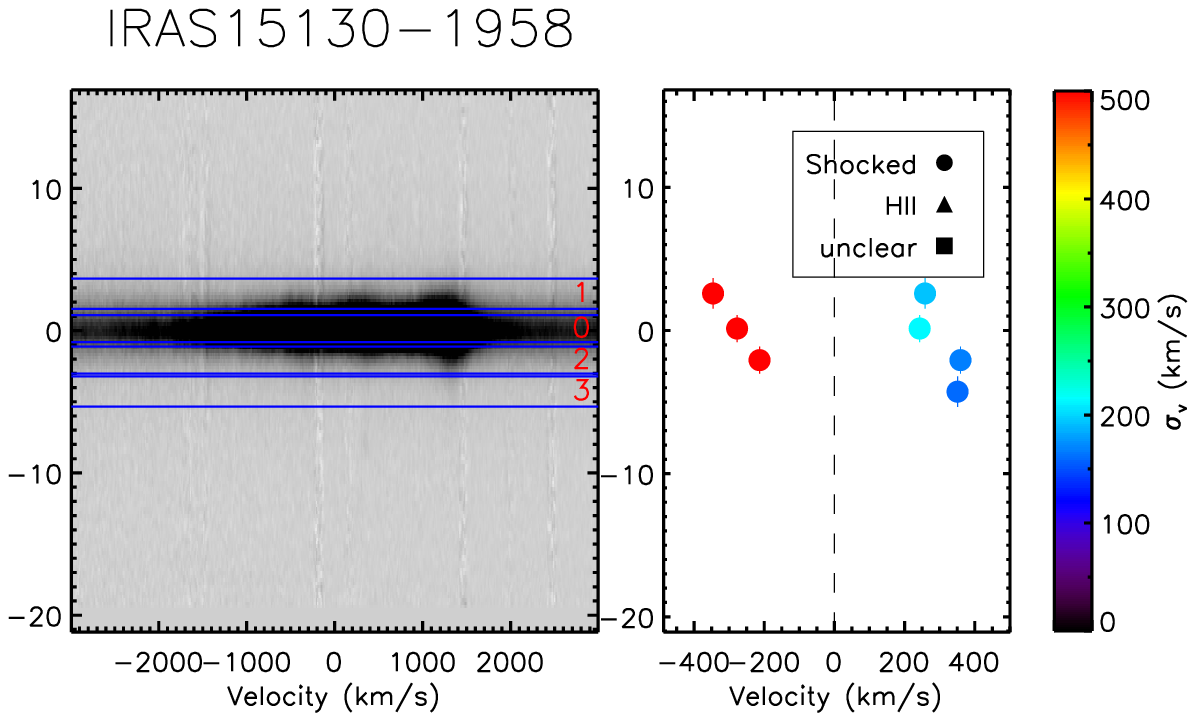, keepaspectratio=true, width=\linewidth}
\epsfig{file=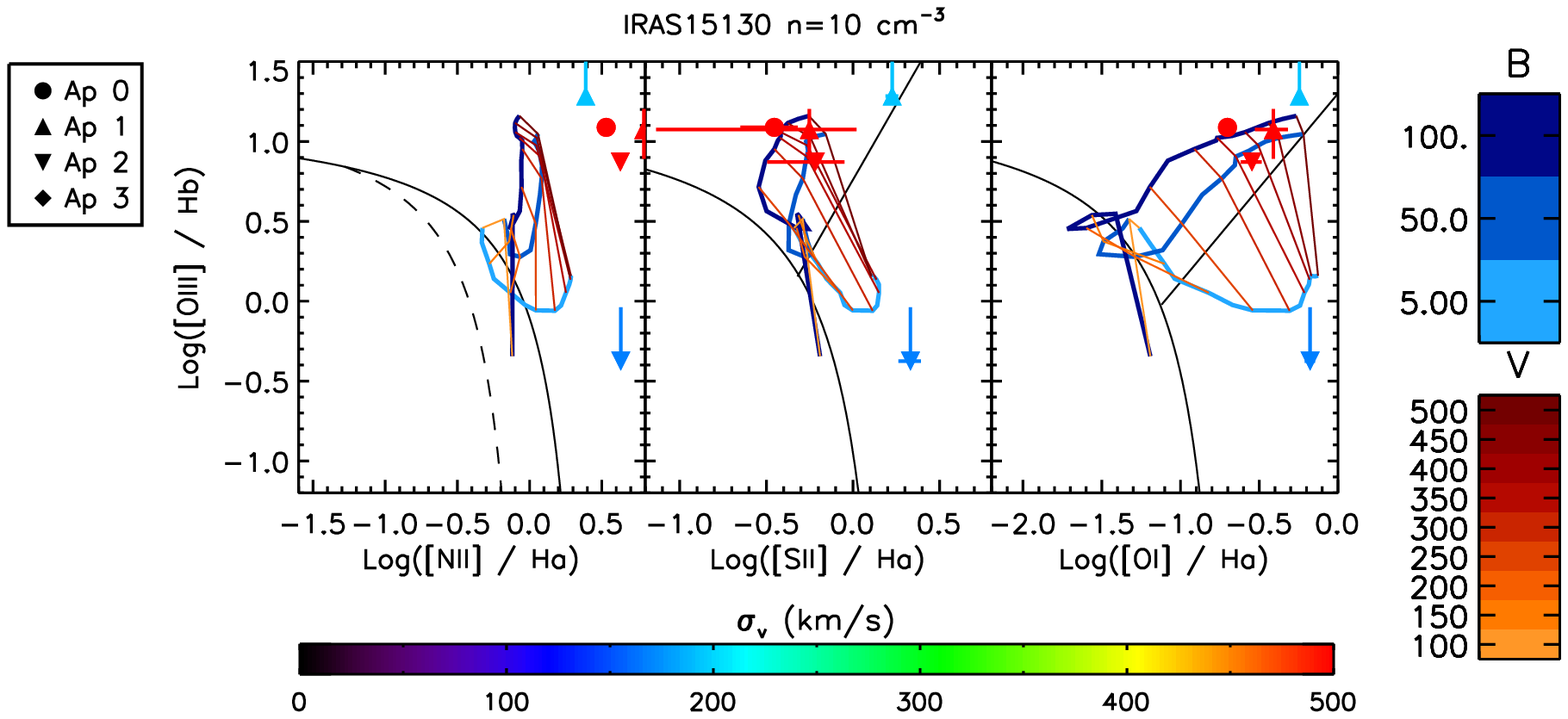, keepaspectratio=true, width=\linewidth}
\caption{\label{fig:i15130_bbc}
}
\end{figure}
\clearpage


\begin{figure}[h]
\centering
\epsfig{file=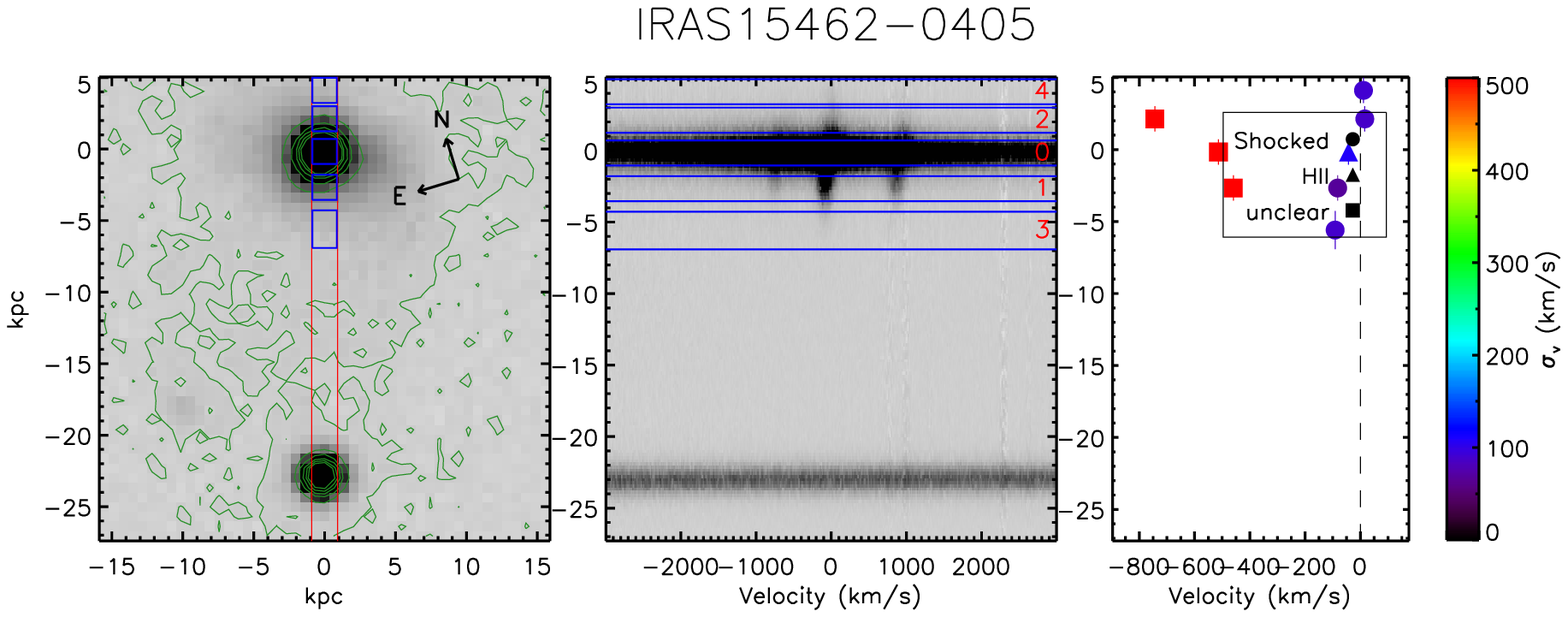, keepaspectratio=true, width=\linewidth}
\epsfig{file=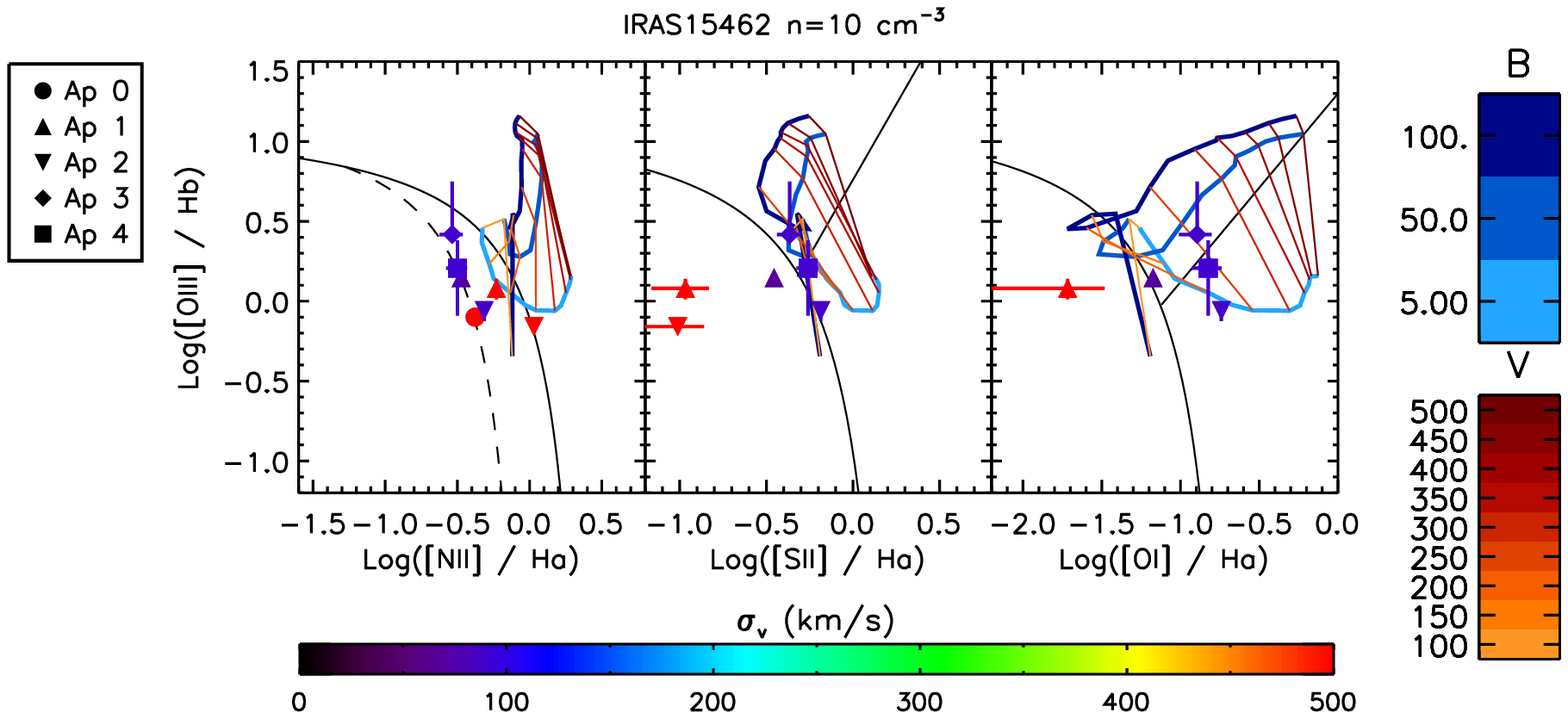, keepaspectratio=true, width=\linewidth}
\caption{\label{fig:i15462_bbc}
}
\end{figure}
\clearpage

\begin{figure}[h]
\centering
\epsfig{file=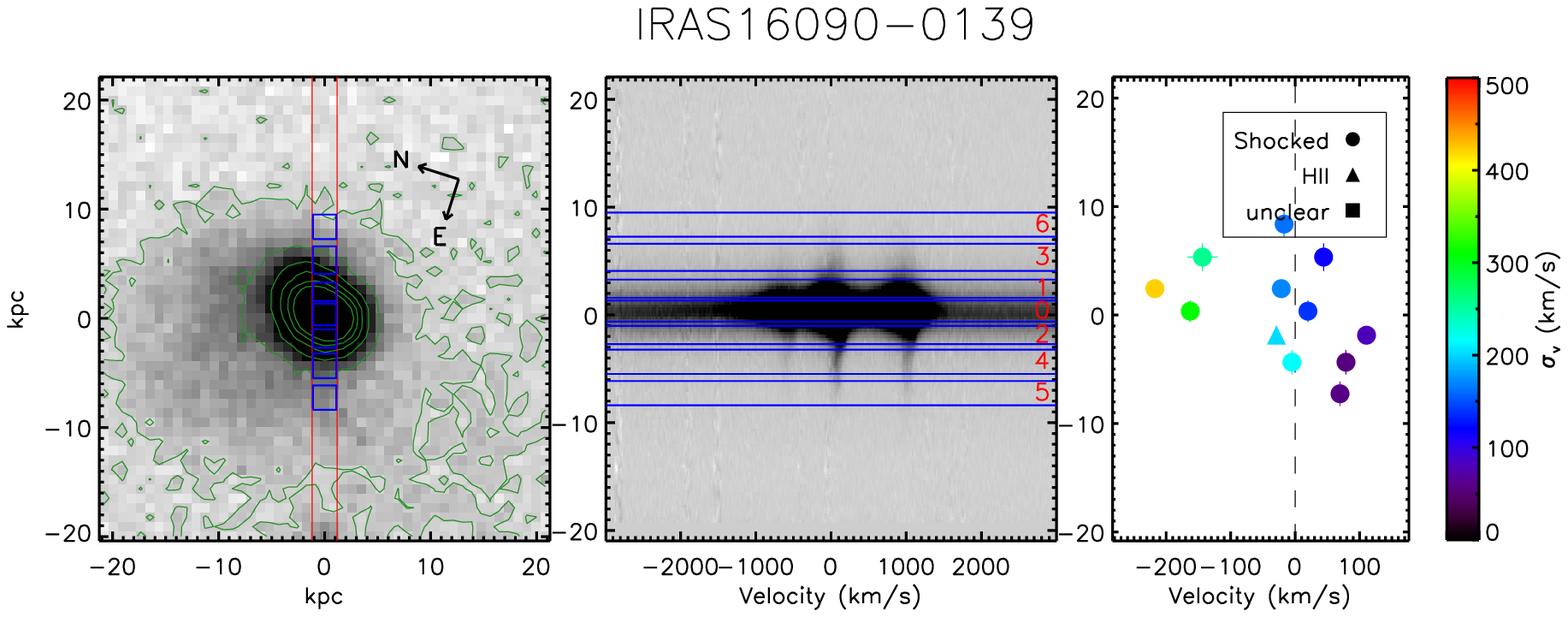, keepaspectratio=true, width=\linewidth}
\epsfig{file=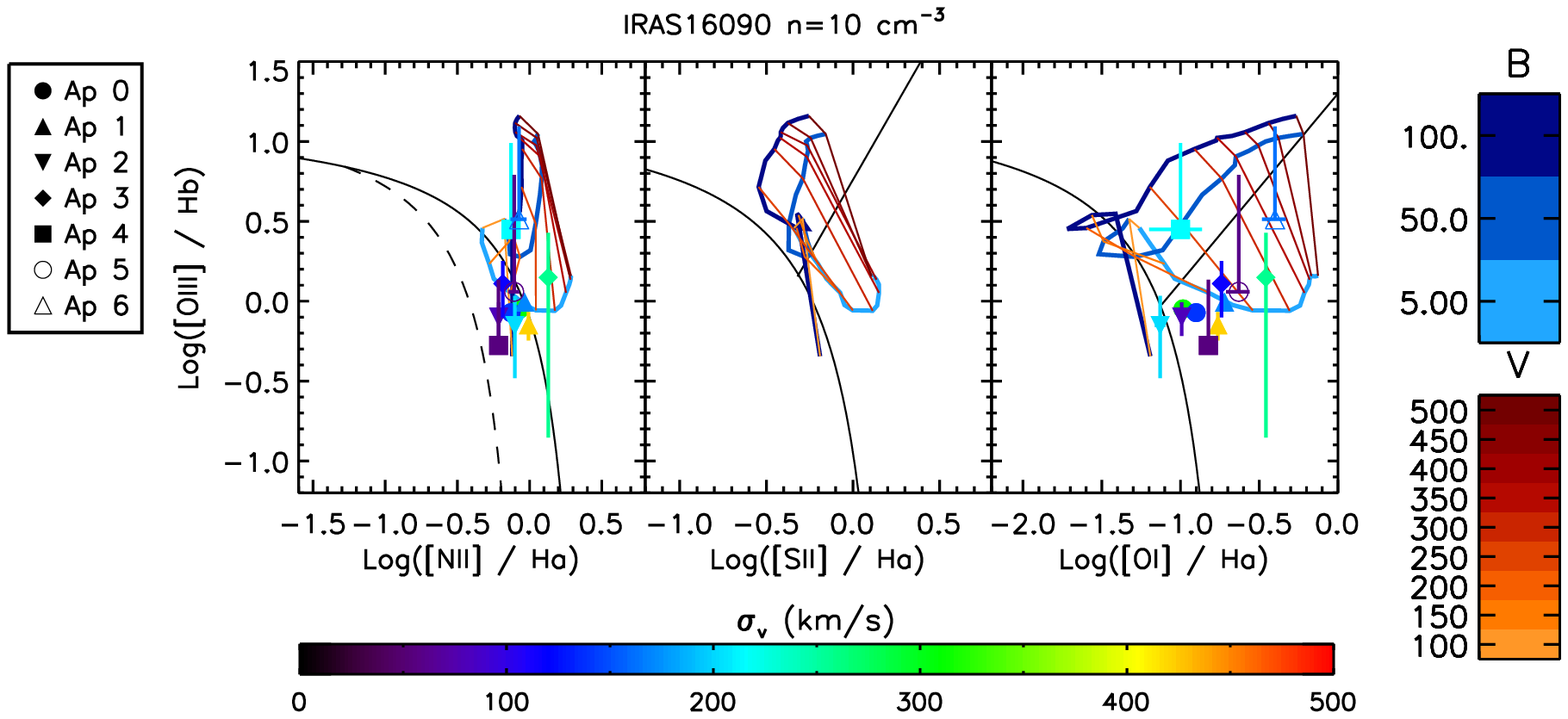, keepaspectratio=true, width=\linewidth}
\caption{\label{fig:i16090_bbc}
}
\end{figure}
\clearpage

\begin{figure}[h]
\centering
\epsfig{file=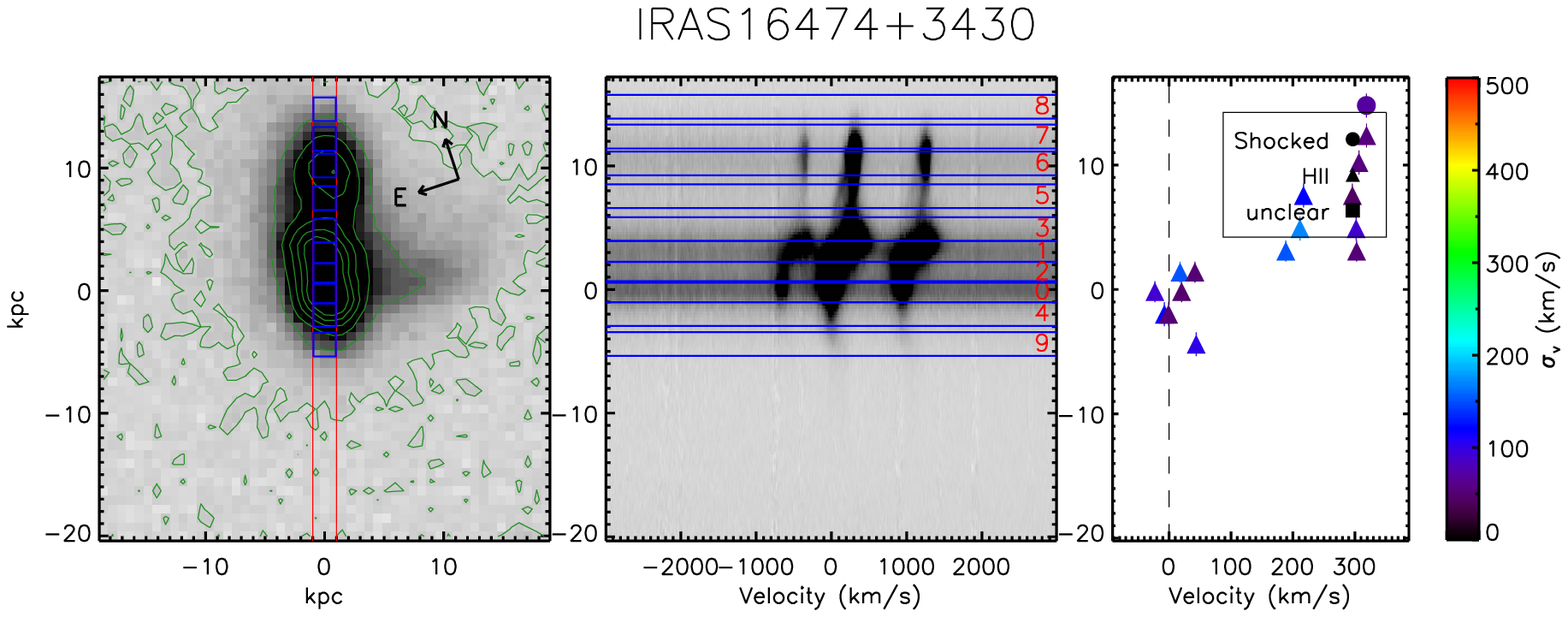, keepaspectratio=true, width=\linewidth}
\epsfig{file=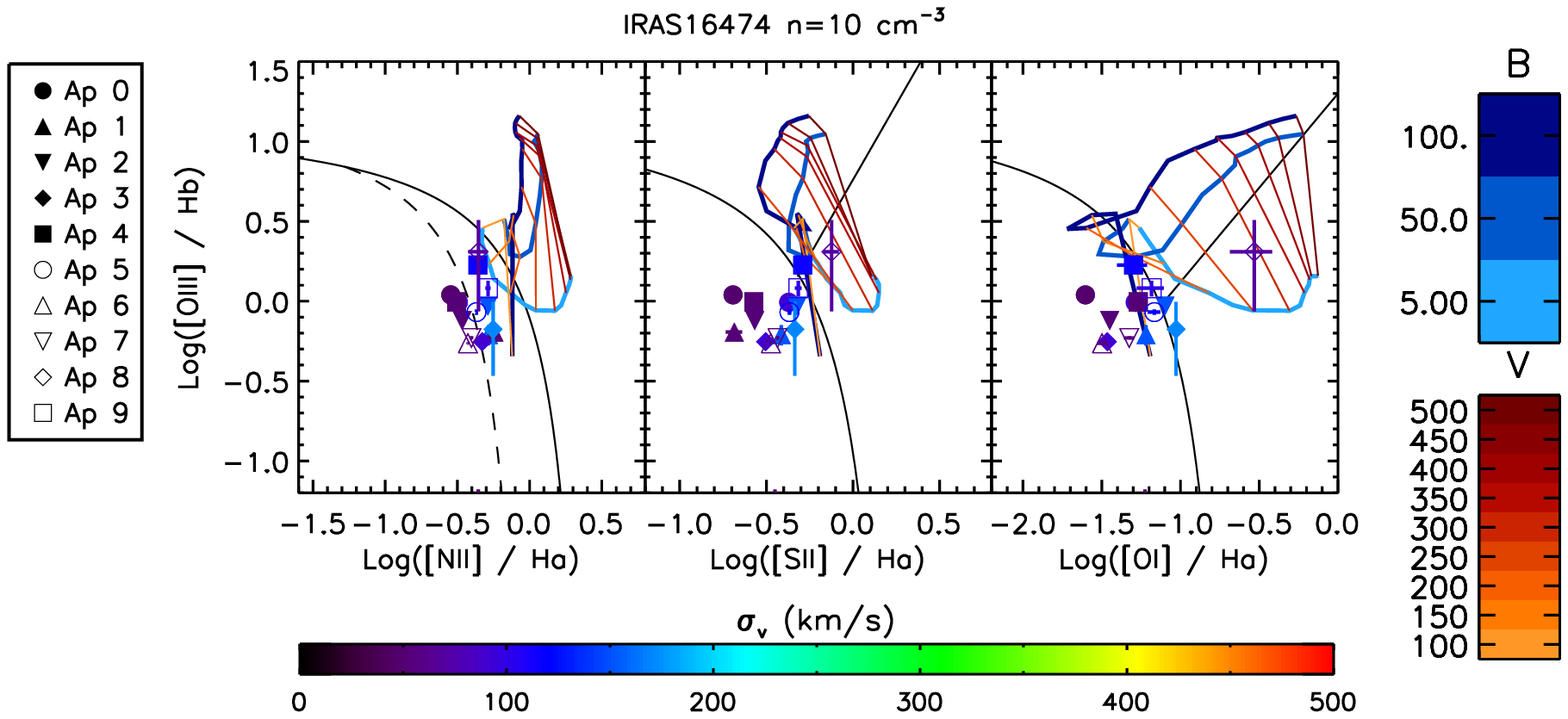, keepaspectratio=true, width=\linewidth}
\caption{\label{fig:i16474_bbc}
}
\end{figure}
\clearpage

\begin{figure}[h]
\centering
\epsfig{file=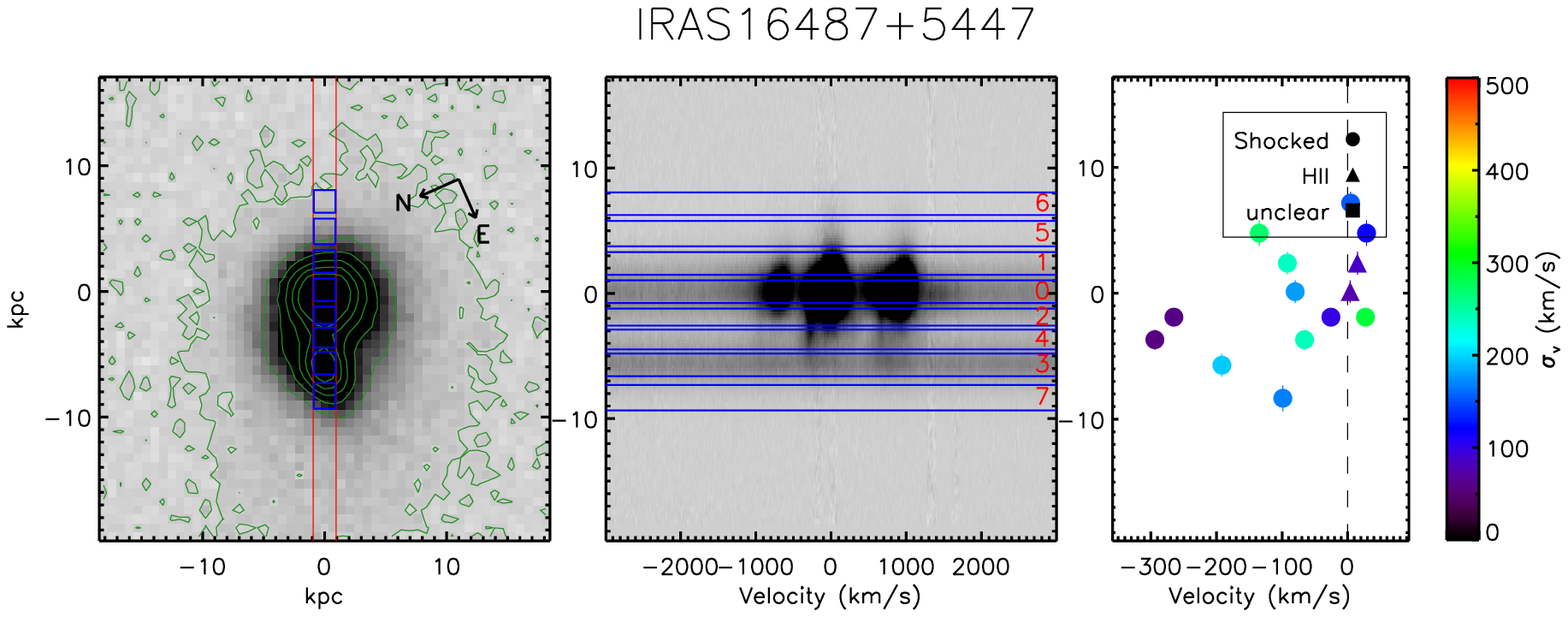, keepaspectratio=true, width=\linewidth}
\epsfig{file=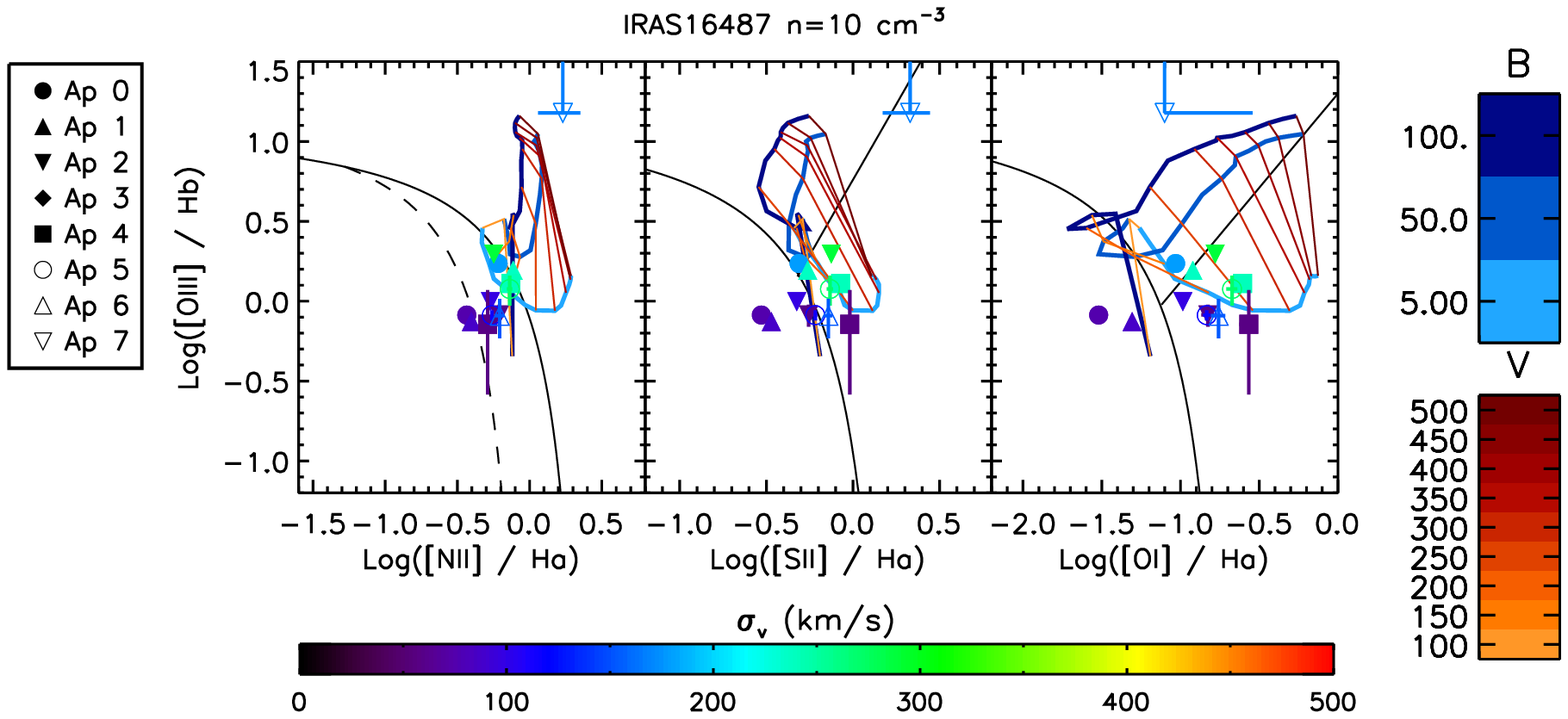, keepaspectratio=true, width=\linewidth}
\caption{\label{fig:i16487_bbc}
}
\end{figure}
\clearpage

\begin{figure}[h]
\centering
\epsfig{file=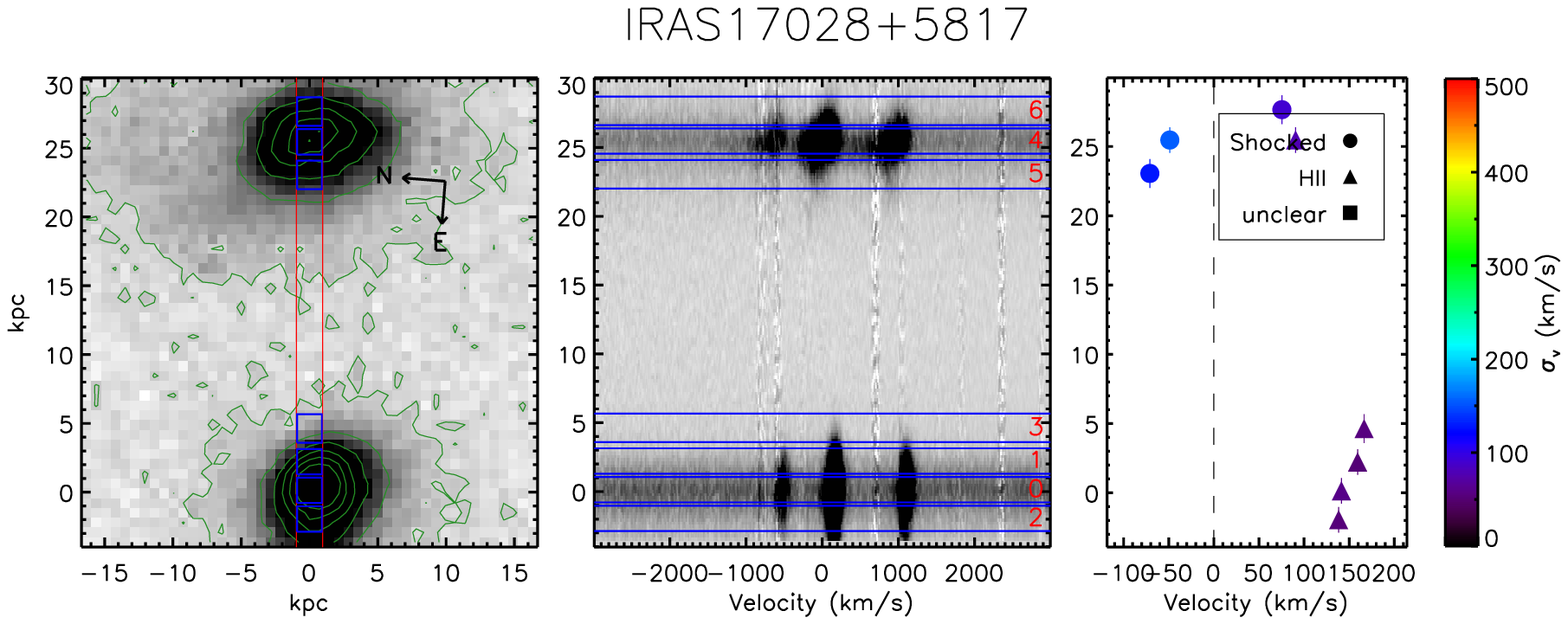, keepaspectratio=true, width=\linewidth}
\epsfig{file=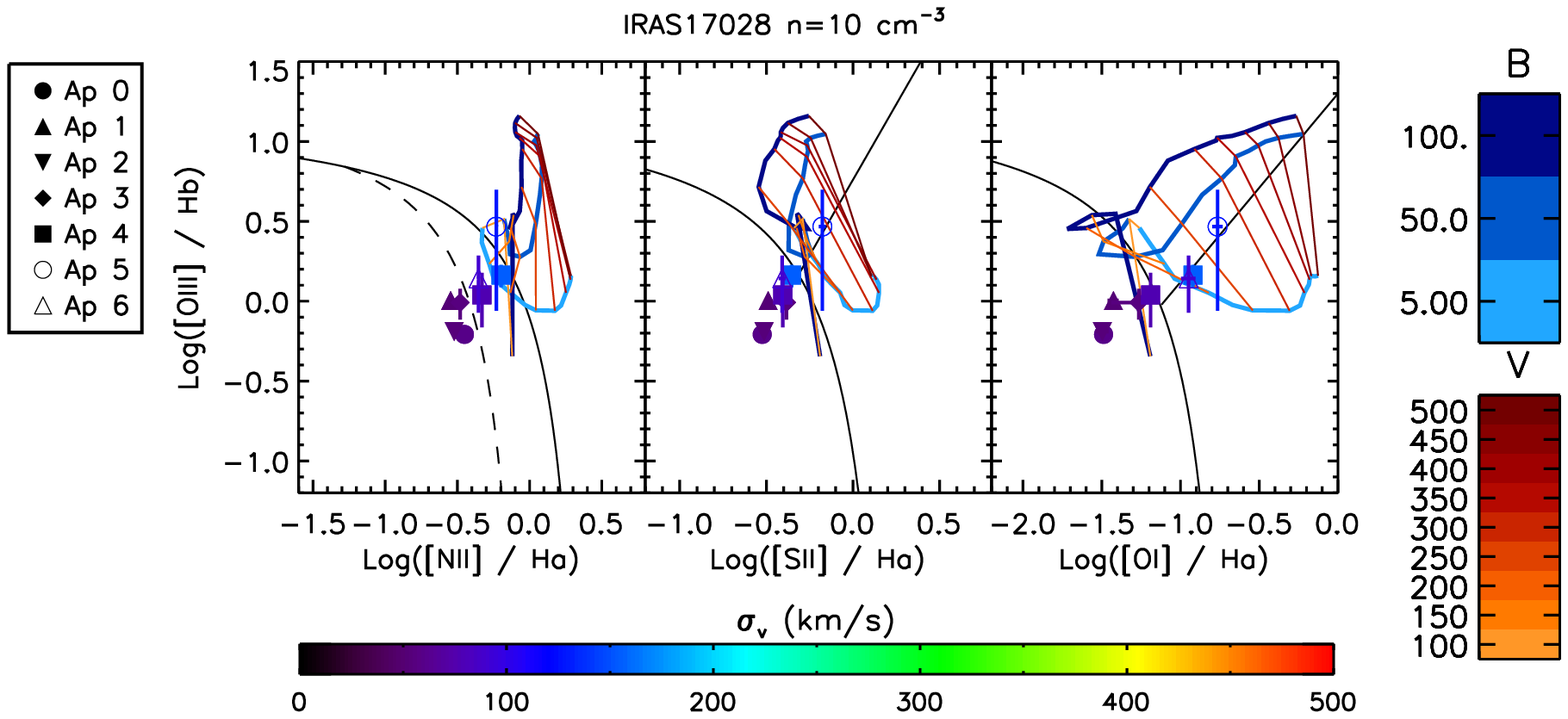, keepaspectratio=true, width=\linewidth}
\caption{\label{fig:i17028_bbc}
}
\end{figure}
\clearpage

\begin{figure}[h]
\centering
\epsfig{file=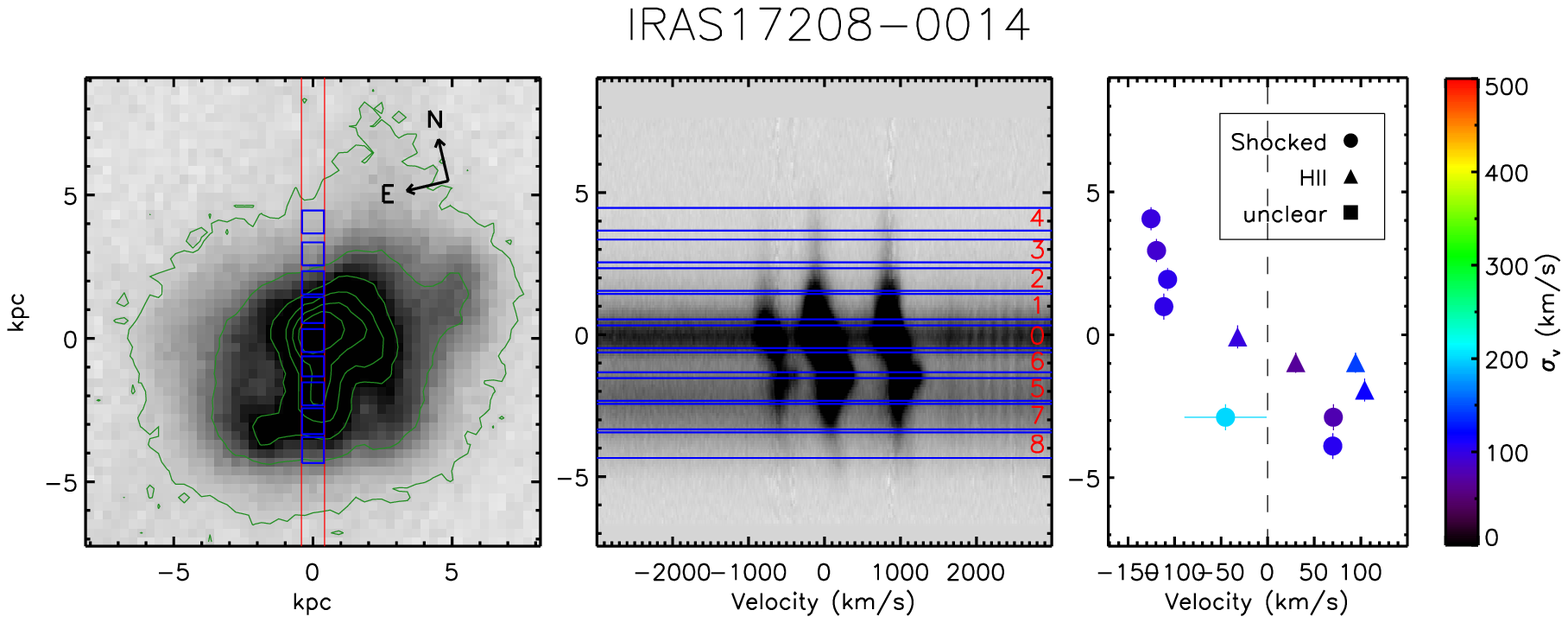, keepaspectratio=true, width=\linewidth}
\epsfig{file=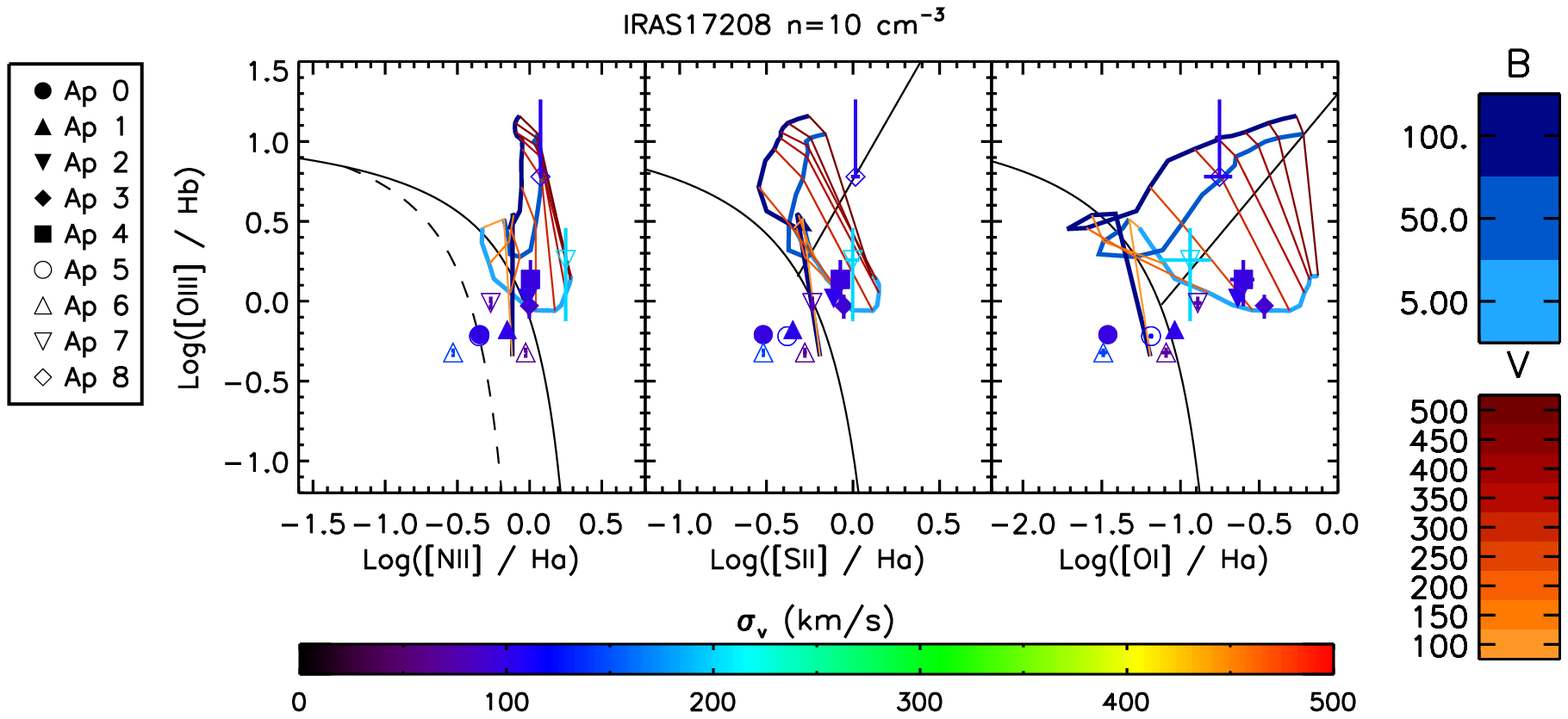, keepaspectratio=true, width=\linewidth}
\caption{\label{fig:i17208_bbc}
}
\end{figure}
\clearpage

\begin{figure}[h]
\centering
\epsfig{file=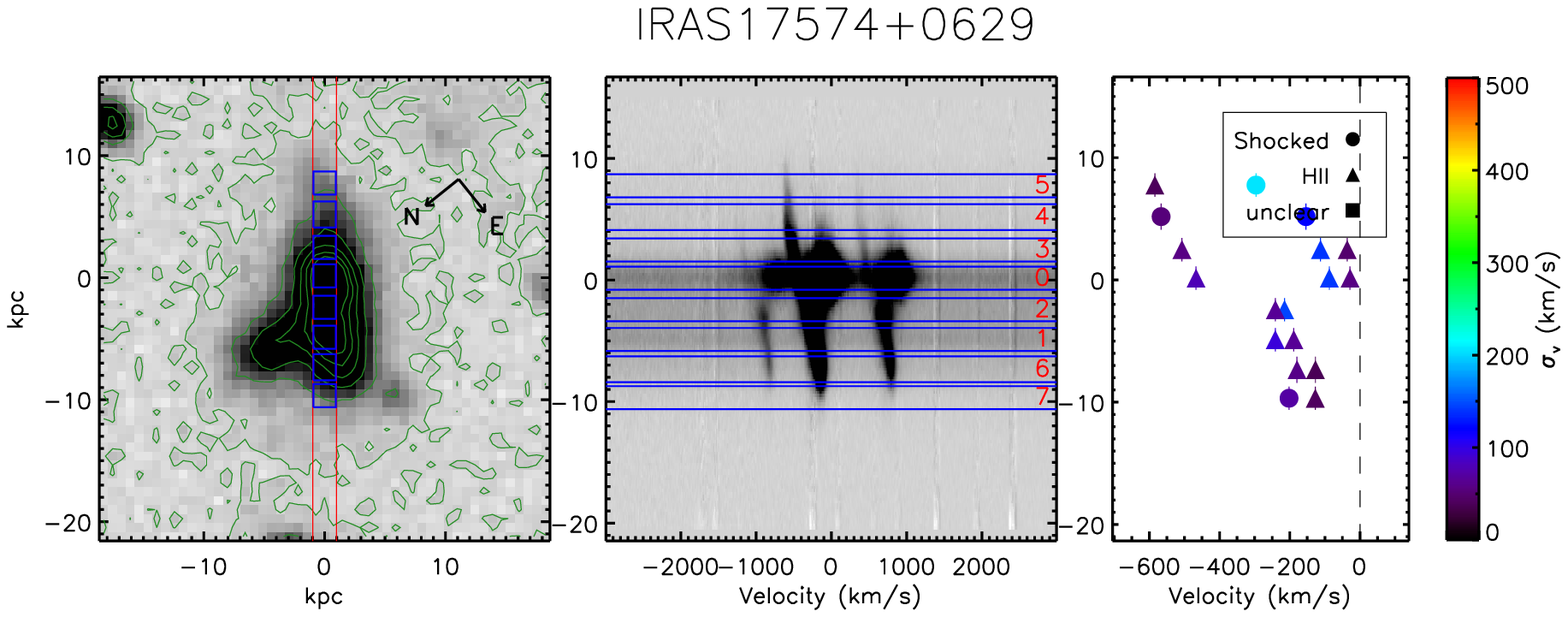, keepaspectratio=true, width=\linewidth}
\epsfig{file=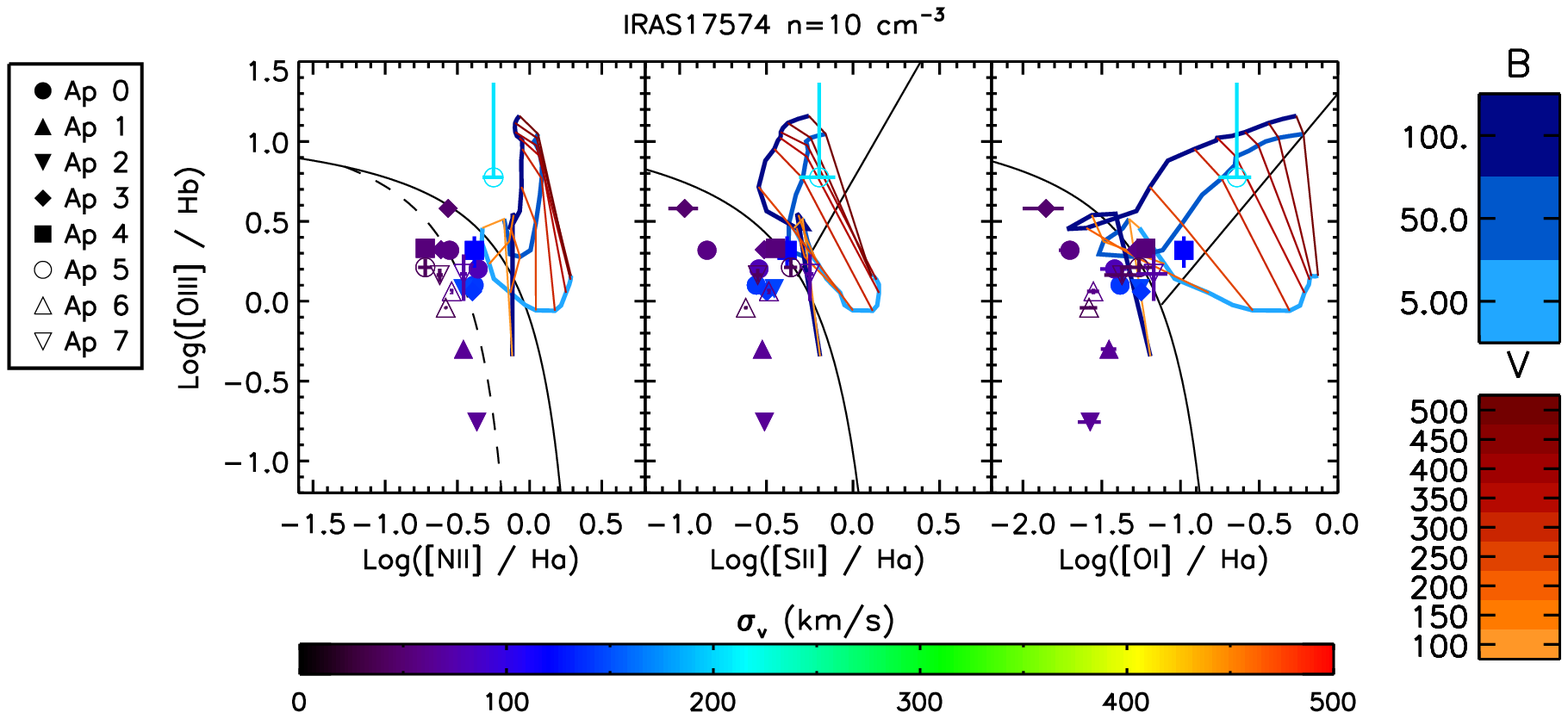, keepaspectratio=true, width=\linewidth}
\caption{\label{fig:i17574_bbc}
}
\end{figure}
\clearpage

\begin{figure}[h]
\centering
\epsfig{file=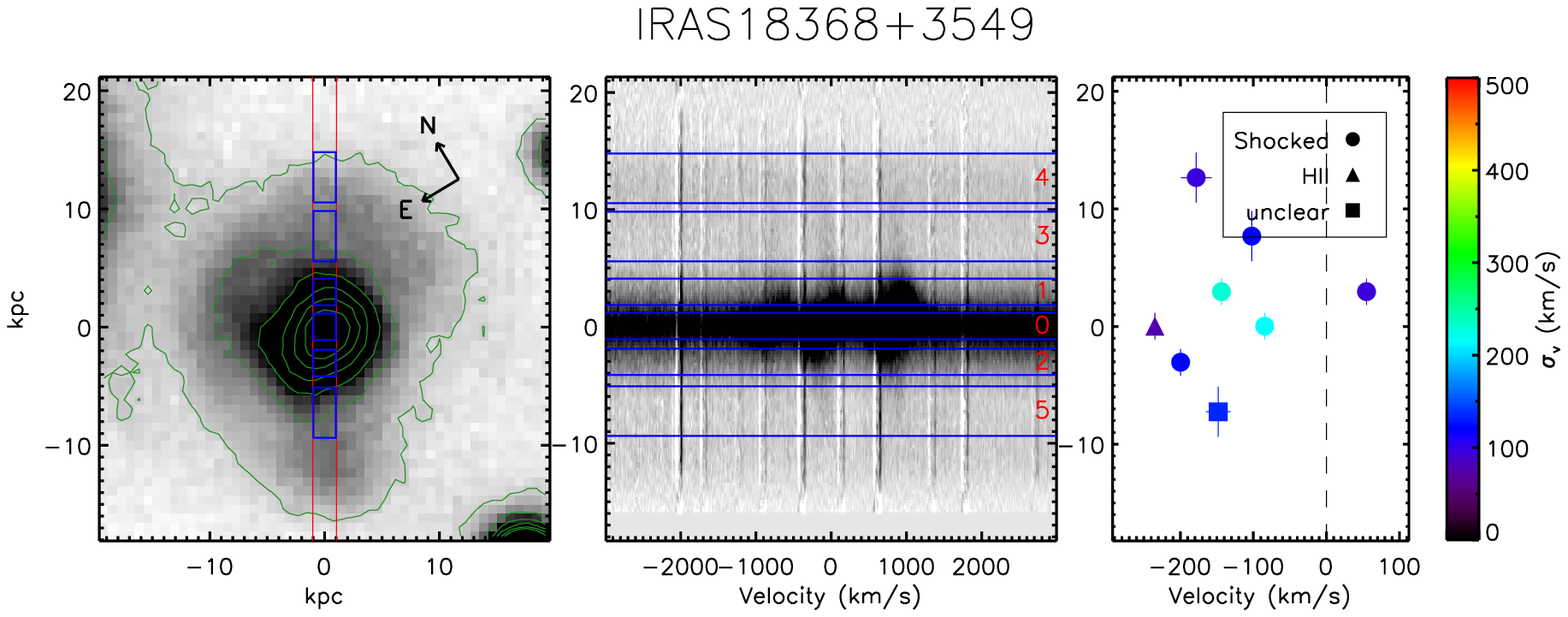, keepaspectratio=true, width=\linewidth}
\epsfig{file=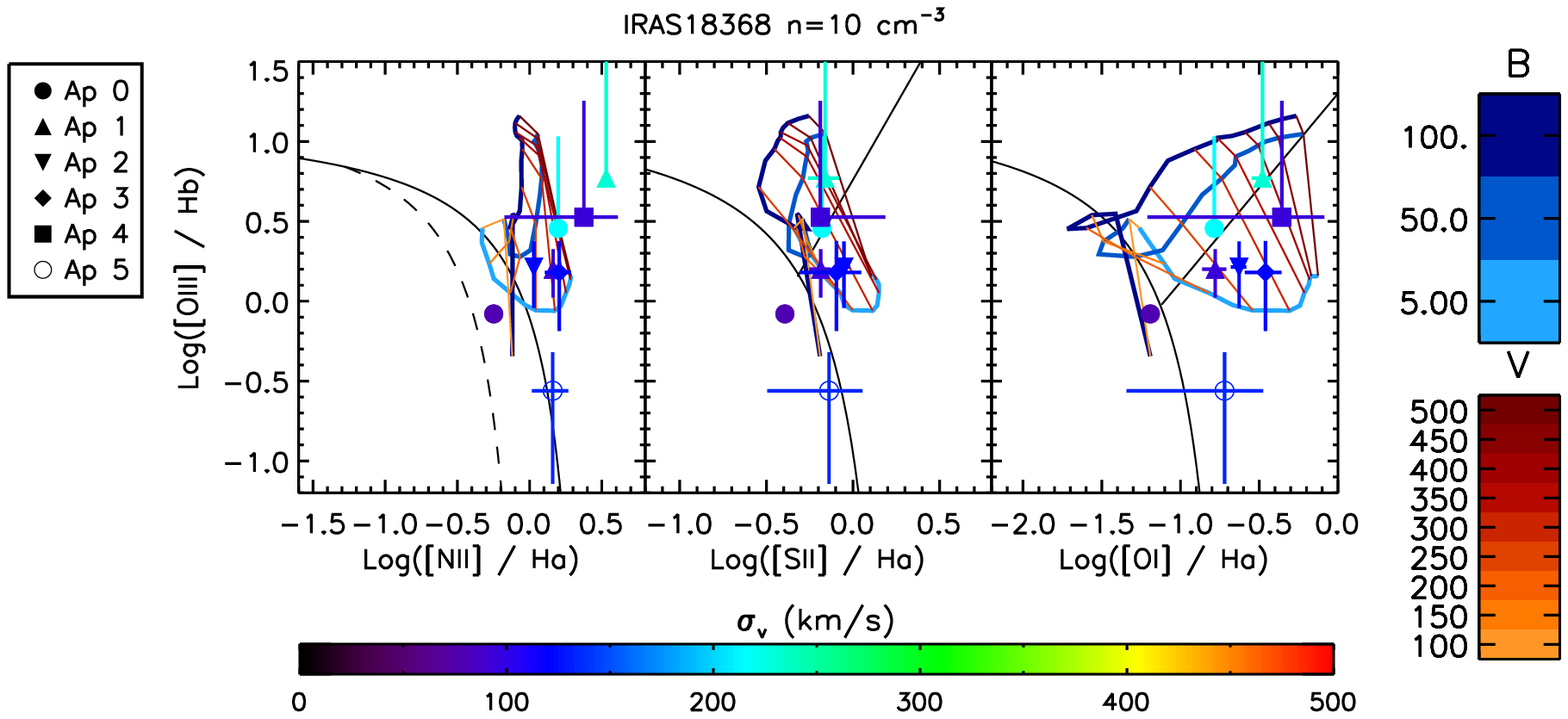, keepaspectratio=true, width=\linewidth}
\caption{\label{fig:i18368_bbc}
}
\end{figure}
\clearpage

\begin{figure}[h]
\centering
\epsfig{file=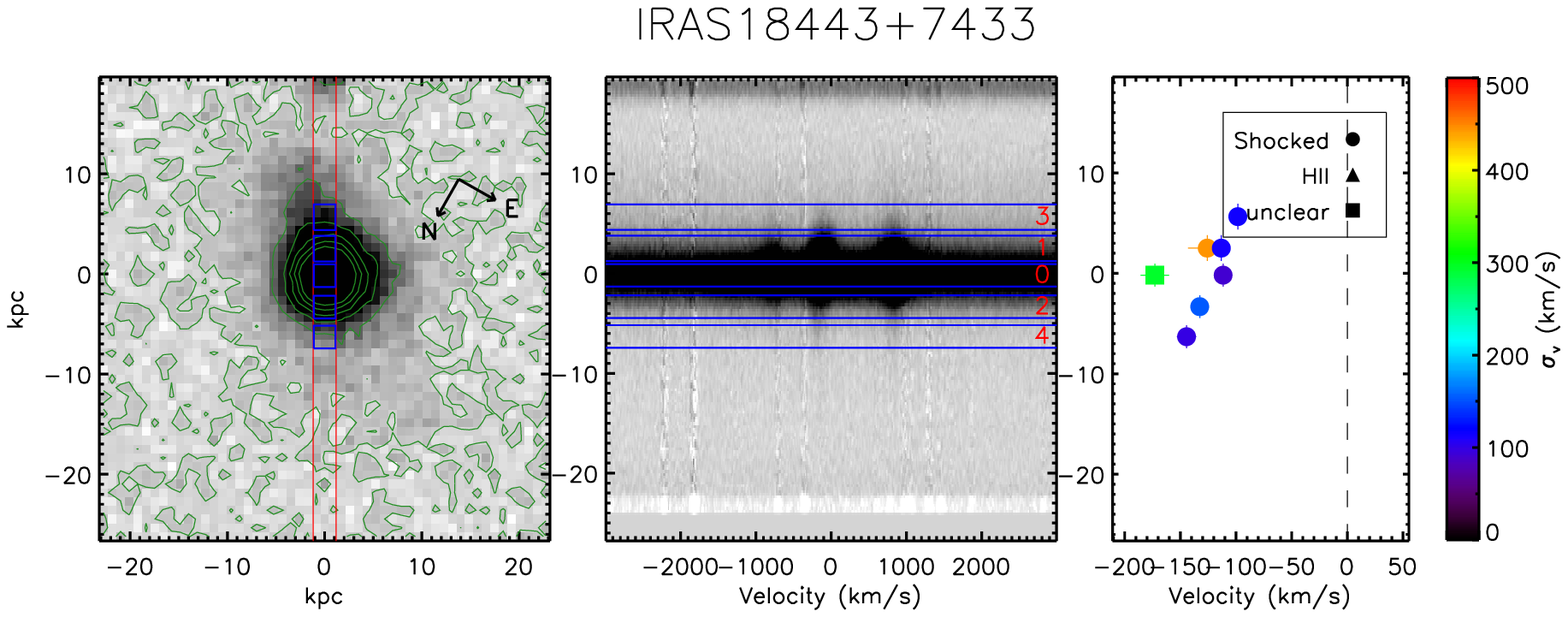, keepaspectratio=true, width=\linewidth}
\epsfig{file=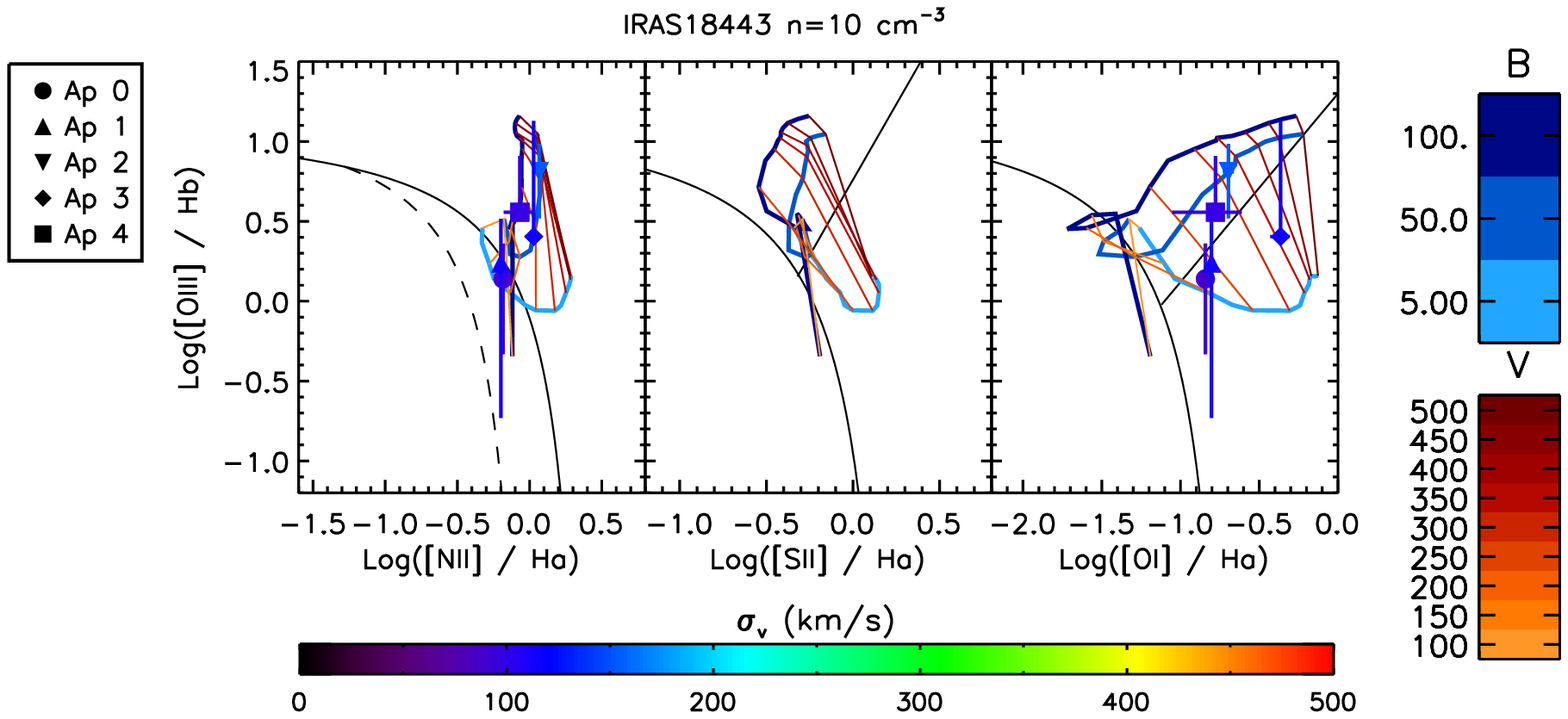, keepaspectratio=true, width=\linewidth}
\caption{\label{fig:i18443_bbc}
}
\end{figure}
\clearpage

\begin{figure}[h]
\centering
\epsfig{file=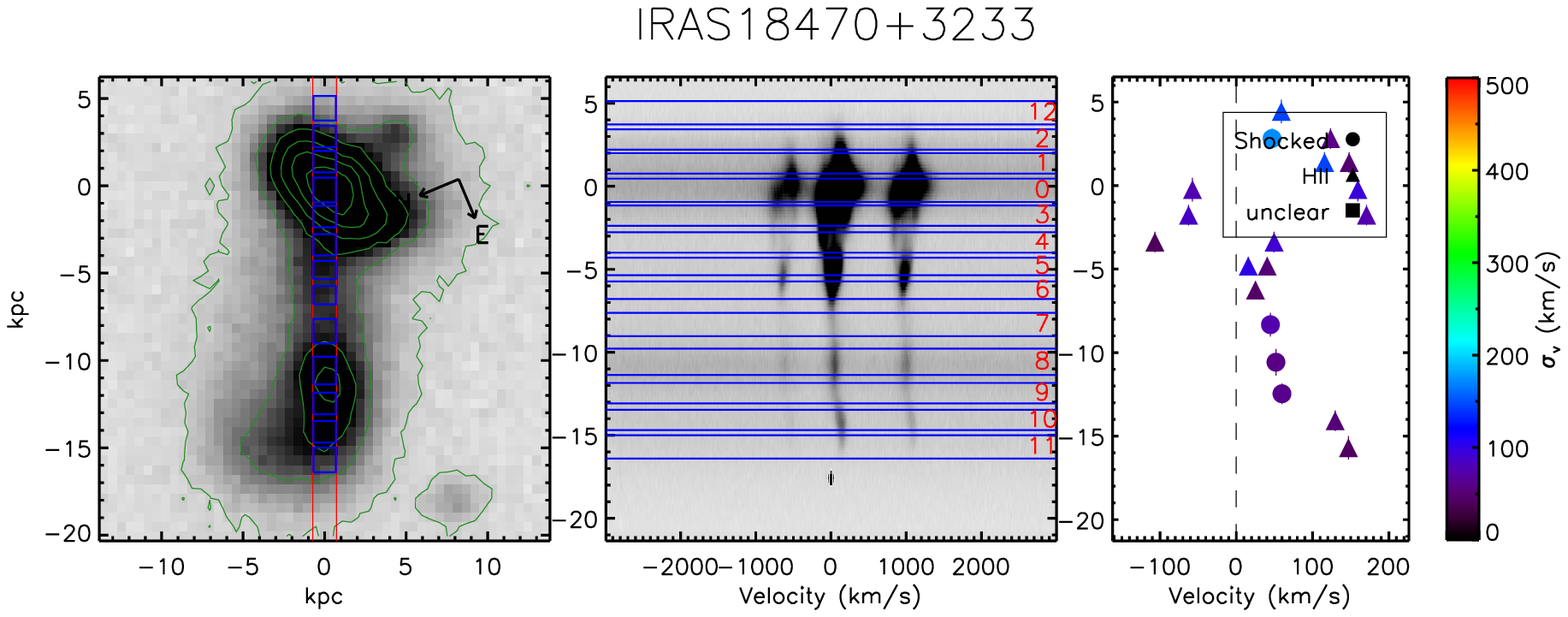, keepaspectratio=true, width=\linewidth}
\epsfig{file=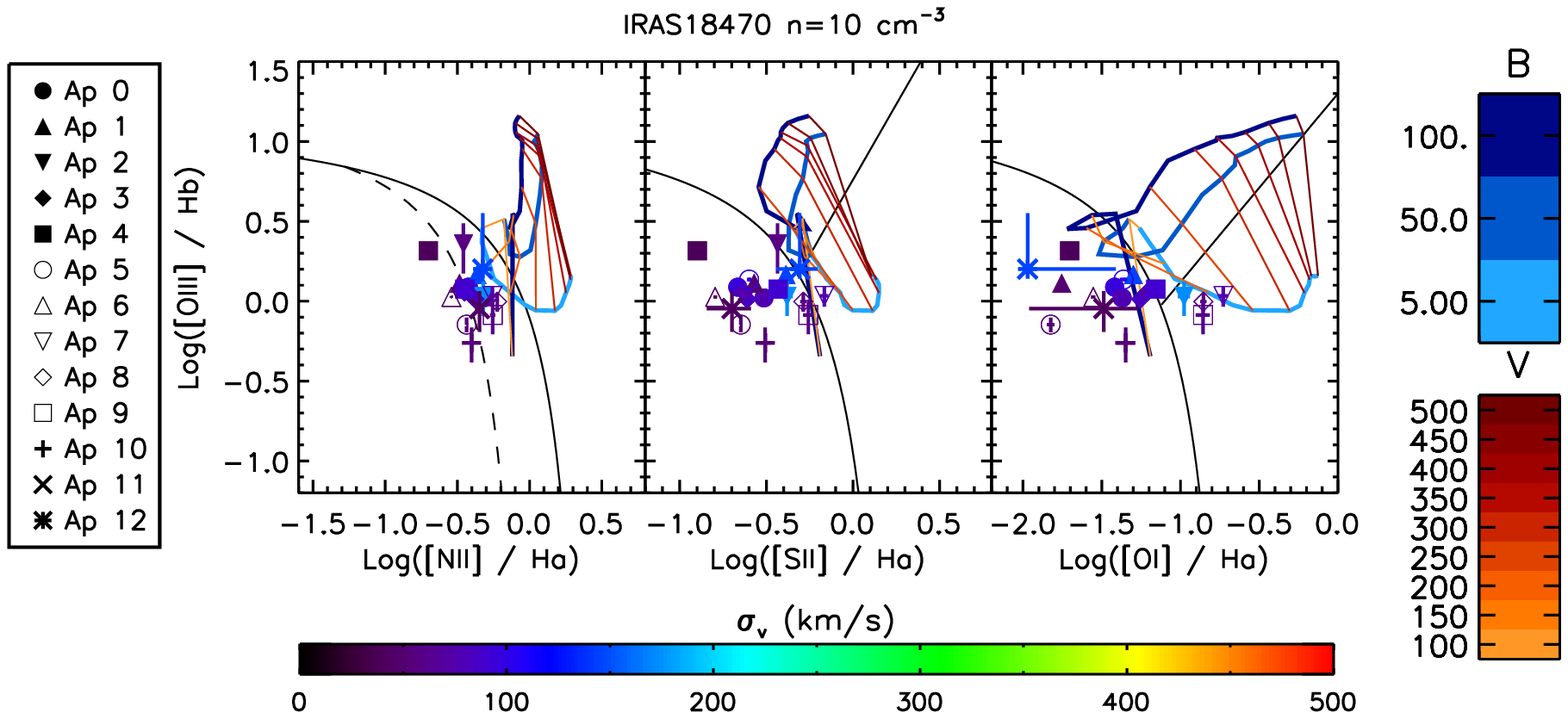, keepaspectratio=true, width=\linewidth}
\caption{\label{fig:i18470_bbc}
}
\end{figure}
\clearpage

\begin{figure}[h]
\centering
\epsfig{file=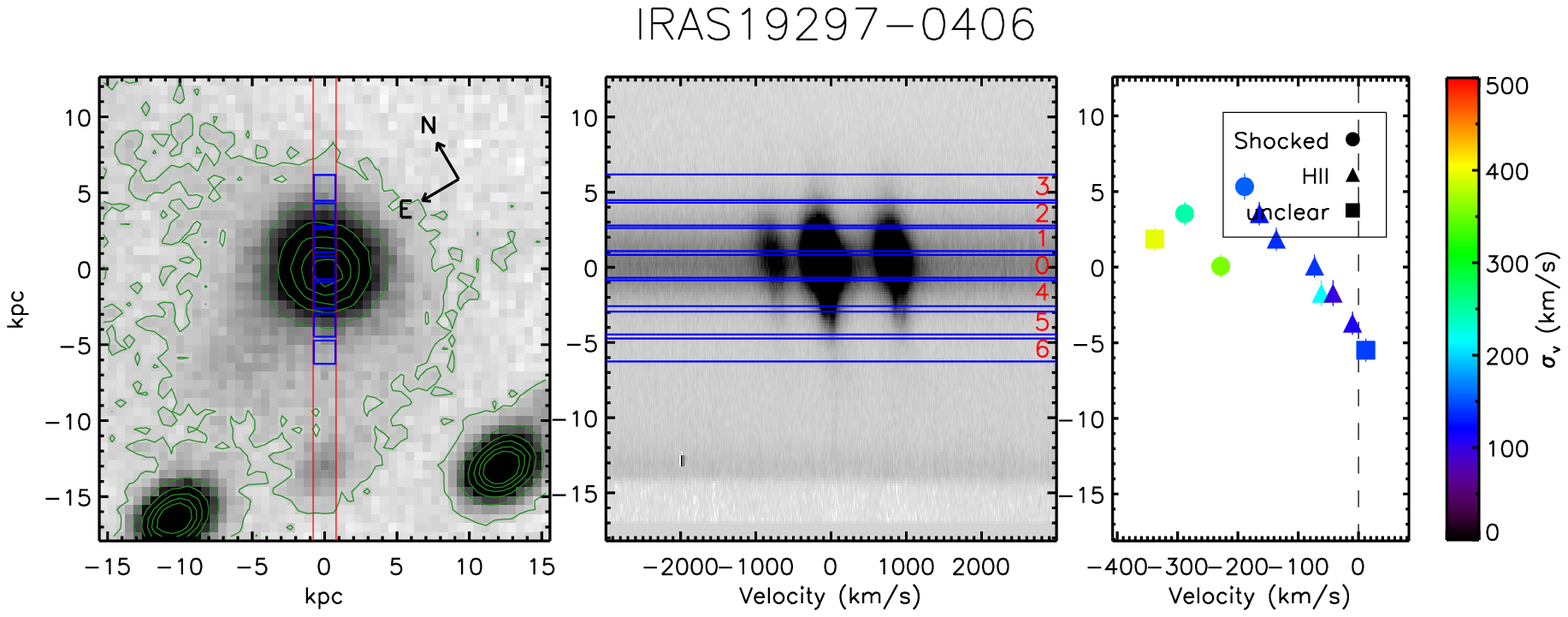, keepaspectratio=true, width=\linewidth}
\epsfig{file=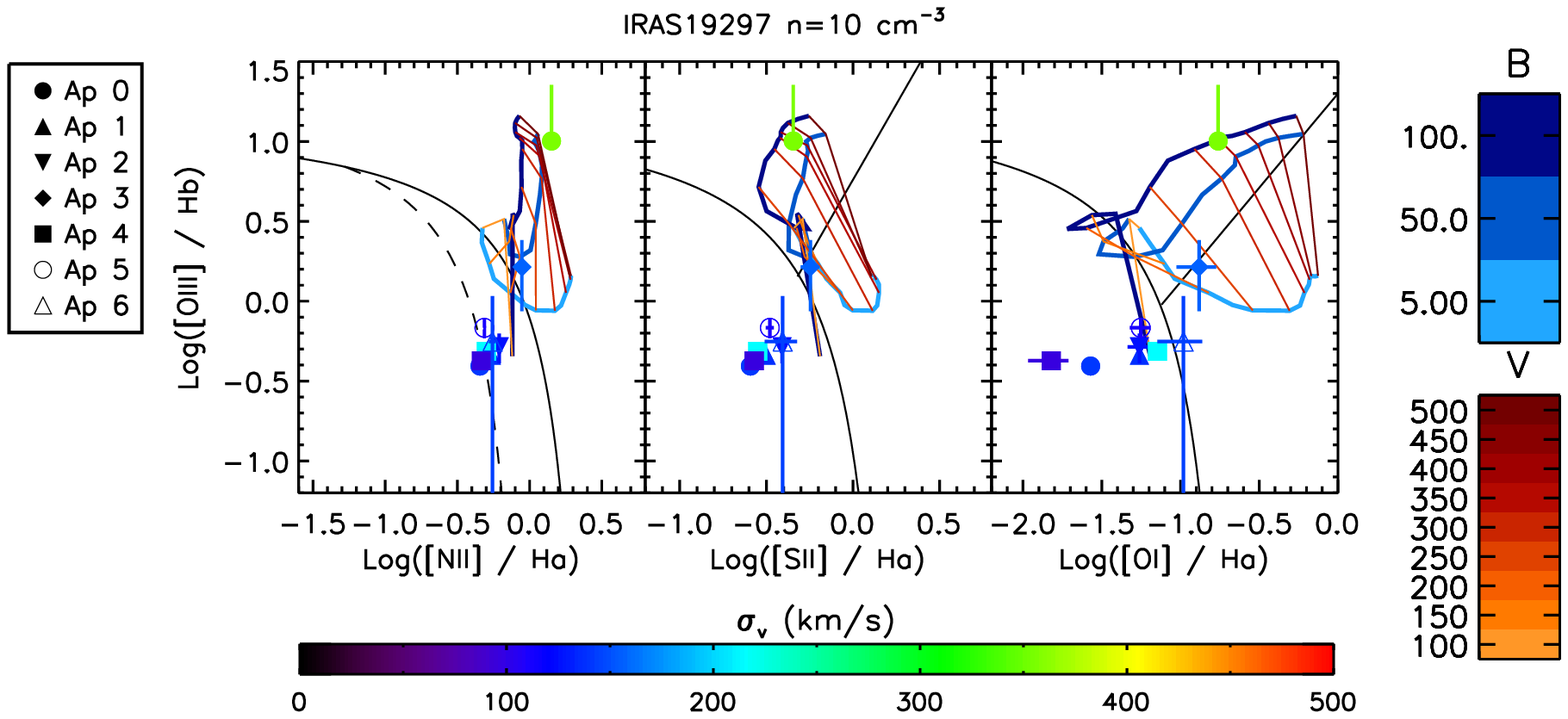, keepaspectratio=true, width=\linewidth}
\caption{\label{fig:i19297_bbc}
}
\end{figure}
\clearpage

\begin{figure}[h]
\centering
\epsfig{file=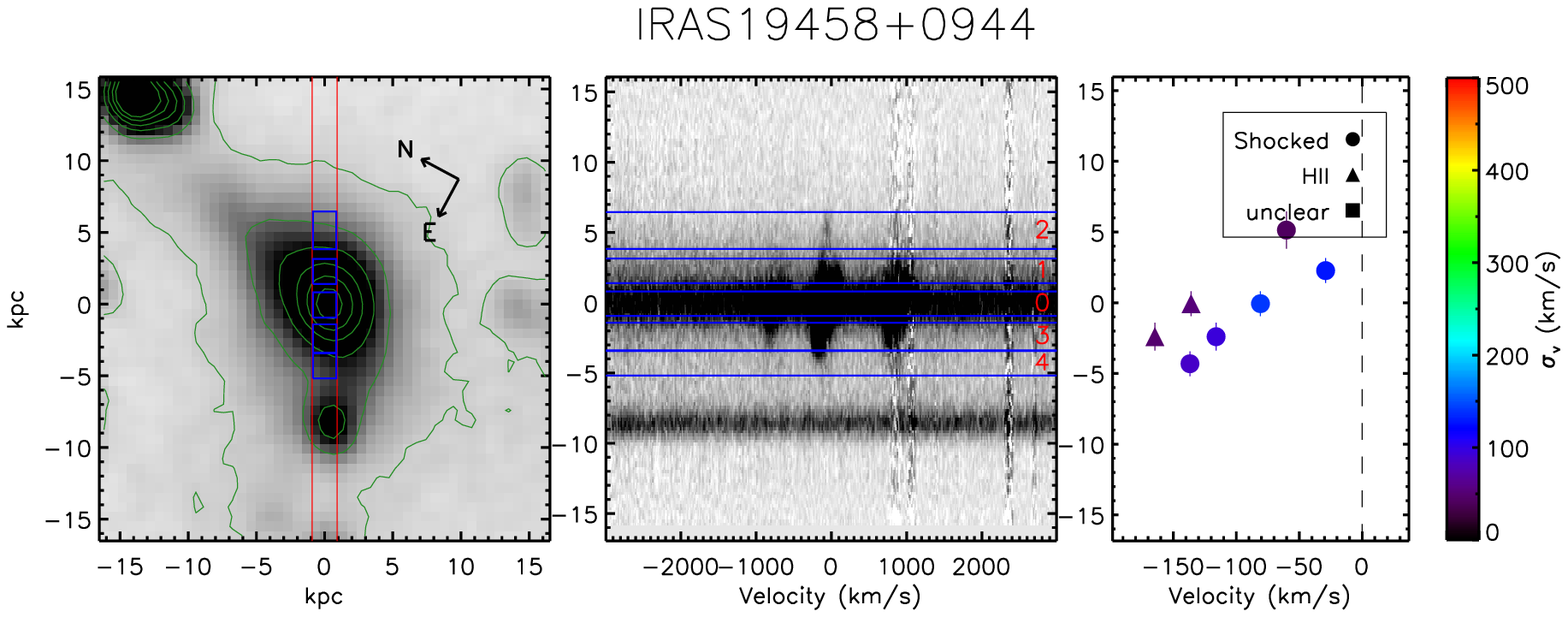, keepaspectratio=true, width=\linewidth}
\epsfig{file=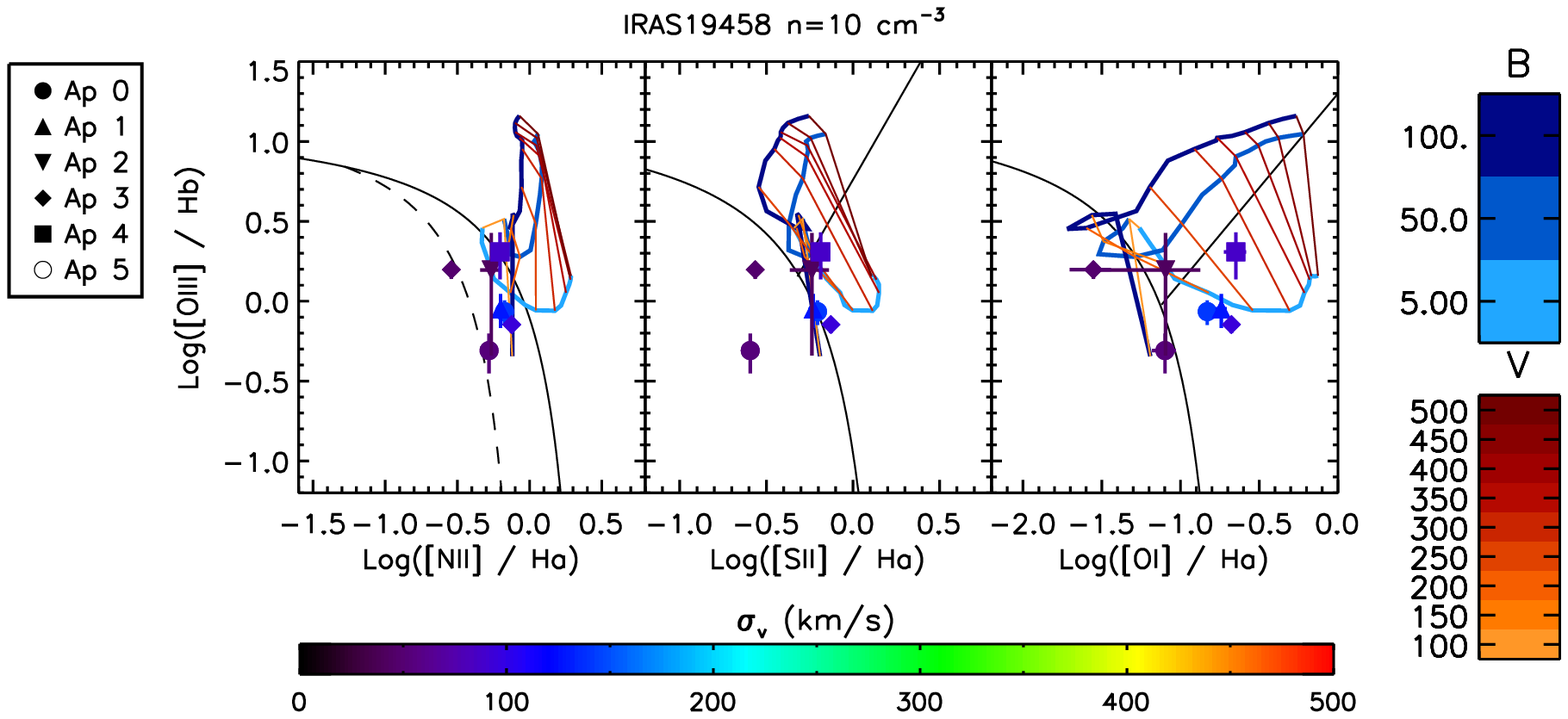, keepaspectratio=true, width=\linewidth}
\caption{\label{fig:i19458_bbc}
}
\end{figure}
\clearpage

\begin{figure}[h]
\centering
\epsfig{file=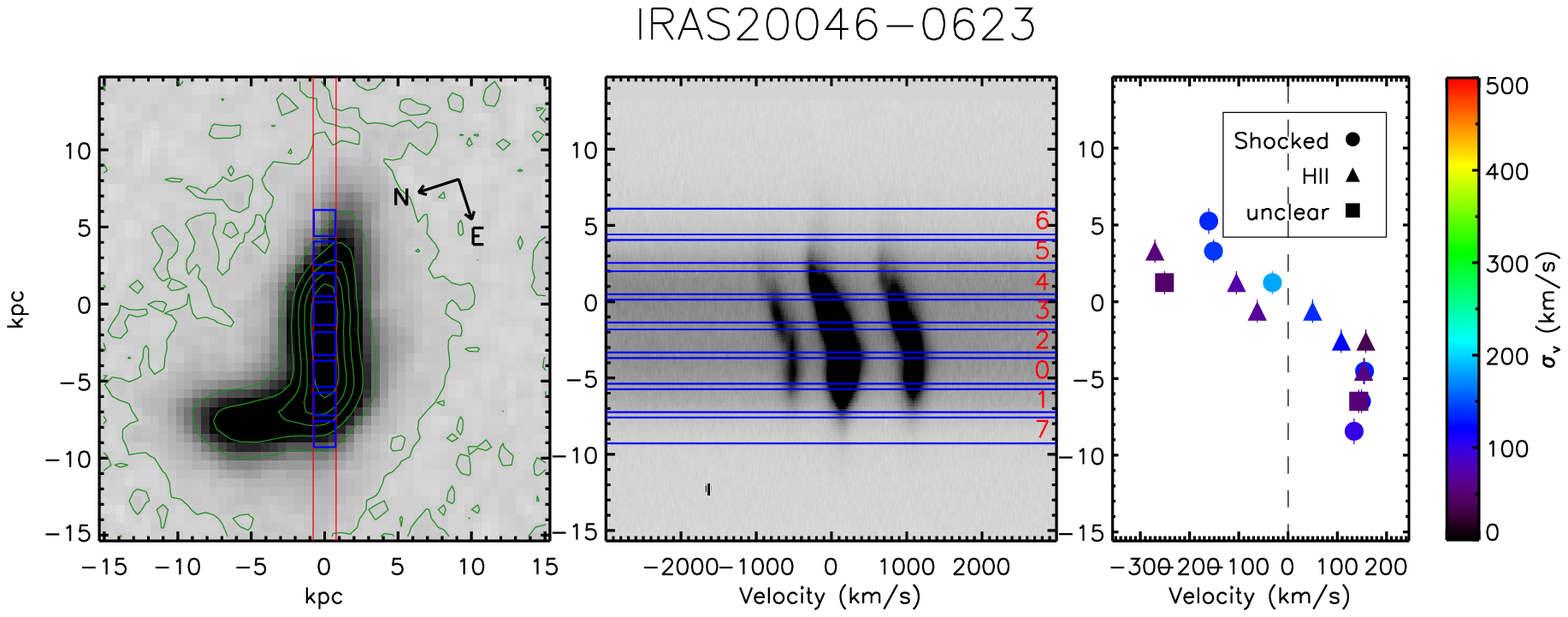, keepaspectratio=true, width=\linewidth}
\epsfig{file=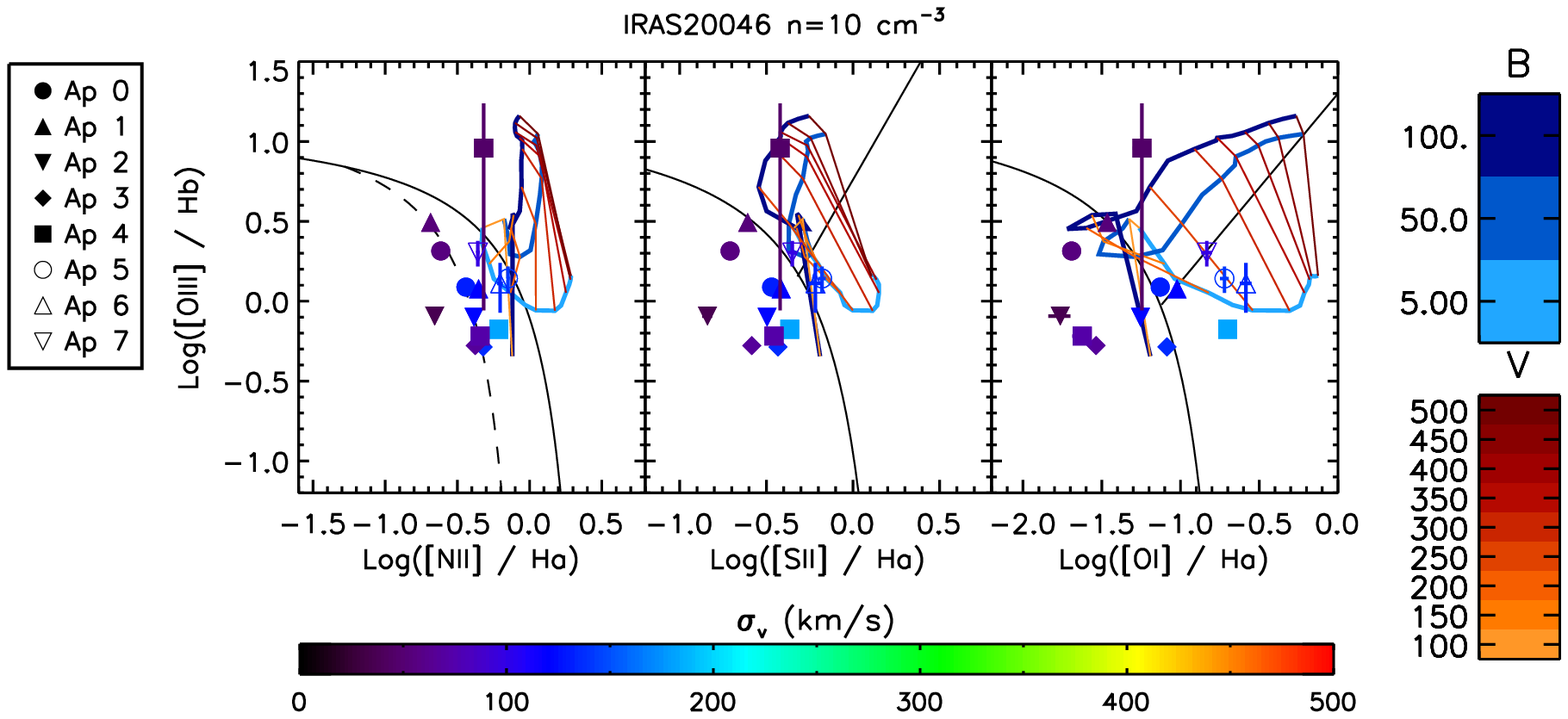, keepaspectratio=true, width=\linewidth}
\caption{\label{fig:i20046_bbc}
}
\end{figure}
\clearpage

\begin{figure}[h]
\centering
\epsfig{file=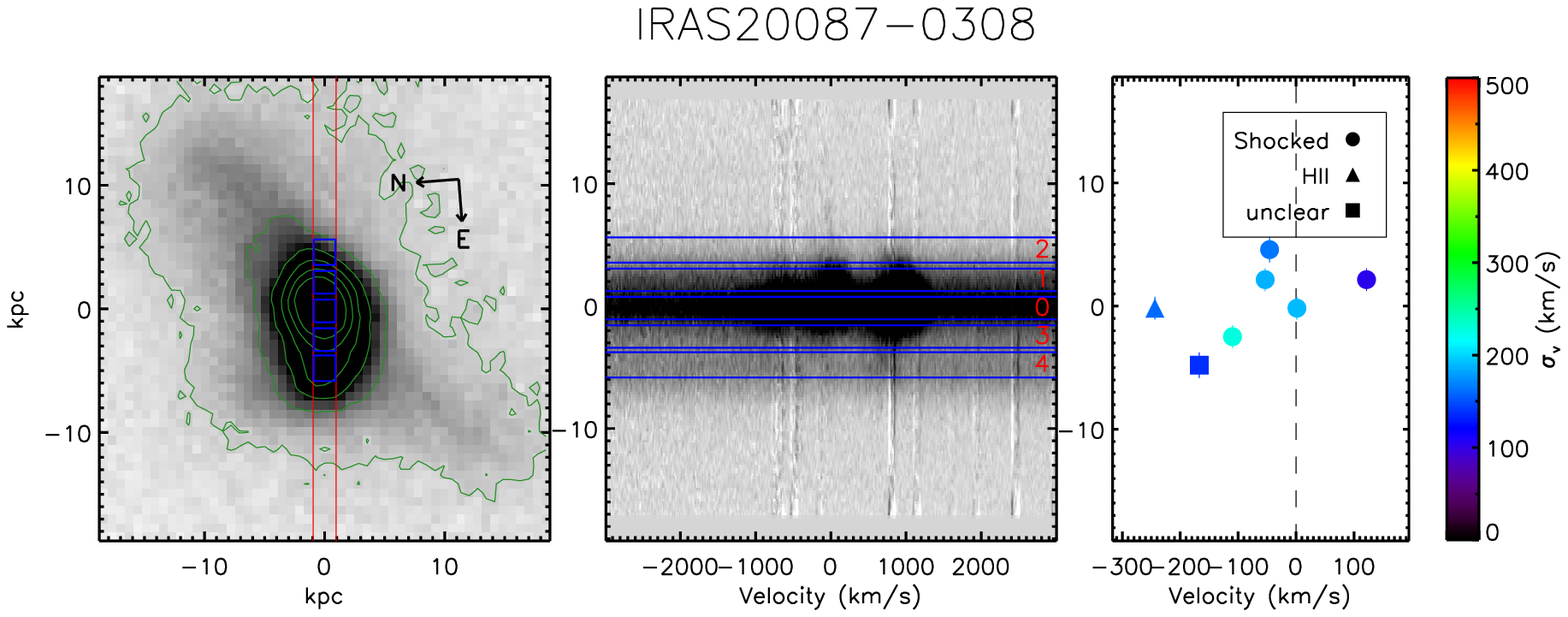, keepaspectratio=true, width=\linewidth}
\epsfig{file=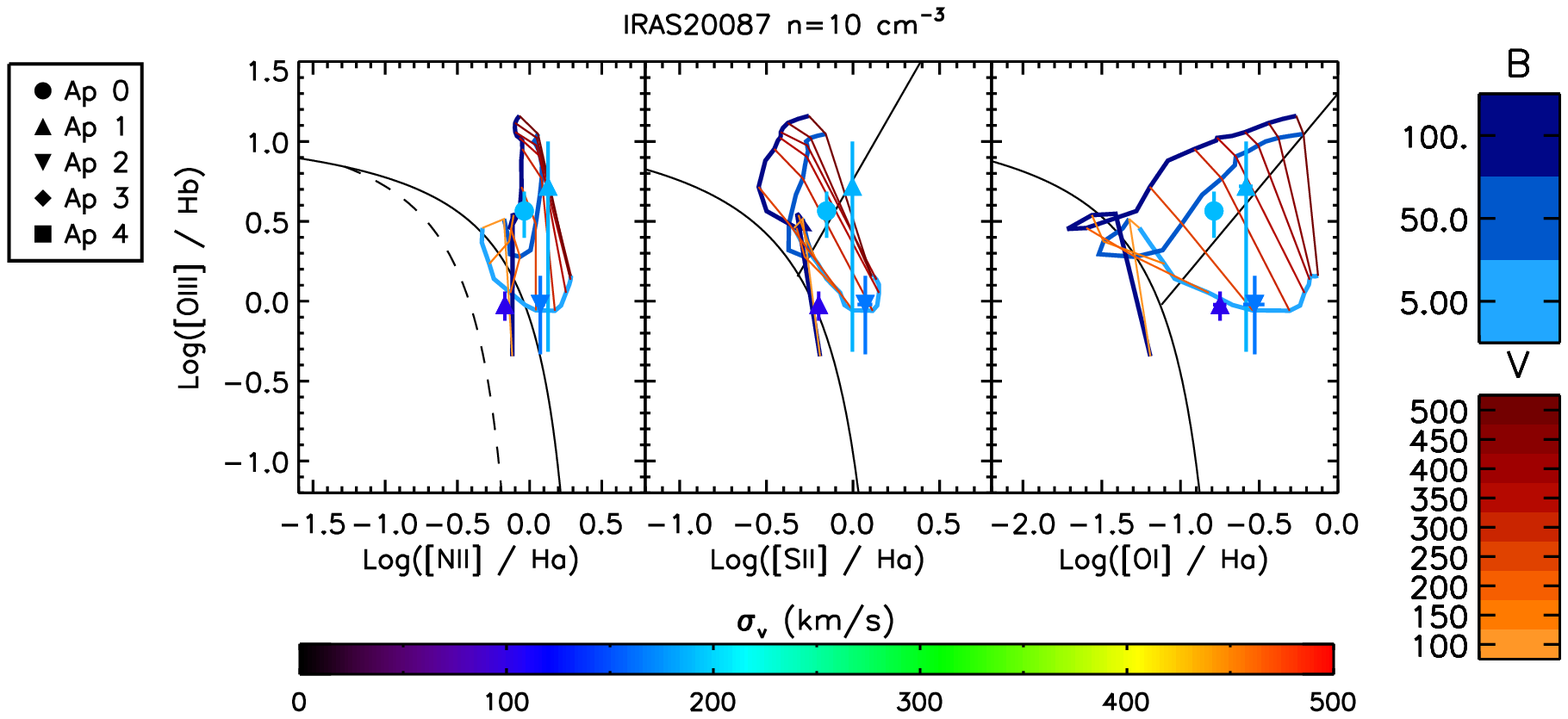, keepaspectratio=true, width=\linewidth}
\caption{\label{fig:i20087_bbc}
}
\end{figure}
\clearpage

\begin{figure}[h]
\centering
\epsfig{file=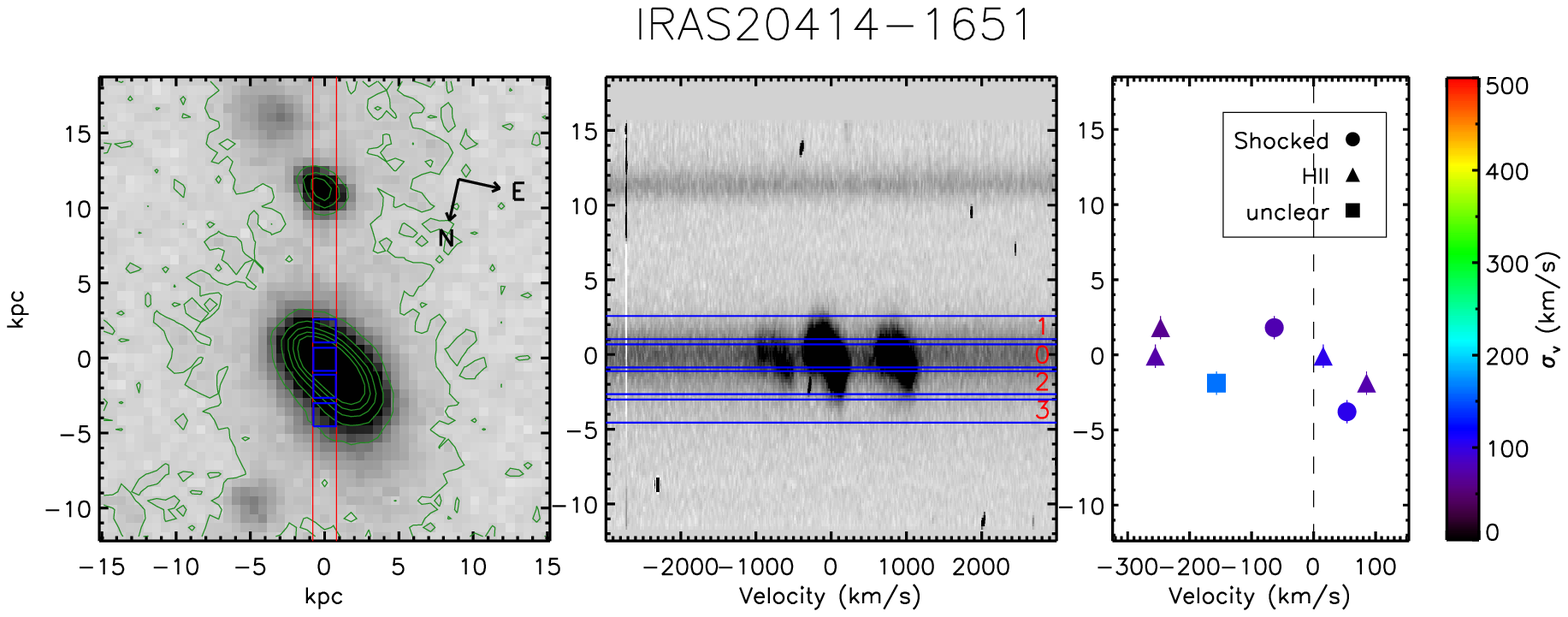, keepaspectratio=true, width=\linewidth}
\epsfig{file=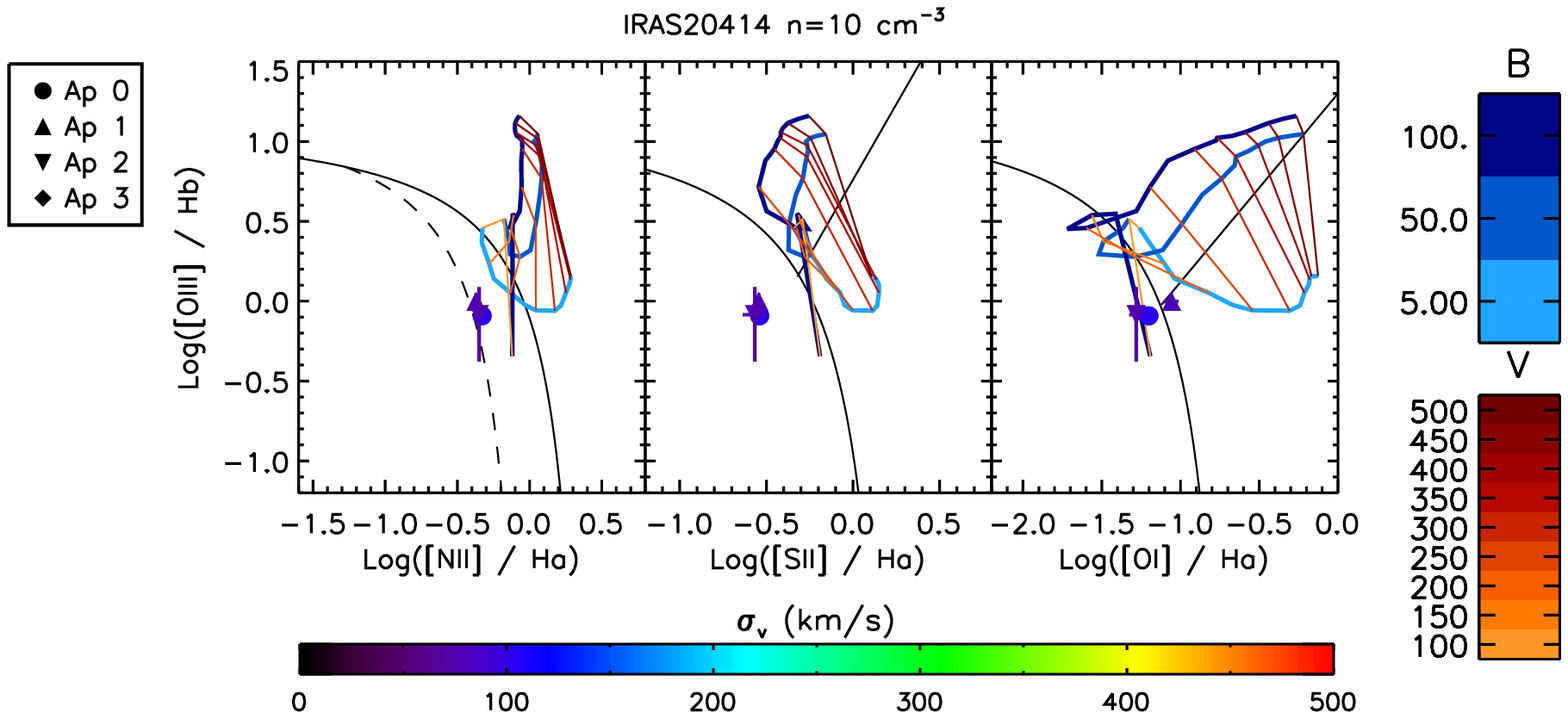, keepaspectratio=true, width=\linewidth}
\caption{\label{fig:i20414_bbc}
}
\end{figure}
\clearpage

\begin{figure}[h]
\centering
\epsfig{file=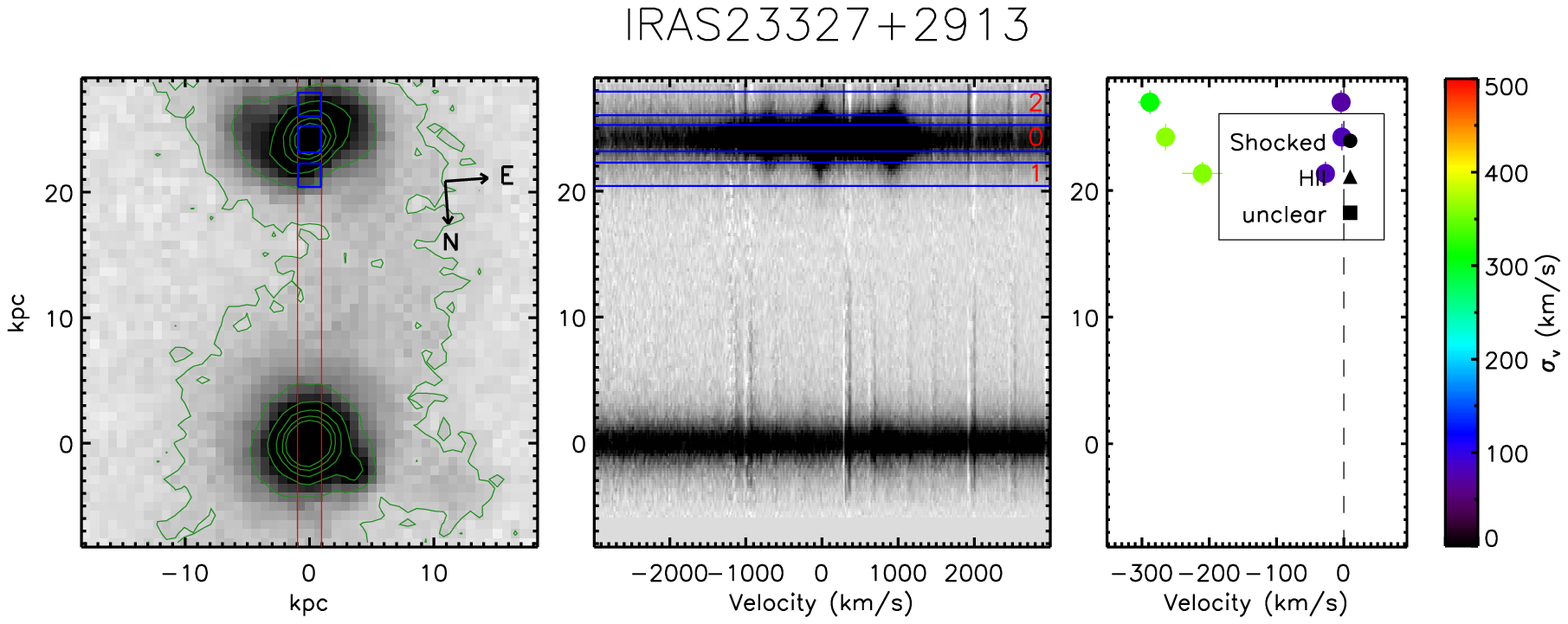, keepaspectratio=true, width=\linewidth}
\epsfig{file=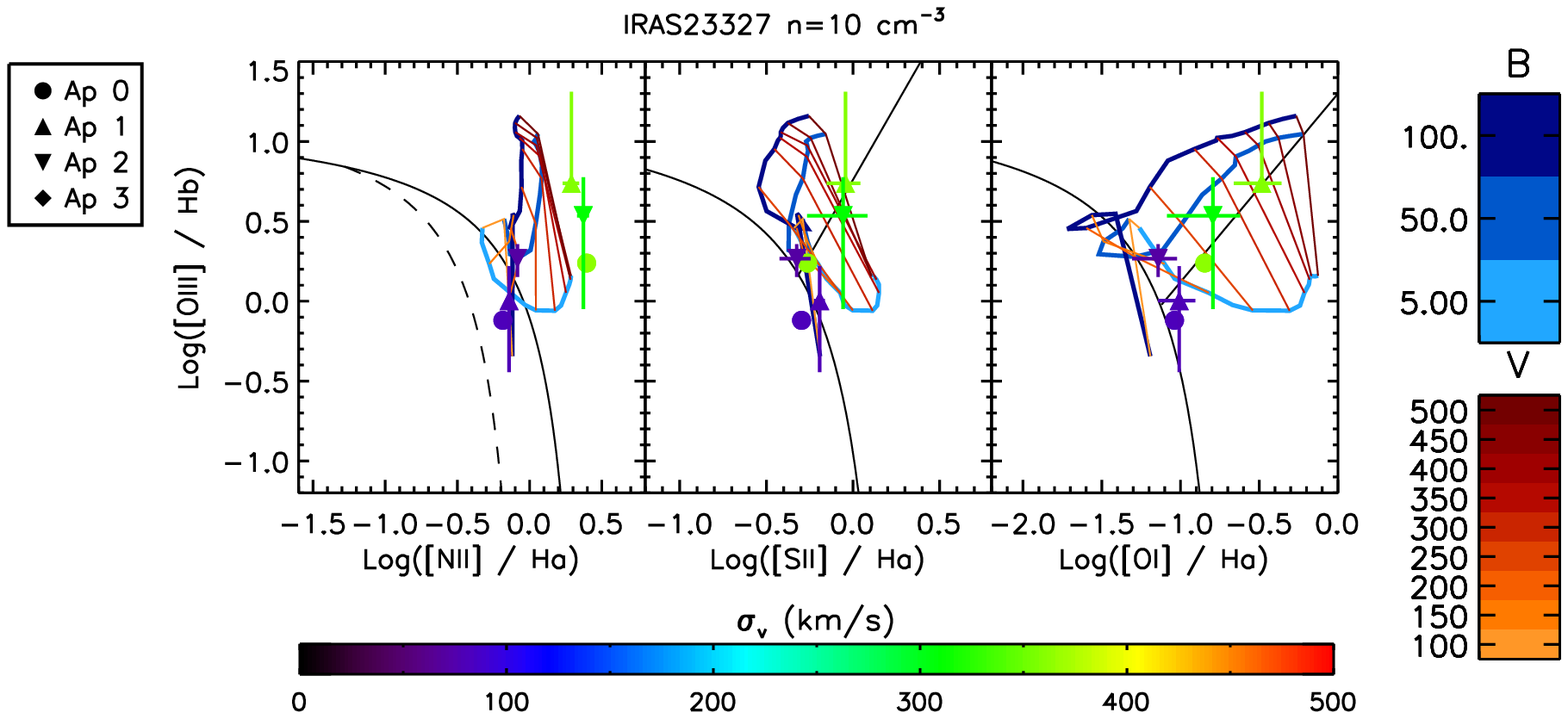, keepaspectratio=true, width=\linewidth}
\caption{\label{fig:i23327_bbc}
}
\end{figure}
\clearpage

\begin{figure}[h]
\centering
\epsfig{file=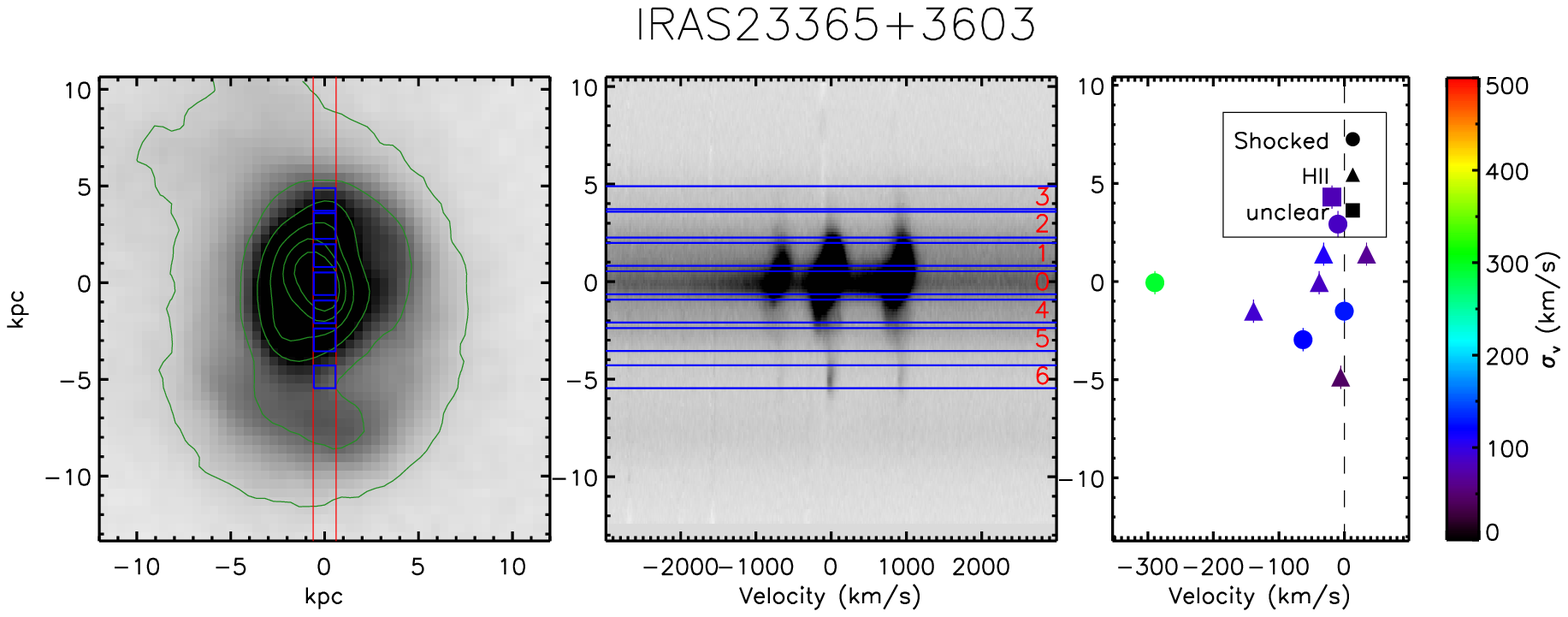, keepaspectratio=true, width=\linewidth}
\epsfig{file=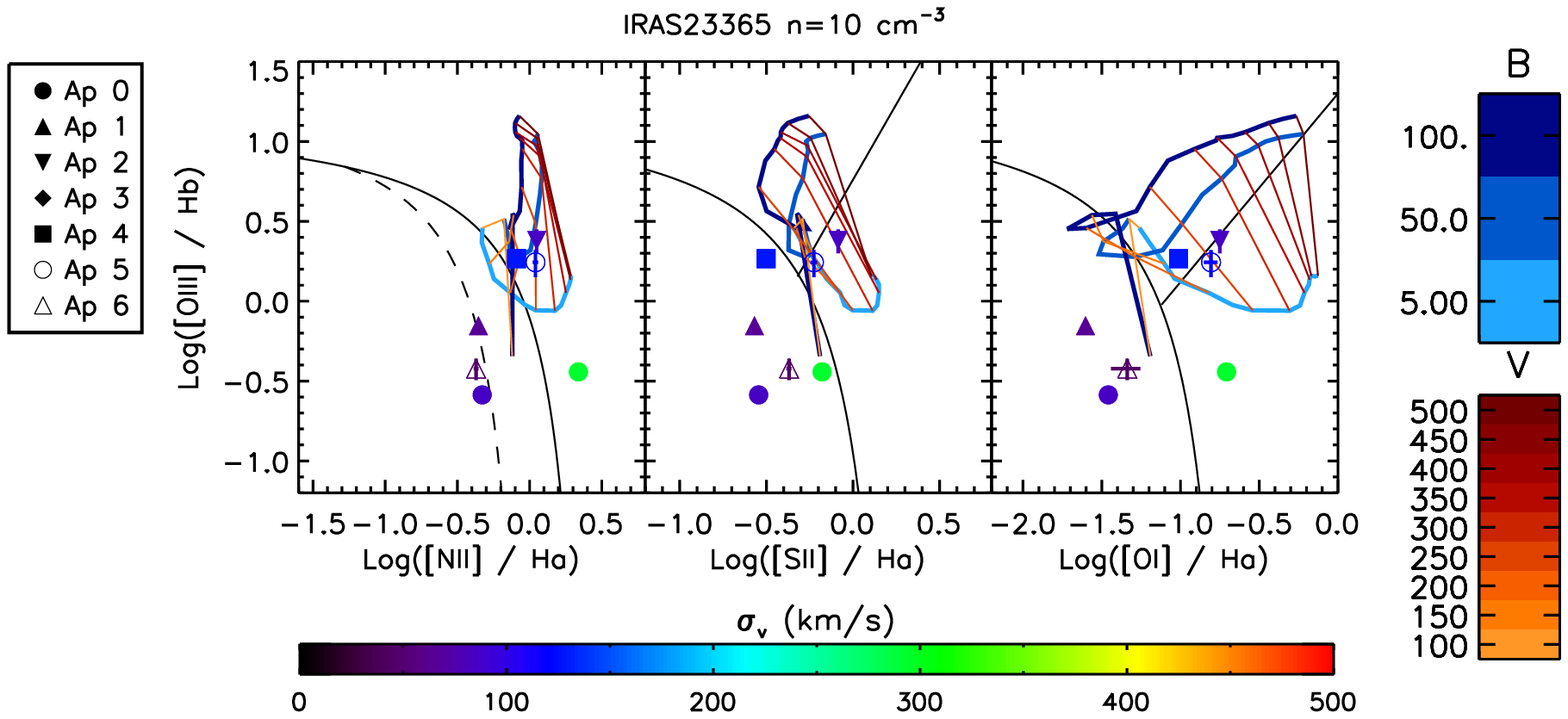, keepaspectratio=true, width=\linewidth}
\caption{\label{fig:i23365_bbc}
}
\end{figure}
\clearpage


\clearpage
\section{Appendix B}
\setcounter{figure}{0}

In this appendix we include the emission line profiles for the transitions identified by the 
	apertures figure set of Appendix A.  
	The measured fluxes and estimated shock velocities from the line ratios are presented in 
		Appendix C. 

\begin{figure}[h]
\centering
\epsfig{file=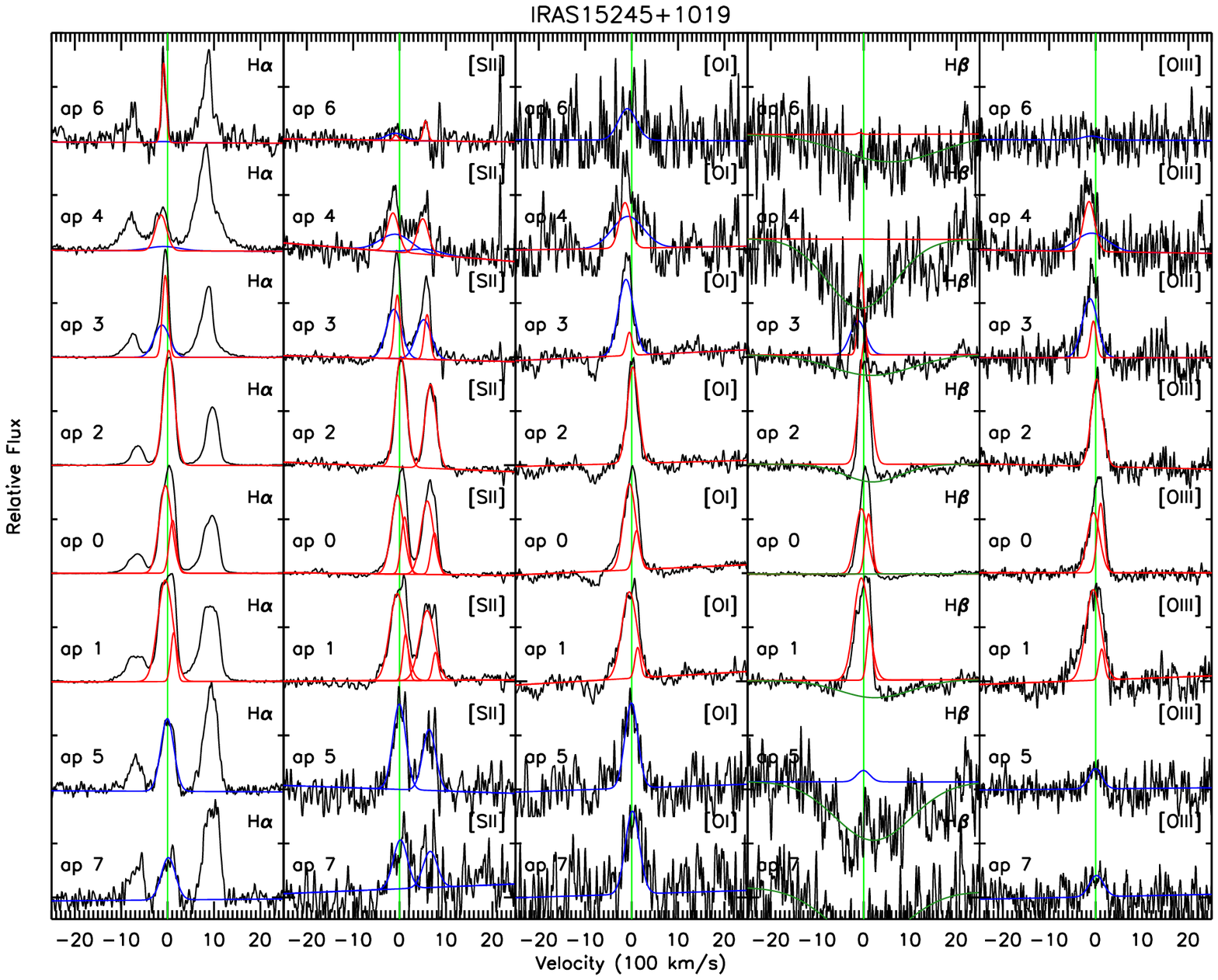, keepaspectratio=true, width=0.9\linewidth}
\caption{\label{fig:i15245_prof}
\scriptsize
Line profiles for each measured transition (horizontal axis) as a function of aperture
(vertical axis). Each cell is labeled with the transition and and 
aperture number identified in Fig. \ref{fig:i15245_bbc}.  The scales have been adjusted 
to illustrate the line profile in each transition. The red and blue lines identify the 
different gaussian components from the resulting simultaneous fit.  The dark green line represents 
the \Hb\ absorption component. For the left column, 
only the components of the \Ha\ transition are displayed to minimize confusion;
the [\ion{N}{2}] kinematics are the same as the other 
transitions, while the amplitudes can vary. The full sample is included in the online version of 
the article.
}
\end{figure}

\begin{figure}[h]
\centering
\epsfig{file=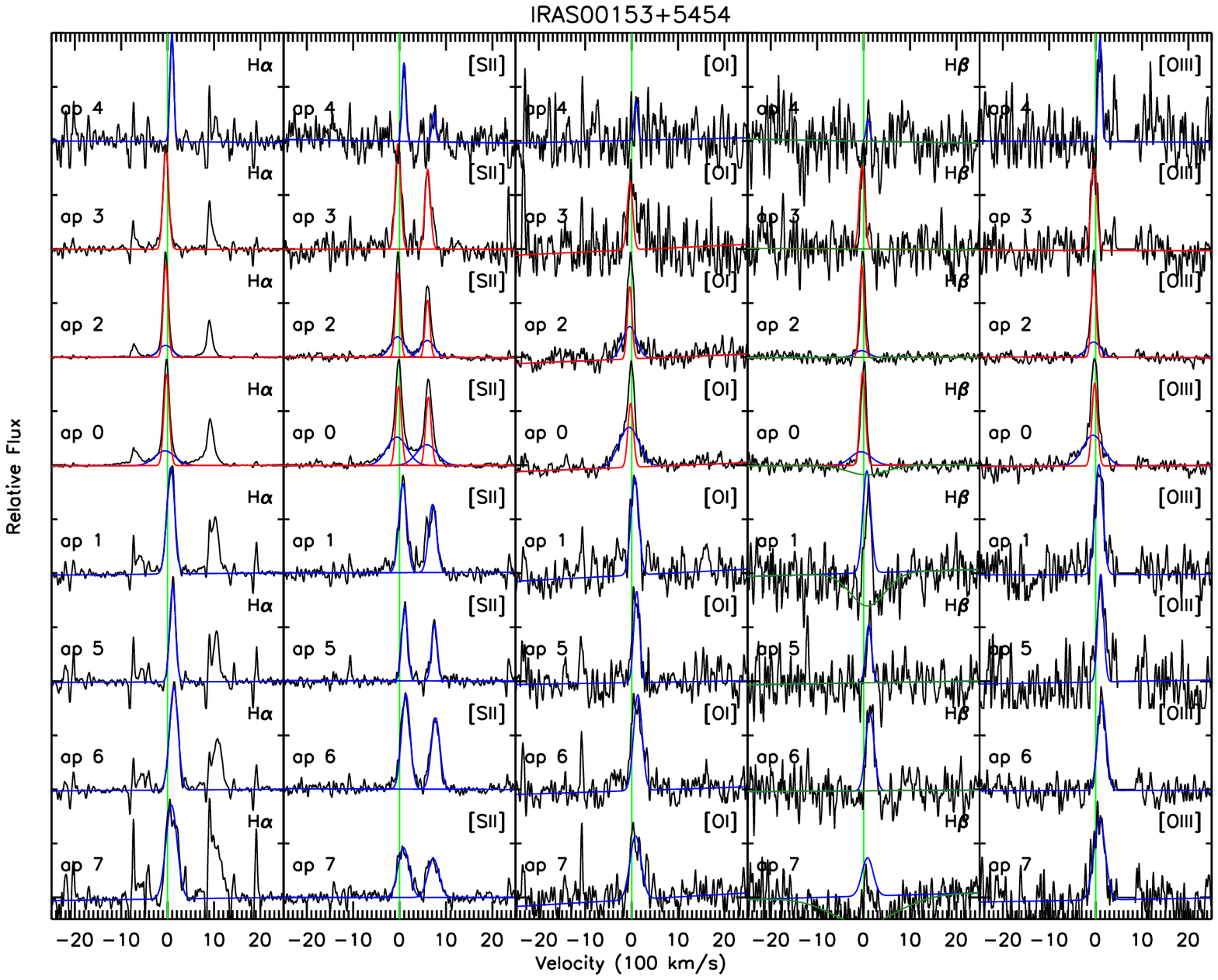, keepaspectratio=true, width=0.9\linewidth}
\caption{\label{fig:i00153_prof}
}
\end{figure}
\clearpage

\begin{figure}[h]
\centering
\epsfig{file=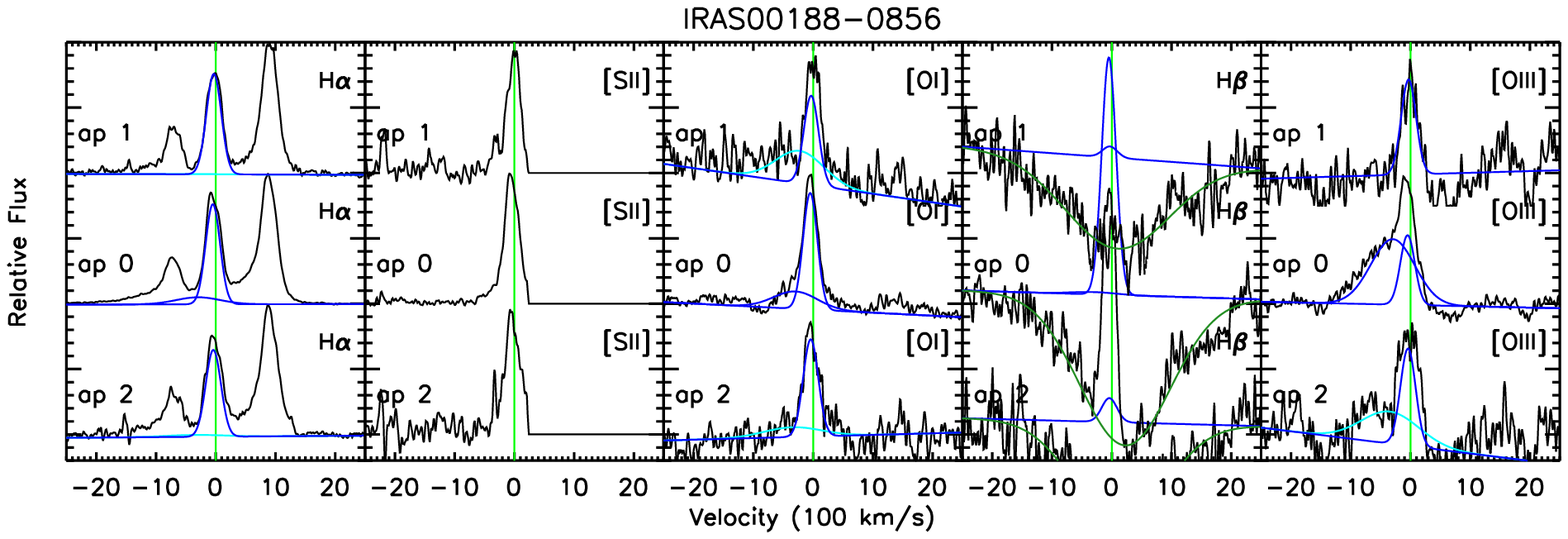, keepaspectratio=true, width=0.9\linewidth}
\caption{\label{fig:i00188_prof}
}
\end{figure}
\clearpage

\begin{figure}[h]
\centering
\epsfig{file=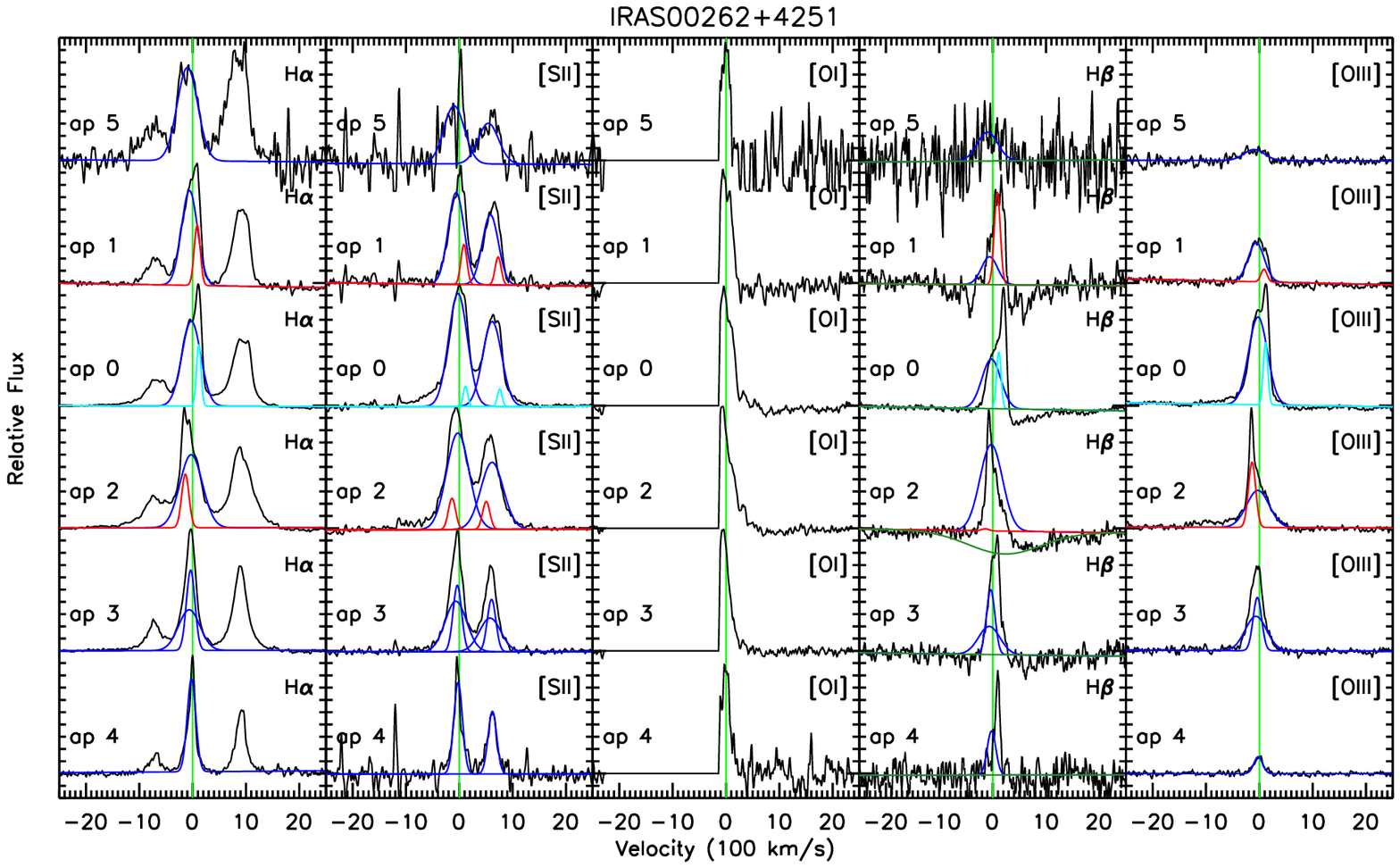, keepaspectratio=true, width=0.9\linewidth}
\caption{\label{fig:i00262_prof}
}
\end{figure}
\clearpage

\begin{figure}[h]
\centering
\epsfig{file=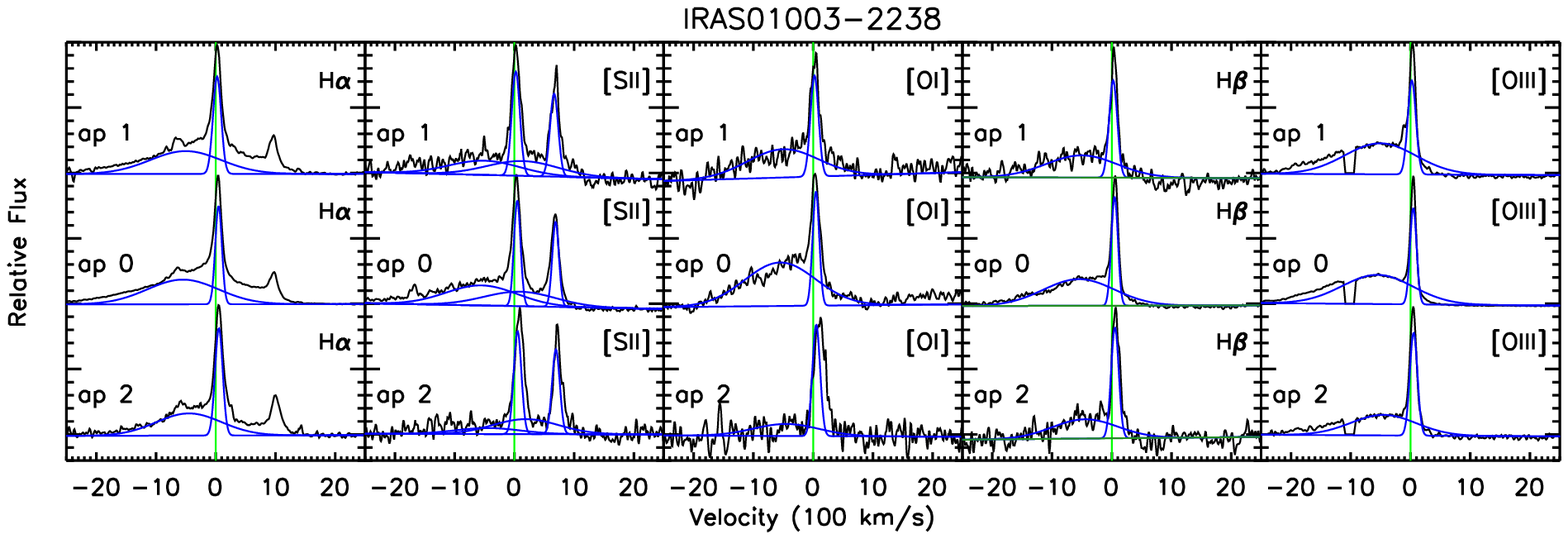, keepaspectratio=true, width=0.9\linewidth}
\caption{\label{fig:i01003_prof}
}
\end{figure}
\clearpage

\begin{figure}[h]
\centering
\epsfig{file=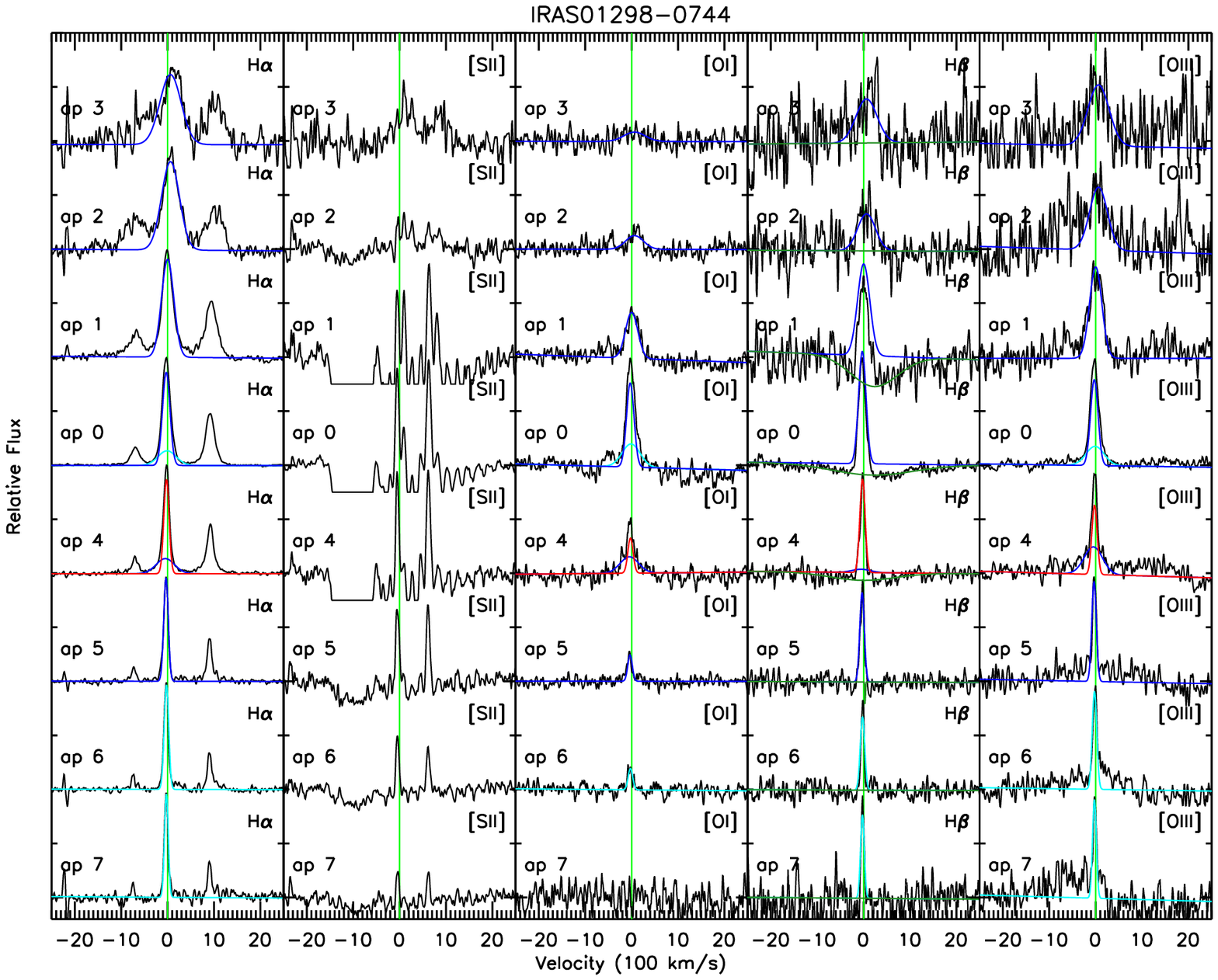, keepaspectratio=true, width=0.9\linewidth}
\caption{\label{fig:i01298_prof}
}
\end{figure}
\clearpage

\begin{figure}[h]
\centering
\epsfig{file=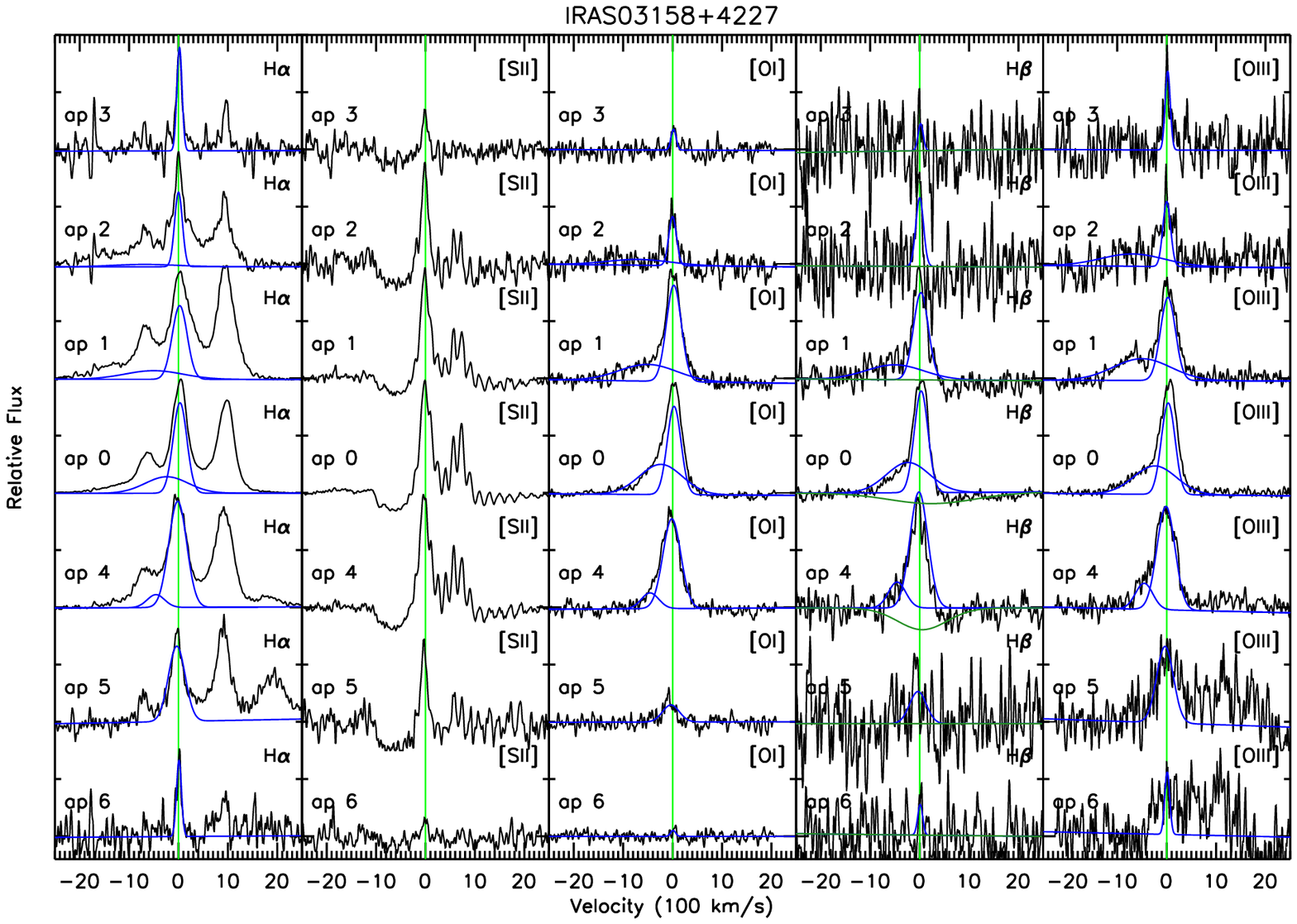, keepaspectratio=true, width=0.9\linewidth}
\caption{\label{fig:i03158_prof}
}
\end{figure}
\clearpage

\begin{figure}[h]
\centering
\epsfig{file=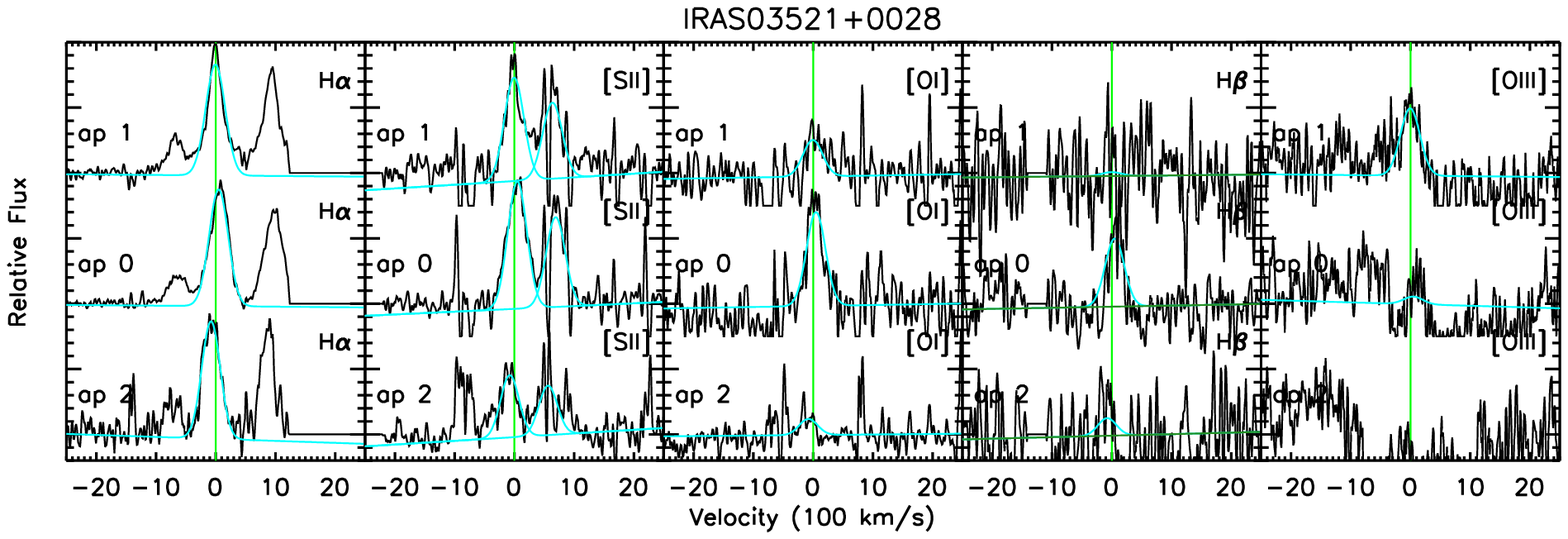, keepaspectratio=true, width=0.9\linewidth}
\caption{\label{fig:i03521_prof}
}
\end{figure}
\clearpage

\begin{figure}[h]
\centering
\epsfig{file=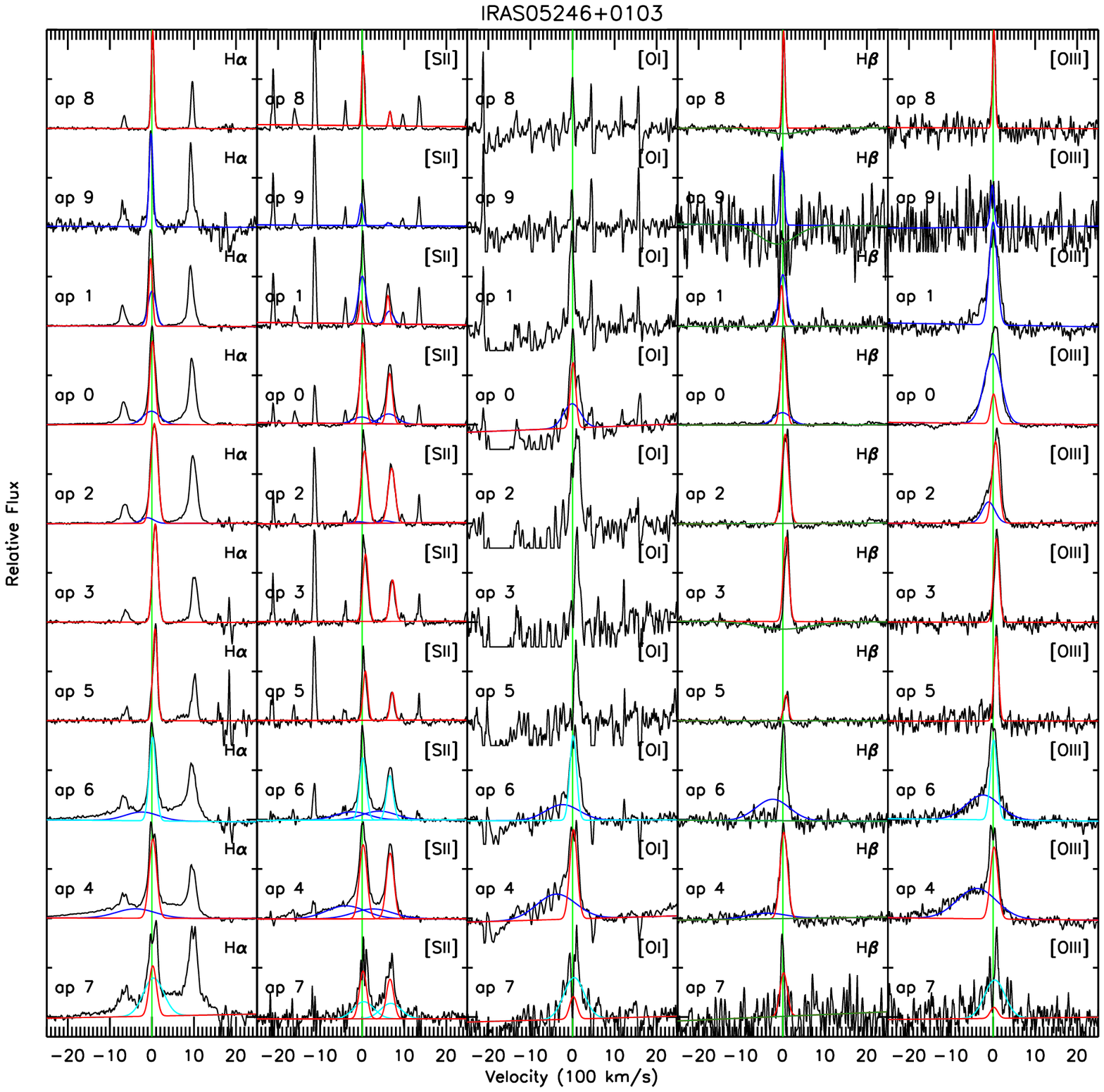, keepaspectratio=true, width=0.9\linewidth}
\caption{\label{fig:i05246_prof}
}
\end{figure}
\clearpage

\begin{figure}[h]
\centering
\epsfig{file=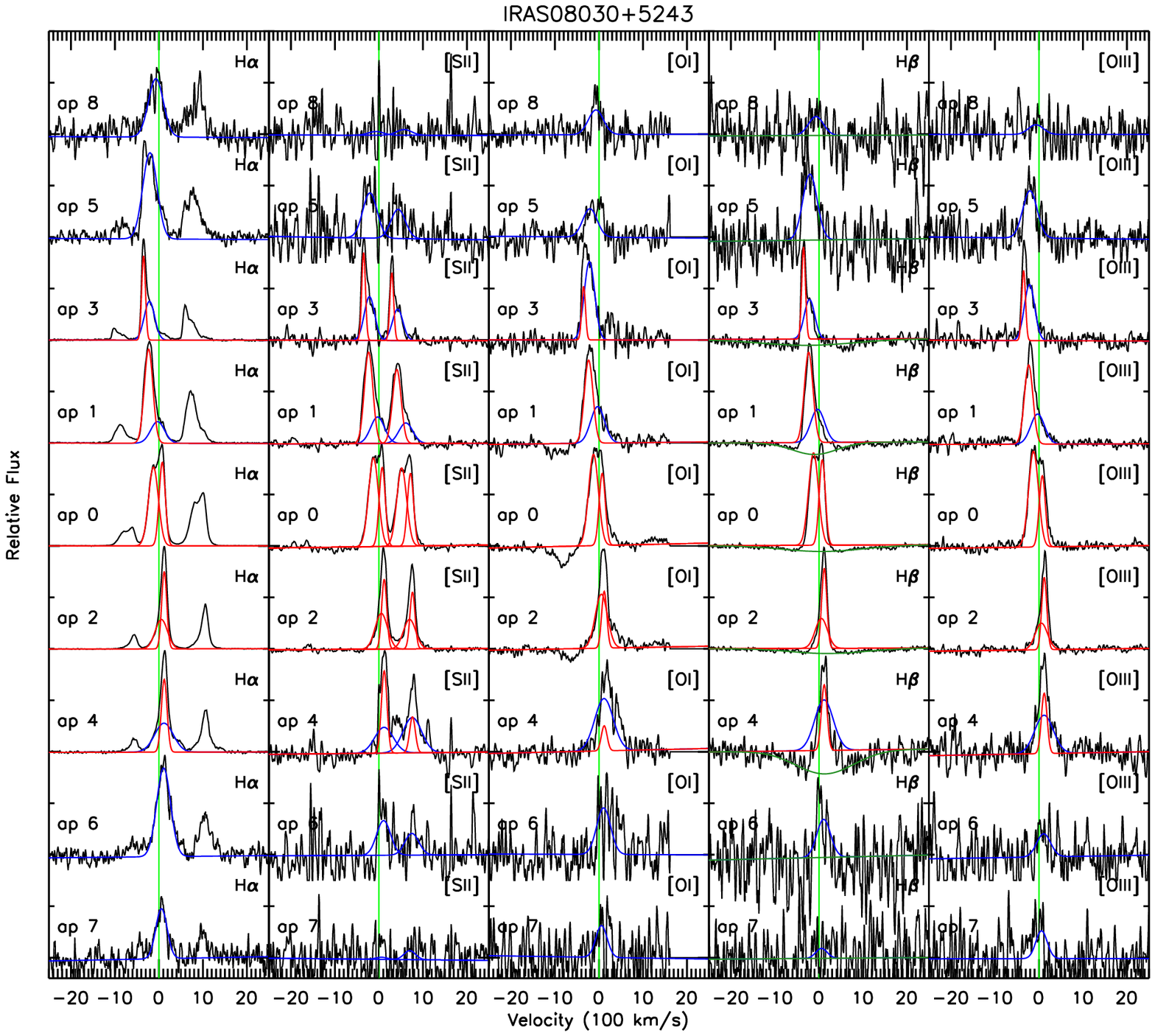, keepaspectratio=true, width=0.9\linewidth}
\caption{\label{fig:i08030_prof}
}
\end{figure}
\clearpage

\begin{figure}[h]
\centering
\epsfig{file=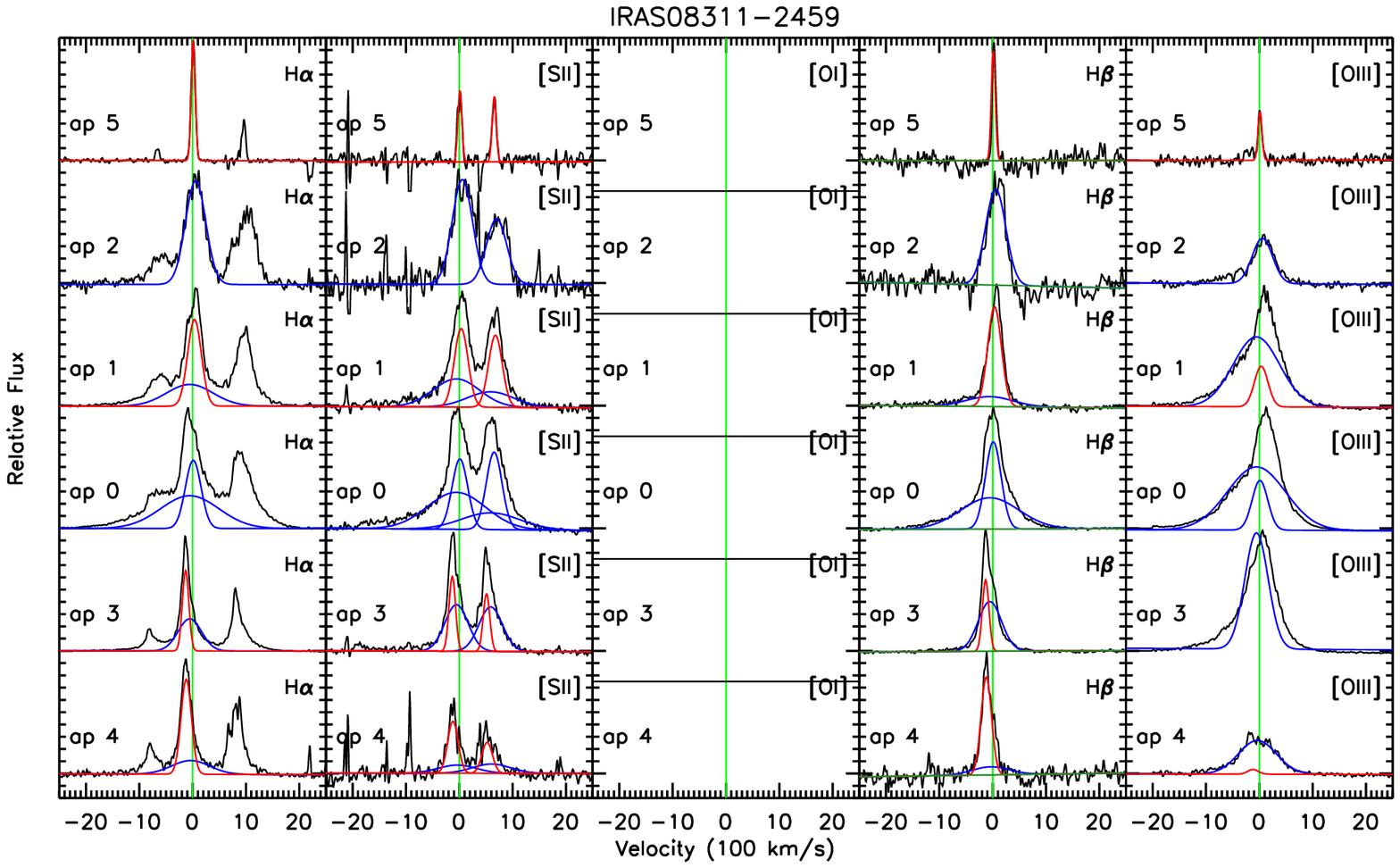, keepaspectratio=true, width=0.9\linewidth}
\caption{\label{fig:i08311_prof}
}
\end{figure}
\clearpage

\begin{figure}[h]
\centering
\epsfig{file=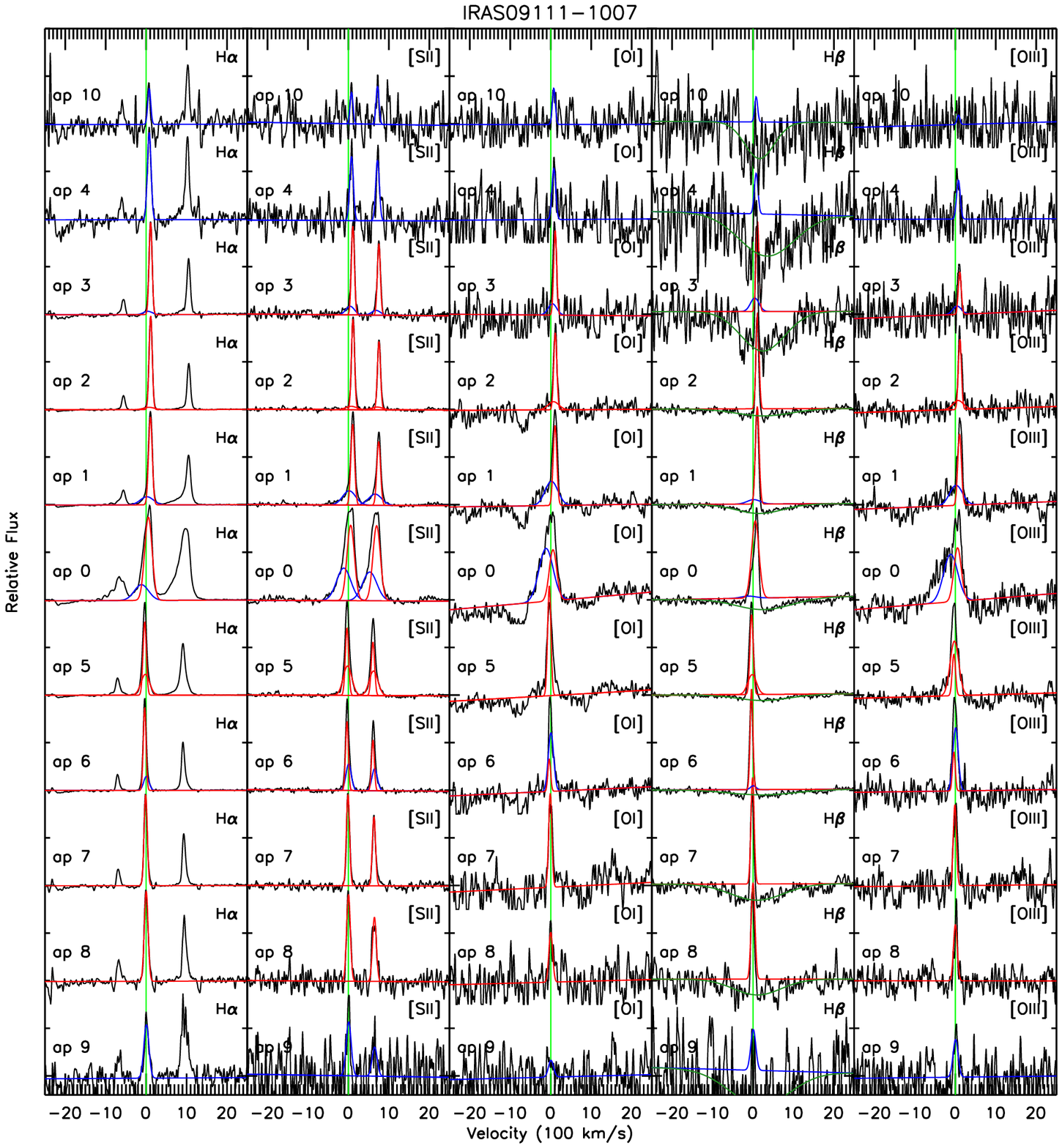, keepaspectratio=true, width=0.9\linewidth}
\caption{\label{fig:i09111_prof}
}
\end{figure}
\clearpage

\begin{figure}[h]
\centering
\epsfig{file=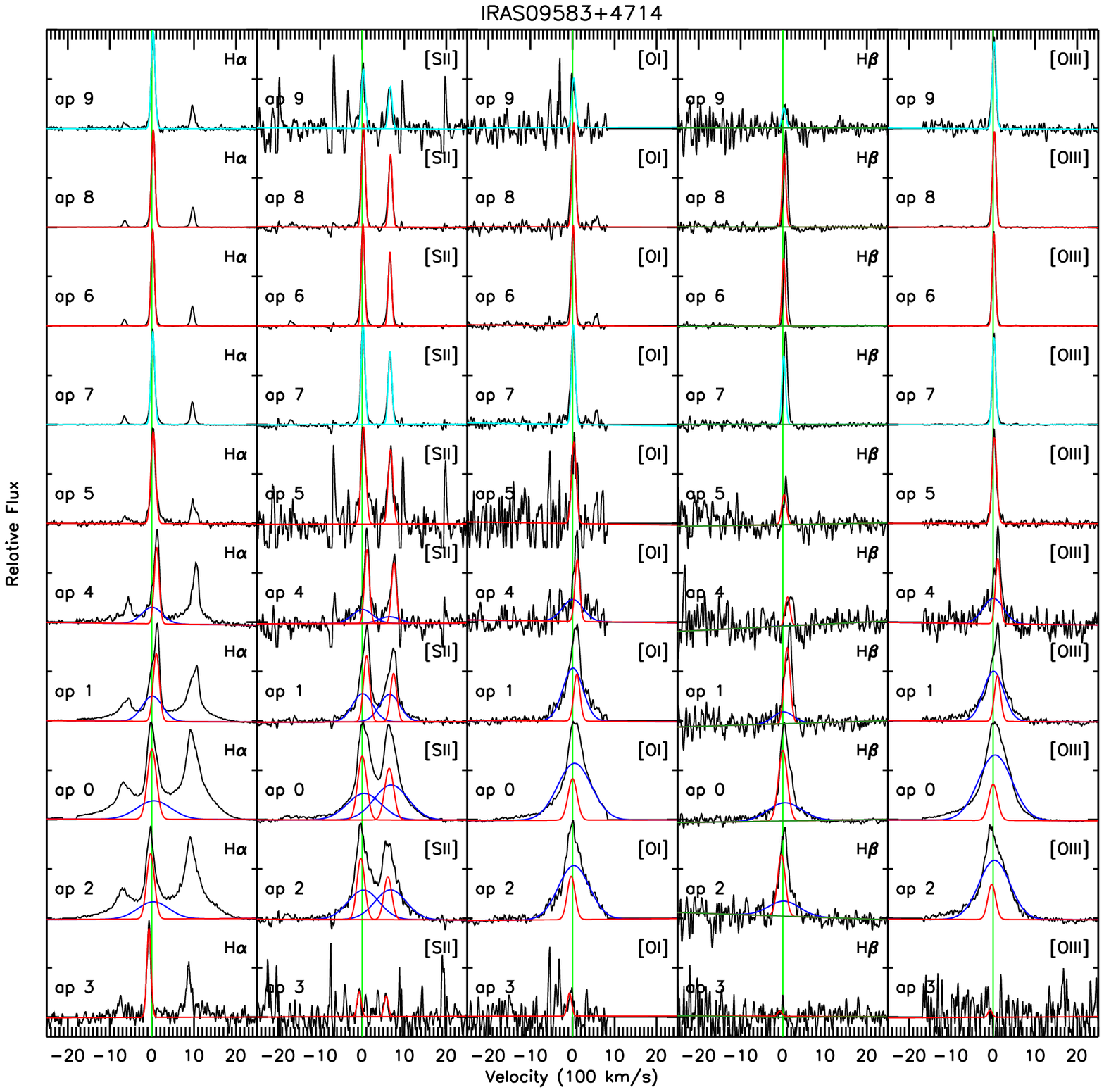, keepaspectratio=true, width=0.9\linewidth}
\caption{\label{fig:i09583_prof}
}
\end{figure}
\clearpage

\begin{figure}[h]
\centering
\epsfig{file=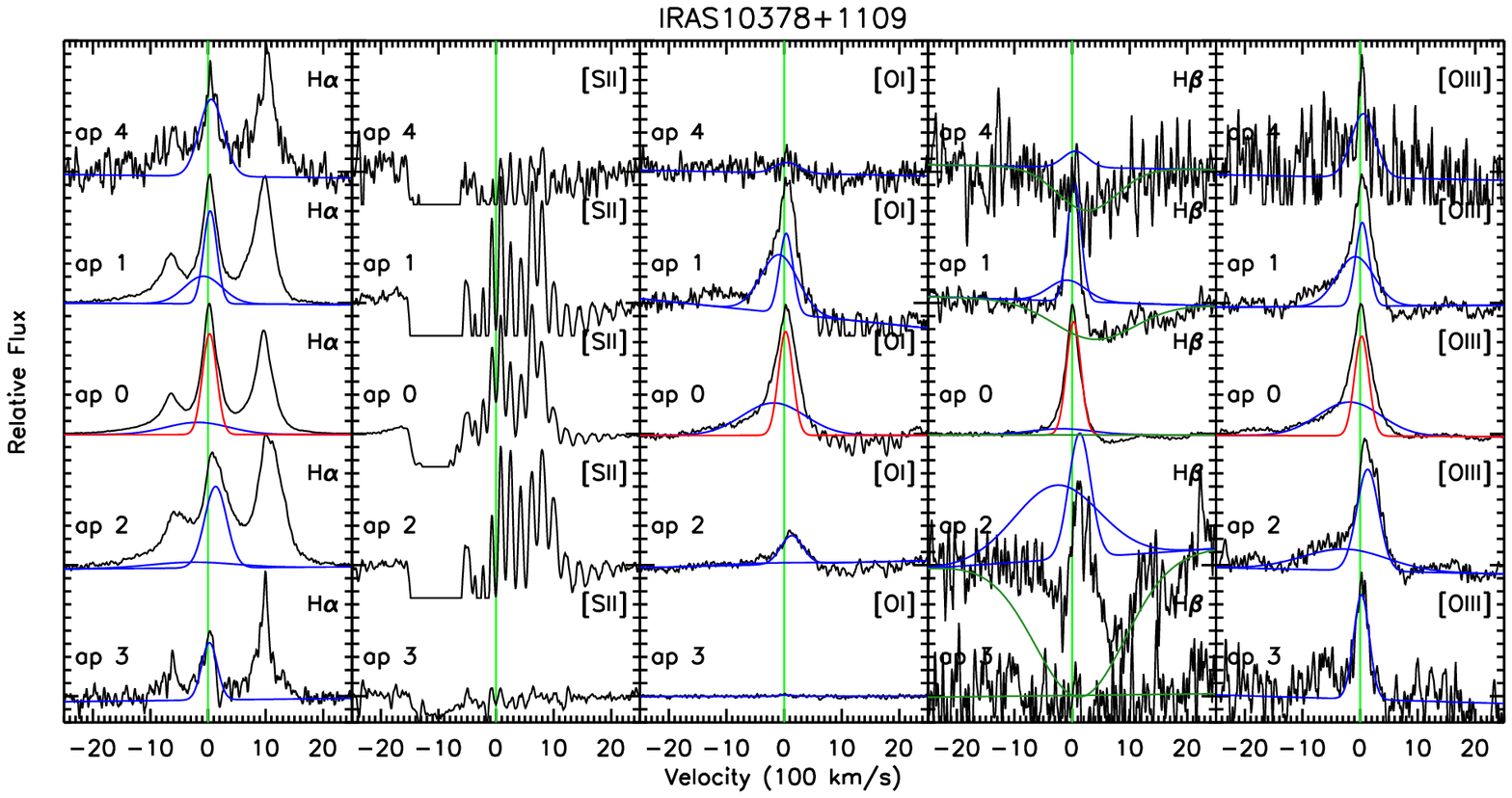, keepaspectratio=true, width=0.9\linewidth}
\caption{\label{fig:i10378_prof}
}
\end{figure}
\clearpage

\begin{figure}[h]
\centering
\epsfig{file=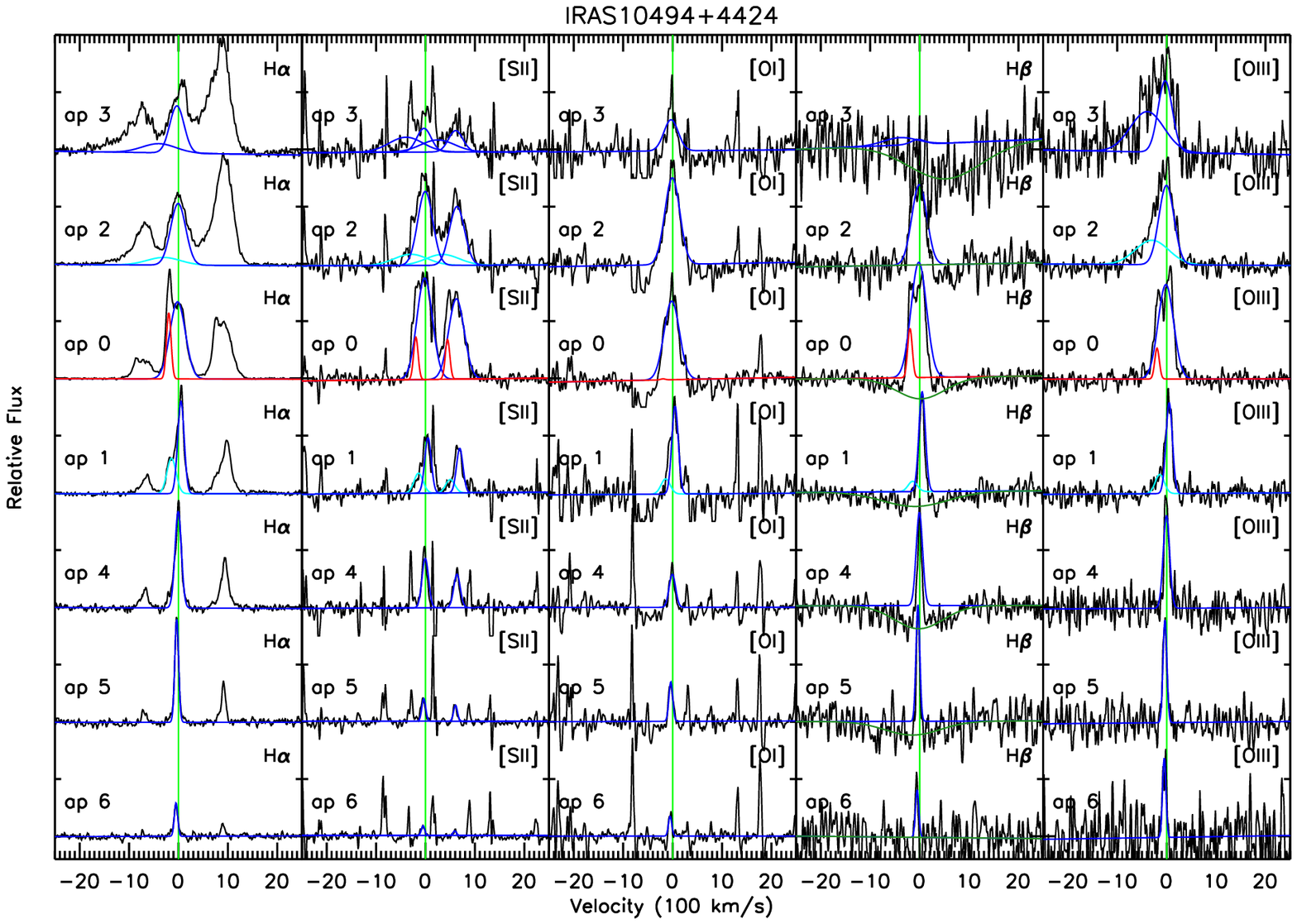, keepaspectratio=true, width=0.9\linewidth}
\caption{\label{fig:i10494_prof}
}
\end{figure}
\clearpage

\begin{figure}[h]
\centering
\epsfig{file=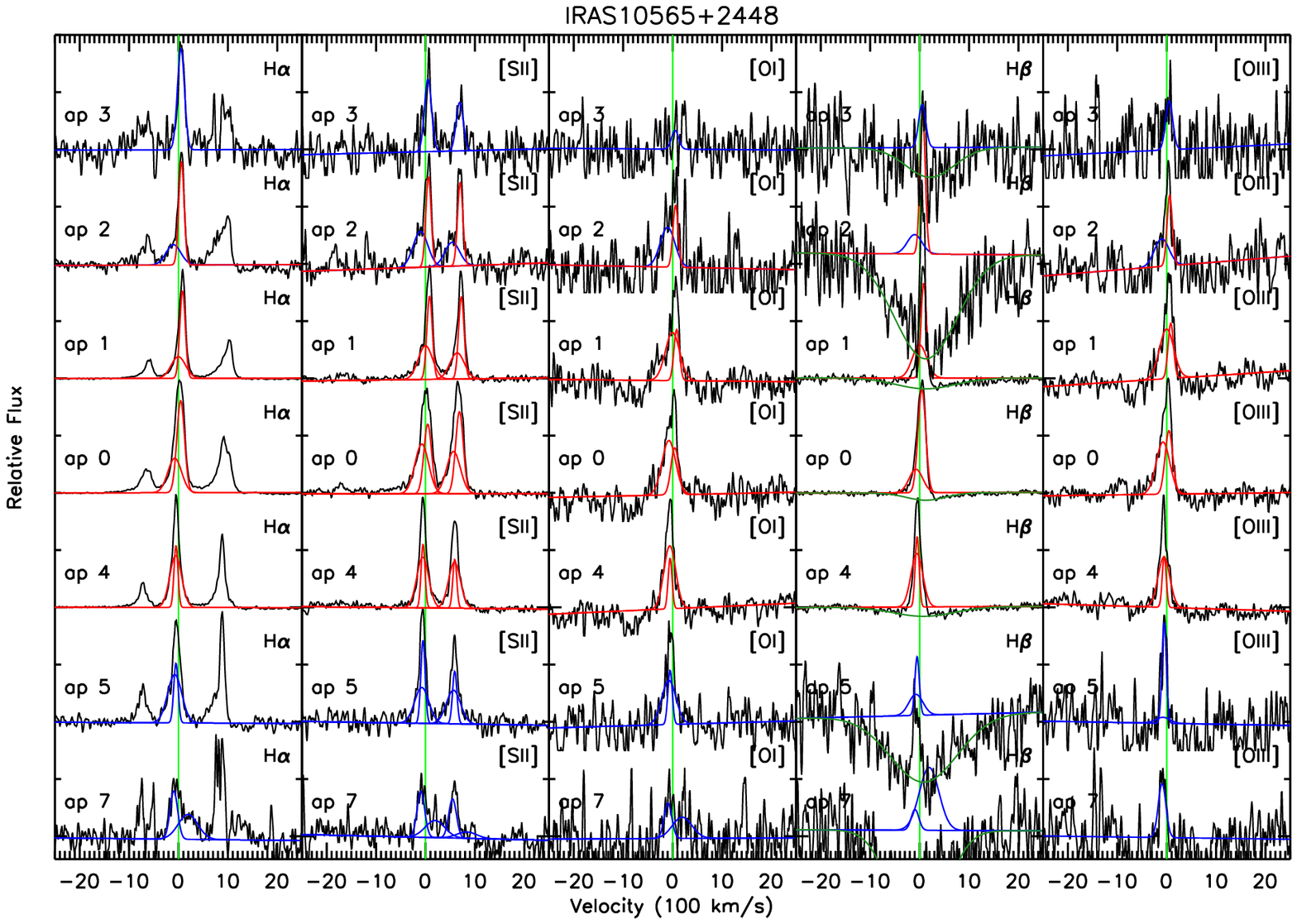, keepaspectratio=true, width=0.9\linewidth}
\caption{\label{fig:i10565_prof}
}
\end{figure}
\clearpage

\begin{figure}[h]
\centering
\epsfig{file=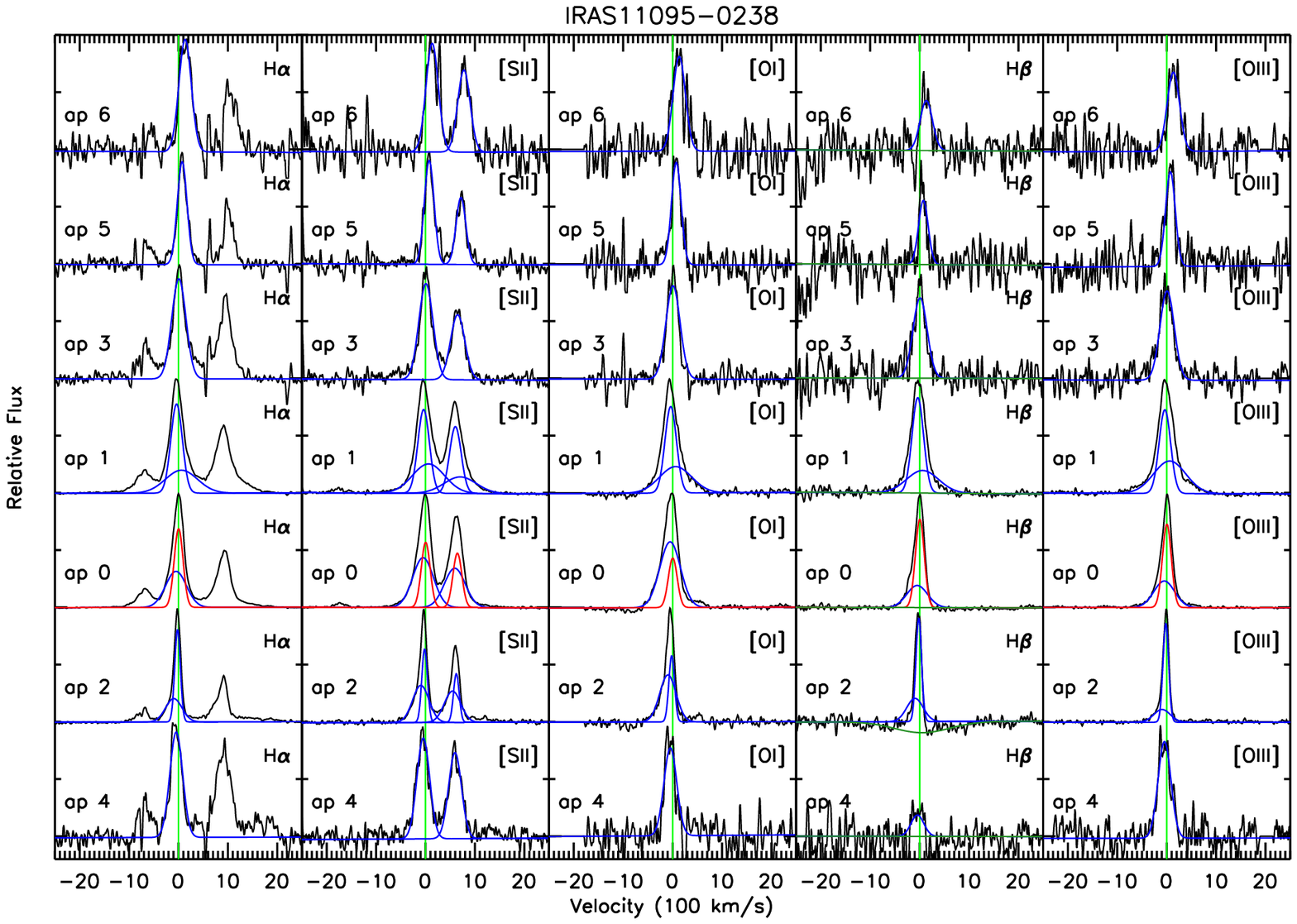, keepaspectratio=true, width=0.9\linewidth}
\caption{\label{fig:i11095_prof}
}
\end{figure}
\clearpage

\begin{figure}[h]
\centering
\epsfig{file=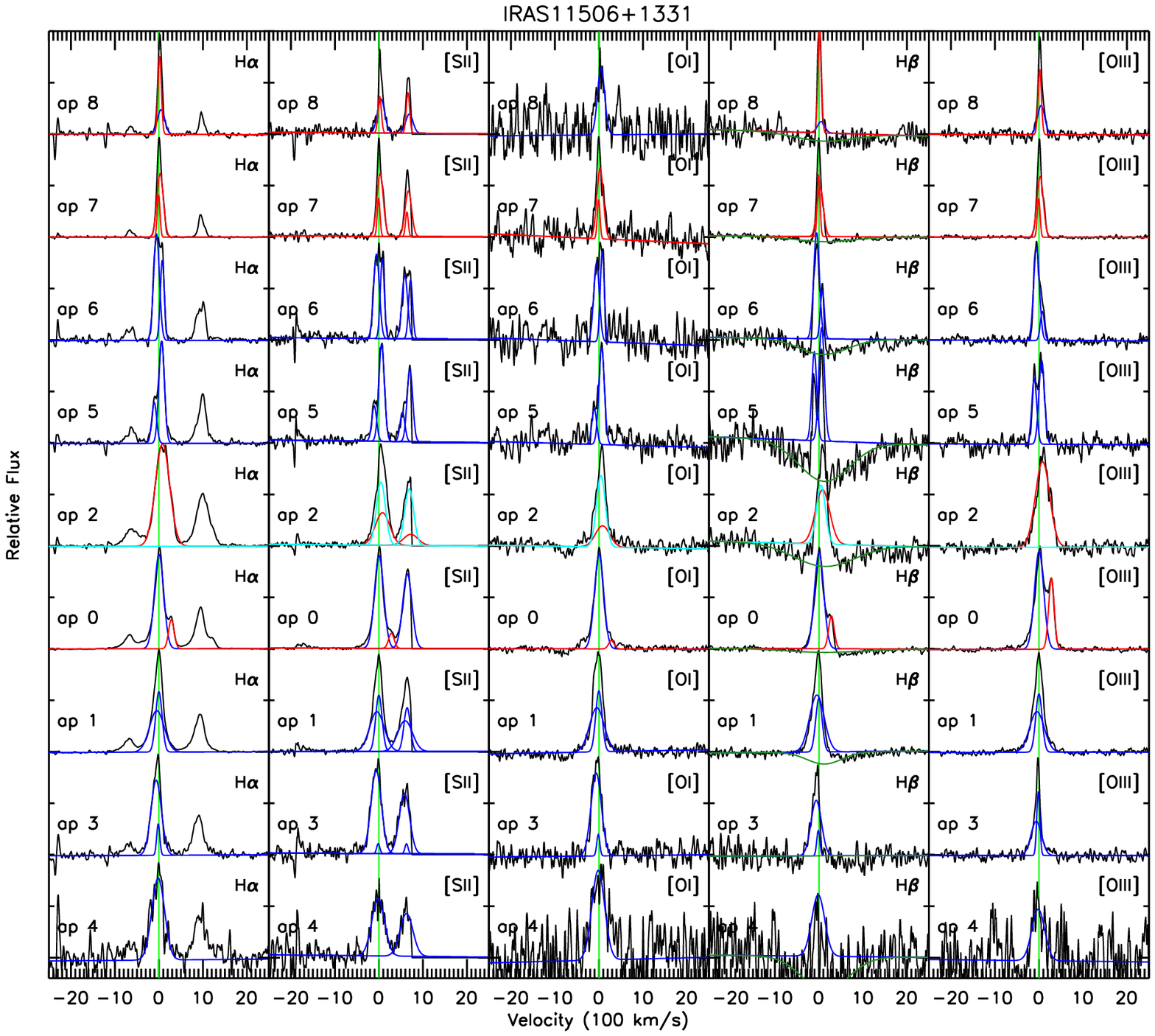, keepaspectratio=true, width=0.9\linewidth}
\caption{\label{fig:i11506_prof}
}
\end{figure}
\clearpage

\begin{figure}[h]
\centering
\epsfig{file=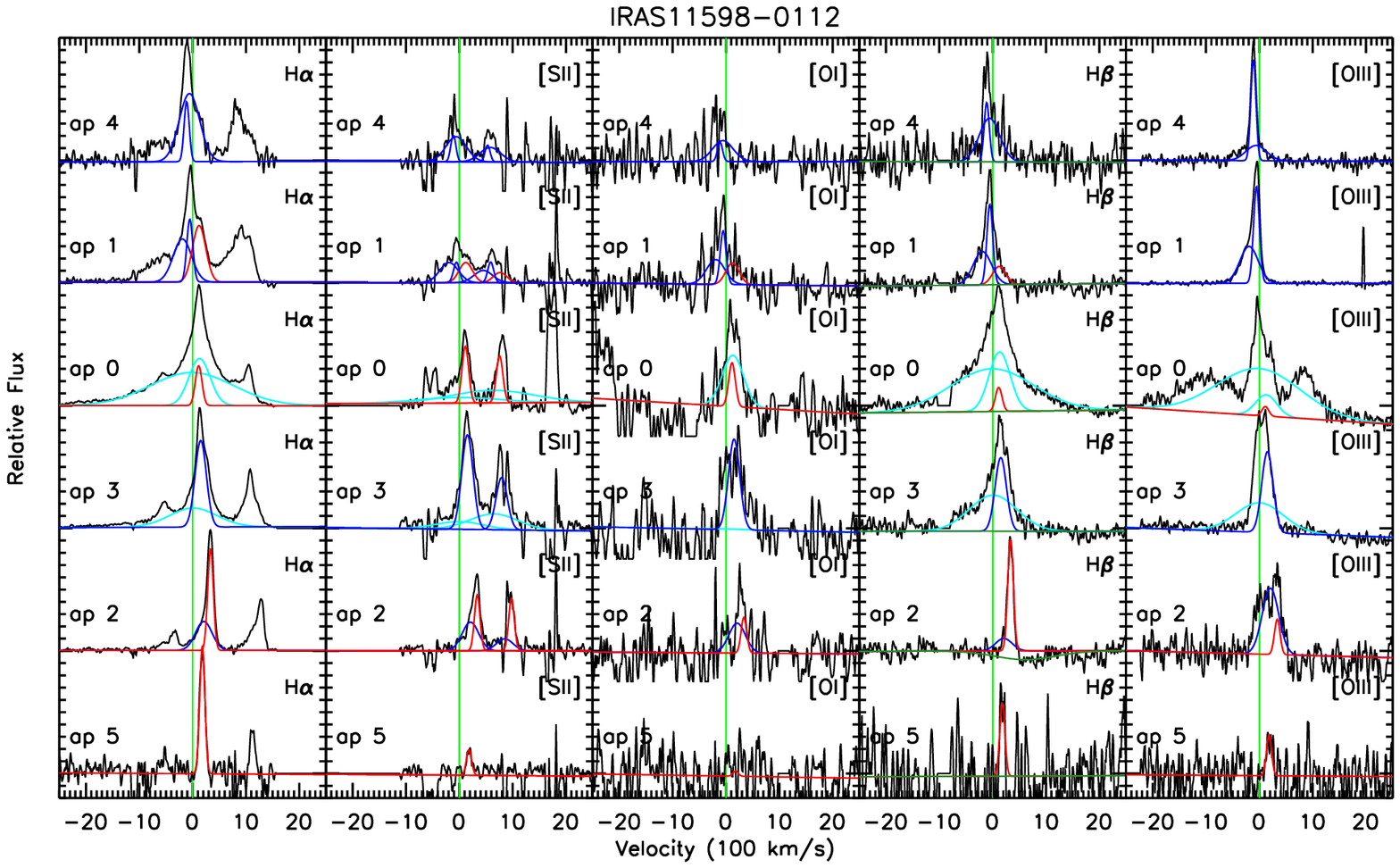, keepaspectratio=true, width=0.9\linewidth}
\caption{\label{fig:i11598_prof}
}
\end{figure}
\clearpage

\begin{figure}[h]
\centering
\epsfig{file=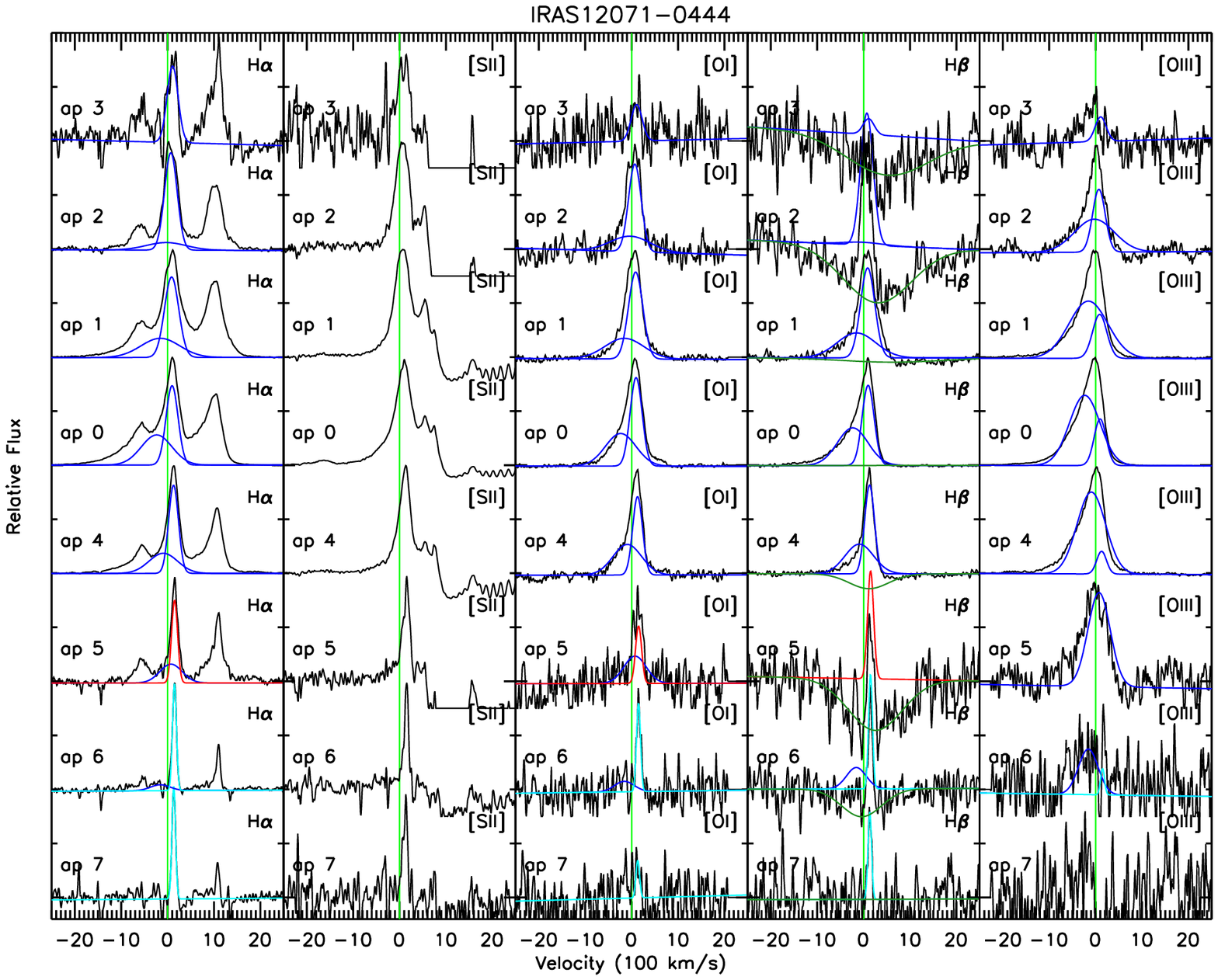, keepaspectratio=true, width=0.9\linewidth}
\caption{\label{fig:i12071_prof}
}
\end{figure}
\clearpage

\begin{figure}[h]
\centering
\epsfig{file=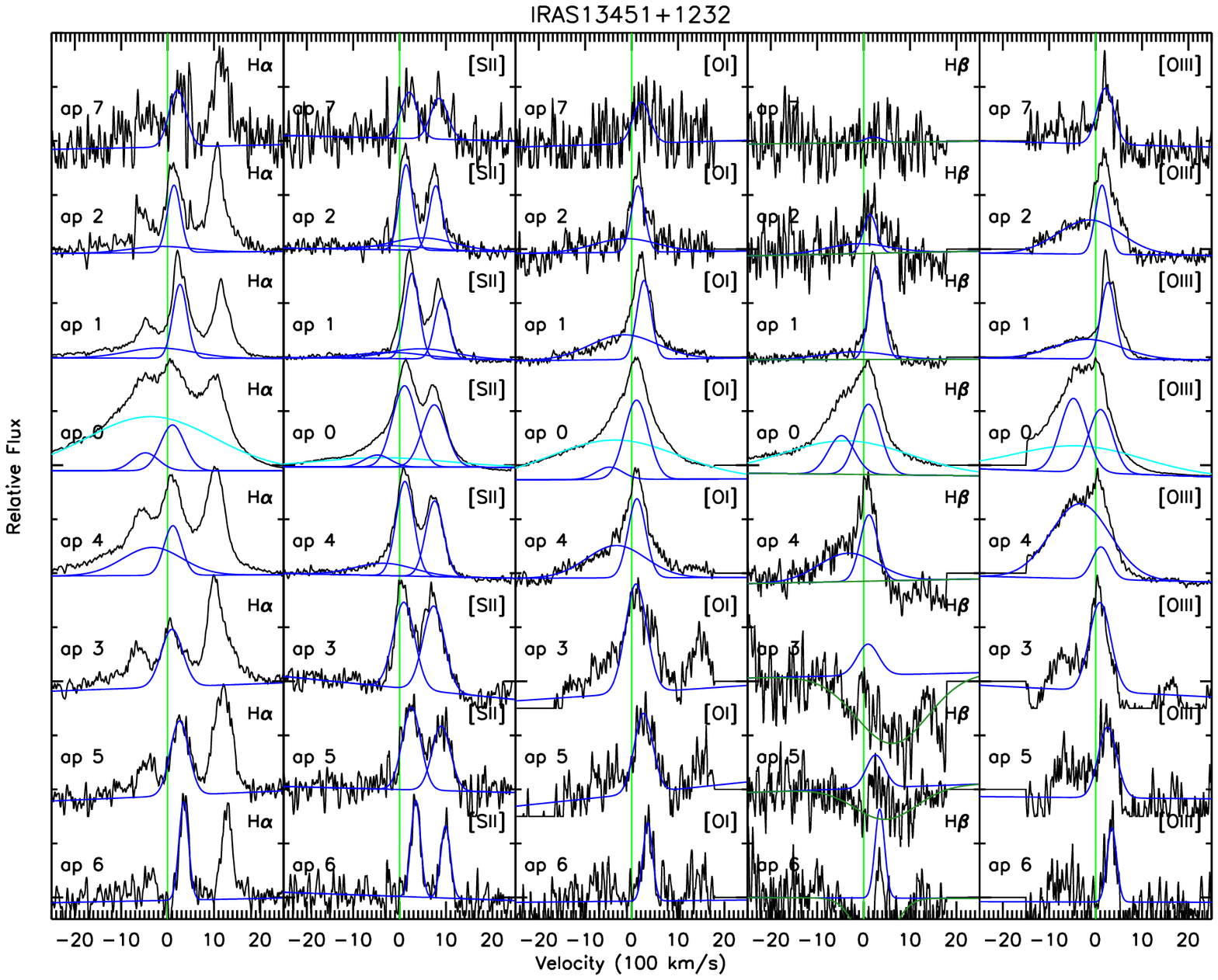, keepaspectratio=true, width=0.9\linewidth}
\caption{\label{fig:i13451_prof}
}
\end{figure}
\clearpage

\begin{figure}[h]
\centering
\epsfig{file=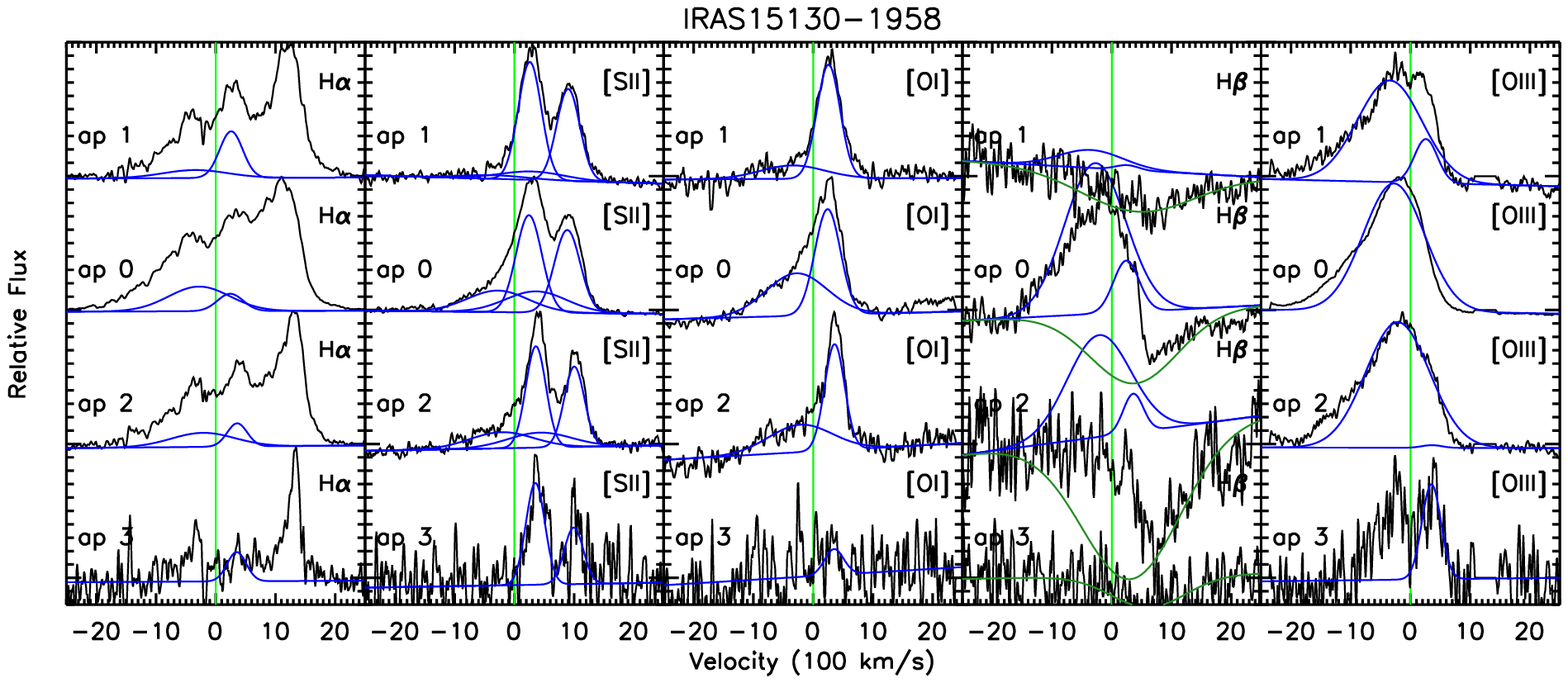, keepaspectratio=true, width=0.9\linewidth}
\caption{\label{fig:i15130_prof}
}
\end{figure}
\clearpage


\begin{figure}[h]
\centering
\epsfig{file=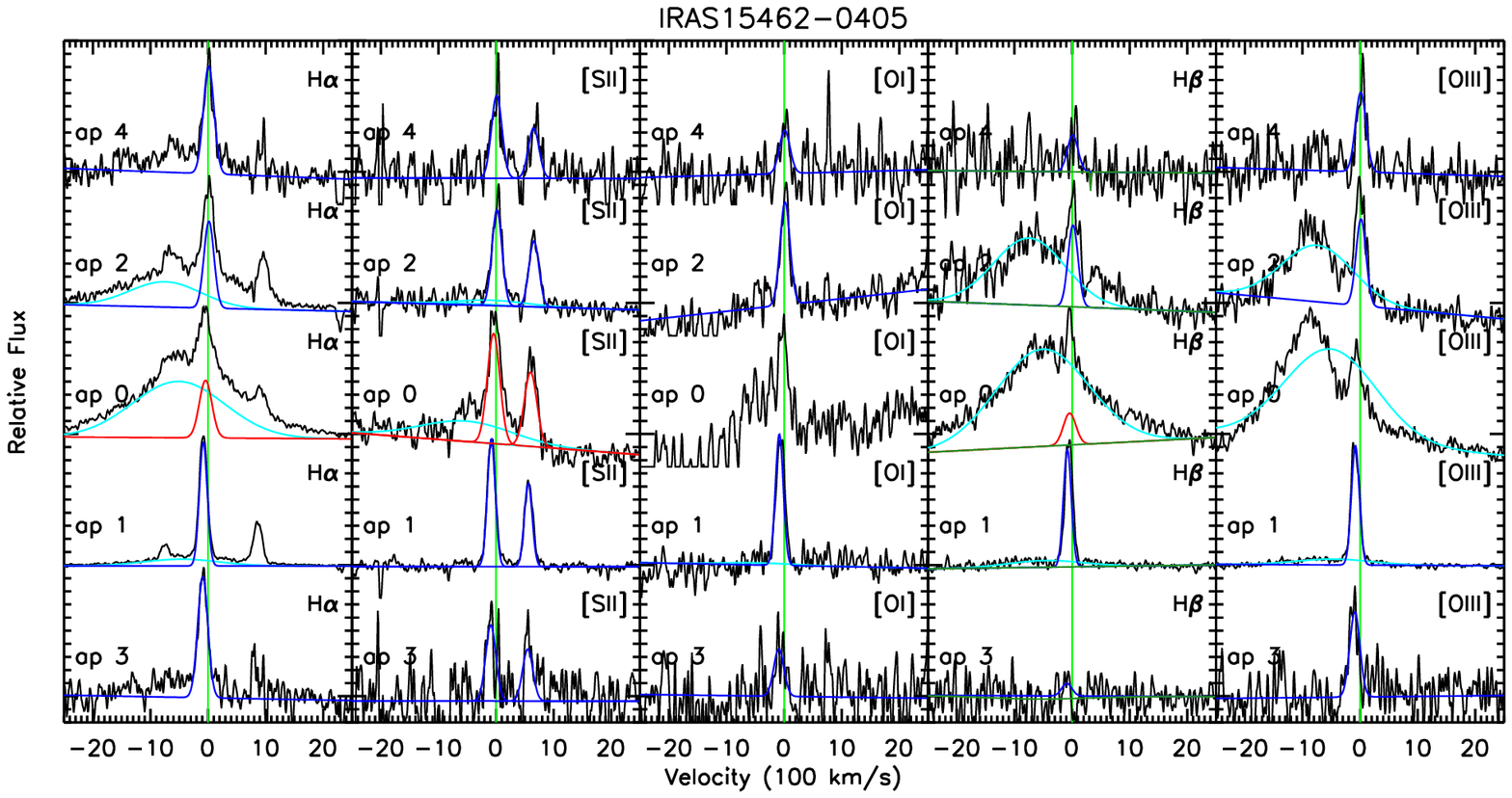, keepaspectratio=true, width=0.9\linewidth}
\caption{\label{fig:i15462_prof}
}
\end{figure}
\clearpage

\begin{figure}[h]
\centering
\epsfig{file=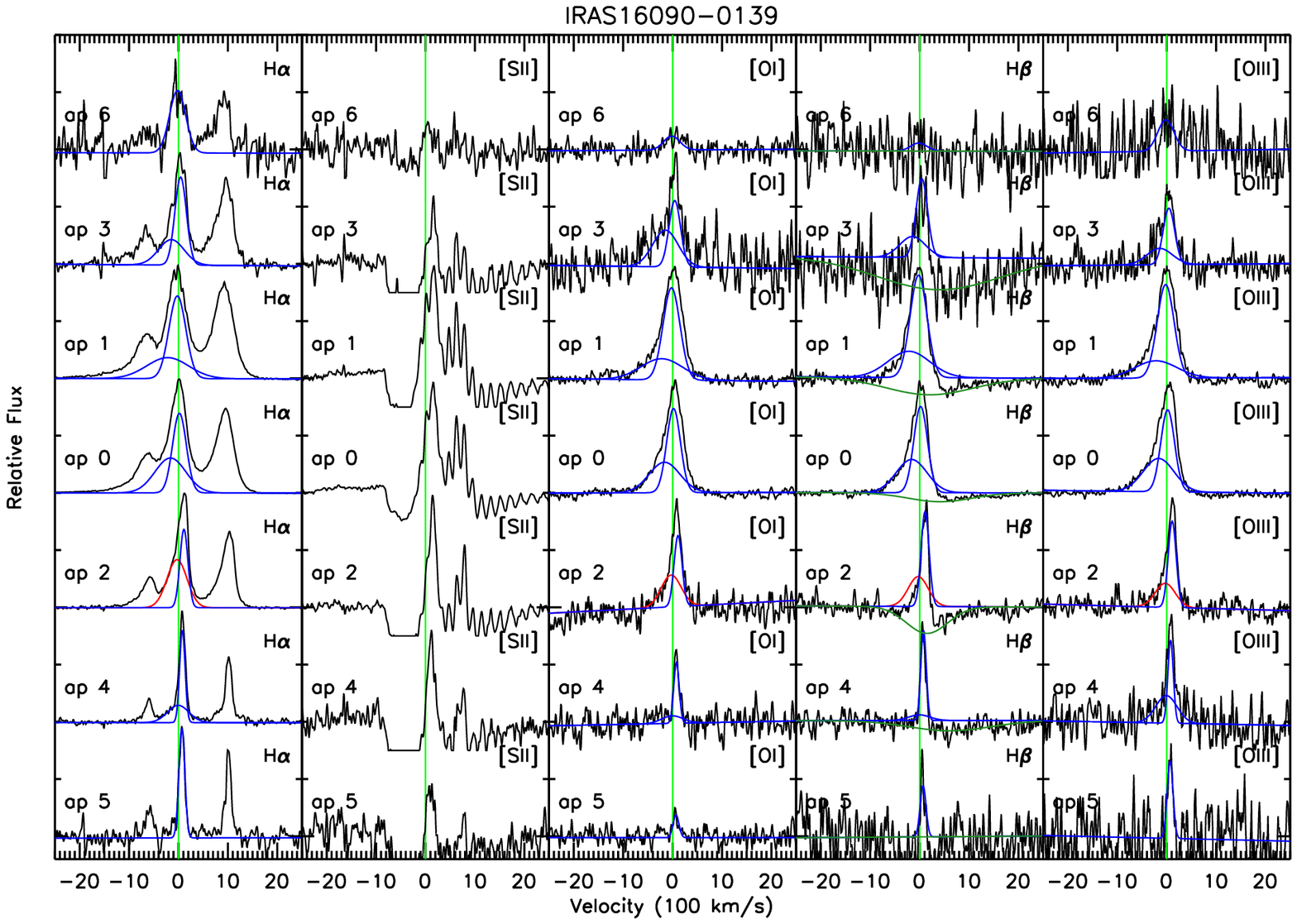, keepaspectratio=true, width=0.9\linewidth}
\caption{\label{fig:i16090_prof}
}
\end{figure}
\clearpage

\begin{figure}[h]
\centering
\epsfig{file=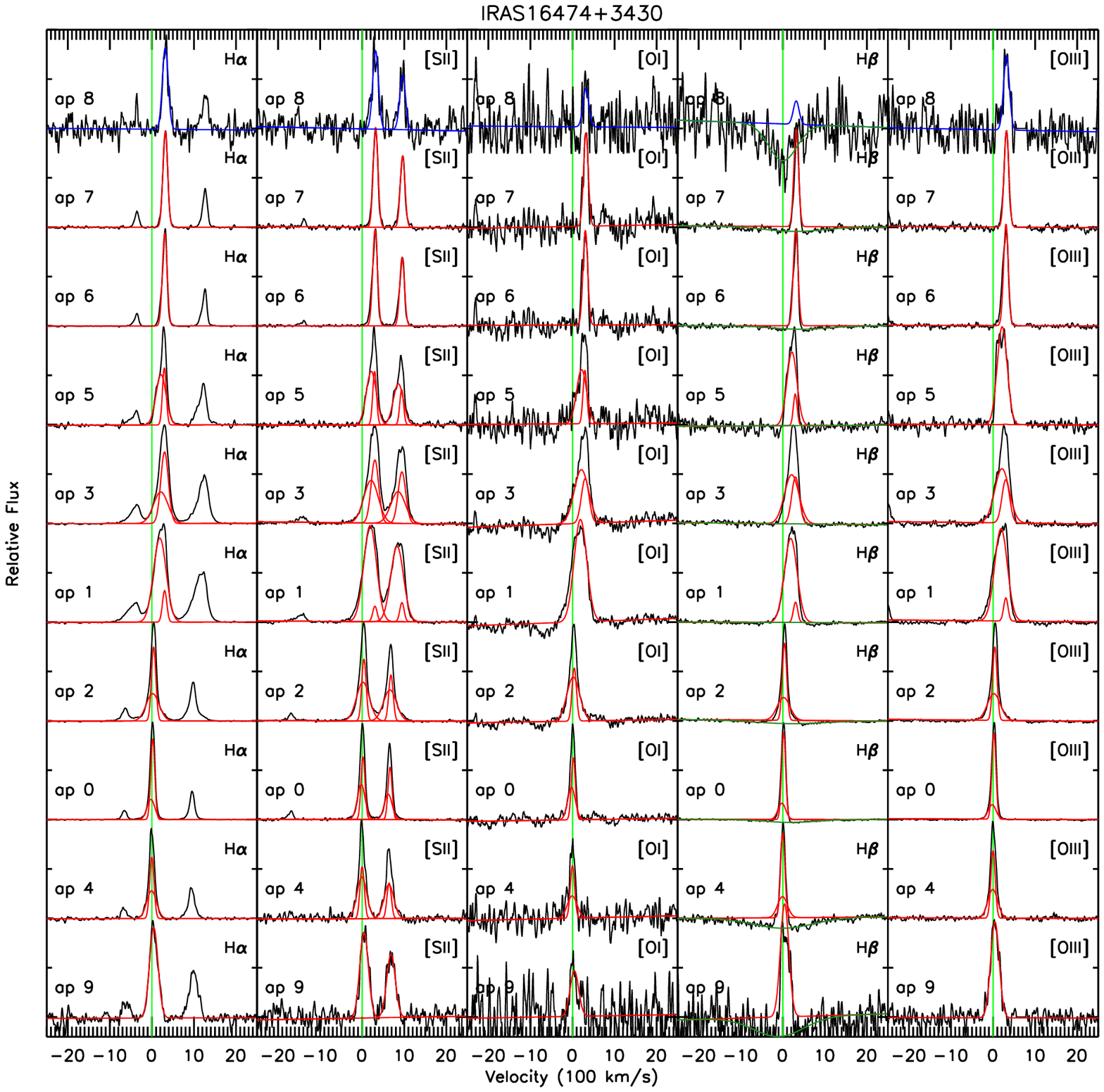, keepaspectratio=true, width=0.9\linewidth}
\caption{\label{fig:i16474_prof}
}
\end{figure}
\clearpage

\begin{figure}[h]
\centering
\epsfig{file=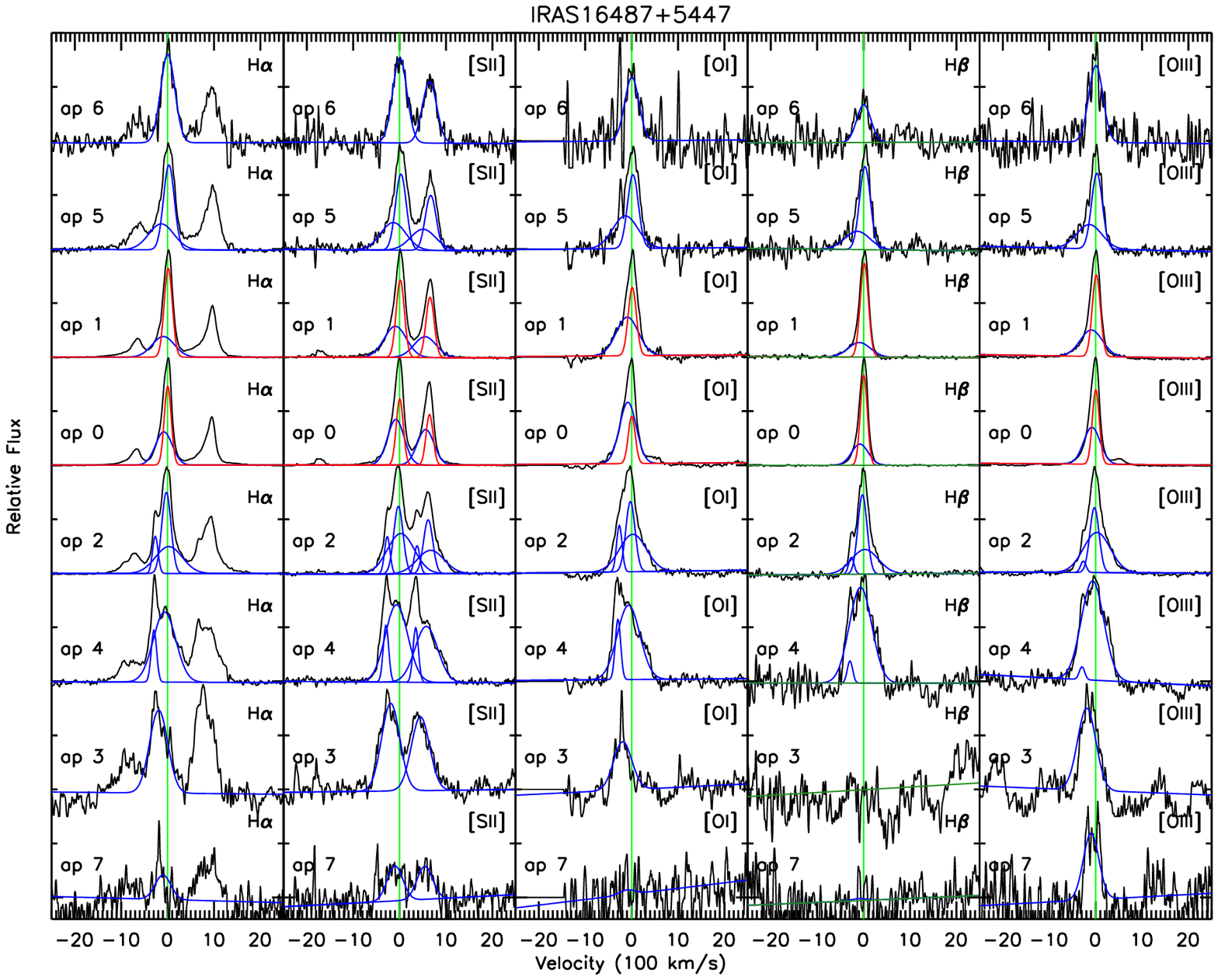, keepaspectratio=true, width=0.9\linewidth}
\caption{\label{fig:i16487_prof}
}
\end{figure}
\clearpage

\begin{figure}[h]
\centering
\epsfig{file=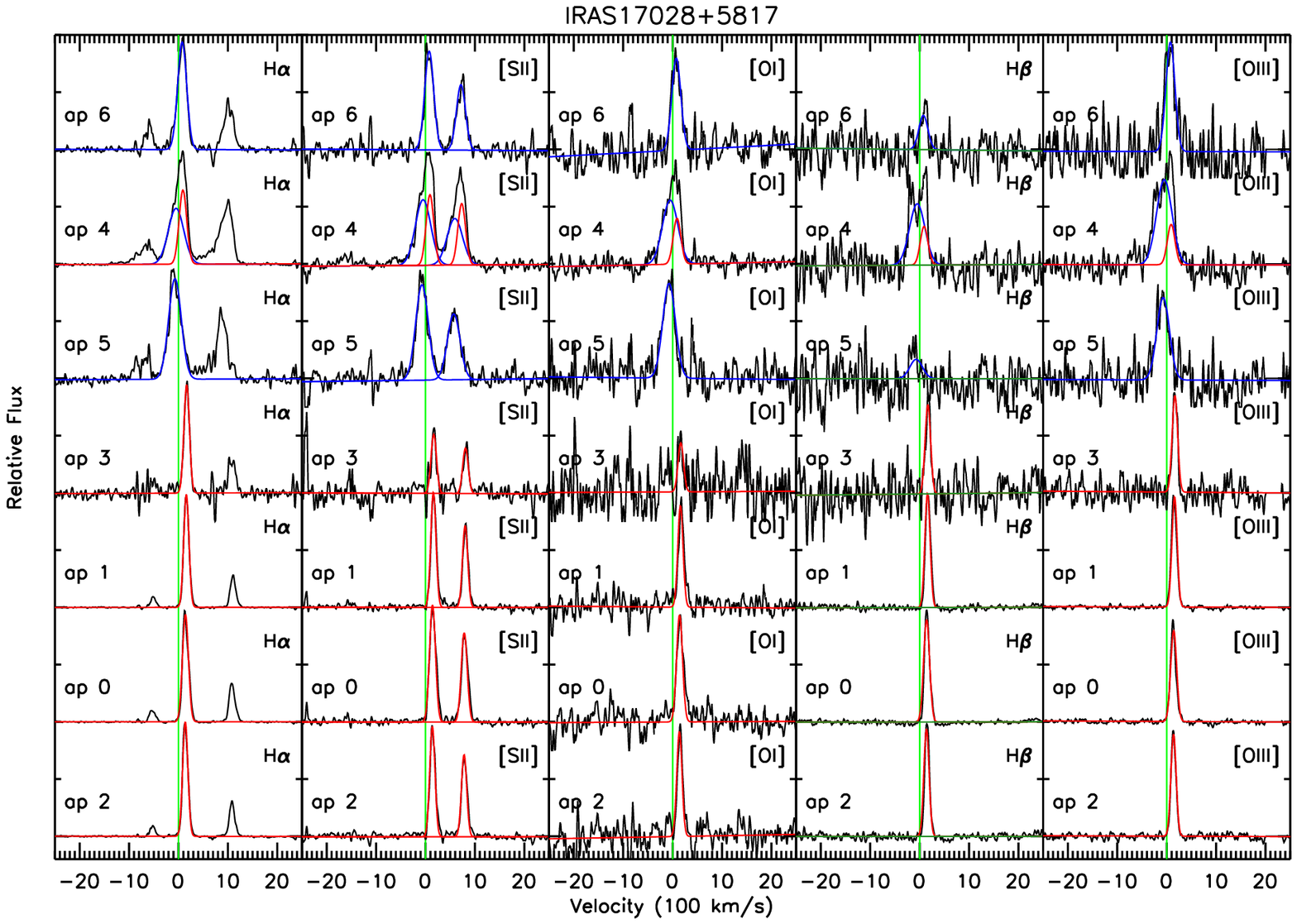, keepaspectratio=true, width=0.9\linewidth}
\caption{\label{fig:i17028_prof}
}
\end{figure}
\clearpage

\begin{figure}[h]
\centering
\epsfig{file=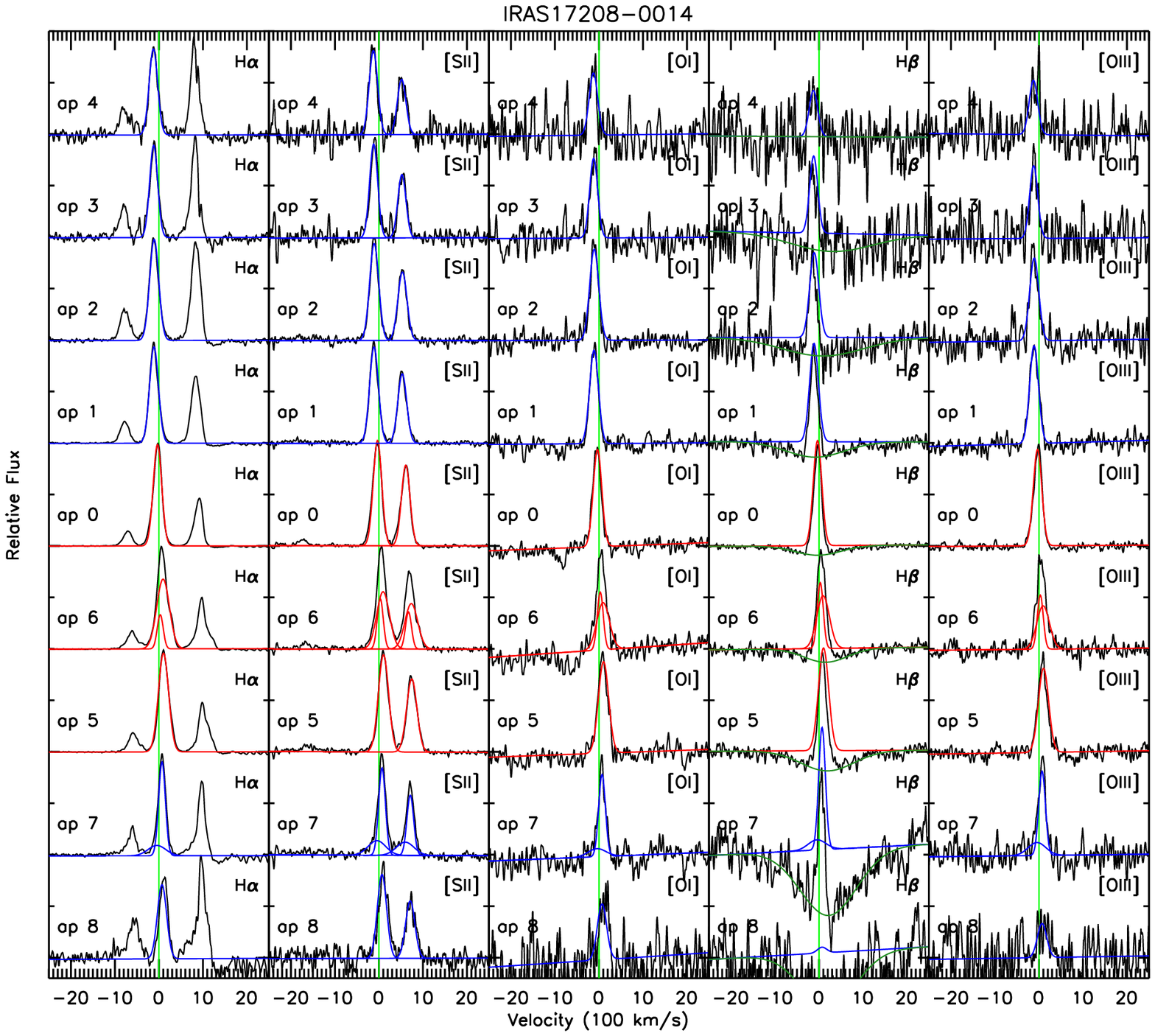, keepaspectratio=true, width=0.9\linewidth}
\caption{\label{fig:i17208_prof}
}
\end{figure}
\clearpage

\begin{figure}[h]
\centering
\epsfig{file=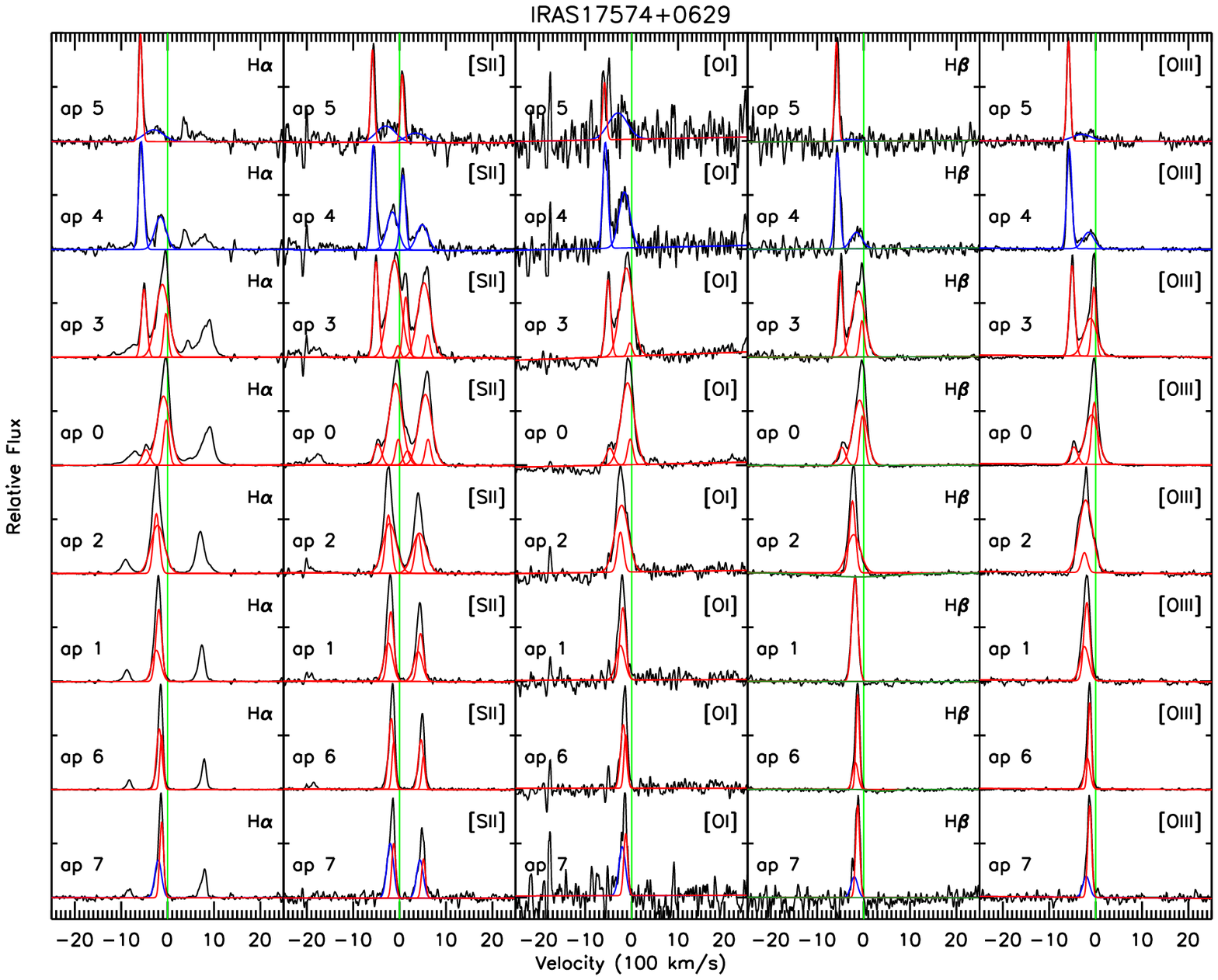, keepaspectratio=true, width=0.9\linewidth}
\caption{\label{fig:i17574_prof}
}
\end{figure}
\clearpage

\begin{figure}[h]
\centering
\epsfig{file=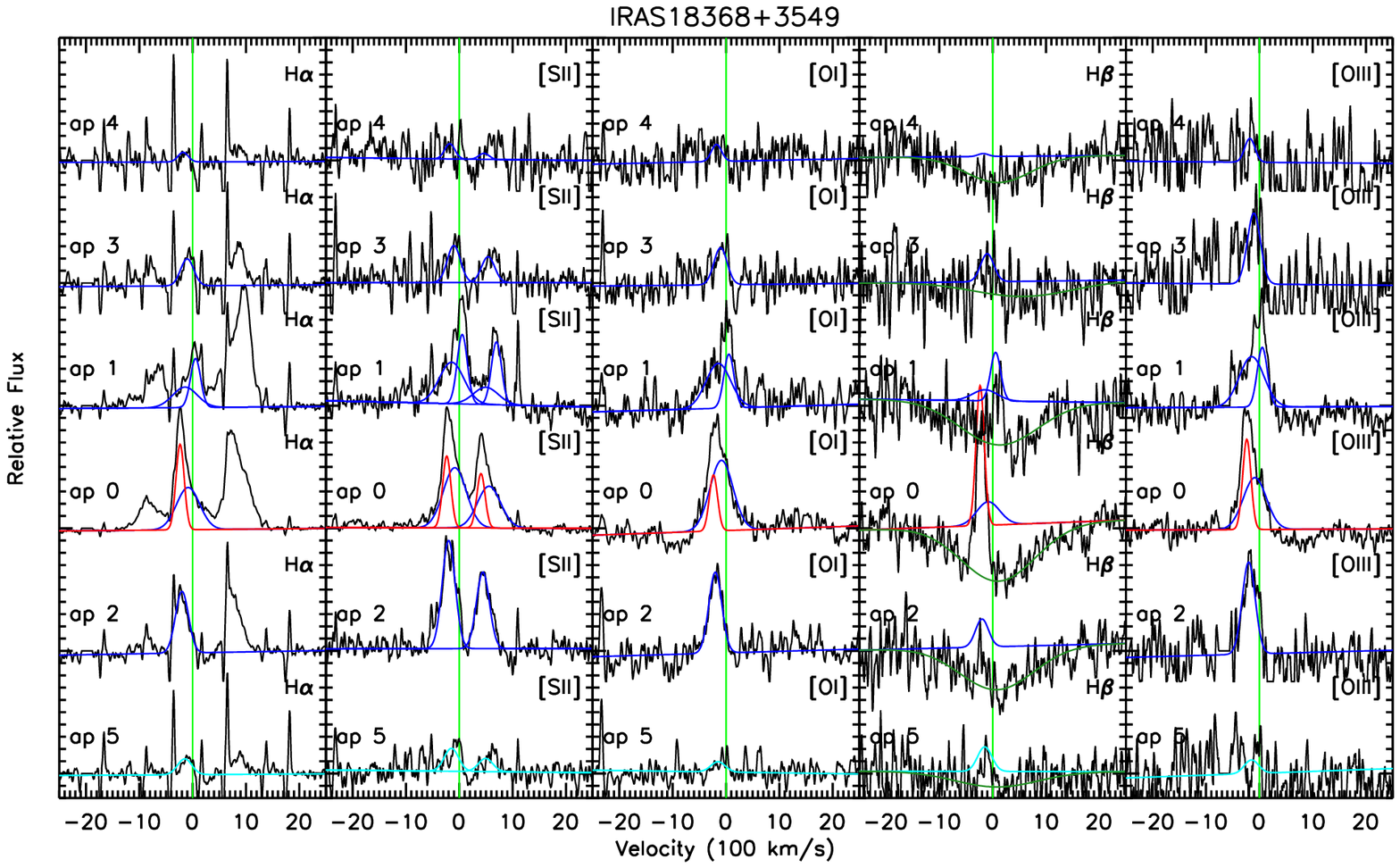, keepaspectratio=true, width=0.9\linewidth}
\caption{\label{fig:i18368_prof}
}
\end{figure}
\clearpage

\begin{figure}[h]
\centering
\epsfig{file=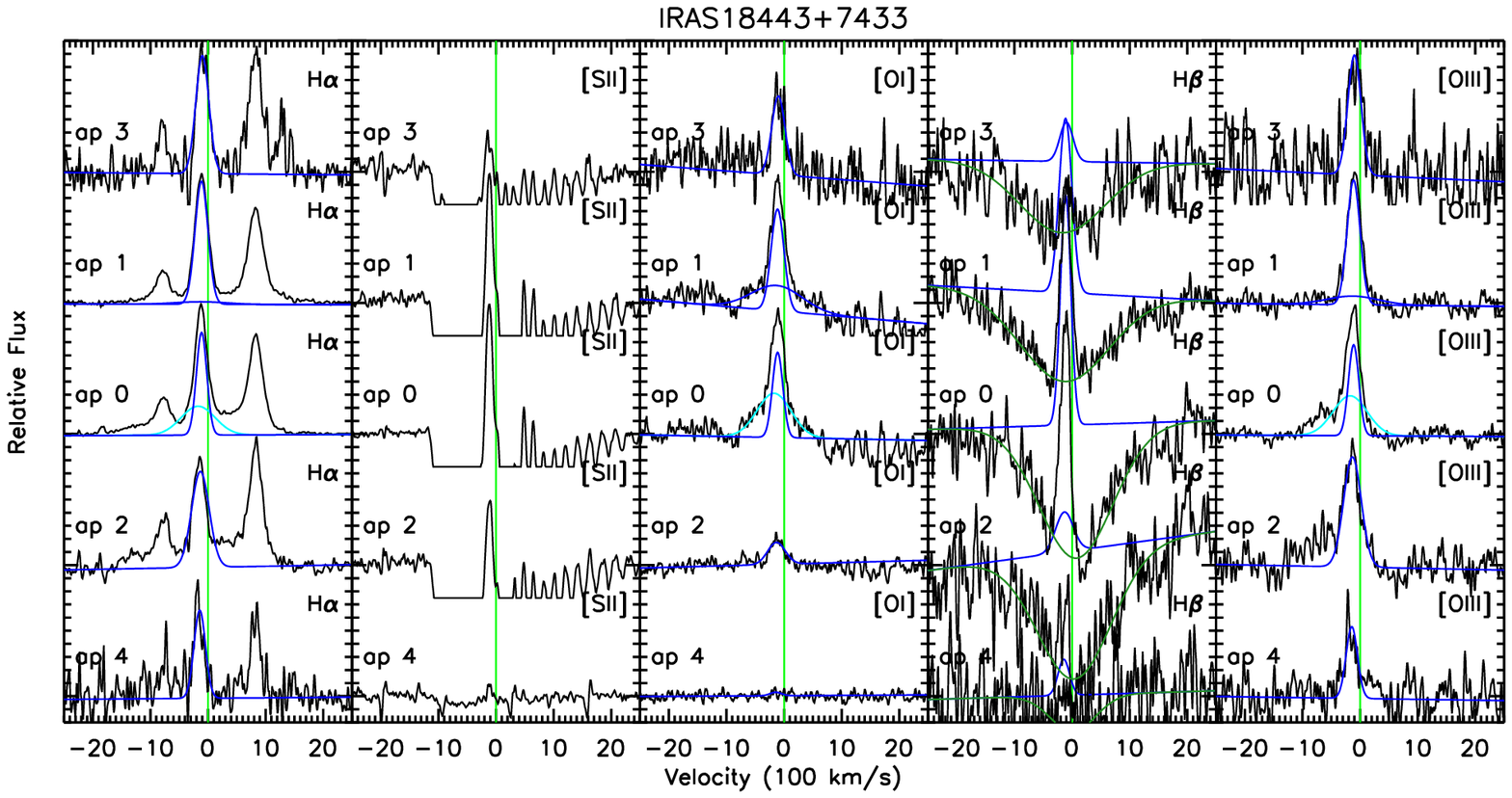, keepaspectratio=true, width=0.9\linewidth}
\caption{\label{fig:i18443_prof}
}
\end{figure}
\clearpage

\begin{figure}[h]
\centering
\epsfig{file=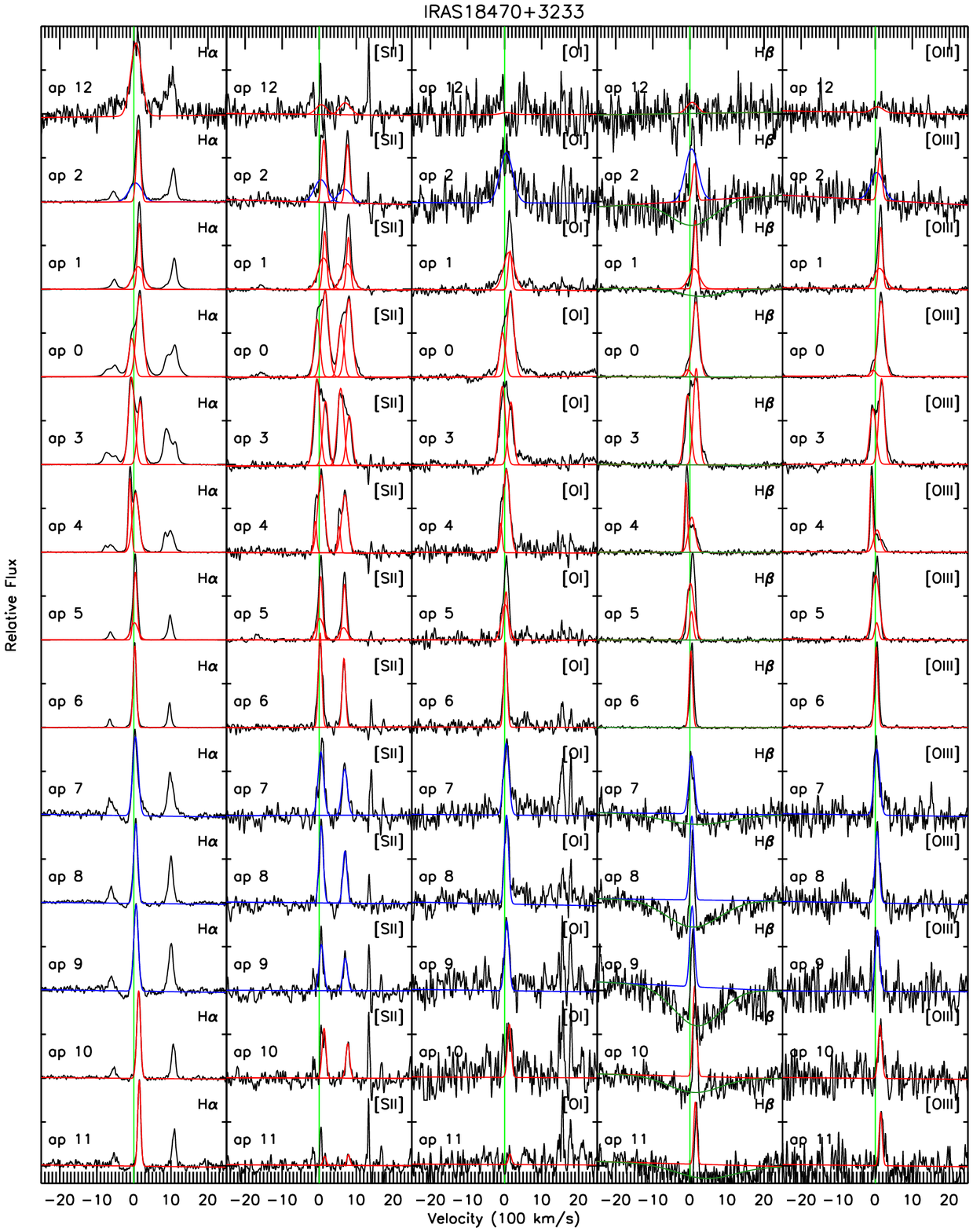, keepaspectratio=true, width=0.9\linewidth}
\caption{\label{fig:i18470_prof}
}
\end{figure}
\clearpage

\begin{figure}[h]
\centering
\epsfig{file=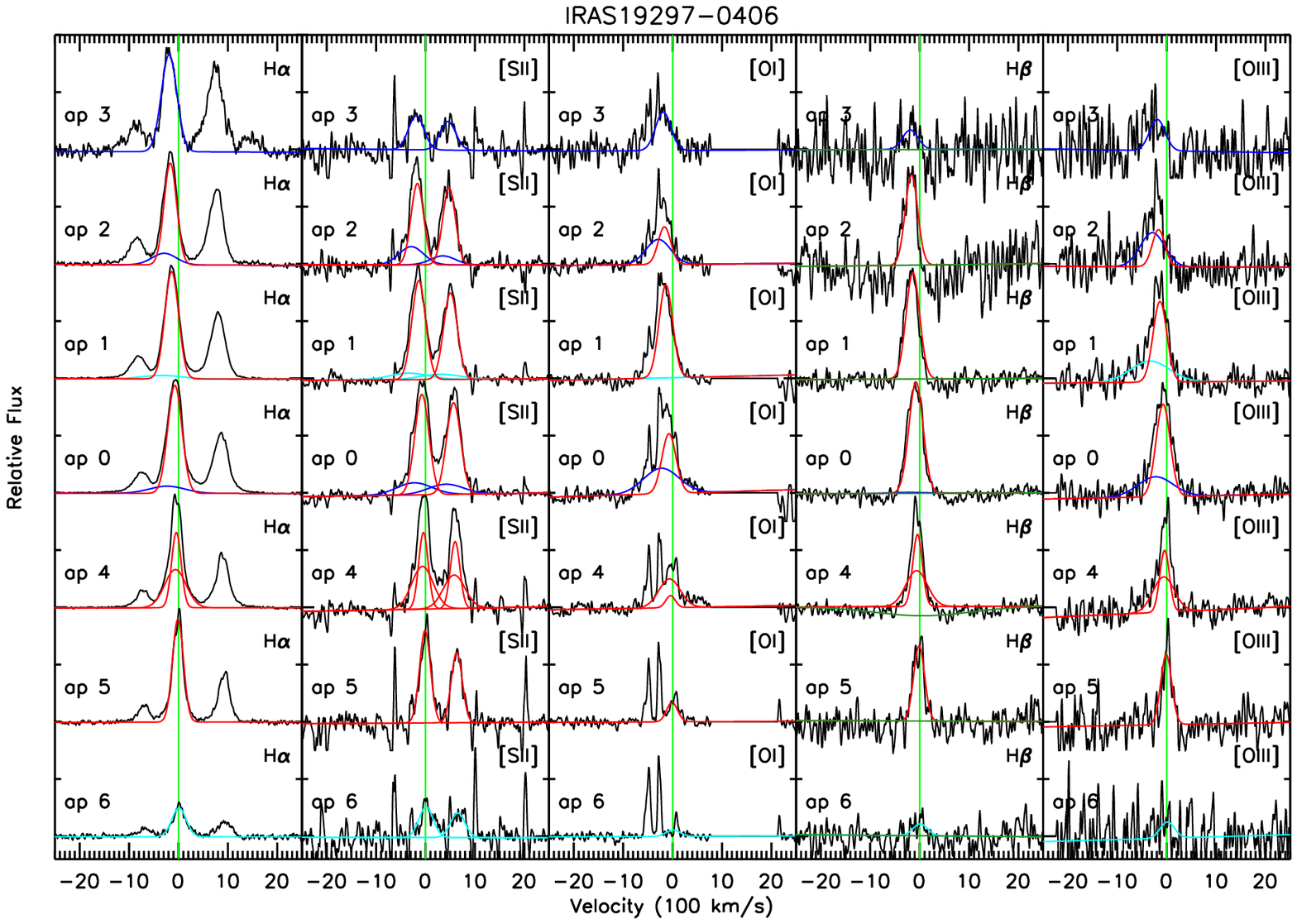, keepaspectratio=true, width=0.9\linewidth}
\caption{\label{fig:i19297_prof}
}
\end{figure}
\clearpage

\begin{figure}[h]
\centering
\epsfig{file=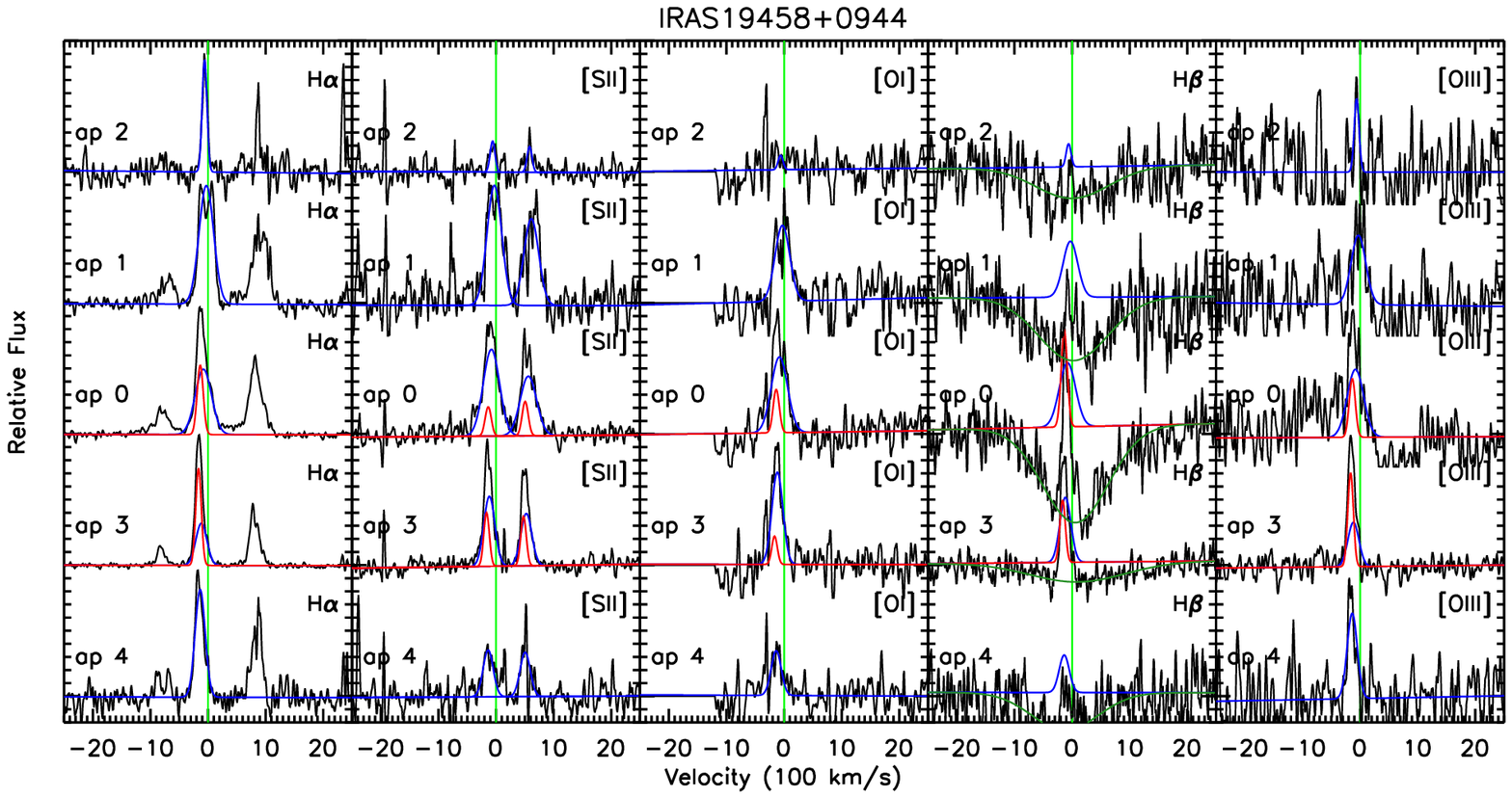, keepaspectratio=true, width=0.9\linewidth}
\caption{\label{fig:i19458_prof}
}
\end{figure}
\clearpage

\begin{figure}[h]
\centering
\epsfig{file=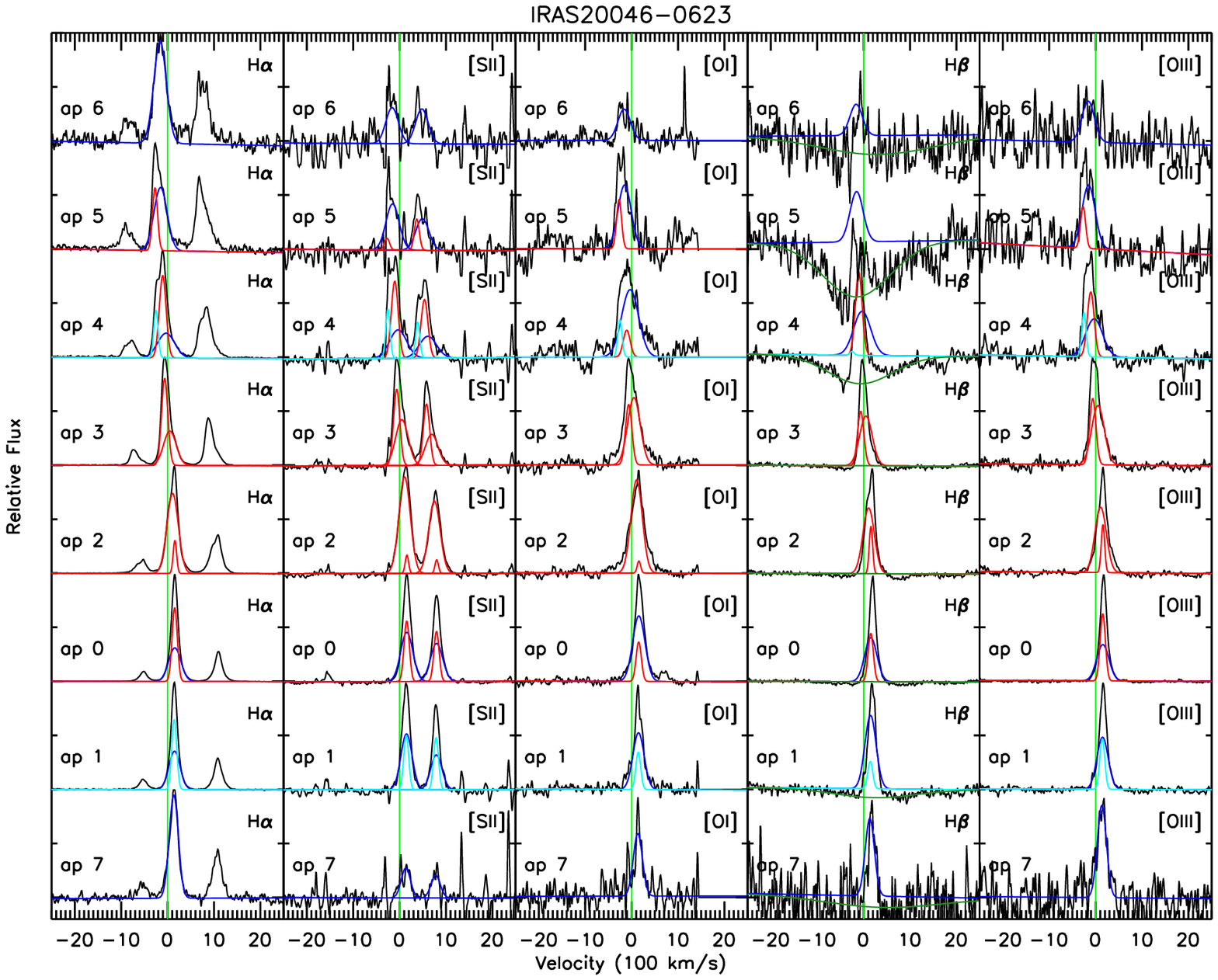, keepaspectratio=true, width=0.9\linewidth}
\caption{\label{fig:i20046_prof}
}
\end{figure}
\clearpage

\begin{figure}[h]
\centering
\epsfig{file=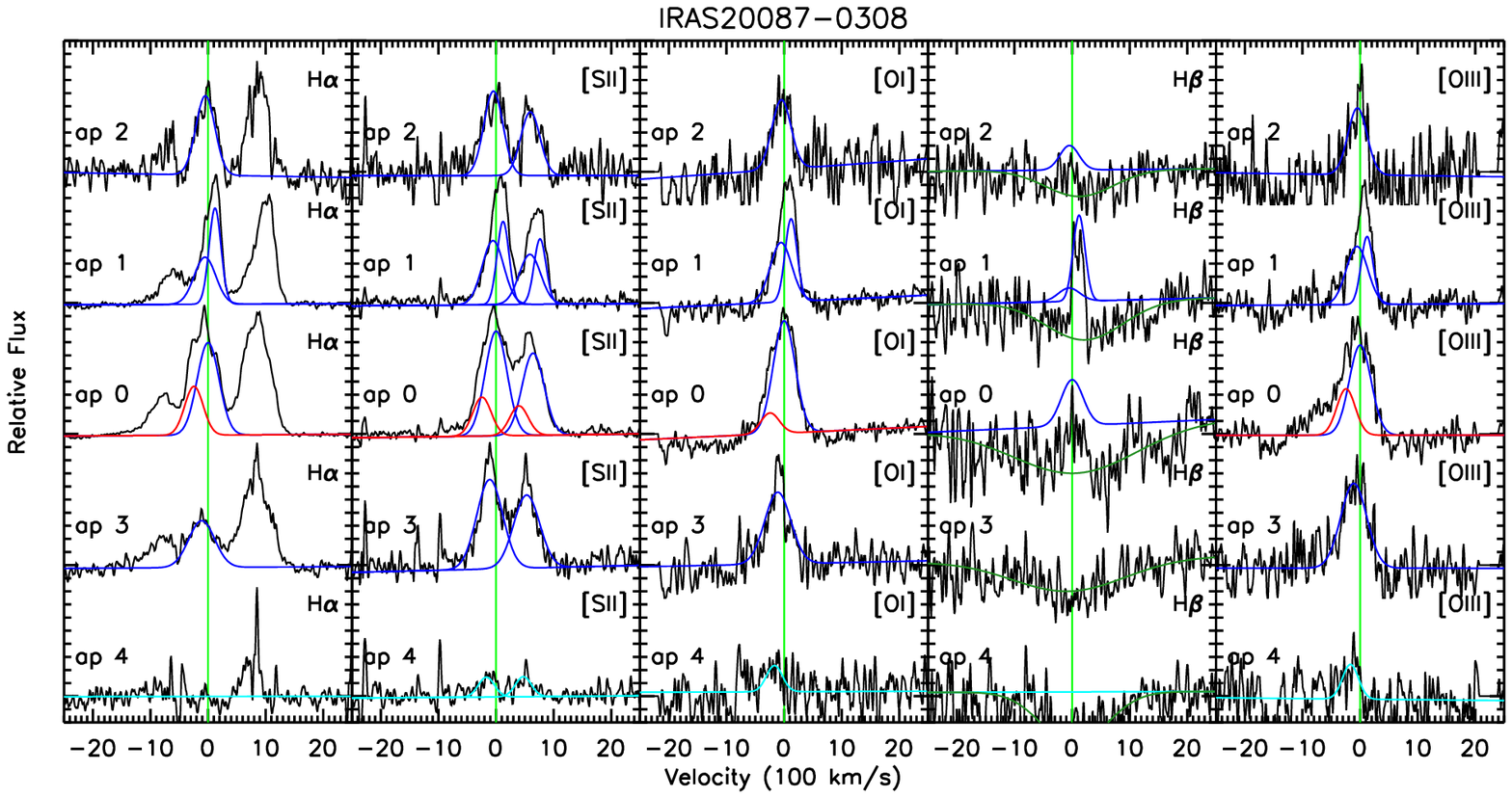, keepaspectratio=true, width=0.9\linewidth}
\caption{\label{fig:i20087_prof}
}
\end{figure}
\clearpage

\begin{figure}[h]
\centering
\epsfig{file=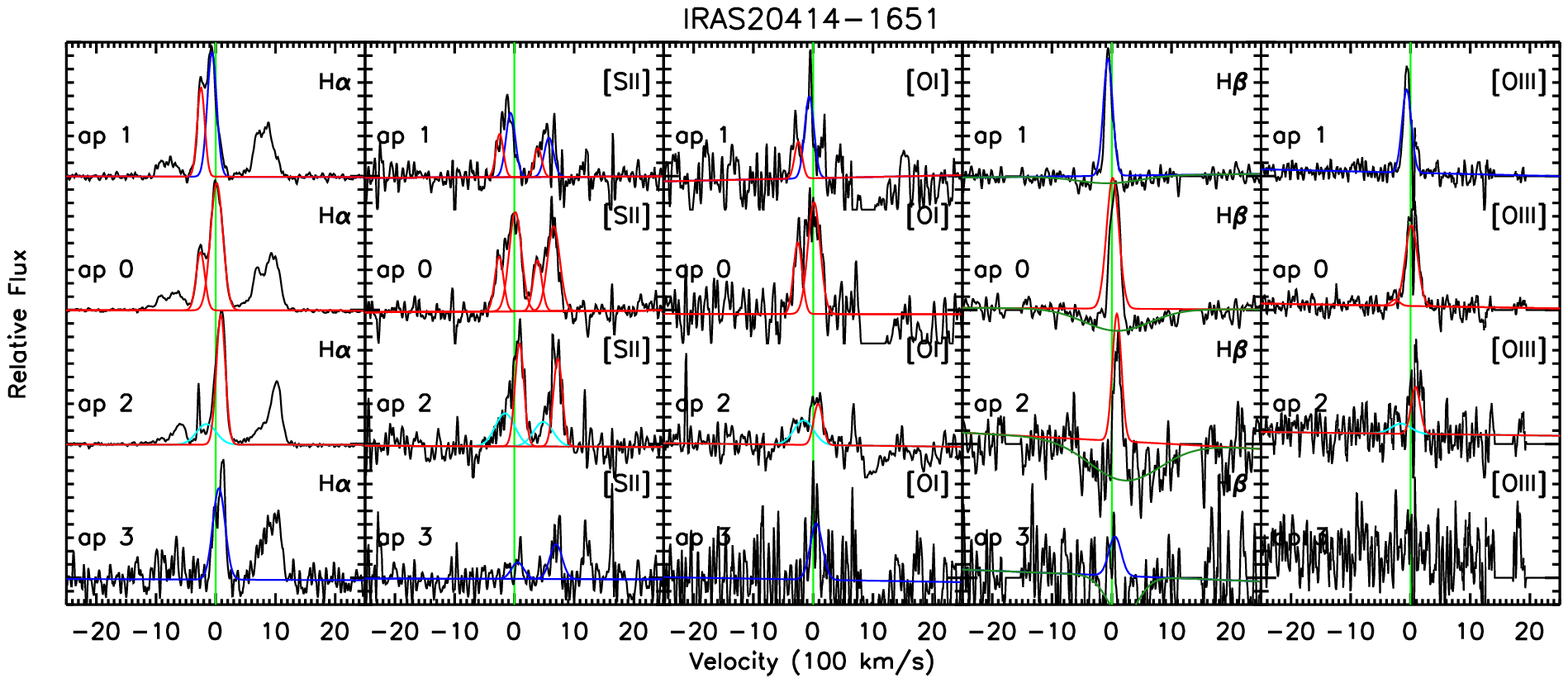, keepaspectratio=true, width=0.9\linewidth}
\caption{\label{fig:i20414_prof}
}
\end{figure}
\clearpage

\begin{figure}[h]
\centering
\epsfig{file=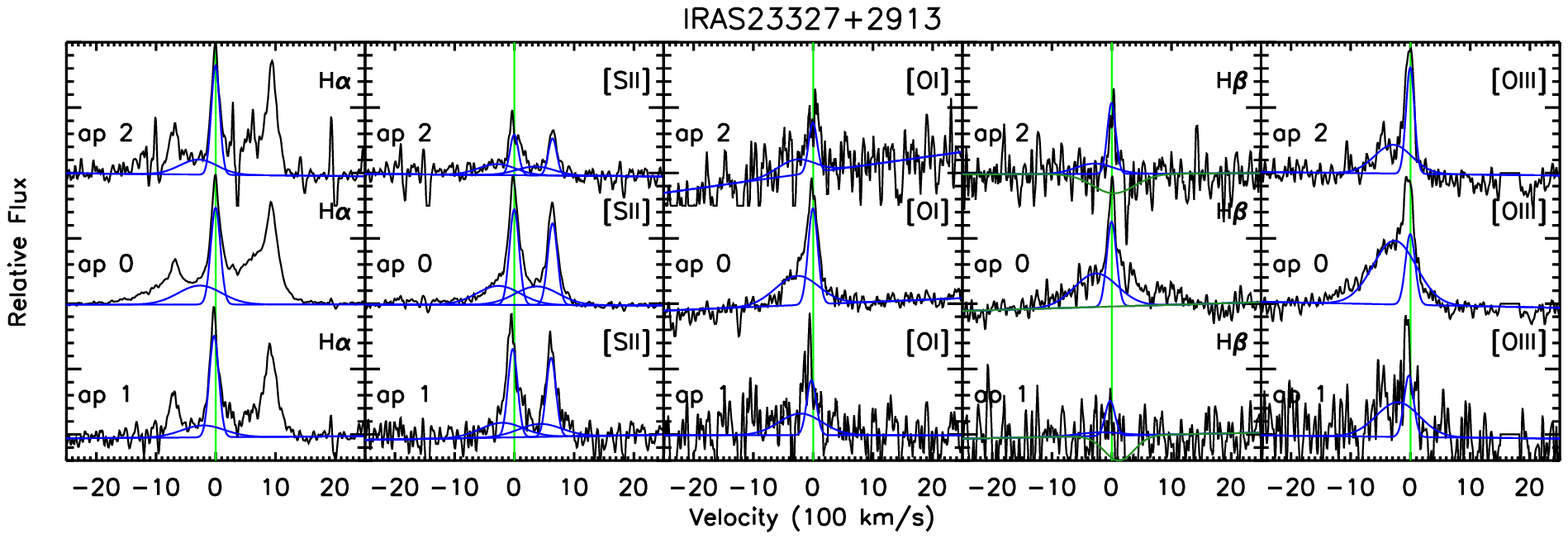, keepaspectratio=true, width=0.9\linewidth}
\caption{\label{fig:i23327_prof}
}
\end{figure}
\clearpage

\begin{figure}[h]
\centering
\epsfig{file=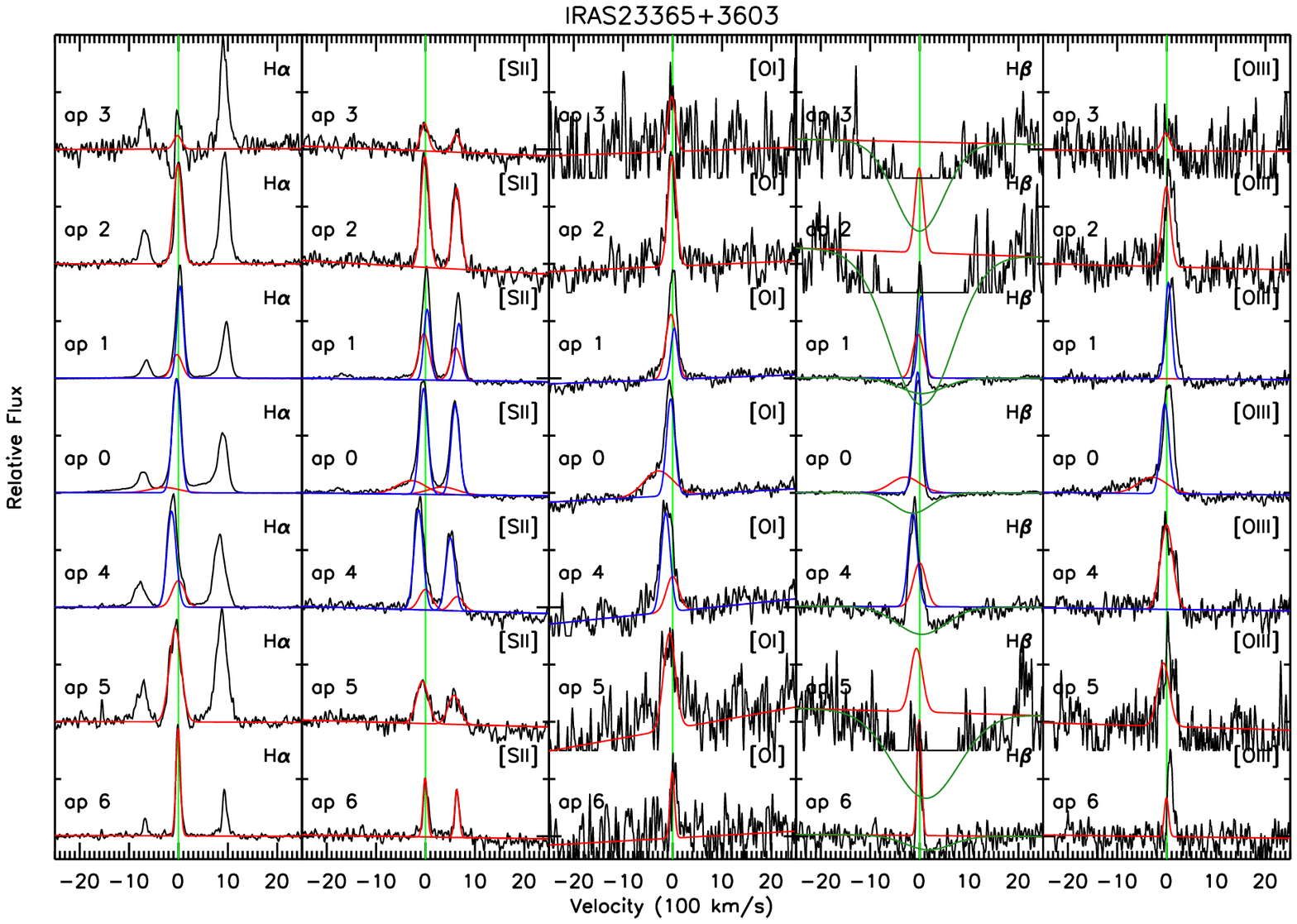, keepaspectratio=true, width=0.9\linewidth}
\caption{\label{fig:i23365_prof}
}
\end{figure}
\clearpage


\begin{landscape}
\section{Appendix C}
\LongTables
\begin{deluxetable}{c c c c c c c c c c c c c}
\tablewidth{0pt}
\tablecaption{ \label{tab:flux} Component Velocities and Fluxes}
\tabletypesize{\tiny}
\tablehead{
 	IRAS 	& Ap\#  & ex & $r-r_{nuc}$ & ap size & v & $\sigma_v$ & $F_{H\alpha}$  & $F_{[NII]}$          	& $F_{[SII]}$          	& $F_{[OI]}$           	& $F_{H\beta}$         & $F_{[OIII]}$          \\
 		        	&		&    & (kpc)       & (kpc) & (\kms) & (\kms)		& (F$\times10^{-17}$) & (F$\times10^{-17}$) & (F$\times10^{-17}$) & (F$\times10^{-17}$) & (F$\times10^{-17}$) & (F$\times10^{-17}$)  \\
	(1)	        	& (2)	& (3)& (4)	    	  & (5)        & (6)	                	& (7)                  & (8)	              		& (9)	             	& (10)	            		& (11)                 & (12) &(13)                 
}
\startdata
00153+5454 & 	0 & 	s & 	0.40 & 	1.9 & 	-56.5 $\pm$ 2.3 & 	216. $\pm$ 2.4 & 	29.9 $\pm$ 0.29 & 	20.6 $\pm$ 0.29 & 	24. $\pm$ 0.76 & 	5.96 $\pm$ 0.27 & 	3.7 $\pm$ 0.49 & 	5. $\pm$ 5.6 \\
00153+5454 & 	0 & 	h & 	0.40 & 	1.9 & 	-20.7 $\pm$ 0.19 & 	51.59 $\pm$ 0.27 & 	51.01 $\pm$ 0.15 & 	20.1 $\pm$ 0.16 & 	20. $\pm$ 0.40 & 	2.6 $\pm$ 0.14 & 	6.94 $\pm$ 0.26 & 	3.51 $\pm$ 0.16 \\
00153+5454 & 	1 & 	s & 	-2.0 & 	1.9 & 	69.6 $\pm$ 0.86 & 	81.0 $\pm$ 0.87 & 	14.6 $\pm$ 0.16 & 	7.89 $\pm$ 0.16 & 	9.3 $\pm$ 0.52 & 	1.9 $\pm$ 0.14 & 	2. $\pm$ 0.34 & 	1.9 $\pm$ 0.19 \\
00153+5454 & 	2 & 	s & 	2.7 & 	1.9 & 	-44.1 $\pm$ 2.5 & 	150 $\pm$ 3.8 & 	14.5 $\pm$ 0.23 & 	7.32 $\pm$ 0.23 & 	9.3 $\pm$ 0.48 & 	1.7 $\pm$ 0.19 & 	1. $\pm$ 0.46 & 	2.0 $\pm$ 0.24 \\
00153+5454 & 	2 & 	h & 	2.7 & 	1.9 & 	-39.91 $\pm$ 0.20 & 	41.6 $\pm$ 0.32 & 	38.92 $\pm$ 0.14 & 	12.3 $\pm$ 0.14 & 	12.1 $\pm$ 0.29 & 	1.3 $\pm$ 0.11 & 	6.70 $\pm$ 0.28 & 	3.94 $\pm$ 0.15 \\
00153+5454 & 	3 & 	h & 	5.0 & 	1.9 & 	-35.0 $\pm$ 0.84 & 	51.7 $\pm$ 0.85 & 	8.93 $\pm$ 0.13 & 	3.19 $\pm$ 0.14 & 	3.53 $\pm$ 0.27 & 	0.44 $\pm$ 0.098 & 	1.4 $\pm$ 0.27 & 	1.2 $\pm$ 0.16 \\
00153+5454 & 	4 & 	s & 	12. & 	3.1 & 	93.7 $\pm$ 1.8 & 	26. $\pm$ 1.8 & 	2.0 $\pm$ 0.091 & 	0.43 $\pm$ 0.092 & 	0.72 $\pm$ 0.15 & 	0.2 $\pm$ 0.11 & 	0.3 $\pm$ 0.27 & 	0.66 $\pm$ 0.17 \\
00153+5454 & 	5 & 	s & 	5.8 & 	1.9 & 	110. $\pm$ 1.1 & 	56.0 $\pm$ 1.1 & 	7.76 $\pm$ 0.14 & 	3.93 $\pm$ 0.15 & 	4.9 $\pm$ 0.44 & 	1.1 $\pm$ 0.11 & 	1.0 $\pm$ 0.27 & 	0.88 $\pm$ 0.16 \\
\enddata
\tablecomments{Column 1: IRAS galaxy. 
Column 2: aperture number.
Column 3: excitation classification: h denotes a component with HII-like line ratios, s represents components with line ratios indicative of shock, u is an unclear component where the errors in the flux ratio make the measurement consistent with either interpretation. 
Column 4: distance of sampled region from associated nucleus in kpc. 
Column 5: size of the selected aperture in kpc. 
Column 6: Line of sight velocity of fitted spectral component in \kms. 
Column 7: Velocity dispersion of fitted spectral component in \kms. 
Column 8: Flux of \Ha\ emission in the fitted spectral component in units of 10$^{-17}$ erg s$^{-1}$ cm$^{-2}$. 
Column 9: Flux of [\ion{N}{2}] emission in the fitted spectral component in 
units of 10$^{-17}$ erg s$^{-1}$ cm$^{-2}$. 
Column 10: Flux of [\ion{S}{2}] emission in the fitted spectral component in 
units of 10$^{-17}$ erg s$^{-1}$ cm$^{-2}$. 
Column 11: Flux of [\ion{O}{1}] emission in the fitted spectral component in 
units of 10$^{-17}$ erg s$^{-1}$ cm$^{-2}$. 
Column 12: Flux of \Hb\ emission in the fitted spectral component in 
units of 10$^{-17}$ erg s$^{-1}$ cm$^{-2}$. 
Column 13: Flux of [\ion{O}{3}] emission in the fitted spectral component in 
units of 10$^{-17}$ erg s$^{-1}$ cm$^{-2}$.}
\end{deluxetable}
\clearpage
\end{landscape}

\clearpage
\LongTables
\begin{deluxetable}{l l l l l l l}
\tablewidth{0pt}
\tabletypesize{\scriptsize}
\tablecaption{\label{tab:vshock} Shock Velocities for Identified Shock Components}
\tablehead{
 	IRAS	& Ap\#	& v                 & $\sigma_v$       & v$_{shock}$([\ion{N}{2}]) & v$_{shock}$([\ion{S}{2}]) & v$_{shock}$([\ion{O}{1}]) \\
 		        & 	& (\kms) & (\kms) & (\kms)   & (\kms)   & (\kms)  \\
	(1)	        & (2)	& (3)	            & (4)	       & (5)                & (6)                & (7)               \\
}
\startdata
00153+5454 & 0 & -56.5 $\pm$ 2.3 & 216. $\pm$ 2.4 & 150 $\pm$ 50 & 250 $\pm$ 50 & 250 $\pm$ 25 \\
00153+5454 & 1 & 69.6 $\pm$ 0.86 & 81.0 $\pm$ 0.87 & 50 $\pm$ 50 & 100 $\pm$ 25 & 175 $\pm$ 25 \\
00153+5454 & 2 & -44.1 $\pm$ 2.5 & 150 $\pm$ 3.8 & 150 $\pm$ 25 & 200 $\pm$ 50 & 215 $\pm$ 25 \\
00153+5454 & 4 & 93.7 $\pm$ 1.8 & 26. $\pm$ 1.8 & 50 $\pm$ 50 & 100 $\pm$ 100 & 225 $\pm$ 125 \\
00153+5454 & 5 & 110. $\pm$ 1.1 & 56.0 $\pm$ 1.1 & 50 $\pm$ 50 & 100 $\pm$ 25 & 175 $\pm$ 25 \\
00153+5454 & 6 & 134. $\pm$ 0.74 & 87.5 $\pm$ 0.76 & 50 $\pm$ 50 & 100 $\pm$ 25 & 190 $\pm$ 15 \\
00153+5454 & 7 & 84.8 $\pm$ 2.2 & 118. $\pm$ 2.2 & 175 $\pm$ 25 & 200 $\pm$ 25 & 225 $\pm$ 25 \\
00153+5454 & 8 & -30 $\pm$ 12. & 240 $\pm$ 17. & 150 $\pm$ 50 & 300 $\pm$ 200 & 225 $\pm$ 75 \\
\enddata
\tablecomments{
Column 1: IRAS galaxy. 
Column 2: aperture number. 
Column 3: Line of sight velocity of 
the fit component in \kms. 
Column 4: Velocity dispersion of the fit component in \kms. 
Column 5: velocity of shock front measured from the ratios 
log([\ion{O}{3}]/\Ha) vs log([\ion{N}{2}]/\Ha) from shock grids of \cite{Allen:2008p695}. 
Column 6: velocity of shock front measured from the ratios 
log([\ion{O}{3}]/\Ha) vs log([\ion{S}{2}]/\Ha) from shock grids of \cite{Allen:2008p695}.
Column 7: velocity of shock front measured from the ratios log([\ion{O}{3}]/\Ha) vs 
log([\ion{O}{1}]/\Ha) from shock grids of \cite{Allen:2008p695}.}
\end{deluxetable}
\clearpage

\end{document}